\documentclass[12pt,a4paper]{article}
\usepackage{jheppub}
\usepackage[utf8]{inputenc}
\usepackage[T1]{fontenc}
\pdfoutput = 1

\usepackage{main}
\usepackage{aas_macros}
\usepackage{amsmath}
\usepackage{amssymb}
\usepackage{mathrsfs}
\def\dt{{\mathrm d}}
\def\CO{{\cal O}}
\def\e{{\epsilon}}
\def\ads{{\text{AdS}}}
\def\0{{(0)}}
\def\1{{(1)}}
\def\2{{(2)}}
\def\3{{(3)}}
\def\g{{\gamma}}
\def\a{{\alpha}}
\def\O{{\Omega}}
\def\mnn{{\mathbb N}}
\def\mfs{{\mathscr S}}
\def\l{{\lambda}}

\def\far{{\text{far}}}
\def\intt{{\text{int}}}
\def\near{{\text{near}}}

\newcommand{\ttbr}[1]{{(#1)}}
\def\5{{(5)}}
\newcommand{\bs}[1]{{\boldsymbol{#1}}}
\def\tf{{\text{TF}}}

\usepackage{mathabx}
\usepackage{mathtools}
\usepackage{pdfsync}
\usepackage{slashed}
\usepackage[normalem]{ulem}
\usepackage{upgreek}
\usepackage{url}
\usepackage{enumitem}

\usepackage[ragged]{footmisc}
\def\be{\begin{equation}}
\def\ee{\end{equation}}

\renewcommand{\i}{\mathrm{i}}
\renewcommand{\tilde}{\widetilde}

\def\lsim{\mathrel{\rlap{\lower4pt\hbox{\hskip1pt$\sim$}}
    \raise1pt\hbox{$<$}}}
\def\gsim{\mathrel{\rlap{\lower4pt\hbox{\hskip1pt$\sim$}}
    \raise1pt\hbox{$>$}}}
\def\be{\begin{equation}}
\def\ee{\end{equation}}
\def\bea{\begin{eqnarray}}
\def\eea{\end{eqnarray}}

\numberwithin{equation}{section}

\begin{document}
\title{New phases of $\mathcal{N}=4$ SYM at finite chemical potential}
\author[a]{\'Oscar~J.C.~Dias,}
\author[b]{Prahar Mitra}
\author[b]{and Jorge~E.~Santos}
\affiliation[a]{STAG research centre and Mathematical Sciences, University of Southampton,\\ Southampton SO17 1BJ, UK}
\affiliation[b]{Department of Applied Mathematics and Theoretical Physics, University of Cambridge,\\
Wilberforce Road, Cambridge, CB3 0WA, UK}


\emailAdd{ojcd1r13s@soton.ac.uk}
\emailAdd{pmitra@damtp.cam.ac.uk}
\emailAdd{jss55@cam.ac.uk}

\abstract{
We do a systematic search of supergravity solutions that, via the AdS$_5$/CFT$_4$ correspondence, are dual to thermal states in $\mathcal{N}=4$ SYM at finite chemical potential. These solutions dominate the microcanonical ensemble and are required to ultimately reproduce the microscopic entropy of AdS black holes. Using a mix of analytical and numerical methods, we construct and study static charged hairy solitonic and black hole solutions with global AdS$_5$ asymptotics. They are constructed in two distinct consistent truncations of five dimensional gauged supergravity (and can thus be uplifted to asymptotically AdS$_5\times$S$^5$ solutions of type IIB supergravity). In the ``single charge'' truncation which consists of one charged scalar field, hairy black holes exist above a critical charge and merge with the known Behrndt-Cveti\v c-Sabra (BCS) black holes along a curve determined by the onset of superradiance in the latter family. The lowest mass hairy black hole is a singular zero entropy soliton. In the ``two charge'' truncation which consists of a two equal charged scalar fields, hairy black holes exist for all charges and merge with the known BCS black holes along their superradiant onset curve. The lowest mass hairy black hole is a smooth supersymmetric zero entropy soliton. Together with the known phases of the truncation with three equal charges, our findings permit a good understanding of the full phase space of SYM thermal states with three arbitrary chemical potentials.
}

\maketitle


\section{Introduction}

One of the most important conformal field theories (CFTs) in high energy physics is the four dimensional $\mathcal{N}=4$ supersymmetric Yang-Mills (SYM) with gauge group $SU(N)$. The AdS/CFT correspondence was originally formulated using this CFT \cite{Maldacena:1997re} and the holographic dictionary is perhaps best understood in this version of the correspondence  \cite{Gubser:1998bc,Witten:1998qj,Aharony:1999ti}. Despite this, the spectrum of thermal states in this CFT at non-zero chemical potential is still not completely understood.

Maldacena's AdS$_5$/CFT$_4$ duality \cite{Maldacena:1997re} conjectures that classical type IIB superstring theory on $\ads_5\times S^5$ (at equal radii $L$) with string coupling $g_s$ and string length $\ell_s$ is equivalent to $\CN=4$ SYM with gauge group $SU(N)$ and 't Hooft coupling $\l = g_\text{YM}^2 N$. The CFT lives on the conformal boundary of $\ads_5$ -- which in global coordinates is the Einstein Static Universe $\mrr_t \times S^3$ -- and the parameters on the two sides of the duality are identified according to $\l \sim (L/\ell_s)^4$ and $g_s \sim \l/N$. The stringy theory side of this duality is best understood in the low energy limit $\ell_s/L \to 0$ (which suppresses stringy corrections) and at weak coupling $g_s \to 0$ (which suppresses loop corrections) where the theory reduces to classical type IIB supergravity on $\ads_5\times S^5$. In this limit, the CFT is strongly coupled ($\l \to \infty$) and is truncated to the planar sector ($N \to \infty$).

Under the holographic dictionary, thermal states of $\mathcal{N}=4$ SYM with temperature $T$, chemical potentials $\mu_i$ and energies of order $N^2$ living on the Einstein static universe are dual -- in the large $N$ limit -- to black hole solutions with Hawking temperature $T$  and chemical potentials $\mu_i$  of IIB supergravity with global AdS$_5\times S^5$ asymptotics \cite{Witten:1998qj}. Consequently, finding the full phase space of black hole solutions of IIB is mandatory to understand the dynamics and thermodynamics of thermal phases of $\CN=4$ SYM.

The massless bosonic fields of $d=10$ type IIB supergravity are the metric tensor $g_{ab}$, the  dilaton $\Phi$,  the axion $C$, the NS-NS antisymmetric  2-tensor $B_\2$,  the R-R 2-form potential $C_\2$, and the R-R 4-form field $C_\4$ with a 5-form field strength $F_\5 = \dt C_\4$ satisfying a self-duality condition $\star \tilde{F}_\5=\tilde{F}_\5=F_\5- \frac{1}{2} C_\2 \wedge H_\3 + \frac{1}{2} B_\2 \wedge F_\3$ (their fermionic superpartners are a complex Weyl gravitino and a complex Weyl dilatino) \cite{schwarz1983covariant,grana2002gauge}. Solving the associated coupled system of equations of motion (EoM) to find solutions of type IIB with all or some of these fields switched on is usually  a fairly complicated task. A notable exception is the Schwarzschild-$\ads_5\times S^5$ black hole or its rotating partner, the Hawking-Hunter-Taylor-$\ads_5\times S^5$ black hole \cite{Hawking:1998kw} (with two arbitrary angular momenta) which are solutions of the $SL(2,\mrr)$-invariant sector of type IIB where only the metric $g$ and the self-dual 5-form field $F_\5$ are turned on. These are ``simple'' solutions because they are everywhere (not only at the boundary) the direct product of two base spaces $\CM_5 \times S^5$ and have horizon topology $S^3 \times S^5$. Despite their simplicity, these solutions exhibit exceptionally rich thermodynamics: for example, we can have small and large black holes and the latter dominate the canonical ensemble at high temperatures, with a phase transition into a thermal AdS$_5\times S^5$ gas of gravitons at the Hawking-Page temperature \cite{Hawking:1982dh,Witten:1998qj} (dual to a confinement/deconfinement first order phase transition on the SYM \cite{Witten:1998qj}).

Less trivial solutions of type IIB in the $SL(2,\mrr)$-invariant sector which are asymptotically globally $\ads_5\times S^5$ but break the $SO(6)$ symmetry of $S^5$ down to $SO(5)$ have also been found recently. They describe either lumpy black holes with polar deformations along the $S^5$ \cite{Dias:2015pda,Cardona:2020unx} or black holes that localize on the $S^5$ \cite{Dias:2016eto} (the latter have $S^8$ horizon topology). Their existence demonstrates how important it is to find the full phase diagram of asymptotically global AdS$_5\times S^5$ black holes. Indeed, these solutions show that the much loved $SO(6)$-preserving  Schwarzschild-AdS$_5\times S^5$ black holes can be unstable to a localisation on the $S^5$ if their radius (in AdS units) is sufficiently small \cite{Banks:1998dd,Peet:1998cr,Hubeny:2002xn,Buchel:2015gxa} and the localized $SO(5)$-preserving  black holes associated to this Gregory-Laflamme-like instability actually dominate the microcanonical ensemble at small energies. This first-order transition is dual to spontaneous breaking of the $SO(6)$ $R$-symmetry of $\CN = 4$ SYM down to $SO(5)$. In other words, from the viewpoint of a dimensional reduction of IIB along the $S^5$, the localized phases correspond to the condensation of an infinite tower of scalar operators (the VEV of the condensed scalars vanish for the $SO(6)$-preserving states \cite{Dias:2015pda,Dias:2016eto}) with increasing conformal dimension that can be read off using  Kaluza-Klein holographic renormalization \cite{Skenderis:2006uy,Dias:2015pda,Dias:2016eto}.

The above examples invite us to explore even further (and ultimately, in full) the phase space of thermal states of Maldacena's AdS/CFT. Once this is done, we can identify all the relevant saddle points for the thermodynamic partition functions of the theory. For this we can benefit from the fact that, if we are interested in systems with enough symmetry so that the above localization phenomenon does not occur, a dimensional reduction of type IIB supergravity along the $S^5$ yields 5-dimensional $\mathcal{N}=8$ gauged supergravity \cite{Gunaydin:1985cu}. It is believed (although not yet proven) that this is a consistent reduction  of the full IIB supergravity on AdS$_5\times S^5$.\footnote{At the linearised level, the reduction ansatz was given in \cite{Kim:1985ez} and the full non-linear reduction ansatz was conjectured in \cite{Khavaev:1998fb}. However, at the full non-linear level, so far the only complete proofs that the reduction is consistent are for the consistent embedding of the maximal Abelian $U(1)^3$ truncation \cite{Cvetic:1999xp}, the $\mathcal{N} = 4$ gauged $SU(2) \times U(1)$ truncation \cite{Lu:1999bw}, and the scalar truncation in \cite{Cvetic:1999xx,Cvetic:2000eb}. For recent progress on trying to extend the proof to the full theory see \cite{Ciceri:2014wya} were a strategy to establish the proof is outlined.} If so, full information of the 10-dimensional fields $\{g_{ab}, \Phi, C, B_\2, C_\2, C_\4 \}$ is equivalently encoded in the 5-dimensional spectrum of  gauged $\mathcal{N}=8$ supergravity whose field content consists of one graviton, fifteen $SO(6)$ gauge fields, twelve 2-form gauge potentials in the $6+\widebar{6}$ representations of $SO(6)$, 42 scalars in the $1+1+20^\prime+10+\overline{10}$ representations of $SO(6)$ and the fermionic superpartners. But solving for these fields in full generality is still a formidable task. 

Out of the above IIB fields, the most relevant ones for Maldacena's AdS/CFT are the graviton and the self-dual 5-form (since these are the fields that are sourced by D3-branes) and it is well known  that type IIB supergravity itself can be consistently truncated in $d=10$ to this $SL(2,\mrr)$-invariant subsector (in the sense that the system of equations of type IIB closes if we set the other fields to zero). A dimensional reduction of this IIB subsector along the $S^5$ yields the so-called 5-dimensional $SO(6)$ gauged supergravity and it has been established that this is a consistent reduction of the $SL(2,\mrr)$-invariant subsector of IIB supergravity \cite{Cvetic:2000nc}.\footnote{The Kaluza-Klein reduction along $S^5$ is also proven to be a consistent reduction if one further retains the dilaton $\Phi$ and axion $C$ of the type IIB. In $D=5$ the $SO(6)$ gauged supergravity is simply supplemented with an additional $SL(2,\mathbb{R})$ invariant term in the action \cite{Cvetic:2000nc}.} Additionally, the $d=5$ $SO(6)$ gauged supergravity is also a consistent truncation of $d=5$ $\CN=8$ gauged supergravity where we simultaneously set the $1+1+10+\overline{10}$ scalars and the $6+\widebar{6}$ 2-form potentials to zero. The bosons that survive -- namely the graviton, the 15 $SO(6)$ gauge fields $A^{ij}_{(1)}$, and the $20^\prime$ scalars which parameterise the full $SL(6,\mathbb{R})/SO(6)$ submanifold of the complete scalar coset (the scalars are parameterized by a symmetric unimodolar $SO(6)$ tensor $T_{ij}$) -- descend  from the metric and the self-dual 5-form of the original $d=10$ type IIB supergravity. Undoubtedly, we should attempt to find the full phase space of thermal solutions in this consistent reduction of IIB, but even this task is challenging. 

On this long-term programme we can however start by looking into a further consistent truncation of 5-dimensional $SO(6)$ gauged supergravity which is singled out by the $U(1)^3$ Cartan subgroup of $SO(6)$. In this truncation, the non-zero fields in the bosonic sector are the graviton, two neutral real scalar fields $\{\varphi_1, \varphi_2\}$\footnote{It is often convenient to replace the two real scalar fields $\{\varphi_1, \varphi_2\}$ with 3 real scalars $\{X_1,X_2,X_3\}$ subject to the constraint $X_1X_2 X_3=1$.}, 3 complex scalar fields $\{\Phi_1,\Phi_2,\Phi_3 \}$ that are charged under three $U(1)$ gauge field potentials $\{A^1_{(1)},A^2_{(1)},A^3_{(1)}\}$. The most general black hole solution of this theory is expected to have 6 conserved charges: the energy $E$, three $U(1)$ electric charges $\{Q_1,Q_2,Q_3\}$, and two independent angular momenta $\{J_1,J_2\}$ along the two independent rotation planes of AdS$_5$ with $SO(4)$ symmetry. In the holographic dictionary, the dual thermal states in $\CN=4$ SYM have $SU(4)\cong SO(6)$ $R$-charge given by the weight vector $(Q_1,Q_2,Q_3)$ and chemical potentials $\{\mu_1,\mu_2,\mu_3\}$ given by the sources of $\{A^1_{(1)},A^2_{(1)},A^3_{(1)}\}$ \cite{Cvetic:1999xp}. On the other hand, $(J_1,J_2)$ is proportional to a weight vector of the four dimensional rotation group $SO(4)$. In the dual CFT language, one usually works with $J_L \equiv J_1 + J_2$ and $J_R \equiv J_1 - J_2$, which are proportional to the weights with respect to the two $SU(2)$ factors in $SO(4) \sim SU(2)_L \times SU(2)_R$ \cite{Kunduri:2006ek}.

The black hole solutions in this theory when the charged scalar fields $\Phi_{1,2,3}$ vanish -- in which case the theory reduces to $d=5$, $\CN=2$ $U(1)^3$ gauged supergravity coupled to two vector multiplets (or minimal supergravity when the three $U(1)$'s are equal) -- are already fully known. In this case, the most general non-extremal solution with arbitrary $\{E,Q_1,Q_2,Q_3\}$ but zero angular momentum $J_1=J_2=0$ was found by Behrndt-Cveti\v c-Sabra \cite{Behrndt:1998jd}. Solutions with equal angular momenta were found by Cveti\v c-L\"u-Pope \cite{Cvetic:2004ny} which was then generalized  to have all the six charges arbitrary  by Wu \cite{Wu:2011gq}\footnote{The solution of \cite{Wu:2011gq} reduces to previously known black holes of the theory, namely: to \cite{CVETIC2004273} (with arbitrary $Q_1=Q_2=Q_3$ and $J_1=J_2$),  to \cite{Chong:2005hr} (with arbitrary $Q_1=Q_2=Q_3$ and $J_1,J_2$), to \cite{Cvetic:2004ny} (with arbitrary $Q_1,Q_2,Q_3$ and $J_1=J_2$), to \cite{Chong:2005da} (with arbitrary $Q_1=Q_2, Q_3=0$ and $J_1,J_2$), to \cite{MEI200764} (with arbitrary $Q_1=Q_2, Q_3$ and $J_1,J_2$), to \cite{Chong:2005da} (with arbitrary $Q_1=Q_2=0, Q_3$ and $J_1=J_2$), and  to \cite{Chong:2006zx,Wu:2011zzh} (with arbitrary $Q_1=Q_2=0, Q_3$ and $J_1,J_2$).}. In Section \ref{sec:BCS-BHs} we will review the solution of \cite{Behrndt:1998jd} while taking the opportunity to write it in a novel form that is more tailored to study its physical and thermodynamical properties (which we will do here thus filling a gap in the literature).\footnote{In this manuscript, we will only be interested in asymptotically global AdS$_5\times S^5$ supergravity solutions. However, there are also solutions with Poincar\'e AdS$_5\times S^5$ asymptotics that are dual to Coulomb flows or to top-down models of holographic superconductors; see e.g. \cite{Freedman:1999gk,Khavaev:2001yg,Gubser:2009qm,Gauntlett:2009dn,Gubser:2009gp,Gauntlett:2009bh,Bobev:2011rv} and references therein.}

Often, the extremal limit of the non-extremal black holes has conserved charges that match those of the supersymmetric black holes of the theory.\footnote{There are subsectors of the theory where there is no $T\to 0$ limit of the non-extremal black holes. This is e.g. the case when at least one of the charges $Q_{1,2,3}$ is zero as we will find later.} The most general such solution known is a 4-parameter solution with charges $\{E,Q_1,Q_2,Q_3\}$ and angular momenta $\{J_1,J_2\}$ whose mass satisfies the BPS condition $E=Q_1+Q_2+Q_3+J_1/L+J_2/L$, where $L$ is the AdS$_5$ radius. This is described by the Kunduri-Lucietti-Reall supersymmetric black hole  \cite{Kunduri:2006ek}.\footnote{In contains as special cases the previously known supersymmetric black holes of \cite{Gutowski:2004ez} (with arbitrary $Q_1=Q_2=Q_3$ and $J_1=J_2$), of \cite{Gutowski:2004yv} (with arbitrary $Q_1,Q_2,Q_3$ and $J_1=J_2$) and of
\cite{Chong:2005da,Chong:2005hr} (with arbitrary $Q_1=Q_2=Q_3$ and $J_1,J_2$).} Note that although the most general supersymmetric black holes are expected to have  5 independent conserved charges $\{Q_1,Q_2,Q_3,J_1,J_2\}$, the supersymmetric black holes of \cite{Kunduri:2006ek} are just a 4-parameter family of solutions because its 5 parameters have to obey an additional constraint (this adds to the BPS condition that fixes the energy as a function of the other conserved charges).

In this manuscript we want to extend the programme of finding thermal states at finite chemical potential of $\mathcal{N}=4$ SYM     
further and find solutions of 5-dimensional $SO(6)$ gauged supergravity with non-vanishing complex scalars  $\Phi_{1,2,3}$ (hereinafter we refer to these solutions as ``hairy'' black holes since they have non-trivial scalar hair given by the expectation value (VEV) of the spontaneously broken fields $\Phi_{1,2,3}$). These asymptotically global AdS$_5\times S^5$ hairy black holes are dual to thermal states with finite chemical potential and they are relevant because the ``bald'' black holes \cite{Behrndt:1998jd,Cvetic:2004ny,Wu:2011gq} with  $\Phi_{1,2,3}=0$ are unstable to the condensation of these scalars and, as we will find, the novel hairy black holes can dominate some thermodynamic ensembles in windows of the parameter space\footnote{The physical mechanisms that are responsible for the condensation of scalar fields in black hole backgrounds with a Maxwell field or rotation are the superradiant instability and/or the violation of the near-horizon AdS$_2$ Breitenl\"ohner-Freedman (BF) bound \cite{Breitenlohner:1982jf}. This occurs already in solutions of AdS-Einstein(-Maxwell) gravity coupled to a scalar field a.k.a the Abelian Higgs model in AdS which can be seen as a bottom up model for the supergravity physics and hairy black holes we discuss here; see \cite{Gubser:2008px,Hartnoll:2008vx,Hartnoll:2008kx,Basu:2010uz,Dias:2011tj,Dias:2010ma,Dias:2011at} for early discussions of these condensation mechanisms in Abelian Higgs model systems with bound states.}. Moreover, in a phase diagram of asymptotically global AdS$_5\times S^5$ solutions, often these hairy black holes exist in a region between the onset of the condensation instability (where they merge with \cite{Behrndt:1998jd,Cvetic:2004ny,Wu:2011gq}) and the boundary defined by the BPS condition of the system. A special family of these hairy black holes with $\Phi_1=\Phi_2=\Phi_3$ (and $A^1=A^2=A^3$) was already found recently in \cite{Bhattacharyya:2010yg,Markeviciute:2016ivy,Markeviciute:2018yal}. The solutions of \cite{Bhattacharyya:2010yg,Markeviciute:2016ivy} are static hairy black holes with $Q_1=Q_2=Q_3$ and $J_1=J_2=0$ while the equal angular momenta partner solutions with $Q_1=Q_2=Q_3$ and $J_1=J_2$ were found in \cite{Bhattacharyya:2010yg,Markeviciute:2018yal}. In this case, the solutions without hair are literally the Reissner-Nordstr\"om-AdS$_5\times S^5$ (in the static case) and Kerr-Newman-AdS$_5\times S^5$ black holes.

In this manuscript, we shall lift the restriction of $\Phi_1=\Phi_2=\Phi_3$ and $A^1=A^2=A^3$ and construct asymptotically globally AdS$_5\times S^5$ static hairy solutions in two other sectors of the consistent truncation, 1) $A^1=A^2\equiv 0$, $A^3\equiv A$ (the EoM then imply $\Phi_1=\Phi_2\equiv 0$, $\Phi_3\equiv \Phi$), and 2) $A^1=A^2\equiv A$, $A^3\equiv 0$ (the EoM then imply $\Phi_1=\Phi_2\equiv \Phi$, $\Phi_3\equiv 0$). In future work, we plan to extend the analysis to find the rotating partners of these solutions with equal angular momenta. Altogether, the sector studied in \cite{Bhattacharyya:2010yg,Markeviciute:2016ivy,Markeviciute:2018yal} together with the two sectors discussed here gives us a good understanding of the full phase space of hairy solutions with three arbitrary charged fields $\Phi_{1,2,3}$.\footnote{Our asymptotically global AdS$_5\times S^5$ hairy black holes and (most of) our solitons are regular. Truncations of $\mathcal{N}=8$ gauged supergravity that are different from ours and described in \cite{Bobev:2010de} may also have similar hairy black holes and supersymmetric solitons.} 

There are at least three main motivations to undergo the programme advocated above. $(i)$ Firstly, as mentioned before but worth emphasizing, if we are to fully understand and benefit from Maldacena's AdS/CFT correspondence we should (besides formally proving it) find all the thermal solutions and map them into states in the dual $\mathcal{N}=4$ SYM. Only then will we be able to identify the dominant phases (as saddle points) in the relevant thermodynamic ensembles. For example, in this manuscript we will identify new thermal phases with a finite chemical potential that can dominate some thermodynamic ensembles (at least in some regions of the phase space) over already known phases.
$(ii)$ Secondly, the Bekenstein-Hawking entropy of some asymptotically flat black holes has already been reproduced microscopically within string theory and with the help of holographic techniques (notably in \cite{Strominger:1996sh}). Remarkably, such a programme is still lacking for asymptotically AdS black holes though there are several promising recent developments on this front (see \cite{Zaffaroni:2019dhb} and references therein). In order to complete such an initiative, we necessarily need to identify all the black holes of the bulk theory. $(iii)$ Finally, a remarkable puzzle of $SO(6)$ gauged supergravity is that its most general supersymmetric thermal solution known so far has only 4 independent parameters. This is the aforementioned  Kunduri-Luietti-Reall solution \cite{Kunduri:2006ek}. However, given that such asymptotically AdS$_5\times S^5$ black holes are characterized by 6 conserved charges with one of them constrained by the BPS relation  $E=Q_1+Q_2+Q_3+J_1/L+J_2/L$, one should expect that the most general  supersymmetric black hole should be a 5-parameter solution. From the dual CFT perspective, we also expect the most general supersymmetric states to be characterized by 5 parameters. So, what is the missing gravitational parameter? An important observation  is that the Kunduri-Lucietti-Reall solution has no charged scalar hair (\emph{i.e.} $\Phi_{1,2,3}=0$). In the consistently truncated theory with $\Phi_1=\Phi_2=\Phi_3$ (and thus with $Q_1=Q_2=Q_3$) and for the $J_1=J_2$ case, it was found that there are hairy black holes that, in the extremal $T\to 0$ limit  obey the BPS condition and thus fill the BPS surface \emph{beyond} the region where the Kunduri-Luietti-Reall solutions exist \cite{Markeviciute:2018yal}. This suggests that the charged scalar condensate might be the missing gravitational parameter. One would like to extend this proposal to the most general  $SO(6)$ gauged supergravity without particular restrictions on $\{ \Phi_1,\Phi_2,\Phi_3 \}$. Motivated by this conjecture, in the present manuscript, we will construct static hairy solutions within certain consistent truncations of $SO(6)$ gauged supergravity. In future work \cite{DMS2021}, we generalize our analysis to rotating solutions in order to test the conjecture proposed in \cite{Markeviciute:2018yal}

The plan of the manuscript is the following. In Section~\ref{sec:theory} we start by describing the  consistent truncation of type IIB supergravity on $\ads_5 \times S^5$, namely  $SO(6)$ gauged supergravity and its $U(1)^3$ gauged supergravity truncations that we will study (which retain the gauge symmetry associated to the Cartan subgroup of $SO(6)$). We also revisit the Behrndt-Cveti\v c-Sabra black holes of the theory and we take the opportunity to study their thermodynamics (thus filling a gap in the literature). In Section~\ref{sec:SingleQ} we do a consistent search of static hairy solutions (i.e. with finite chemical potential) of the $SO(6)$ gauged supergravity truncation with a single charge while in Section~\ref{sec:TwoQ} we repeat the process but this time for the truncation with two equal charges. We follow a similar exposition plan for both truncations/sections, as it is best clear from the table of contents. Indeed, in the first subsection of both sections, we start by setting up the ansatz\"e and boundary conditions of the boundary value problem that we need to solve. Next, in the second subsection, we find the thermodynamic quantities of the truncated system using holographic renormalization. Before finding the hairy black hole of the system, in the third subsection, we first find the hairy supersymmetric solitons. In the fourth subsection, we then revisit the ``bald'' BCS black holes of the theory, in the particular truncation at hand, to study their thermodynamic properties (that were not studied previously in the literature). In the fifth subsection, we find that these BCS black holes are unstable to condensation of the charged scalar field of $SO(6)$ gauged supergravity and we find the instability timescale. In the  sixth subsection, we find directly the onset of this scalar condensation instability which also marks the merger of the hairy and bald black holes of the theory in a phase diagram of solutions. In the seventh and eighth subsections we collect the results (about BCS and hairy black holes and solitons) of the previous subsections to finally mount the phase diagram of static solutions of the truncated theory, first (seventh subsection) in the microcanonical ensemble (that hairy black holes can dominate) and then in the grand-canonical ensemble (eighth subsection). Finally, in the ninth subsection, we present a complementary construction of the hairy solitons and black holes of the truncation using perturbation theory (with the perturbation parameters being the charged scalar condensate and in addition, for the black holes, the horizon radius). These perturbative results are a very good approximation to the numerical solutions for small energies and charges and nearby the instability onset where the ``bald'' BCS and hairy black holes merge.
We have three appendices. In Appendix \ref{app:holo-renorm}, we apply {\it mutatis mutandis} the holographic renormalization procedure of Bianchi-Freedman-Skenderis \cite{Bianchi:2001de,Bianchi:2001kw} to compute the holographic stress tensor and current of the $U(1)^3$ gauged supergravity $-$ i.e. of the truncation of  $SO(6)$ gauged supergravity that retains a $U(1)^3 \cong SO(6) / \mzz_2^3$ gauge symmetry (associated to the Cartan subgroup of $SO(6)$) with associated gauge fields $\{A_\1^K\}$ $-$ with all sources turned on. 
Finally, in Appendices~\ref{App:PerturbativeSingleQ} and \ref{App:PerturbativeTwoQ} we give details of the perturbative analysis done in Sections~\ref{sec:PerturbativeSingleQ} and~\ref{sec:PerturbativeTwoQ} for the single charge and two equal charge truncations, respectively.

\section{A consistent truncation of $SO(6)$ gauged supergravity}\label{sec:theory}

\subsection{Truncating $SO(6)$ down to $U(1)^3$ gauged supergravity}\label{sec:theory2}

As discussed in the introduction, five-dimensional ${\cal N}=8$ gauged supergravity is expected to be a consistent truncation of type IIB supergravity on $\ads_5 \times S^5$ \cite{Gunaydin:1985cu}. The bosonic field content of this theory consists of one graviton, fifteen $SO(6)$ gauge fields, twelve 2-form gauge potentials in the $6+\widebar{6}$ representations of $SO(6)$, and 42 scalars in the $1+1+20^\prime+10+\overline{10}$ representations of $SO(6)$ \cite{Gunaydin:1985cu}. This theory admits a further truncation -- the $SO(6)$ gauged supergravity \cite{Cvetic:2000nc} -- that retains only the metric, the scalars in the $20^\prime$  that parameterise the full $SL(6,\mathbb{R})/SO(6)$ submanifold of the complete scalar coset (that are  parametrized by a symmetric unimodolar $SO(6)$ tensor $T_{ij}$), and the 15 Yang-Mills fields $A^{ij}_{(1)}$. It is a consistent reduction of the $SL(2,\mrr)$ invariant sector of 10-dimensional IIB supergravity (which contains only the metric and the self-dual 5-form field) along the $S^5$. The action for  $SO(6)$ gauged supergravity is given by \cite{Cvetic:2000nc}
\begin{equation}\label{ActionSO6sugra}
\begin{split}
S &=\frac{1}{16 \pi G_5}
\int \dt^5 x \sqrt{-g} \bigg[ R - \frac{1}{4}T_{ij}^{-1}\left(D_a T_{jk}\right)T_{kl}^{-1}\left(D^a T_{li}\right)-\frac{1}{8}T_{ik}^{-1}T_{jl}^{-1}F^{ij}_{a b}(F^{kl})^{a b} - V\\
&- \frac{1}{192} \ve^{abcde} \varepsilon_{i_1 \cdots i_6} \left(F_{ab}^{i_1 i_2}F_{cd}^{i_3 i_4} A_{e}^{i_5 i_6} - \frac{1}{L} F_{ab}^{i_1 i_2} A_c^{i_3 i_4} A_d^{i_5 j}A_e^{j i_6} + \frac{2}{5L^2} A_a^{i_1 i_2} A_b^{i_3 j} A_c^{j i_4 }A_d^{ i_5 k}A_e^{ k i_6}\right)\! \bigg]
\end{split}
\end{equation}
where $R$ is the Ricci scalar, $V$ is the potential associated to the scalar fields described by the symmetric  $SO(6)$ tensors $T_{ij}$ with unit determinant, $(i,j,k,\cdots)$ denote the $SO(6)$ vector indices, $(a, b,\cdots)$ are the spacetime indices, $\varepsilon^{abcde}$ denotes the spacetime Levi-Civita tensor with $\ve^{01234} \equiv - \frac{1}{\sqrt{-g}}$, $\varepsilon_{i_1 \cdots i_6}$ is the Levi-Civita tensor with $\varepsilon_{i_1 \cdots i_6}\equiv 1$, and \begin{equation}\label{ActionSO6sugra2}
\begin{split}
V &=\frac{1}{2}\big[2 T_{ij} T_{ij} - (T_{ii})^2\big],\\
F_\2^{ij} &= \dt A_\1^{ij} + A_\1^{ik} A_\1^{kj},\\
D_a T_{ij} &= \partial_a T_{ij} + A_a^{ik}T_{kj} + A_a^{jk}T_{ik}.
\end{split}
\end{equation}
Using the holographic dictionary, we can relate the 5-dimensional Newton's constant with the rank $N$ of the gauge group $SU(N)$ of the dual ${\cal N}=4$ SYM theory and with the $\ads_5$  radius $L$  as\footnote{In detail, matching the low energy limit of string theory with type IIB supergravity in the Einstein frame one finds that the 10-dimensional Newton constant is $G_{10}=8\pi^6 g_s^2\ell_s^8$ where $g_s$ and $\ell_s$ are the string coupling and string length respectively. The t'Hooft coupling is $\l=g_{\text{YM}}^2N$ where for $p$-branes the YM coupling is given by $g_{\text{YM}}^2=(2\pi)^{p-2}g_s\ell_s^{p-3}$. For $p=3$, the equivalence between the 3-brane and D3-brane charges requires that $\l=\frac{L^4}{2\ell_s^4}$ and thus $G_{10}=\frac{\pi^4 L^8}{2N^2}$. Finally, the 5-dimensional Newton constant $G_5$ is obtained by dimensional reduction of the 10d theory so $G_5=\frac{G_{10}}{\Omega_5 L^5} = \frac{\pi L^3}{2 N^2}$ ($\Omega_5=\pi^3$ is the volume of a unit $S^5$ and $L$ is the radius of the $S^5$ and of the $\ads_5$ due to the 10-dimensional EoM).}
\begin{equation}\label{G5}
G_5= \frac{\pi L^3}{2 N^2}.
\end{equation}

In this manuscript, we consider a further consistent truncation of  \eqref{ActionSO6sugra}. To describe this, it is convenient  to use a complex basis for the $SO(6)$ vector indices that appear summed in \eqref{ActionSO6sugra}, as was done in the simpler truncation of \cite{Bhattacharyya:2010yg}.
Let $\{x_j\}$ $(j = 1,\cdots,6)$ denote $SO(6)$ Cartesian directions and introduce the complex coordinates 
\begin{equation}
    \begin{split}
        z_K = x_{2K - 1} + i x_{2K}, \qquad \bar z_K = x_{2K - 1} - i  x_{2K} , \qquad K = 1,2,3.
    \end{split}
\end{equation}
We now consider a restriction of \eqref{ActionSO6sugra} which preserves a $\mzz_2^\1 \times \mzz_2^\2 \times \mzz_2^\3$ symmetry where $\mzz_2^{(K)}$ denotes a rotation by $\pi$ in the complex $z_K$ plane (under which $z_K,{\bar z}_K \to - z_K,-{\bar z}_K$). The most general field configuration which preserves this symmetry satisfies
\begin{equation}
    \begin{split}
    \label{eq:cons-trunc-1}
        T_{z_J z_K} = T_{{\bar z}_J{\bar z}_K} = T_{z_J{\bar z}_K} = A_\1^{z_J z_K} = A_\1^{ {\bar z}_J {\bar z}_K} = A_\1^{z_J {\bar z}_K} = 0 \quad \forall \quad J \neq K.
    \end{split}
\end{equation}
The remaining non-vanishing fields, namely $T_{z_Kz_K}$ and its conjugate $T_{z_Kz_K}^\dagger = T_{{\bar z}_K {\bar z}_K}$, $T_{z_K {\bar z}_K}$ and $A_\1^{z_K {\bar z}_K}$ can be parameterized as
\begin{equation}\label{Truncation}
 \begin{split}
  T_{z_Kz_K} &= \frac{1}{4}\,X_K \Phi_K, \qquad T_{z_K{\bar z}_K} = \frac{1}{4}\,X_K \sqrt{4+\Phi_K^\dagger \Phi_K}, \qquad A_\1^{{z_K{\bar z_K}}} = 2 i A_\1^K.
 \end{split}
\end{equation}
The scalar fields $X_K$ satisfy the unimodularity constraint $X_1X_2X_3=1$ and thus effectively describe two real scalar fields $\varphi_1$ and $\varphi_2$,
\begin{equation}\label{Truncation2}
 X_1=e^{-\frac{1}{\sqrt{6}}\varphi_1-\frac{1}{\sqrt{2}}\varphi_2}, \qquad 
 X_2=e^{-\frac{1}{\sqrt{6}}\varphi_1+\frac{1}{\sqrt{2}}\varphi_2},\qquad 
 X_3=e^{\sqrt{\frac{2}{3}} \varphi_1}.
\end{equation}
Equations \eqref{eq:cons-trunc-1}--\eqref{Truncation2} describe the most general configuration which is invariant under the discrete $\mzz_2^3$ symmetry described above. Consequently, this is a consistent truncation of the system \eqref{ActionSO6sugra}.\footnote{We also verify that  \eqref{eq:cons-trunc-1} is consistent with the EoM derived from \eqref{ActionSO6sugra}.} More precisely, it is a truncation which retains a $U(1)^3 \cong SO(6) / \mzz_2^3$ gauge symmetry (this is the Cartan subgroup of $SO(6)$) with associated gauge fields $\{A_\1^K\}$. The matter content consists of 2 neutral scalars  $\{\varphi_1,\varphi_2\}$ and three complex scalar fields $\{\Phi_K\}$ that are charged under the $U(1)$'s gauge fields. All 5 scalars have mass $m^2L^2=-4$ and thus saturate the AdS$_5$ Breitenl\"ohner-Freedman (BF) bound \cite{Breitenlohner:1982jf}.\footnote{For  the neutral scalar field  $\varphi_K$,  given the relative normalization of its kinetic and potential contributions to the action, the mass is $m^2=\partial^2_{\varphi_K}V\big|_{\varphi_K=0}$. On the other hand, taking into account the  relative normalization of the kinetic and potential contributions in the action for the charged scalar field, the mass of $\Phi_K$ is $m^2=\partial_{\Phi_K}\partial_{\Phi^\dagger_K}(8V)\big|_{\Phi_K=\Phi^\dagger_K=0}$.} Under the AdS/CFT dictionary these fields are dual to operators
of conformal dimension $\Delta=\frac{d}{2} + \sqrt{\frac{d^2}{4} + m^2 L^2 } = 2$.
Additionally, the 3 complex scalars  $\{\Phi_K\}$ have electric charge $q L=2$.

Substituting \eqref{eq:cons-trunc-1}--\eqref{Truncation2} into the EoM derived from the  action \eqref{ActionSO6sugra}, we can check that the resulting EoM for the dynamical fields can be derived from the following effective action
\begin{equation}
\begin{split}\label{OurCTaction}
S &= \frac{1}{16 \pi G_5} \int \dt^5 x \sqrt{-g} \bigg\{  R - V
-\frac{1}{2} \sum_{K=1}^{2}(\nabla\varphi_K)^2
-\frac{1}{4} \sum_{K=1}^{3}\frac{1}{X_K^2}\left(F_\2^K\right)^2\\
&\qquad -\frac{1}{8} \sum_{K=1}^{3} \bigg[(D_a\Phi_K)(D^a\Phi_K)^\dagger -  \frac{(\nabla \l_K)^2}{4(4+\l_K)} \bigg]\bigg\} - \frac{1}{16 \pi G_5} \int  F_\2^1 \wedge F_\2^2  \wedge A_\1^3 \, , 
\end{split}
\end{equation}
where we have defined (with no Einstein summation convention over $K=1,2,3$)
\begin{equation}\label{OurCTaction2}
\begin{split}
D_a\Phi_K &\equiv \partial_a\Phi_K - i\, \frac{2}{L} A^K_a \Phi_K\,, \\
F^K_{a b} &\equiv \partial_a A^K_b - \partial_b A^K_a\,, \\
\l_K &\equiv \Phi_K\Phi^\dagger_K,\end{split}
\end{equation}
and the scalar potential for the truncation \eqref{Truncation} is 
\begin{equation}\label{OurCTaction3}
\begin{split}
V &=\frac{1}{2L^2}\bigg[ X_1^2\l_1+X_2^2\l_2+X_3^2\l_3-\frac{2}{X_1}\sqrt{4+\l_2}\sqrt{4+\l_3}\\
&\qquad \qquad \qquad \qquad -\frac{2}{X_2}\sqrt{4+\l_1}\sqrt{4+\l_3}-\frac{2}{X_3}\sqrt{4+\l_1}\sqrt{4+\l_2}\bigg].
\end{split}
\end{equation}
Extremization of the action \eqref{OurCTaction} yields the field equation for the graviton
\begin{equation}\label{GravEOM}
R_{a b} - \frac{1}{2}g_{a b}R =   T^{\varphi}_{a b}+\frac{1}{2}T^{A}_{a b}+ \frac{1}{8}T^{\Phi}_{ab} - \frac{1}{2} g_{ab} V ,
\end{equation}
where  
\begin{equation}\label{GravEOM2}
\begin{split}
T^{\varphi}_{a b} =& \frac{1}{2} \sum_{K=1}^{2} \left( \nabla_a \varphi_K \nabla_b \varphi_K 
-\frac{1}{2}g_{a b}  (\nabla_c \varphi_K)(\nabla^c \varphi_K) \right), \\
T^{A}_{a b} = & \sum_{K=1}^{3}\frac{1}{X_K^2}\left( 
 F^K_{ac}(F^K)_b{}^c  - \frac{1}{4} g_{a b} (F^K_\2)^2 \right), \\
T^{\Phi}_{a b} =& \frac{1}{2}\sum_{K=1}^{3} 
\bigg[ (D_a\Phi_K)\left( D_b\Phi_K \right)^\dagger + (D_ b\Phi_K) \left(D_a\Phi_K\right)^\dagger -g_{a b}  (D_c \Phi_K)(D^c \Phi_K)^\dagger  \\
&\qquad \qquad -\frac{1}{2\big(4 + \l_K\big)} \bigg( \nabla_a \l_K\nabla_b\l_K - \frac{1}{2}g_{a b}\nabla_c \l_K \nabla^c\l_K\bigg)\bigg],
\end{split}
\end{equation}
and the equations of motion for the other fields
\begin{eqnarray}\label{ScalarsGaugeEOM}
  && \Box \,\varphi_1 -\frac{1}{2\sqrt{6}} \bigg[ \frac{(F_\2^1)^2}{X_1^2}+ \frac{(F_\2^2)^2}{X_2^2}-2 \frac{(F_\2^3)^2}{X_3^2}  \bigg] -\frac{\partial V}{\partial \varphi_1}=0,\nonumber \\
  &&  \Box\, \varphi_2 -\frac{1}{2\sqrt{2}} \bigg[ \frac{(F_\2^1)^2}{X_1^2}- \frac{(F_\2^2)^2}{X_2^2} \bigg] -\frac{\partial V}{\partial \varphi_2}=0,\nonumber \\
  &&  \dt \left(\frac{1}{X_1^2}\star F_\2^1\right)+F_\2^2  \wedge F_\2^3=-\star J^1_\1, \nonumber \\
  && \dt \left(\frac{1}{X_2^2}\star F_\2^2\right)+F_\2^3  \wedge F_\2^1=-\star J^2_\1, \\
  &&  \dt \left(\frac{1}{X_3^2}\star F_\2^3\right)+F_\2^1  \wedge F_\2^2=-\star J^3_\1,\nonumber \\
  && D^a D_a \Phi_K+\bigg[ \frac{\nabla_a\l_K\nabla^a\l_K}{4\big(4 + \l_K\big)^2} 
 - \frac{\Box \l_K}{2\big(4 + \l_K\big)}  
 -\frac{8}{\Phi_K}\frac{\partial V}{\partial\Phi_K^\dagger}
  \bigg] \Phi_K=0,\quad (K=1,2,3),\nonumber \\
  && D^a D_a \Phi_K^\dagger+\bigg[ \frac{\nabla_a \l_K\nabla^a\l_K}{4\big(4 + \l_K\big)^2} - \frac{\Box \l_K}{2\big( 4 + \l_K \big)}  
 -\frac{8}{\Phi_K^\dagger}\frac{\partial V}{\partial\Phi_K} \bigg] \Phi_K^\dagger=0,\quad (K=1,2,3),\nonumber 
\end{eqnarray}
with $\Box=g^{ab}\nabla_a\nabla_b$ and $J^K_\1=\frac{i}{4L}[\Phi_K^\dagger (D\Phi_K) - \Phi_K (D\Phi_K)^\dagger]$ is the electric current, $\dt $ is the exterior derivative,  $\star$ is the Hodge dual and we use the differential form conventions listed in appendix of \cite{Dias:2019wof}.

There are three special cases of the consistent truncation \eqref{OurCTaction} of $SO(6)$ gauged supergravity where the field equations simplify considerably, namely:
\begin{enumerate}[label=\Roman*)]
\item \textbf{Truncation with three equal charges:} The action \eqref{OurCTaction} admits a $S_3$ permutation symmetry which acts on $K$ index of all the fields. The most general $S_3$-invariant field configuration satisfies $A^1=A^2=A^3 \equiv A$, $\Phi_1 = \Phi_2=\Phi_3 \equiv \Phi$ and $X_1=X_2=X_3=1$ or equivalently $\varphi_1 = \varphi_2 = 0$. This truncation can also be obtained directly from \eqref{ActionSO6sugra} by restricting to $SO(3)$-invariant field configurations (instead of $\mzz_2^3$).\footnote{More precisely, think of $T_{ij}$ as a $3 \times 3$ matrix where each entry is a $2\times 2$ matrix. Consider the $SO(3) \subset SO(6)$ which acts on this $3\times3$ matrix. $SO(3)$-invariant field configurations are proportional to the identity (Schur's Lemma) so we can decompose $T_{ij} = \CT_{2\times 2} \otimes {\mathbb I}_{3\times 3}$. The unimodularity condition on $T$ then implies that $\CT$ is a unimodular $2\times2$ symmetric matrix and such a matrix can be parameterized in terms of one complex field $\Phi$. The same restriction on the gauge field implies $A^{ij} = \CA_{2\times2} \otimes {\mathbb I}_{3\times 3}$ and the antisymmetric $2\times2$ matrix $\CA$ is parameterized by one d.o.f. $A$.\label{footnote:construnc3}}

\item \textbf{Truncation with a single charge:} The action \eqref{OurCTaction} has a $\mzz_2 \times \mzz_2$ symmetry where the first factor is the permutation group of $K=1,2$ and the second factor is a discrete $U(1)$ transformation on $K=1$ (or equivalently on $K=2$), namely a rotation by $\pi$ which maps $\Phi_1 \to - \Phi_1$. Field configurations which are invariant under this symmetry satisfy $\Phi_1 = \Phi_2 = 0$, $\Phi_3 \equiv \Phi$,  $A^1=A^2=B$, $A^3=A$ and $X_1=X_2$ or equivalently $\varphi_1 \equiv \varphi$, $\varphi_2 = 0 $. In this sector, the action has an enhanced $\mzz_2$ symmetry under which $B \to - B$ and we can further consistently truncate to $B=0$. This truncation can also be obtained directly from \eqref{ActionSO6sugra} by restricting to $SO(4)$-invariant field configurations.\footnote{Here, $SO(4) \subset SO(6)$ acts on the top left $4\times 4$ minor matrix of the $6\times 6$ matrix $T_{ij}$.}

\item \textbf{Truncation with two equal charges:} The action \eqref{OurCTaction} has a $\mzz_2\times \mzz_2$ symmetry where the first $\mzz_2$ corresponds to the permutation group of $K=1,2$ and the second $\mzz_2$ is a discrete $U(1)$ transformation on $K=3$, namely a rotation by $\pi$ which maps $\Phi_3 \to - \Phi_3$. Field configurations which are invariant under this symmetry satisfy $A^1=A^2=A$, $A^3=B$, $\Phi_1=\Phi_2\equiv \Phi$, $\Phi_3 = 0$ and $X_1=X_2$ or equivalently $\varphi_1 \equiv \varphi$, $\varphi_2 = 0$. This truncation can also be obtained directly from \eqref{ActionSO6sugra} by restricting to $SO(2)$-invariant field configurations.\footnote{In the language of footnote \ref{footnote:construnc3}, $SO(2) \subset SO(6)$ acts on the top left $2\times2$ minor matrix of the $3\times 3$ matrix $T_{ij}$.}

Static solutions (which is the topic of this manuscript) have an additional time-reversal symmetry which sets the Chern-Simons term in \eqref{OurCTaction} to zero. In this case, the action has an enhanced $\mzz_2$ symmetry under which $B \to - B$ and we can further consistently truncate to $B=0$.

\end{enumerate}
Static asymptotically AdS$_5\times S^5$ hairy black hole and solitonic solutions of the first theory where already studied in detail using perturbation theory in \cite{Bhattacharyya:2010yg} and a full numerical analysis was done in \cite{Markeviciute:2016ivy}. This has been further extended to include angular momenta $J_1=J_2$ in \cite{Bhattacharyya:2010yg,Markeviciute:2018yal}.
In this manuscript, we construct the static asymptotically AdS$_5\times S^5$  hairy black hole and solitonic solutions in the second and third truncations (in future work, we plan to extend this study to include angular momenta $J_1=J_2$ \cite{DMS2021}) using perturbation theory and a full numerical analysis. Truncation II) is studied in section~\ref{sec:SingleQ} and truncation III) is studied in section~\ref{sec:TwoQ}. Altogether, the case studied in \cite{Bhattacharyya:2010yg,Markeviciute:2016ivy,Markeviciute:2018yal} along with the two cases discussed in this manuscript provides us with a good overview of the full phase space of hairy solutions for the case with three arbitrary charged fields $\Phi_{1,2,3}$. 

Before discussing the hairy solutions of our consistent truncations II) and III), it will be useful to review the known static black hole solutions of the theory \eqref{OurCTaction} without the charged condensate: the ``bald'' BCS black holes. We do this in the next subsection.

\subsection{Behrndt-Cveti\v c-Sabra black hole solutions of $SO(6)$ gauged supergravity}\label{sec:BCS-BHs}

When the charged scalar fields vanish, $\Phi_K=0$ $(K=1,2,3)$, the consistent truncation of $SO(6)$ gauged supergravity described by the action \eqref{OurCTaction} reduces to $U(1)^3$ gauged five-dimensional $\mathcal{N} = 2$ supergravity coupled to two vector multiplets. The static black holes of this theory have no scalar hair and are parameterized by an energy $E$ and three electric charges $\{Q_1,Q_2,Q_3\}$ associated to each of the three $U(1)$ gauge fields $A^K$ of the theory. These can be viewed as the ``Reissner-Nordstr\"om-AdS$_5$'' (RNAdS) black holes of the theory (although, in general, they also have  non-trivial neutral scalar fields $\varphi_1$ and $\varphi_2$ supporting them; the exception occurs when $Q_1=Q_2=Q_3$ in which case they are exactly the RNAdS family). 

These static black holes with three arbitrary charges $\{Q_1,Q_2,Q_3\}$ were found by Behrndt-Cveti\v c-Sabra (BCS) \cite{Behrndt:1998jd} (see also \cite{Cvetic:2004ny,Wu:2011gq}). The fields of the BCS black hole solution are given by
\begin{equation} \label{BCS:ansatz}
\begin{split}
\dt s^2 &= ( h_1 h_2 h_3 )^{1/3} \left(-\frac{f}{h_1 h_2 h_3}\dt t^2+  \frac{\dt r^2}{f}+ r^2 \dt \Omega_3^2\right),\\
\varphi_1 &=\frac{1}{\sqrt{6}}\ln \left( \frac{h_1 h_2}{h_3^2} \right)\,,\qquad \varphi_2=
\frac{1}{\sqrt{2}}\ln \left( \frac{h_1}{h_2} \right);\\
A_\1^K &=-\frac{r_0^2}{r^2}\frac{1}{h_K}\,\sinh \delta_K \cosh \delta_K \, \dt t\,;\\
\hbox{where} & \qquad h_K=1+\frac{r_0^2}{r^2}\,\sinh^2 \delta_K\,, \:\:\: (K=1,2,3),\\
& \qquad~~ f = \frac{r^2}{L^2}h_1 h_2 h_3 +1-\frac{r_0^2}{r^2}\,,
\end{split} 
\end{equation}
and $\dt \Omega_3^2$ is the line element of a unit radius $S^3$ and we have chosen the gauge where $r^2 (h_1h_2h_3)^{1/3}$ measures the radius of the $S^3$. Note that we can do a $U(1)$ gauge transformation that takes us to a gauge where at the horizon $\mathcal{H}$ the gauge fields vanish, $A^K |_{\mathcal{H}}=0$. For example, $\tilde{A}^K=A^K-A^K|_{\mathcal{H}}$ with $A^K$  given in \eqref{BCS:ansatz} also describes the BCS black hole. 
 We will work in this latter gauge when presenting the thermodynamic properties of the solution and when we discuss again the BCS black holes in the particular truncations of sections  \ref{sec:SingleQ-BCS} and \ref{sec:TwoQ-BCS}.

It is also useful to note that when $h_1=h_2$ then $\varphi_2=0$ (this is the case of the consistent truncations with a single charge or two equal charges we study in Sections  \ref{sec:SingleQ} and  \ref{sec:TwoQ}, respectively). Moreover, if $h_1=h_2=h_3$ then both neutral scalars vanish,  $\varphi_1=\varphi_2=0$ (this is the case of the theory with three  equal charges studied in \cite{Bhattacharyya:2010yg,Markeviciute:2016ivy,Markeviciute:2018yal}). Because the neutral scalars vanish in this special case, the BCS solutions literally reduce to the RNAdS black holes (and to the Kerr-Newman$-$AdS$_5$ black holes when we include rotation).

It is important to describe the thermodynamic quantities of the BCS black hole in order to study their  competition with the hairy black holes we will find later\footnote{The thermodynamics of BCS black holes with arbitrarily charges and two equal angular momenta was studied in \cite{Cvetic:2005zi}. The phase diagram in the grand canonical ensemble of static BCS black holes with a single charge and three equal charges was discussed in \cite{Henriksson:2019zph}.}.
Defining auxiliary quantities $q_K$ such that $\sinh \delta_K=\frac{\sqrt{2 q_K}}{r_0}$ (the absence of a charged condensate and angular momentum implies that the action has an $A^K \to - A^K$ symmetry which can be used to set $\delta_K \geq 0$ and consequently, $q_K \geq 0$) one can use the condition $f(r_+)=0$, that defines the horizon location ($r=r_+$), to express $r_0$ as a function of $r_+$ and $q_K$ as
 \begin{equation}
 \label{BCS:def-r0}
 \!\!\!
 r_0=\frac{1}{L r_+}\sqrt{r_+^4 \left(L^2+2 q_3+r_+^2\right)+2 q_1 \left(2 q_2+r_+^2\right) \left(2 q_3+r_+^2\right)+2 q_2 r_+^2 \left(2 q_3+r_+^2\right)}\,.
 \end{equation}
The temperature $T$ and the entropy $S$ of the BCS black hole are then:
\begin{eqnarray}\label{BCS:TS}
&& T=\frac{1}{L^2}\frac{r_+^4 \left[L^2+2 (q_1+q_2+q_3)+2 r_+^2\right]-8 q_1 q_2 q_3}{2 \pi  r_+^2 \sqrt{\left(2 q_1+r_+^2\right) \left(2 q_2+r_+^2\right) \left(2 q_3+r_+^2\right)}}\,, \nonumber\\
&& S= \frac{N^2}{L^3}\,\pi \sqrt{\left(2 q_1+r_+^2\right) \left(2 q_2+r_+^2\right) \left(2 q_3+r_+^2\right)}\,.
\end{eqnarray}
An important observation is that when at least one of the electric charges $Q_K\sim  \sinh \delta_K \sim q_K$ vanishes then the BCS does not have an extremal, $T=0$, configuration (since the numerator of $T$ in \eqref{BCS:TS} cannot vanish when the second term $q_1q_2q_3$ is zero).

Using the holographic renormalization method \cite{Bianchi:2001de,Bianchi:2001kw} (which we will describe in detail in Section \ref{sec:ThermoSingleQ} and Appendix \ref{app:holo-renorm})  we find that the energy $E$ (after subtracting the Casimir energy $E_{\hbox{\tiny AdS}_5}= \frac{N^2}{L} \frac{3}{16}$ of the dual $\CN = 4$ SYM on $\mathbb{R}_t \times S^3$) and electric charges $Q_1,Q_2,Q_3$ of the static BCS black hole are given by
\begin{eqnarray}\label{BCS:EQ}
&& E=\frac{N^2}{L^3}\,\frac{r_0^2 }{4} \sum_{K=0}^3\left(s_K^2 + c_K^2 \right),\qquad 
Q_K=  \frac{N^2}{L^3}\,\frac{r_0^2 }{2}  s_K c_K\,,
\nonumber\\
 && \hbox{with} \:\: s_K\equiv  \sinh \delta_K\,, \:\: c_K\equiv \cosh \delta_K \:\:\hbox{and} \:\: K=1,2,3,
\end{eqnarray}
while the chemical potentials $\mu_K$ (that source the operators dual to $A^K$), charge densities $\rho_K $ (\emph{i.e.} the VEVs of the operator dual to $A^K$), and the VEVs of the scalar operators dual to the neutral ($\langle \CO_{\varphi_K} \rangle$) and charged ($\langle \CO_{\Phi_K} \rangle$)  scalar fields are given by:
\begin{subequations}
\begin{eqnarray}\label{BCS:otherVEVs}
 \mu_K&=& \frac{r_0^2 s_K c_K}{ r_0^2 s_K^2+r_+^2} , \qquad  \rho_K = \frac{r^2_0 s_K c_K}{L^4} ; \\
\langle \CO_{\varphi_1} \rangle&=& \frac{r_0^2}{\sqrt{6}L^4}\left( s_1^2+s_2^2-2s_3^2\right),\qquad  \langle \CO_{\varphi_2} \rangle = \frac{r_0^2}{\sqrt{2}L^4}\left( s_1^2-s_2^2 \right);\\
\langle \CO_{\Phi_K} \rangle &=&0\qquad (K=1,2,3)\,.
\end{eqnarray}
\end{subequations}
Note that $\langle \CO_{\Phi_K} \rangle =0$ because the BCS solution has no charge condensate, $\Phi_K=0$. This will not be the case for the hairy black holes that we find in Sections \ref{sec:SingleQ} and \ref{sec:TwoQ} which will have $\Phi_K\neq 0$ and thus scalar hair with expectation value $\langle \CO_{\Phi_K} \rangle \neq 0$ (the sources will be set to zero), at least for one of the $K=1,2,3$.

We can explicitly check that the thermodynamic quantities \eqref{BCS:TS}--\eqref{BCS:otherVEVs}
obey the first law of thermodynamics,
\begin{equation}\label{BCS:FirstLaw}
\dt E= T \dt S + \sum_{K=1}^{3} \mu_K\, \dt Q_K\,.
\end{equation}

From  \eqref{BCS:TS}--\eqref{BCS:otherVEVs} we can also straightforwardly compute the Gibbs free energy which is useful to study the grand-canonical ensemble
\begin{align}\label{BCS:HelmoltzGibbs}
\begin{split}
G&=E - T S -\mu Q\,.
\end{split}
\end{align}

\section{Consistent truncation with $A^1=A^2\equiv 0, A^3\equiv A$}\label{sec:SingleQ}

\subsection{Setup the problem: Ansatz\"e and boundary conditions}\label{sec:AnsatzSingleQ}

We will denote this theory with $A_\1^1=A_\1^2\equiv 0, A_\1^3\equiv A_\1$ and $\Phi_1=\Phi_2\equiv 0, \Phi_3\equiv \Phi$ as the truncation with a single charge of  action \eqref{OurCTaction}.  Motivated by the ansatz  \eqref{BCS:ansatz} we used for the BCS black hole, to find the static and spherically symmetric hairy solutions of this sector we find convenient to use the ansatz: 
\begin{equation} \label{SingleQ:ansatz}
\begin{split}
d s^2&=h^{1/3}\left(-\frac{f}{h}\dt t^2+  \frac{\dt r^2}{g}+ r^2 \dt \Omega_3^2\right);\\
\varphi_1&=-\sqrt{\frac{2}{3}}\ln h\,,\qquad \varphi_2=0;\\
A_\1^1&=0\,,\qquad A_\1^2= 0\,,\qquad  A_\1^3= A_\1 = A_t \dt t\,;\\
\Phi_1&=0\,,\qquad \Phi_2= 0\,,\qquad \Phi_3=\Phi_3^\dagger= \Phi \,;\\
\end{split} 
\end{equation}
where $\dt \Omega_3^2$ is the line element of a unit radius $S^3$ and we have chosen the gauge where $h^{1/3} r^2$ measures the radius of the $S^3$. Moreover, we have fixed the $U(1)$ gauge freedom by taking $\Phi_3= \Phi$ to be real, which implies that the gauge field vanishes on the horizon $r=r_+$ (given by the largest root of $f$), i.e. $A_t|_{r=r_+} \!\!=0$. Note that the neutral scalar is determined by $h$. The full solution is determined in terms of five functions of the radial coordinate, namely $\{h(r),f(r),g(r),A_t(r),\Phi(r)\}$. Plugging this ansatz into the field equations \eqref{GravEOM}--\eqref{ScalarsGaugeEOM} we find that the system closes if the following equations are satisfied:
\begin{equation} \label{SingleQ:EoM}
\begin{split}
0=& L^2 r g h  \left(\Phi ^2+4\right) \left(r h'+6 h\right) f' 
+h^3 r^2 \left(\Phi ^2+4\right) \left(g h L^2 \left(A_t'\right)^2-A_t^2 \Phi ^2\right)
\\
&+f  r^2 \left(\Phi ^2+4\right) \left(-g L^2 \left(h'\right)^2-8 h^3+h \Phi ^2\right) \\
& +f h^2 \left[g L^2 \left(-r^2 \left(\Phi '\right)^2+12 \Phi ^2+48\right)-4 \left(\Phi ^2+4\right) \left(3 L^2+2 r^2 \sqrt{\Phi ^2+4}\right)\right],
\\
0=&L^2 r f h  \left(\Phi ^2+4\right)  \left(r h'+6 h\right)g'
+h^3 r^2 \left(\Phi ^2+4\right) \left[A_t^2 \Phi ^2-g h L^2 \left(A_t'\right)^2\right] \\
& f r \left(\Phi ^2+4\right) \left[h' \left(g L^2 r h'+4 (g-1) h L^2-2 h r^2 \sqrt{\Phi ^2+4}\right)-h r \left(16 h^2+\Phi ^2\right)\right]\\
& +f h^2 \left[g L^2 \left(r^2 \left(\Phi '\right)^2+12 \Phi ^2+48\right)-4 \left(\Phi ^2+4\right) \left(r^3 h'+3 L^2+r^2 \sqrt{\Phi ^2+4}\right)\right],
\\
0=&h''-\frac{\left(h'\right)^2}{h}+\frac{h'}{L^2 r g} \left(L^2(g+2)+2 h r^2+r^2 \sqrt{\Phi ^2+4}\right)
+\frac{4 h^2-2 h \sqrt{\Phi ^2+4}+\Phi ^2}{L^2g}+\frac{h^3 \left(A_t'\right)^2}{f},\\
0=& L^2 r  f g h \left(\Phi ^2+4\right) \left(r h'+6 h\right) A_t'' + g h^4 L^2 r^2 \left(\Phi ^2+4\right) \left(A_t'\right)^3  - f r \Phi ^2 \left(\Phi ^2+4\right) \left(r h'+6 h\right) A_t \\
& +A_t'\bigg\{
f h r \left(\Phi ^2+4\right) \bigg[2 L^2 h'+r \left(r \sqrt{\Phi ^2+4} h'+2 h r h'+4 h^2-2 h \sqrt{\Phi ^2+4}+\Phi ^2\right)\bigg]\\
& + L^2 f g  \bigg[\left(\Phi ^2+4\right) \left(r^2 \left(h'\right)^2+13 h r h'+18 h^2\right)-h^2 r^2 \left(\Phi '\right)^2\bigg]
-A_t^2 h^3 r^2 \Phi ^2 \left(\Phi ^2+4\right)
\bigg\}, \\
0= & L^2 g h \Phi'' -\frac{L^2 g h \Phi  \left(\Phi '\right)^2}{\Phi ^2+4}  
+\frac{h \Phi '}{r} \left(g L^2+2 h r^2+2 L^2+\frac{4 r^2}{\sqrt{\Phi ^2+4}}+\frac{r^2 \Phi ^2}{\sqrt{\Phi ^2+4}}\right)\\
  & +\left(\frac{1}{f}A_t^2 h^2 \left(\Phi ^2+4\right)+4 h \sqrt{\Phi ^2+4}-\Phi ^2-4\right)\Phi . \\
\end{split} 
\end{equation}
This is a system of two first order ODEs for $\{f',g'\}$ plus three second order ODEs for $\{h'',A_t'',\Phi''\}$.

To solve this coupled system of nonlinear ODEs we must impose relevant physical boundary conditions. 
Consider first the asymptotic boundary at $r=\infty$. Since we have two first order plus three second order ODEs we have, \`a priori, $2+6=8$ free UV parameters, some of which will be fixed by boundary conditions. Naturally, we demand that our solutions are asymptotically AdS$_5$ with the normalization for the Killing vector field $\partial_t$ chosen to be $|\partial_t|_{r\to\infty}=-1$. This requires that we impose as boundary condition that $f|_{r\to\infty}=\frac{r^2}{L^2}$ (it then follows from the EoM that $g|_{r\to\infty}=\frac{r^2}{L^2}$). The asymptotic value of the gauge field is  the chemical potential, $A_t |_{r\to\infty}=\mu$, which we leave free. On the other hand, the neutral and charged scalar fields, $\varphi_1$ and $\Phi$, both have mass $m^2L^2=-2$ and thus saturate the BF bound in AdS$_5$. Therefore, $\varphi_1 |_{r\to\infty} \sim s_\varphi\, \frac{L^2 \log r}{r^2} -\sqrt{\frac{2}{3}}h_2 \frac{L^2}{r^2} + \cdots$ and
$\Phi |_{r\to\infty} \sim s_\Phi\, \frac{L^2 \log r}{r^2} +\varepsilon \frac{L^2}{r^2} + \cdots$ where  $ s_{\varphi}$ and $s_\Phi$  are the sources for the operators (both with conformal dimension $\D=2$) dual to $\varphi_1$ and $\Phi$, respectively, and $h_2 \sim \langle \CO_{\varphi_1} \rangle$ and  $\varepsilon = \langle \CO_\phi \rangle$ are their VEVs. We are interested on solutions dual to CFT states that are not sourced, so we set $s_{\varphi}=0$ and $s_\Phi=0$ as  boundary conditions. 
After imposing these boundary conditions, that fix 3 of the 8 UV parameters, a Frobenius analysis off the asymptotic boundary yields the asymptotic expansion,
\begin{align}\label{SingleQ:expansionUV}
\begin{split}
g(r)&\simeq \frac{r^2}{L^2}+\left(1+h_2\right)+f_2 \frac{L^2}{r^2}+\CO\left(L^4r^{-4}\right),\\
f(r)&\simeq \frac{r^2}{L^2}+\left(1+h_2\right)+f_2 \frac{L^2}{r^2}+\CO\left(L^4r^{-4}\right),\quad \\
A_t(r)&\simeq \mu+\rho \frac{L^2}{r^2}+\CO\left(L^4r^{-4}\right),\quad \\
h(r)&\simeq  1+h_2 \frac{L^2}{r^2}+\CO\left(L^4r^{-4}\right)\\
\Phi(r)&\simeq \varepsilon \frac{L^2}{r^2} +\CO\left(L^4r^{-4}\right)\,,
\end{split}
\end{align}
where $\{h_2,f_2,\mu,\rho,\varepsilon\}$ are the 5 free UV parameters that are not fixed by boundary conditions or by the EoM. Essentially, $h_2$ and $\varepsilon$  give the VEVs of $\varphi_1$ and $\Phi$, respectively, $f_2$ is related to the mass of the solution and $\mu$ and $\rho$ are the chemical potential and charge density of the gauge field $A_t$.

In addition to the boundary conditions at the asymptotic boundary, we must also impose boundary conditions in the interior of the spacetime. Consider first the case of black holes with a Killing horizon generated by the Killing vector $\partial_t$ (geometries without horizons -- namely solitons -- are discussed later). In this case, the inner boundary of our integration domain is the horizon which we will take to be located at $r=r_+>0$. Again, \`a priori the number of free IR parameters is given by the order (\emph{i.e.} 8) of the ODE system. Some of these are however fixed by requiring regularity of the solution at the horizon. To find the constraints imposed by this regularity, it is enlightening to note that our coupled ODE system can effectively be rewritten as a system of 4 second order ODEs. And the horizon is a regular singular point with 
degeneracy 2 (\emph{i.e.} the indicial root is 2). Thus, the 4 functions have a pair of independent solutions where one of them is proportional to $\ln(r-r_+)$ and the other is a regular power law of $r-r_+$. Demanding regularity  at the horizon eliminates the logarithm terms and we are left with 4 free IR parameters. Since we have a (non-extremal) horizon at $r=r_+$, $f$ and $g$ vanish linearly at the horizon. Moreover, we work in the $U(1)$ gauge where $\Phi$ is real and $A_t$ vanishes linearly at $r=r_+$. Altogether these conditions define the parameter $r_+$ and impose horizon regularity.

We will solve equations \eqref{SingleQ:EoM} with the above boundary conditions either numerically (at full nonlinear level) or within perturbation theory. The details of the perturbative construction is discussed in Section \ref{sec:PerturbativeSingleQ}. When solving the ODE system numerically, the above boundary conditions can be imposed in practice if we introduce the field redefinitions  
\begin{align}\label{SingleQ:FieldRedef}
\begin{split}
g& =\frac{r^2}{L^2}\left(1-\frac{r_+^2}{r^2}\right)q_1\,,\qquad
f=\frac{r^2}{L^2}\left(1-\frac{r_+^2}{r^2}\right)q_1 q_2\,,\\
A_t &=\left(1-\frac{r_+^2}{r^2}\right)q_3\,, \qquad
h=q_4\,, \qquad
\Phi =2\frac{r_+^2}{r^2} q_5 \,\sqrt{2+\frac{r_+^4}{r^4} q_5^2}\,,
\end{split}
\end{align}
and look for solutions $q_j$, ($j=1,2,\cdots 5$) that are everywhere smooth (not to be confused with the parameters $q_K$ in the BCS solution). Note that the peculiar redefinition of $\Phi$ in terms of $q_5$ was introduced to avoid square root terms of the form $\sqrt{4+\Phi^2}$ in the EoM; see \eqref{SingleQ:EoM}.
For the numerical search of the hairy  solutions it is also convenient to introduce the compact coordinate and adimensional horizon radius,
\begin{equation}\label{SingleQ:compactRadius}
y=1-\frac{r_+^2}{r^2}\,,\qquad y_+=\frac{r_+}{L}\,,
\end{equation}
where $y$ ranges between $y=0$ (\emph{i.e.} $r=r_+$) and $y=1$ (\emph{i.e.} $r\to \infty$). 

We can now specify the boundary conditions for the auxiliary fields $q_j$. Demanding that our solutions are asymptotically AdS$_5$ at $y=1$ requires that $q_{1}(1)=1=q_{2}(1)$. The EoM then require that $q_{4}(1)=1$. We will find useful to construct lines of solutions that have constant electric charge $Q$. Later, in \eqref{SingleQ:EQ}, we will find that $Q$ is a function of $q_{3}(1)$ and $q_{3}'(1)$, $Q=\frac{N^2}{L}\frac{1}{2}y_+^2(q_{3}'+q_{3})|_{y=1}$. To introduce $Q$ in our numerical code as an input parameter (that will allow us to run lines of constant $Q$) we thus use this condition to give a mixed boundary condition for $q_3$. Finally, the EoM require that $q_5$ also satisfies a  mixed boundary condition. Altogether, we impose the boundary conditions at the asymptotically AdS$_5$ boundary ($y=1$):  
\begin{align}\label{SingleQ:BCq}
\begin{split}
& q_1\big|_{y=1}= 1\,, \quad q_2\big|_{y=1} = 1\,, \quad q_3'\big|_{y=1}= - q_3\big|_{y=1}+\frac{2Q}{y_+^2}\frac{L}{N^2}\,, \\  
 & q_4\big|_{y=1} = 1\,, \quad q_5'\big|_{y=1}=\left( \frac{q_3^2}{y_+^2}-q_4'\right)q_5\big|_{y=1} 
\end{split}
\end{align}
On the other hand, at the horizon ($y=0$) the derived boundary conditions from the EoM are that 
$q_1$ must obey the Dirichlet $q_1\big|_{y=0}=1+\frac{1}{y_+^2}+\left(q_4+q_5^2\right)\big|_{y=0}$ and $q_{2,3,4,5}$ must obey mixed boundary conditions which are not enlightening to display.

We now discuss our numerical strategy to find the nonlinear solutions of our boundary-value problem. As mentioned earlier, after imposing the asymptotic and horizon boundary conditions we have 5 free UV parameters and 4 free IR parameters. It follows that our black hole solutions depend on $5-4=1$ parameter plus the dimensionless horizon radius $y_+$, \emph{i.e.} a total of 2 parameters. We can take these parameters to be e.g. the dimensionless electric charge $Q L/N^2$ and the dimensionless radius $y_+=r_+/L$ (the latter is related to the temperature and entropy of the solutions; see \eqref{SingleQ:TS}). 

We solve our boundary-value problem using a Newton-Raphson algorithm. For the numerical grid discretization we use a pseudospectral collocation with a Chebyshev-Lobatto grid and the Newton-Raphson linear equations are solved by LU decomposition. These methods are reviewed and explained in detail in the review \cite{Dias:2015nua} and used in a similar context e.g.~\cite{Dias:2015pda,Dias:2016eto,Dias:2017uyv,Dias:2017opt,Bena:2018vtu,Bea:2020ees}. Our solutions have analytical polynomial expansions at all the boundaries of the integration domain and thus the pseudospectral collocation guarantees that the numerical results have exponential convergence with the number of grid points. We further use the first law to check our numerics. In the worst cases, our solutions satisfy these relations with an error that is smaller than $0.1\%$. As a final check of our  full nonlinear numerical results, we compare them against the perturbative expansion results of Section \ref{sec:PerturbativeSingleQ}. 

As usual, to initiate the Newton-Raphson algorithm one needs an educated seed. The hairy black holes merge with the BCS black holes when the condensate $\Phi(q_5)$ vanishes (see later Section~\ref{sec:OnsetSingleQ}). Therefore, it is natural to expect that the BCS solution with a  small $q_5$ perturbation can be used as seed for the solution near the merger. Actually, in  Section \ref{sec:PerturbativeSingleQ}, we will find the hairy black hole solution in perturbation solution for small values of the charged condensate $q_5(1)$ and of the horizon radius $y_+$. This is an even better seed for the Newton-Raphson code. To scan the 2-dimensional parameter space of hairy black holes we can fix $Q$ and run the numerical code for several values of $y_+$. Alternatively, we can generate lines of constant $y_+$ parametrized by $Q$. In practice, we will mainly use the former strategy since this will allow us to densely fill the phase space along constant-$Q$ families of solutions that span between the two boundaries of the 2-dimensional triangular shaped $Q$-$M$ phase space where hairy black holes exist. Indeed, constant-$Q$ families depart from the merger line and end up at zero entropy $E=Q$ solutions with finite dimensionless temperature $T L$. Once we have the numerical solutions $q_j(y)$, the thermodynamic quantities are read straightforwardly from the expressions \eqref{SingleQ:FGexpansion2b}--\eqref{SingleQ:holoCurrent} and \eqref{SingleQ:EQ}--\eqref{BCS:HelmoltzGibbs} that we will find in the next subsection.

\subsection{Thermodynamic quantities using holographic renormalization}\label{sec:ThermoSingleQ} 

To determine the thermodynamic quantities of the solutions we implement the holographic renormalization procedure. Our solutions are asymptotically $\ads_5$ solutions with scalar fields $\varphi_1$ and $\Phi$ that both have mass $m^2 L^2=-4$, \emph{i.e.} they saturate the BF bound in AdS$_5$ and thus have conformal dimension $\Delta=2$. In these conditions, the holographic renormalization procedure to find the holographic stress tensor $\CT_{\mu\nu}$, holographic current $\CJ_\mu$ (we use Greek indices $\mu,\nu,\ldots$ to denote the boundary coordinate indices), and expectation values $\langle {\cal O}_{\varphi} \rangle$ and $\langle {\cal O}_{\Phi} \rangle$ of the operators dual to the scalar fields $\varphi_1$ and $\Phi$  was developed in \cite{Bianchi:2001de,Bianchi:2001kw}. The details for this procedure applied to the $U(1)^3$ gauged supergravity with \emph{all} sources turned on is presented in Appendix \ref{app:holo-renorm}. In this section, we simplify the results of that section to the single charge truncation.

We start by introducing the Fefferman-Graham (FG) radial coordinate $z$ that is such that the asymptotic boundary is at $z=0$ and $g_{zz}=L^2/z^2$ and $g_{z\mu}=0$ (with $\mu=t,\xi_j)$ at all orders in a Taylor expansion about $z=0$ ($\xi_j$ being the 3 coordinates that parametrize the $S^3$). It follows that the radial coordinate $y$ is given as a function of the FG coordinate $z$ as
 \begin{align}
 \label{SingleQ:FGcoord}
y=& 1-y_+^2\frac{z^2}{L^2} - \frac{y_+^2}{6}\Big( 3-2y_+^2 q_4'(1)\Big)\frac{z^4}{L^4}  
- \frac{y_+^2}{144} \bigg[  27-12 y_+^2 \Big(3+2 q_4'(1) \Big) \nonumber \\
&+ 4y_+^4\left(  
\frac{9 q_1''(1)}{2}+2 q_4'(1)^2+9 q_4'(1)+3 q_5(1)^2-9
\right) \bigg]\frac{z^6}{L^6} 
+\CO(z^8/L^8).
\end{align}    
The expansion of the gravitational around the boundary up to the order that contributes to the thermodynamic quantities is then
\begin{eqnarray}\label{SingleQ:FGexpansion1}
&& \dt s^2 = \frac{L^2}{z^2}\left[ \dt z^2 + \dt s^2_{\partial}+z^2 \,\dt s^2_\2+z^4\, \dt s^2_\4+\CO(z^6)\right],  \\[2mm]
&& \hbox{with}\nonumber\\[2mm]
&& \dt s^2_{\partial}=g_{\mu\nu}^\0\dt x^\mu \dt x^\nu = -\dt t^2+L^2 \dt \Omega_3^2,  \nonumber\\[2mm]
&& \dt s^2_\2=g_{\mu\nu}^\2\dt x^\mu \dt x^\nu =-\frac{1}{2 L^2}\!( \dt t^2+L^2 \dt \Omega_3^2 )\,, \nonumber\\[2mm]
&&  \dt s^2_\4= g_{\mu\nu}^\4\dt x^\mu \dt x^\nu \nonumber\\
&& \hspace{0.2cm} =- \frac{1}{144 L^4}
\bigg[ 9-36 y_+^2 \Big(3-2 q_4'(1)\Big)+4 y_+^4 \left(\frac{27}{2}\,q_1''(1)-2 q_4'(1)^2+27 q_4'(1)+21 q_5(1)^2-27\right) \!\! \bigg]\dt t^2\nonumber\\
&& \hspace{0.2cm}+ \frac{1}{144 L^2} \bigg[9+12 y_+^2 \Big(3-2 q_4'(1)\Big)  - 4 y_+^4 \left(\frac{9}{2}\,q_1''(1)+2 q_4'(1)^2+9 q_4'(1)+15 q_5(1)^2-9\right) \bigg] \dt \Omega_3^2 \nonumber
\,;  
\end{eqnarray}
and the relevant expansion of the gauge and scalar fields around the boundary is
\begin{eqnarray}\label{SingleQ:FGexpansion2} 
A_t &=& a_\0 +a_\2  \,z^2+ \CO(z^4) \,, \qquad \mu=q_3(1), \quad
a_\2 =-\frac{y_+^2}{L^2}\Big(q_3'(1)+q_3(1)\Big);
\\[2mm] 
\varphi_1 &=&  \tilde{\varphi}_\0 \,z^2+ \CO(z^4),\qquad  \tilde{\varphi}_\0= \sqrt{\frac{2}{3}} \,\frac{y_+^2}{L^2}\,q_4'(1) \,; \label{SingleQ:FGexpansion2b} 
\\[2mm] 
\Phi &=&  \tilde{\Phi}_\0 \,z^2+ \CO(z^4),\qquad  \tilde{\Phi}_\0= 2 \sqrt{2}\,\frac{y_+^2}{L^2}\,q_5(1) \,.\label{SingleQ:FGexpansion2c} 
\end{eqnarray}
The remaining holographic quantities can now be computed using the holographic renormalization procedure of Bianchi-Freedman-Skenderis \cite{Bianchi:2001de,Bianchi:2001kw} as done in Appendix~\ref{app:holo-renorm} 
(In particular, see \eqref{app:HoloStressTensor}--\eqref{app:TraceHoloStressTensor} and \eqref{app:VEVcurrent}--\eqref{app:VEVscalars}).\footnote{\label{footHoloRen}Note however that we use different conventions for the Riemann curvature, that is to say, with respect to \cite{Bianchi:2001de,Bianchi:2001kw} our action \eqref{OurCTaction} has the opposite relative sign between the Ricci scalar ${\cal R}$ and the scalar fields' kinetic terms  $\left( \nabla \varphi \right)^2$ and $\left( D \Phi \right)^2$. Further note that our relative  normalization factor in the action \eqref{OurCTaction} between the Ricci scalar and the scalar kinetic terms differs from  \cite{Bianchi:2001de,Bianchi:2001kw}: compare e.g. our action \eqref{OurCTaction} with (2.1) of \cite{Bianchi:2001de}.}
In short, one needs to compute the renormalized on-shell action $S_{\rm ren}$ (with all the source terms included) and then, the expectation value of an operator dual to a particular bulk field can be obtained taking the variation of $S_{\rm ren}$  w.r.t. the source while setting the sources to their Dirichlet value in the end. 

Using \eqref{SingleQ:FGexpansion2b}--\eqref{SingleQ:FGexpansion2c}, it is a simple exercise to compute the expectation values $ \langle {\cal O}_{\varphi} \rangle$ and $\langle {\cal O}_{\Phi} \rangle$  for the operators $ {\cal O}_{\varphi}$ and  ${\cal O}_{\Phi}$  dual to the neutral scalar $\varphi_1$ and charged scalar $\Phi$
\begin{equation}\label{SingleQ:vevs}
\langle {\cal O}_{\varphi} \rangle=\frac{N^2}{\pi^2}\tilde{\varphi}_\0\qquad\text{and}\qquad \langle {\cal O}_{\Phi} \rangle=\frac{N^2}{\pi^2}\tilde{\Phi}_\0\,,
\end{equation}
respectively. Recall that these operators have conformal dimension $\Delta=2$ and in our solutions we have killed the sources of these operators; see discussion above \eqref{SingleQ:expansionUV}. On the other hand, the source $\mu$ (\emph{i.e.} the chemical potential that is given by the boundary value of $A_t$)\footnote{The chemical potential associated to a gauge field $A_a$ of a black hole is given by $\mu=A_a k^a |_{\infty} - A_a k^a |_{\mathcal{H}}$ where $k$ is the Killing horizon generator, \emph{i.e.} $|k|^2|_{\mathcal{H}}=0$.} and the charge density $\rho = - \frac{1}{\sqrt{-g_\0}}\frac{\delta S_{\rm ren}}{\delta a_\0} = - \frac{L}{8\pi G_5}\,a_\2$ of the dual operator to $A_t$ are:
\begin{equation}\label{SingleQ:holoCurrent}
\mu= q_3(1)\,,\qquad \rho=\frac{N^2}{4 \pi ^2 L^4} y_+^2 \Big(q_3'(1)+q_3(1)\Big),
\end{equation}
where we use \eqref{G5} to write $G_5$ in terms of the SYM quantities. Finally, the expectation value of the holographic stress tensor is given by:
\begin{align}\label{SingleQ:holoT}
\begin{split}
 \langle \CT_{\mu\nu} \rangle
=& \frac{N^2}{2 \pi ^2} \bigg[ g_{\mu\nu}^\4
-\frac{1}{8}g_{\mu\nu}^\0\left( \tr{ g^\2}^2 - \tr { ( g^\2 )^2 }  \right)
-\frac{1}{2} \left(g^\2\right)_{\mu\nu}^2   \\
& \qquad \qquad \qquad \qquad +\frac{1}{4} g_{\mu\nu}^\2 \tr{ g^\2 }  + \frac{1}{12} g_{\mu\nu}^\0 \tilde{\varphi}_\0^2 + \frac{1}{48} g_{\mu\nu}^\0 \tilde{\Phi}_\0^2 \bigg]\,,
\end{split}
\end{align}
where the metric components $g_{\mu\nu}^\0$,  $g_{\mu\nu}^\2$, $g_{\mu\nu}^\4$ and $\langle {\cal O}_{\varphi} \rangle$ can be read directly from \eqref{SingleQ:FGexpansion1}--\eqref{SingleQ:FGexpansion2} and, for the consistent truncation of  \eqref{OurCTaction} analysed in this section,  one has $\langle {\cal O}_{\Phi_1} \rangle=\langle {\cal O}_{\Phi_2} \rangle=0$ and $\langle {\cal O}_{\Phi_3} \rangle\equiv \langle {\cal O}_{\Phi} \rangle$ is given by \eqref{SingleQ:FGexpansion2c}.

The trace of the  expectation value yields the expected Ward identity associated to the conformal anomaly\footnote{The trace anomaly in a theory with the field content of \eqref{OurCTaction} has three possible sources of anomaly \cite{Bianchi:2001de,Bianchi:2001kw} (see Appendix \ref{app:holo-renorm} for details). The first comes from terms of the type $\phi_\0 \langle {\cal O}_{\phi} \rangle$ for a scalar field $\phi$ with source $\phi_\0$ and VEV  $\langle {\cal O}_{\phi} \rangle$. In our case such terms vanish because we set the sources of $\varphi_1$ and $\Phi$ to zero. For the same reason the holographic scalar field anomalies $\mathcal{A}_{\varphi}$ and $\mathcal{A}_{\Phi}$ also vanish in our system. A second source of the anomaly is the gauge field $\CA_{\text{gauge}}$ which has the form $(F^\0)^2$. In our case, the gauge field on the boundary has a vanishing field strength (it is pure gauge) so this contribution vanishes. We are left with the gravitational conformal anomaly $\mathcal{A}_{\rm grav}$, given by the right-hand-side of \eqref{SingleQ:traceAnomaly} that is responsible for the fact that the trace of the holographic stress tensor is non-vanishing.}
\begin{equation} \label{SingleQ:traceAnomaly}
\langle \CT^\mu{}_\mu \rangle=\frac{N^2}{2 \pi ^2 L^3} \frac{1}{16} \left(R^{\0\mu\nu} R_{\mu\nu}^\0 -\frac{1}{3} (R^\0)^2 \right),
\end{equation}
where $R^{\0}_{\mu\nu}$ and $R^\0$ are the Ricci tensor and scalar, respectively, of the boundary metric $g_{\mu\nu}^\0$ as defined in \eqref{SingleQ:FGexpansion1}. Note that this gravitational conformal anomaly is a consequence of the fact that the conformal boundary is the Einstein Static Universe $\mathbb{R}_t\times S^3$, \emph{i.e.} that our solutions are asymptotically global AdS$_5$ (not planar AdS$_5$). Furthermore, we can confirm that the expectation value of the holographic stress tensor is conserved, \emph{i.e.}\footnote{The conservation of the holographic stress tensor is spoiled by the the scalar and gauge field sources as shown in Appendix \ref{app:holo-renorm}. But in our case such terms vanish because we set the sources of $\varphi$ and $\Phi$ to zero.} 
\begin{equation}
(\nabla^\0)^\mu \avg{\CT_{\mu\nu} } =0 \,.
\end{equation}
From \eqref{SingleQ:holoT}  we can read the energy of our solutions. This is done by pulling-back $\langle \CT_{\mu\nu} \rangle$ to a 4-dimensional spatial hypersurface  $\Sigma_t$, with unit normal $n$ and induced metric $\sigma^{\mu\nu}=g^{\0\mu\nu} +n^\mu n^\nu$, and contracting it with the Killing vector $\xi=\partial_t$ that generates time translations. More concretely, the integral  $M=-\int_{\Sigma_t}\sqrt{\sigma} \avg{\CT_{\mu\nu} }  \xi^\mu n^\nu$ gives the desired energy.
This energy contains a contribution from the AdS$_5$ background, $E_{\hbox{\tiny AdS}_5}= \frac{N^2}{L} \frac{3}{16}$, which is the well known Casimir energy of the dual AdS$_5$ $\mathcal{N} = 4$ SYM on $\mathbb{R}_t \times S^3$. We define our final energy with this Casimir energy removed, $E=M-E_{\hbox{\tiny AdS}_5}$, in which case the BPS condition for the system reads $E=Q$.
On the other hand, from \eqref{SingleQ:holoCurrent}  we can compute the electric charge of our solutions.\footnote{Of course we can also compute the electric charge using the standard ADM formula associated to the gauge field equation \eqref{ScalarsGaugeEOM}, namely $Q_K=\frac{1}{16 \pi G_5}\int_{\Sigma_t} \frac{1}{X_K^2} \star F_\2^K$. Note that the Chern-Simmons terms vanish for the static solutions of our theory.} We do the pulling-back of the holographic current $\avg{ \CJ_\mu } = (\rho,0,\cdots,0)$ to the aforementioned 4-dimensional spatial hypersurface  $\Sigma_t$ with normal $n$ and integrate it, $Q=\int_{\Sigma_t}\sqrt{\sigma} n^\mu \avg{ \CJ_\mu } $.
So, for our solutions \eqref{SingleQ:ansatz} with the field redefinitions \eqref{SingleQ:FieldRedef}, the energy and electric charge  are given by
\begin{eqnarray}\label{SingleQ:EQ}
E&=&\frac{N^2}{L}\,\frac{y_+^2 }{4}\left[ 
 3-2q_4'(1)  +3 y_+^2 \left( 1-\frac{1}{2}\,q_1''(1)-q_4'(1)-q_5(1)^2 \right) \right],
\nonumber\\
Q&=&  \frac{N^2}{L}\,\frac{y_+^2}{2} \Big(q_3'(1)+q_3(1)\Big).
\end{eqnarray}
The temperature $T$ and the entropy $S$ of the hairy black holes can be read simply from the surface gravity and the horizon area of the solutions  \eqref{SingleQ:ansatz}, respectively:
\begin{eqnarray}\label{SingleQ:TS}
&& T=\frac{1}{L} \frac{y_+}{2\pi} \frac{q_1(0) q_2(0)} {\sqrt{q_4(0)}}\,, \qquad S= N^2 \pi y_+^3\sqrt{q_4(0)}\,.
\end{eqnarray}

These thermodynamic quantities must obey the first law of thermodynamics \eqref{BCS:FirstLaw}
where, for the theory of this subsection, one has $Q_1=Q_2=0$, $\mu_1=\mu_2=0$ and $Q_3\equiv Q$, $\mu_3 \equiv \mu$, and thus  \eqref{BCS:FirstLaw} reads simply $\dt E= T \dt S +\mu \dt Q$.

From \eqref{SingleQ:EQ}--\eqref{SingleQ:TS} and \eqref{BCS:HelmoltzGibbs} we can also straightforwardly compute the Gibbs free energy $G=E - T S -\mu Q$ which is useful to study the grand-canonical ensemble.

\subsection{Hairy supersymmetric solitons}\label{sec:ansatzSingleQ-Solitons}

So far we have discussed the setup of the boundary-value problem for black hole solutions (i.e. solutions with horizons). However, the single charge sector of \eqref{OurCTaction} also has supersymmetric solitonic solutions. Some of these have connections to hairy black holes in the limit where the temperature of the latter reaches its minimum value. For this reason, it is important to find these solutions before proceeding with further discussions of hairy black holes.
 
These solitons are still described by the field ansatz\"e \eqref{SingleQ:ansatz} and thus satisfy the field equations \eqref{SingleQ:EoM}. However, because they are supersymmetric, instead of solving  \eqref{SingleQ:EoM} one can solve directly the Killing spinor equations, which are first order ODEs. At the end of the day we find that any supersymmetric solution of the consistent truncation of \eqref{OurCTaction} with a single charge  can be described by the ansatz \eqref{SingleQ:ansatz} with the  fields $\{f,g,A_t, \varphi_1,\Phi\}$ given as a function of $h$ as
\begin{equation}\label{SingleQ:SUSYfunctions}
\begin{split}
g&=f=1+\frac{r^2}{L^2}\,h \, , \\
A_t &= \frac{1}{h}\,, \\
\varphi_1 &= -\sqrt{\frac{2}{3}}\ln h\,, \\
\Phi &=\sqrt{ (2h+ rh' )^2-4} \,,
\end{split}
\end{equation}
where $h(r)$ satisfies a second order ODE,
 \begin{equation}\label{SingleQ:solitonODE}
L^2 r\left(1+ \frac{r^2}{L^2}\,h\right)h''  +r^3 \left(h'\right)^2+\left(7 h r^2+3 L^2\right)h'  -4 r \left(1-h^2\right)
=0.
\end{equation}
Doing a Frobenius analysis of this ODE at the asymptotic boundary we conclude that we must have $h|_{r\rightarrow\infty}=1$ which also ensures that the fields $f$ and $g$ are asymptotically AdS$_5$ and $\varphi_1$ and $\Phi$ are normalizable (i.e. the scalar field's sources are zero); see also discussion of \eqref{SingleQ:expansionUV}.  
Next consider the behaviour of the soliton solutions of \eqref{SingleQ:solitonODE} at the origin, $r=0$.
Assuming that at $r=0$ these solutions behave as
\begin{equation}\label{SingleQ:solitonOrigin}
h\big|_{r\to 0} = \frac{h_\a}{r^\a}\,, 
\end{equation}
for arbitrary constants $\a$ and $h_\a$, a Frobenius analysis of \eqref{SingleQ:solitonODE} yields two possible solutions for the exponent: $\a=0$ and $\a=2$. So, we have two distinct families of supersymmetric solitons. The family with $\a=0$ is clearly regular at the origin while the family with $\a=2$ is irregular at $r=0$. Both solutions can be found analytically solving \eqref{SingleQ:solitonODE}. 

For $\a=0$, the regular soliton is a $1$-parameter family of solutions (parametrized by $q$) described by \eqref{SingleQ:ansatz}  with
\begin{align}\label{SingleQ:RegularSoliton}
\begin{split}
h&= \sqrt{1+2 (1+2 q)\frac{L^2}{r^2}+\frac{L^4}{r^4}}-\frac{L^2}{r^2}\,, \\
\end{split}
\end{align}
where $Q=\frac{N^2}{L} q$ is the electric charge of the solution and, as required by the BPS condition, its energy is $E=Q$ (after subtracting the Casimir contribution). Both are computed from the holographic stress tensor $\avg{ \CT_{\mu\nu} }$ using holographic renormalization \cite{Bianchi:2001de,Bianchi:2001kw} as described for the hairy black hole in Section \ref{sec:ThermoSingleQ}. The chemical potential $\mu$, charge density $\rho$, and expectation values of the operators dual to $\varphi_1$ and $\Phi$ are similarly determined by
\begin{equation}\label{SingleQ:RegularSolitonHoloQuantities}
\begin{split}
&\mu =1\,, \qquad \rho = \frac{N^2}{L^4}\frac{q}{2 \pi ^2} \,; \\
&\avg{ \CO_\varphi } = -\frac{2}{L^2} \sqrt{\frac{2}{3}} q\,\frac{N^2}{\pi^2} \,; \\
&\avg{ \CO_\Phi  } = \frac{1}{L^2}\,4 \sqrt{q (q+1)}\frac{N^2}{\pi^2} \,.
\end{split}
\end{equation}
Note in particular that $\mu=1$ as expected for a supersymmetric solution. It is also easy to verify that the soliton satisfies the first law $\dt E=\mu \dt Q=\dt Q$ as it should.

For $\a=2$, the singular soliton is a $2$-parameter family of solutions (parameterized by $C_1$ and $C_2$) as
\begin{equation}\label{SingleQ:SingularSoliton}
h= \sqrt{1+C_2 \frac{L^2}{r^2}+(1+C_1)\frac{L^4}{r^4}}-\frac{L^2}{r^2}\,.
\end{equation}
Doing holographic renormalization one finds that
\begin{align}\label{SingleQ:SingularSolitonHoloQuantities}
\begin{split}
&E=Q=\frac{N^2}{L}\,\frac{1}{4}\left(C_2-2\right)\,;\\
& \mu=1\,, \qquad \rho=\frac{N^2}{L^4}\frac{C_2-2}{8 \pi ^2}\,;\\
&\avg{ \CO_{\varphi} }  = -\frac{1}{L^2} \frac{C_2-2}{\sqrt{6}}\frac{N^2}{\pi^2}\,;\\
&\avg{ \CO_{\Phi} } = \frac{1}{L^2}\, \sqrt{C_2^2-4(C_1+1)}\frac{N^2}{\pi^2}\,.
\end{split}
\end{align}
Again, as expected for a supersymmetric soliton one has $E=Q$ and $\mu=1$. The regular soliton is obtained from this solution by setting $C_1=0$ and $C_2 = 2(1+2q)$.

It is enlightening to compare our soliton spectrum of the consistent truncation of \eqref{OurCTaction} with a single charge ($Q_1=Q_2=0, Q_3=Q$) with the one for the consistent truncation of \eqref{OurCTaction} with three equal charges, $Q_1=Q_2=Q_3\equiv Q$ that was studied in \cite{Bhattacharyya:2010yg,Markeviciute:2016ivy}. In the latter case, there are four (not two)
families of solitons.  We still have the regular 1-parameter soliton with $\a=0$ and a singular 2-parameter soliton with $\a=2$. But, unlike the current single charge truncation, the regular soliton of \cite{Bhattacharyya:2010yg,Markeviciute:2016ivy} has a Chandrasekhar limit. That is, it exists from $E=Q=0$ all the way up to a critical $E=Q=Q_c$ where the central density $h_{\a=0}\to \infty$, with $h_\a$ defined in \eqref{SingleQ:solitonOrigin}. There is then a third singular soliton family with $\a=1$ that departs from $E=Q=Q_c$ where $h_{\a=1}\to 0$ and extends all the way to $E=Q\to \infty$. Finally, precisely at $E=Q=Q_c$ (and only at this point) there is a fourth soliton with $\a=2/3$ and  $h_{\a=2/3}=1$. So this is a singular 0-parameter solution that exists at the merger of the regular 1-parameter soliton family with $\a=0$ with the singular 1-parameter soliton with $\a=1$. Returning back to the single charge case, we can view the absence of the singular solitons with $\a=1$ and $\a=2/3$ as due to the fact that the regular 1-parameter soliton with $\a=0$ now extends from $E=Q=0$ to arbitrary large $E=Q\to \infty$ without a Chandrasekhar limit. 

Although we have not attempted to prove this, we believe that whenever the three charges $Q_1, Q_2,Q_3$ are non-zero, the solitonic spectrum of the system should be similar to the one of the case $Q_1= Q_2=Q_3$ just reviewed. That is, we should have a total of 4 soliton families with physical properties and relations between them in the phase diagram that are similar to the ones described above for $Q_1= Q_2=Q_3$. On the other hand, when at least one of the electric charges is zero, there should exist only 2 families of supersymmetric solitons: a regular 1-parameter soliton without Chandrasekhar limit, and a singular 2-parameter soliton. This is certainly true in the case $Q_1=Q_2=0$ discussed in this section. It will also be the case of the theory with $Q_3=0$ that we will describe in Section \ref{sec:ansatzTwoQ-Solitons}. 

\subsection{Behrndt-Cveti\v c-Sabra black holes}\label{sec:SingleQ-BCS}

In Section \ref{sec:BCS-BHs} we have presented the most general static BCS black hole with three different charges $Q_K$ $(K=1,2,3)$. When $Q_1=Q_2=0$ and $Q_3\equiv Q$, this is a solution of the consistent truncation \eqref{OurCTaction} with $A^1=A^2\equiv 0, A^3\equiv A$ and no charged condensate, $\Phi_K=0$ $(K=1,2,3)$. For our physical discussions of the hairy black holes of the theory with $\Phi_1=\Phi_2\equiv 0, \Phi_3\equiv \Phi$ it will be enlightening to rewrite the BCS black hole  \eqref{BCS:ansatz} for the particular case with $A^1=A^2\equiv 0, A^3\equiv A$  (and thus $\delta_1=\delta_2=0, \delta_3\neq 0 \Rightarrow h_1=h_2=1$ and  $\varphi_2=0$) in the form of the ansatz \eqref{SingleQ:ansatz} with the field redefinitions \eqref{SingleQ:compactRadius} and compact radial coordinate \eqref{SingleQ:FieldRedef}. This is because the hairy black hole family merges with the BCS at the onset of the scalar condensation instability (Section~\ref{sec:OnsetSingleQ}).

Without further delay, set 
\begin{equation}\label{BCS:singleQ-def-q}
\delta_1=\delta_2=0, \qquad \sinh \delta_3\equiv \frac{2 q}{r_0}
\end{equation}
in  \eqref{BCS:ansatz}, where from \eqref{BCS:def-r0} $r_0$ can be written as a function of the horizon radius $r_+$ and the charge parameter $q$ as $r_0=\frac{r_+}{L}\sqrt{r_+^2+L^2+2 q}$. Further introduce the dimensionless quantities $y_+=r_+/L$ and $\tilde{q}=q /L^2$ and choose a $U(1)$ gauge such that the gauge field vanishes at the horizon $\mathcal{H}$. Then one finds that the BCS black hole with a single charge sourced by $A^3$  is described $-$ via \eqref{SingleQ:ansatz}, \eqref{SingleQ:compactRadius}  and \eqref{SingleQ:FieldRedef} $-$  by the functions
\begin{align}\label{BCS:SingleQ}
\begin{split}
& q_1= 1+(1-y)\frac{y_+^2+2 \tilde{q}+1}{y_+^2}\,, \\
&q_2 = 1\,, \\
&q_3= \frac{\sqrt{2} \sqrt{\tilde{q}} \, y_+^2 \sqrt{y_+^2+1}}{ \left[2 \tilde{q} (1-y)+y_+^2\right]\sqrt{y_+^2+2 \tilde{q}}}\,, \\  
&q_4 =1+ (1-y)\frac{2 \tilde{q}}{y_+^2}\,, \qquad q_5=0\,.
\end{split}
\end{align}
With \eqref{BCS:SingleQ}, the thermodynamic quantities for the BCS black hole with a single charge can now be read directly from the expressions \eqref{SingleQ:FGexpansion2b}--\eqref{SingleQ:holoCurrent} and \eqref{SingleQ:EQ}--\eqref{BCS:HelmoltzGibbs} that were found in the previous subsection. We find that the energy (after subtracting the Casimir energy), electric charges, chemical potentials and expectation values of the scalar fields are given by:
\begin{eqnarray}\label{BCSsingleQ:EQ}
E&=&\frac{N^2}{L}\,\frac{1}{4}\Big[3y_+^2(1+y_+^2)+ 2 \tilde{q} \left(2+3y_+^2\right) \Big],
\nonumber\\
Q_1&=&Q_2=0\,,\qquad Q_3= \frac{N^2}{L}\,\frac{\sqrt{\tilde{q}}}{\sqrt{2}}\, \sqrt{y_+^2+1} \,\sqrt{y_+^2+2 \tilde{q}}\,;\\
 \mu_1&=&\mu_2=0\,, \qquad \mu_3=\frac{\sqrt{2} \sqrt{ \tilde{q} } \sqrt{y_+^2+1}}{\sqrt{y_+^2+2  \tilde{q}}}\,;\\
\langle \CO_{\varphi_1} \rangle&=& -\frac{2}{L^2}\sqrt{\frac{2}{3}} \, \tilde{q}\frac{N^2}{\pi^2}\,;\qquad  \langle \CO_{\varphi_2} \rangle =0\,;\qquad \langle \CO_{\Phi_K} \rangle =0\:\:\:(K=1,2,3)\,;
\end{eqnarray}
and the temperature $T$ and the entropy $S$ of the single charged BCS black hole are:
\begin{eqnarray}\label{BCSsingleQ:TS}
&& T=\frac{1}{L} \,\frac{2 y_+^2+2 \tilde{q}+1}{2 \pi  \sqrt{y_+^2+2 \tilde{q}}}\,, \nonumber\\
&& S= N^2\,\pi  y_+^2 \sqrt{y_+^2+2 \tilde{q}}\,.
\end{eqnarray}
As it could not be otherwise, these quantities agree with \eqref{BCS:EQ}--\eqref{BCS:TS} in the appropriate limit.
An important conclusion that follows from \eqref{BCSsingleQ:TS} is that there is no extremal configuration (\emph{i.e.} with $T\to 0$) in the single charged BCS family.

\subsection{The scalar condensation instability of Behrndt-Cveti\v c-Sabra black holes}\label{sec:GrowthSingleQ}
Having introduced all the necessary machinery to compute thermodynamic quantities, we now turn our attention to the issue of dynamical stability of the single charged BCS with respect to charged scalar field condensation. This is a question that can be addressed by analysing linear perturbations of the sixth equation in (\ref{ScalarsGaugeEOM}) about the single charged BCS background. The resulting linear equation takes the following form:
\begin{equation}
D_a D^a \Phi+4\,e^{\frac{\varphi_1}{\sqrt{6}}}\left(2-e^{3\frac{\varphi_1}{\sqrt{6}}}\right)\Phi=0\,.
\label{eq:linear_single}
\end{equation}

We expect the dominant instability to be in the s-wave channel, so we take $\Phi$ to be spherically symmetric. Since the single charged BCS is static, we can further expand perturbations into Fourier modes of the form
\begin{equation}
\Phi(t,r)=\widehat{\Phi}_{\omega}(r)\,e^{-i\omega t}\,,
\label{eq:scalarpartialwave}
\end{equation}
which introduces the frequency $\omega$ of the modes.
We now solve for the eigenpair $\{\widehat{\Phi}_{\omega},\omega\}$ subject to appropriate boundary conditions that we discuss next.

At the horizon, we demand regularity in ingoing Eddington-Finkelstein coordinates $(v,r)$,
\begin{equation}
\dt v=\dt t+\frac{\dt r}{f(r)}\,,
\end{equation}
which in turn imposes
\begin{equation}
\widehat{\Phi}_{\omega}(r)\approx \left(1-\frac{r_+}{r}\right)^{-\i \frac{h(r_+)^{1/2}}{f^\prime(r_+)}\omega}\left[C^+_0+C^+_1\left(1-\frac{r_+}{r}\right)+\ldots\right]\,,
\end{equation}
near $r=r_+$ (where $C^+_0$ and $C^+_1$ are constants). At the conformal boundary, we choose standard quantisation for the scalar field $\Phi$, which fixes the asymptotic behaviour of $\widehat{\Phi}_{\omega}$ to be
\begin{equation}
\widehat{\Phi}_\omega = \left(\frac{r_+}{r}\right)^2\left[C_0^-+C_1^-\left(\frac{r_+}{r}\right)^2+\ldots\right]
\end{equation}
as $r\to+\infty$ (where $C^-_0$ and $C^-_1$ are constants).

To solve for $\widehat{\Phi}_\omega$ we change to a variable $q_{\omega}$ which is regular everywhere (and thus also at $r=r_+$ and near the conformal boundary),
\begin{equation}
\widehat{\Phi}_\omega(r)=\left(1-\frac{r_+}{r}\right)^{-\i \frac{h(r_+)^{1/2}}{f^\prime(r_+)}\omega}\left(\frac{r_+}{r}\right)^2q_{\omega}(r)\,,
\label{eq:change}
\end{equation}
and introduce a compact coordinate
\begin{equation}
r=\frac{r_+}{\sqrt{1-y}}\,
\label{eq:compact_linear}
\end{equation}
so that the conformal boundary is located at $y=1$ and the black hole event horizon at $y=0$. The boundary conditions for $q_{\omega}(y)$ that follow from imposing standard boundary conditions at the conformal boundary and smoothness across the future event horizon are of the Robin type. They can be read (as derived boundary conditions) from the equation for $q_{\omega}(y)$ assuming that $q_{\omega}(y)$ admits a regular Taylor expansion at $y=0$ and $y=1$. These turn out to be too lengthy to present here.

\begin{figure}[th]
\centering \includegraphics[width=\textwidth]{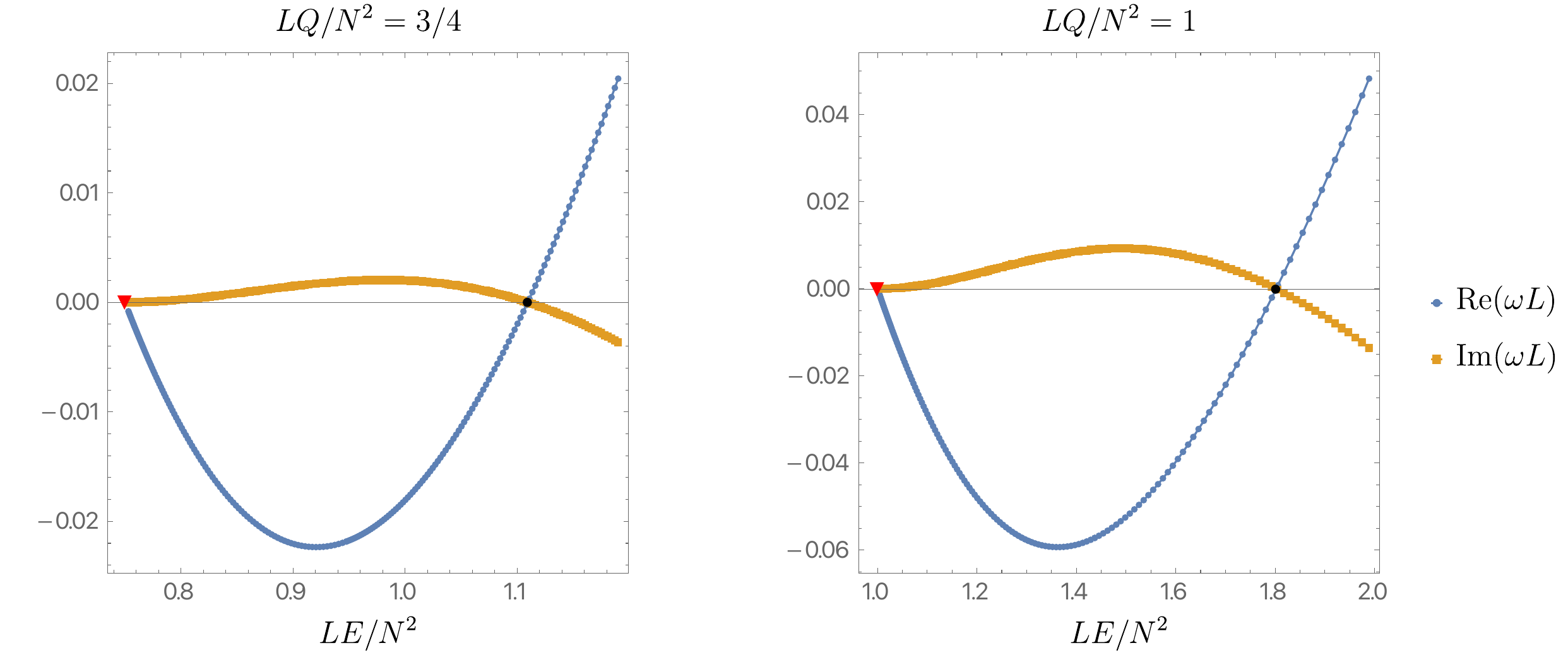}
\caption{\label{figs:growth_rate} The real (blue disks) and imaginary (orange squares) of $\omega$ as a function of the energy $L E/N^2$ for $L Q/N^2=0.75$ (left panel) and $L Q/N^2=1$ (right panel). The inverted red triangle pinpoint the supersymmetric bound for given value of $L Q/N^2$ and the black disk describes the onset of the instability.}
\end{figure}

In Fig.~\ref{figs:growth_rate} we plot the real (blue disks) and imaginary (orange squares) parts of the frequency $\omega$ as a function of the energy $L E/N^2$ for $L Q/N^2=0.75$ (left panel) and $L Q/N^2=1$ (right panel). The inverted red triangles show the supersymmetric bound for the given value of $L Q/N^2$ and the black disk describes the onset of the instability, which was determined using a strategy that we will outline in Section \ref{sec:OnsetSingleQ}. The agreement between the code that searches for the onset directly, and our calculation of the quasinormal mode spectrum $\{\widehat{\Phi}_\omega,\omega\}$ is reassuring. Note that in order for an onset to exist for a static solution, it must be the case that $\mathrm{Re}(\omega)=\mathrm{Im}(\omega)=0$ at the onset. Fig.~\ref{figs:growth_rate} shows that the BCS black hole is unstable (since $\mathrm{Im}(\omega L)>0$) in the region $E_{\mathrm{SUSY}}(Q)<E(Q)<E_{\mathrm{onset}}(Q)$ for fixed $Q$. This is the region of moduli space where we expect the hairy black holes to play a role. We will confirm this picture shortly.
Finally, we note that $\mathrm{Re}(\omega)$ is really not a gauge invariant quantity. Indeed, under $U(1)$ transformations $A\to A+\dt \chi$, with $\chi$ a smooth function, the charged scalar transforms as $\Phi\to e^{\frac{2\,i}{L}\chi}\,\Phi$. In particular, if we choose $\Phi$ as in~\eqref{eq:scalarpartialwave} and consider a class gauge transformations of the form $\chi = \chi^\0t$ with $\chi^\0$ constant and real, these induce a change in $\omega$ as $\omega\to\omega -\frac{2}{L}\chi^\0$\,. We often choose to regard $A$ as a smooth one-form, in which case this gauge freedom is chosen to that $A_t=0$ at the black hole event horizon. This is the gauge we used when computing $\omega$ shown in Fig.~\ref{figs:growth_rate}.

\subsection{Onset of scalar condensation instability}\label{sec:OnsetSingleQ}
Having shown that the scalar is unstable, we now turn out attention to the systematic study of the instability onset as a function of the BCS energy $E$ and charge $Q$. Our starting point is still~\eqref{eq:linear_single}), but we now choose $\Phi$ to be spherically symmetric and exhibit no time dependence. We still perform the same change of variable as in ~\eqref{eq:change} (with $\omega=0$) and introduce a compact coordinate as in ~\eqref{eq:compact_linear}. The resulting equation can be written as
\begin{subequations}
\begin{equation}
L_2\big(y;\tilde{\l}\big)\,q_0^{\prime\prime}(y)+L_1\big(y;\tilde{\l}\big)\,q_0^\prime(y)+L_0\big(y;\tilde{\l}\big)\,q_0(y)=0\,,
\label{eq:quadratic_single}
\end{equation}
with
\begin{equation}
L_2\big(y;\tilde{\l}\big)=(1-y) y \left(1+2 \tilde{Q}\right) \left[\tilde{\l } \left(2 (1-y) \tilde{Q}-y+2\right)-y+1\right]^2\,,
\end{equation}
\begin{multline}
L_1\big(y;\tilde{\l}\big)=\left(1+2 \tilde{Q}\right) \left\{\tilde{\l } \left[2 (1-y) \tilde{Q}-y+2\right]-y+1\right\} \bigg\{2 \tilde{\l } \left(1+\tilde{Q}\right)
\\
-y \left[\tilde{\l } \left(8 \tilde{Q}-6 y \tilde{Q}-3
   y+6\right)-3 y+4\right]+1\bigg\}\,,
\end{multline}
and 
\begin{multline}
L_0\big(y;\tilde{\l}\big)=-\tilde{\l }^2 \left[y^2 \left(1+2 \tilde{Q}\right)^3-y \left(3+4 \tilde{Q}\right) \left(1+\tilde{Q}\right)^2+2 \left(1+\tilde{Q}\right) \left(4 \tilde{Q}^2+2 \tilde{Q}+1\right)\right]
\\
-\tilde{\l }
   \left[2 y^2\left(1+2\tilde{Q}\right)^2-2 y \tilde{Q} \left(9+8 \tilde{Q}\right)+8 \tilde{Q} \left(1+\tilde{Q}\right)-5 y+3\right]-(1-y)^2 \left(1+2 \tilde{Q}\right)\,,
\end{multline}
\end{subequations}%
where we defined $\tilde{\l}=y_+^2$ and $\tilde{Q}=\tilde{q}/y_+^2$. Equation (\ref{eq:quadratic_single}) appears to be a quadratic eigenvalue problem in $\tilde{\l}$ for a given value of $\tilde{Q}$. However, in order to show that this is the case, boundary conditions have to be supplied for $q_0(y)$. These can be obtained around the regular singular points $y=0,1$ by demanding that $q_0$ is smooth there. The boundary condition at $y=0$ turns out to be
\begin{subequations}
\begin{equation}
\left(1+2 \tilde{Q}\right) \left[1+2 \left(1+\tilde{Q}\right) \tilde{\l }\right] q^\prime_0(0)-\left\{1+\tilde{\l }+2 \tilde{Q} \left[1+\tilde{\l }\left(1+2 \tilde{Q}\right) \right]\right\} q_0(0)=0\,,
\end{equation}
while for $y=1$ we find
\begin{equation}
\tilde{\l } \left(1+2 \tilde{Q}\right) q_0^\prime(1)-2 \tilde{Q} \left[1+2 \left(1+\tilde{Q}\right) \tilde{\l }\right] q_0(1)=0\,.
\end{equation}
\end{subequations}%
In Fig.~\ref{fig:2D_single} (solid blue line) we will show the onset curve $Q_{\mathrm{onset}}(E)$, which was determined by the procedure outlined above.

\subsection{Phase diagram in the microcanonical ensemble}\label{sec:ResultsSingleQmicro}
\begin{figure}[th]
\centering \includegraphics[width=0.6\textwidth]{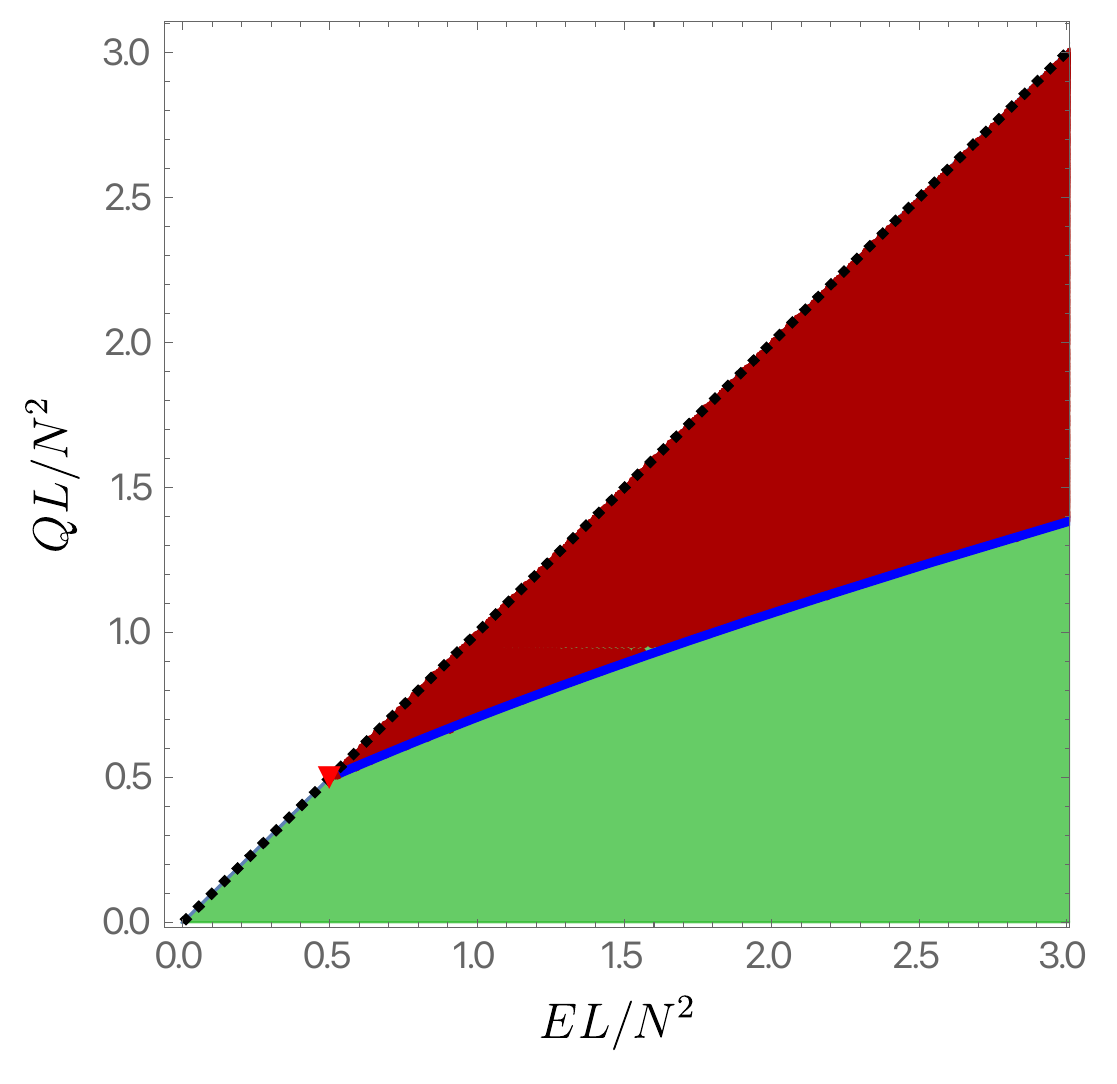}
\caption{\label{fig:2D_single} Phase diagram of singly charged solutions. The supersymmetric bound $E=Q$ is represented as a black dashed curve. The regular 1-parameter supersymmetric soliton is described by this curve. The BCS black holes exist below this BPS line in the $0\leq Q< E$ region. The inverted red triangle at the BPS curve marks the point $(L E/N^2,L Q/N^2)=(1/2,1/2)$. The onset curve of the scalar condensation instability of the BCS is represented as the solid blue line that starts at the inverted red triangle. The hairy black holes exist in the dark red region between the onset and the BPS curves. This phase diagram is also reproduced analytically (for small $E$ and $Q$) by a non-interacting model discussed in section \ref{singleQ:nonintmodel}.}
\end{figure}

We will start by discussing the microcanonical ensemble. Here the state variables are the energy $E$ and charge $Q$ and the relevant thermodynamic potential is the entropy $S$. Dominant phases have the largest $S$ at fixed $E$ and $Q$. The system we will study will involve two phases: the hairy black holes and the BCS black hole. 
At the onset of the BCS condensation instability (analyzed in the previous subsection), the BCS and hairy black holes have the same $E$, $Q$ and $S$ and the transition between the two families is second order\footnote{That is to say, the entropy (or Gibbs free energy) has continuous first derivative across the transition, but the second derivative jumps discontinuously.} (see solid blue line curve in Fig.~\ref{fig:2D_single}).

In Fig.~\ref{fig:2D_single} we plot the phase diagram of the solutions we found. The horizontal axis is $L E/N^2$, whereas the vertical axis is labeled by $L Q/N^2$. We draw the supersymmetric bound $Q=E$ as a thick black dashed line. The regular 1-parameter supersymmetric soliton \eqref{SingleQ:RegularSoliton} is described by this line, starting at $Q=E=0$ and extending for arbitrarily large values. The singular 2-parameter supersymmetric soliton is also described by this line  but we can choose one of its parameters $-$ namely, $C_2=4$ in \eqref{SingleQ:SingularSolitonHoloQuantities} $-$ to have it starting at $(L E/N^2,L Q/N^2)=(1/2,1/2)$ (the red inverted triangle) and extending for arbitrarily large $E=Q$. The BCS black holes exist for any $0\leq Q< E$ i.e. below the supersymmetric thick black dashed line (at the BPS line the BCS black hole approaches a singular solution with $S=0$ that, in the $E-Q$ diagram of Fig.~\ref{fig:2D_single} coincides with the supersymmetric solitons). It is important to note that single charge BCS black holes can get arbitrarily close to saturating the BPS bound. This is unlike the two charge BCS black holes that we will analyse later in section~\ref{sec:TwoQ}. The \emph{limiting} single charge BCS black hole family that saturates the BPS bound is, of course, singular. Still in Fig.~\ref{fig:2D_single}, the scalar condensation onset curve (determined using the method outlined in section \ref{sec:OnsetSingleQ}) is represented as a solid blue line; BCS black holes above this line are unstable. Hairy black holes exist in the dark red region between the supersymmetric bound and the onset curve, which is precisely where the BCS black holes are  unstable (see section \ref{sec:GrowthSingleQ}). Unlike the two charge case that we will discuss later in section~\ref{sec:TwoQ}, in the single charge system the hairy solutions \emph{always} coexist with the BCS black holes, i.e. the dark red region is on top of a green region in Fig.~\ref{fig:2D_single}.

Interestingly enough, the hairy black holes do not exist for arbitrarily small values of $Q$ or $E$ (unlike the two equal charge case of section~\ref{sec:TwoQ}). Indeed, we find that all hairy single charged black hole solutions must have $L Q/N^2>1/2$ and $L E/N^2>1/2$; see the inverted red triangle with $(L E/N^2,L Q/N^2)=(1/2,1/2)$ in Fig.~\ref{fig:2D_single}. This peculiarity along with the fact that the supersymmetric limit of the BCS black hole is a singular soliton makes the perturbative scheme presented in section \ref{sec:PerturbativeSingleQ} considerably more intricate than the two charge case (that will be discussed in section~\ref{sec:PerturbativeTwoQ}). 

\begin{figure}[th]
\centering \includegraphics[width=0.6\textwidth]{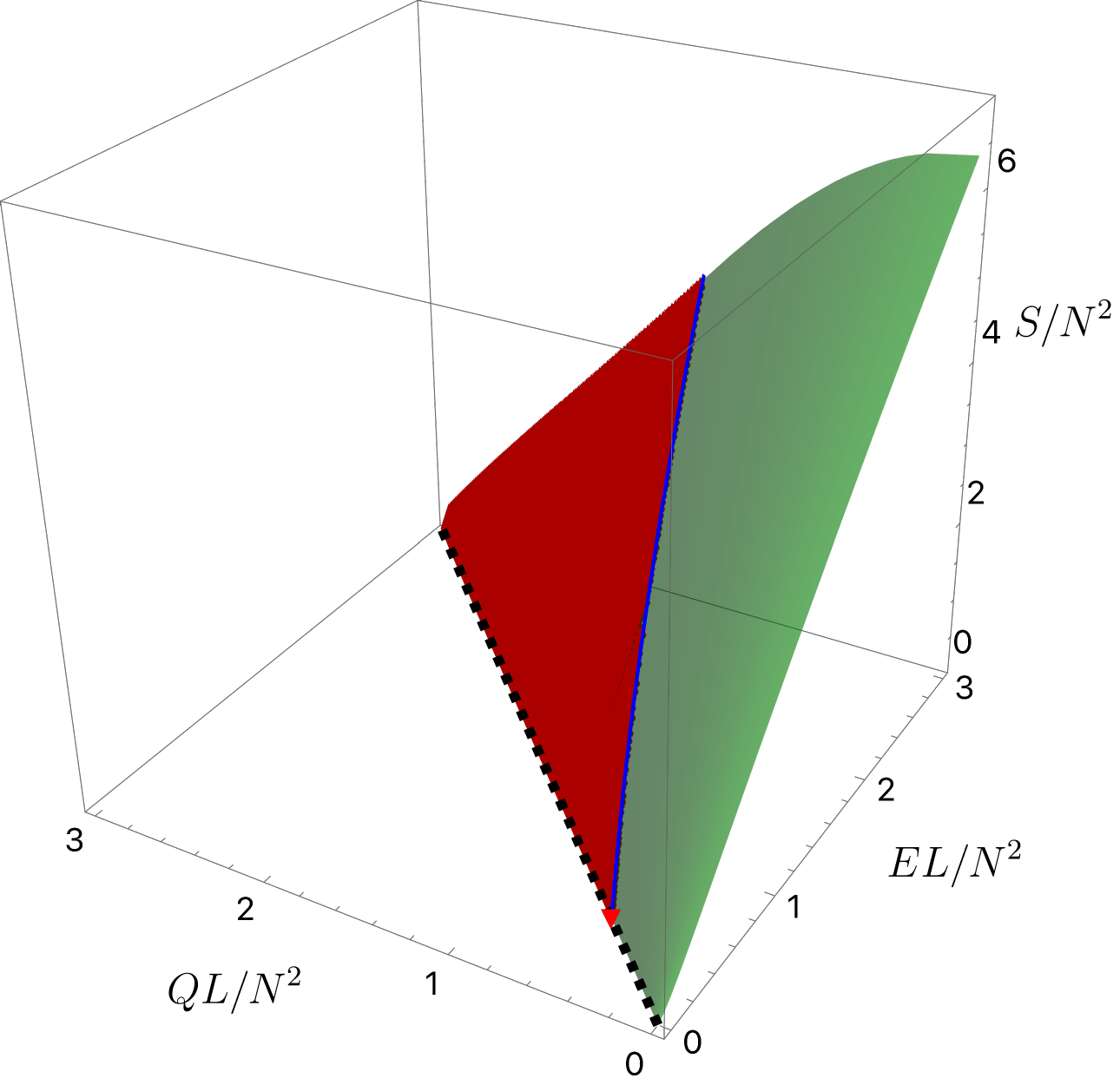}
\caption{\label{fig:3D_single} Microcanonical phase diagram for the single charged system: the entropy $S/N^2$ as a function of $L Q/N^2$ and $L E/N^2$ using the same colour coding as in Fig.~\ref{fig:2D_single}. Although not easy to see, the dark red surface is always above the green surface in the $E-Q$ region where they coexist except at the BPS bound (dashed black line) and at the solid blue line (onset curve), where the two families merge in a second order phase transition.}
\end{figure}

We now address the issue of phase dominance in the microcanonical ensemble. In Fig.~\ref{fig:3D_single} we show a three-dimensional plot of the entropy $S/N^2$ as a function of $L Q/N^2$ and $L E/N^2$ using the same colour coding as in Fig.~\ref{fig:2D_single}. We find that in $E-Q$ region where the hairy black holes coexist with the BCS black holes, the hairy black holes always have a \emph{larger} entropy and are thus dominant in the microcanonical ensemble. This suggests that, as expected, the hairy black holes should be the endpoint of the dynamical instability of BCS black holes uncovered in section \ref{sec:GrowthSingleQ}.

\begin{figure}[th]
\centering \includegraphics[width=0.6\textwidth]{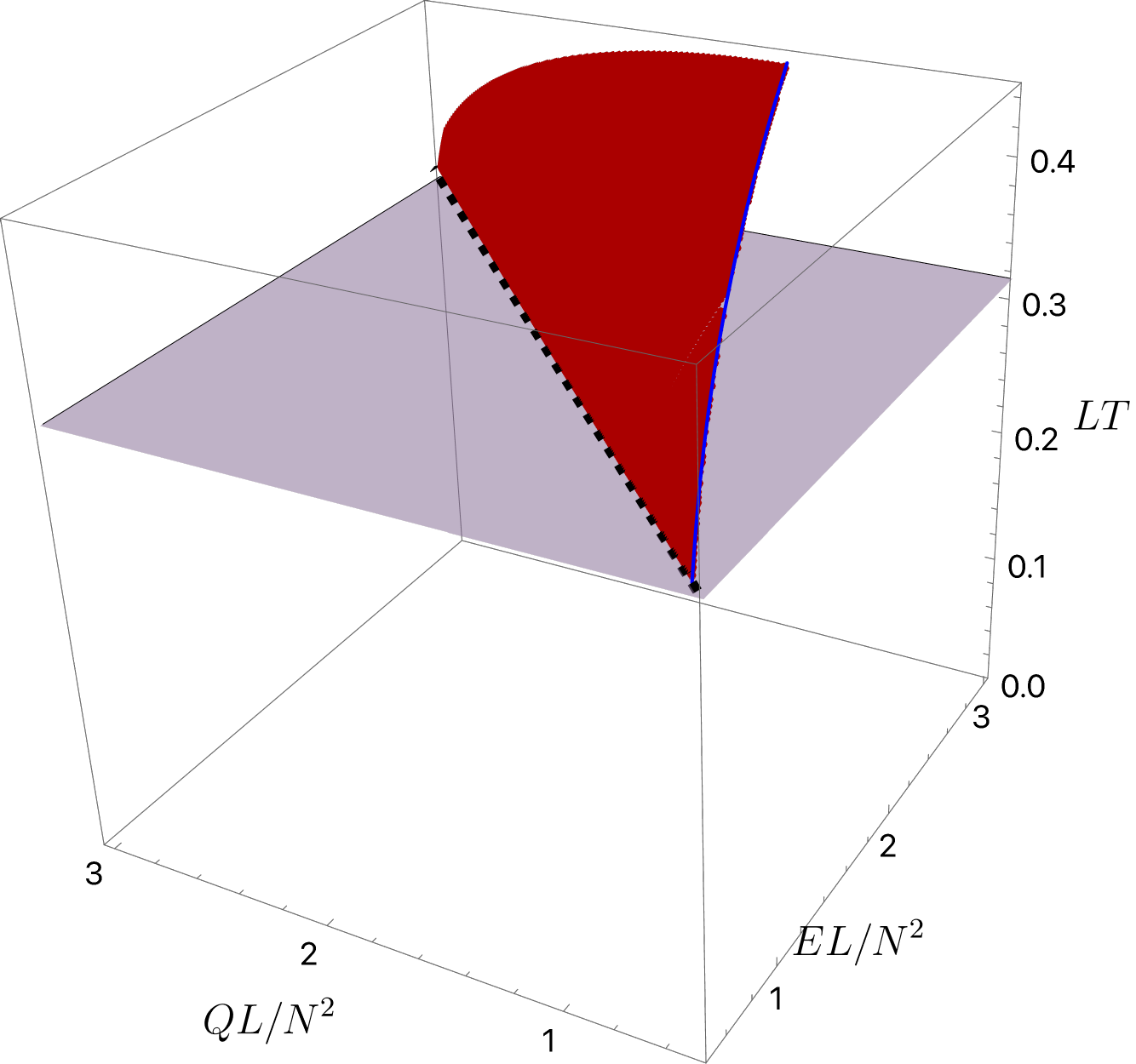}
\caption{\label{fig:T_3D} The temperature $L T$ of the single charge hairy black holes as a function of $L Q/N^2$ and $L E/N^2$ using the same colour coding as in Fig.~\ref{fig:2D_single}. For reference, we also plot the plane $L T = 1/\pi$ in purple: the hairy black hole temperature approaches it in the supersymmetric limit (dashed black line).}
\end{figure}

It is important to investigate the hairy black hole solutions near the BPS bound $Q=E$. This is the region of moduli space where our numerical schemes struggle the most to find solutions. We have managed to reach $y_+=0.1$, but found very hard to lower $y_+$ below this (solutions should exist all the way down to $y_+\to 0$). Nevertheless, with enough resolution, there are a number of striking features that we can infer. First, we find that the hairy black hole temperature tends to $L T = 1/\pi$ as one approaches the supersymmetric bound $Q=E$. This is best seen in Fig.~\ref{fig:T_3D} where we plot the hairy black hole temperature as a function of $E$ and $Q$. To aid the reader we also plot the plane $L T = 1/\pi$ in purple.
Actually, the temperature $L T = 1/\pi$ also plays an important role in the singly charged BCS black hole. Namely, one can ask what is the \emph{smallest} temperature one can reach with the BCS black holes. This minimum temperature can be reached by letting $\tilde{q}=1/2$ and $y_+=0$ and it turns out to be $L T = 1/\pi$. In this limit, one approaches the supersymmetric singular soliton discussed earlier.

\begin{figure}[th]
\centering \includegraphics[width=0.9\textwidth]{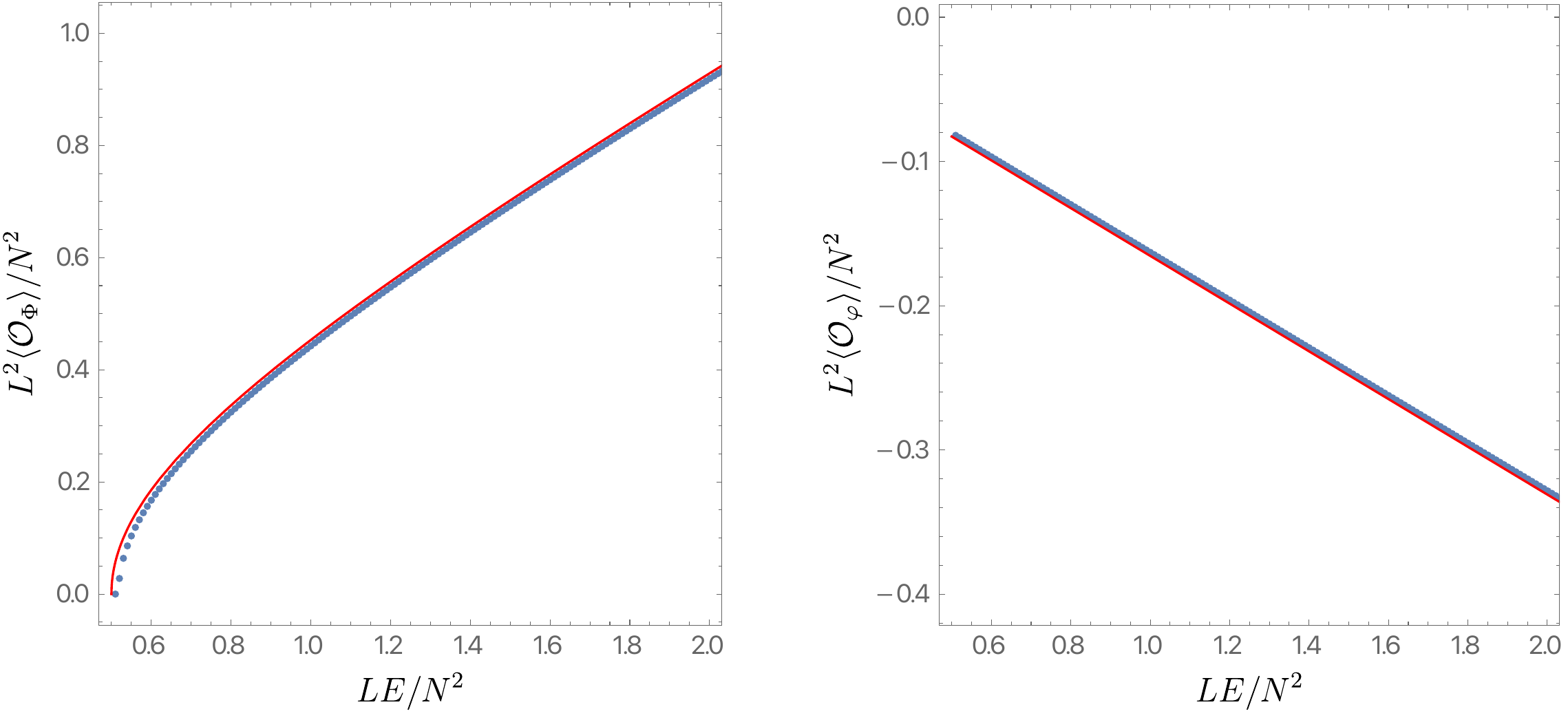}
\caption{\label{fig:comp_sing} Expectation values $L^2\langle {\cal O}_{\Phi} \rangle/N^2$ (left panel) and $L^2\langle {\cal O}_{\varphi} \rangle/N^2$ (right panel) as a function of $L E/N^2$ for the family of single charged hairy black holes with $y_+=0.1$ (the closest family we have to the supersymmetric limit). The blue disks are the numerical data, and the solid red lines are given by \eqref{SingleQ:SingularSolitonHoloQuantities} with $C_2\geq4$ and $C_1=3$.}
\end{figure}

We also monitored the charged and neutral scalar field expectation values $\langle {\cal O}_{\Phi} \rangle$ and $\langle {\cal O}_{\varphi} \rangle$, as defined in \eqref{SingleQ:vevs}, for the line of solutions closest to the supersymmetric bound (these have $y_+=0.1$). Perhaps surprisingly, we find that this curve is very well fit by that of a singular soliton given in \eqref{SingleQ:SingularSolitonHoloQuantities} with $C_1=3$ and $C_2\geq 4$. At the moment we have no understanding why this is the case. Note that in principle we could have $C_1(C_2)$, but it turns out that our best fit yields $C_1=3$. To back our claim, in Fig.~\ref{fig:comp_sing} we plot  $\langle {\cal O}_{\Phi} \rangle$ (left panel) and $\langle {\cal O}_{\varphi} \rangle$ (right panel) as a function of $L E/N^2$. The blue disks are the numerical data, and the solid red lines are given by~\eqref{SingleQ:SingularSolitonHoloQuantities} with $C_2\geq4$ and $C_1=3$. The agreement is striking (specially if we remember that $y_+=0.1$ for the hairy solutions, and we expect the agreement to improve as $y_+$ gets smaller). At the moment we have no analytic understanding of why the solution with $C_1=3$ is preferred over other possible choices.

\subsection{Phase diagram in the grand-canonical ensemble}\label{sec:ResultsSingleQgrand}

We now turn out attention to the grand-canonical ensemble. The state variables are now the temperature $T$ and chemical potential $\mu$. The relevant thermodynamic potential is the Gibbs free energy $G$, and dominant solutions have the lowest free energy. There is now a competition between three phases: thermal AdS, BCS solutions and the hairy black holes. We expect large enough black holes to eventually dominate the ensemble. The question is then which of the black holes will dominate in a given window of $T$ and $\mu$.

\begin{figure}[th]
\centering \includegraphics[width=0.9\textwidth]{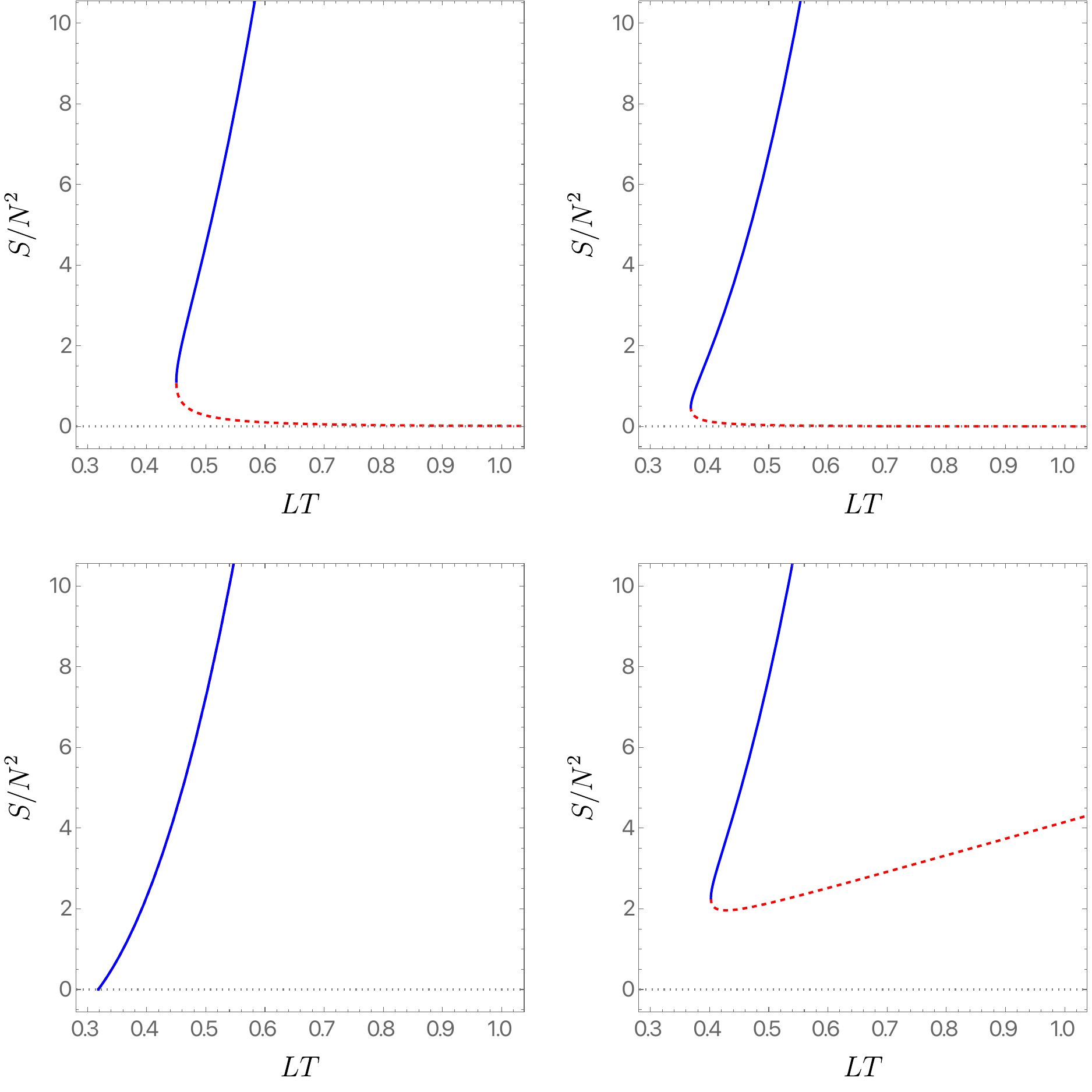}
\caption{\label{fig:entropy_single_BCS} Entropy $S/N^2$ as a function of the temperature $LT$ for singly charged BCS black holes for four different values of $\mu$. Using lexicographic ordering, we have $\mu=0,\, 0.9,\, 1,\, 1.1$, respectively. The solid blue (dashed red) line describes large (small) BCS black holes.}
\end{figure}

We first discuss the thermodynamic properties of the BCS black hole in the grand-canonical ensemble. This has been previously discussed in \cite{Henriksson:2019zph} but the presentation here is more detailed. We first note that for fixed temperature $T L>\frac{1}{\pi}$ and fixed $\mu\neq1$ there are two black hole solutions. These two solutions have distinct entropies. We coin the one with larger entropy the ``large'' BCS black hole, and the one with smaller entropy the ``small'' BCS black hole. Large BCS black holes have an entropy that scales as $S/N^2\simeq \pi^4 (L T^3)$ at large $T$, while small BCS black holes have an entropy that approaches zero as $T\to+\infty$ for $|\mu|<1$ (see top panels of Fig.~\ref{fig:entropy_single_BCS}), and diverge linearly in $T$ as $S/N^2\simeq 2 \pi ^2 L T (\mu ^2-1)$ for $\mu^2>1$ (see bottom-right panel of Fig.~\ref{fig:entropy_single_BCS}). For $\mu=1$ there is a degeneracy, and the small and large black hole branch merge and form a single black hole family (see bottom-left panel of Fig.~\ref{fig:entropy_single_BCS}). Fig.~\ref{fig:entropy_single_BCS} shows a set of snapshots of the entropy $S/N^2$ as a function of the temperature $L T$ computed for different values of $\mu$ that illustrate the aforementioned properties. Using lexicographic ordering, we have $\mu=0,\, 0.9,\, 1,\, 1.1$, respectively.

The small BCS black hole is always locally thermodynamically unstable in the grand-canonical ensemble. To see this, start by recalling that local thermodynamic stability in the grand-canonical ensemble is equivalent to demanding that minus the Hessian of the entropy with respect to the energy $E$ and charge $Q$, 
\begin{subequations}
\begin{equation}
\mathcal{S}_{ij}=-\frac{1}{L^2}\frac{\partial^2 S}{\partial \mathcal{Q}^i\partial \mathcal{Q}^j}
\qquad \hbox{where $\mathcal{Q}=\{E,Q\}$ and $i=1,2$,}
\end{equation}
is positive definite\footnote{One could have equally well investigate positivity of the Hessian of the Gibbs free energy $G$ with respect to $T$ and $\mu$. See for instance \cite{Monteiro:2008wr,Monteiro:2009tc} for details.}. We find
\begin{align}
&\mathcal{S}_{11}=-\frac{4 \pi  \sqrt{2 \tilde{q}+y_+^2} \left[4 \left(\tilde{q}-3\right) \tilde{q}-10 y_+^2 \tilde{q}-2 y_+^4-y_+^2+1\right]}{\left[2 \tilde{q}+2 y_+^2+1\right)^3 \left(\left(6 y_+^2+8\right) \tilde{q}+3
   \left(y_+^4+y_+^2\right)\right]}, \nonumber
\\
&\mathcal{S}_{12}=-\frac{4 \sqrt{2} \pi  \sqrt{\left(y_+^2+1\right) \tilde{q}} \left[6 \left(\left(y_+^2+2\right) \tilde{q}-2 \tilde{q}^2+y_+^4\right)+7 y_+^2+1\right]}{\left(2 \tilde{q}+2 y_+^2+1\right)^3 \left[\left(6
   y_+^2+8\right) \tilde{q}+3 \left(y_+^4+y_+^2\right)\right]},
\\
&\mathcal{S}_{22}=-\frac{4 \pi  \sqrt{2 \tilde{q}+y_+^2} \left[4 \left(6 y_+^2+5\right) \tilde{q}^2-2 \left(6 y_+^4+13 y_+^2+6\right) \tilde{q}-3 \left(y_+^2+1\right) \left(2 y_+^2+1\right){}^2\right]}{\left(2 \tilde{q}+2
   y_+^2+1\right)^3 \left(\left[6 y_+^2+8\right) \tilde{q}+3 \left(y_+^4+y_+^2\right)\right]}.\nonumber
\end{align}
\end{subequations}%
It is a simple exercise to compute the eigenvalues of $\mathcal{S}$ and verify that they are both positive for large BCS black holes, while at least one is negative for small BCS black holes. This shows that small BCS black holes are locally thermodynamically unstable, while large BCS black holes are locally thermodynamically stable in the grand-canonical ensemble.

\begin{figure}[th]
\centering \includegraphics[width=0.6\textwidth]{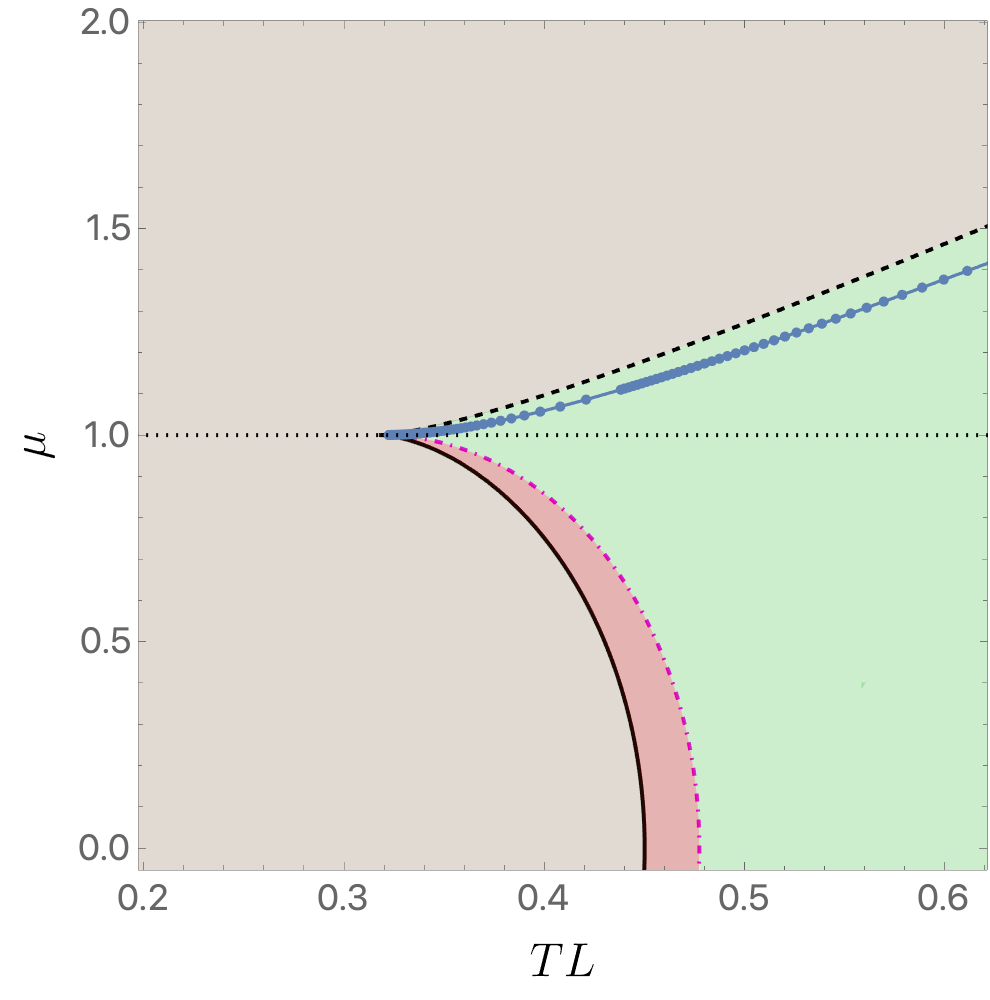}
\caption{\label{fig:2D_Gibbs_BCS} Phase space of singly charged BCS black holes in the grand-canonical ensemble. The light brown region indicates regions where BCS black holes do not exist (thermal AdS dominates the ensemble); the light red region is a region where thermal AdS and large BCS black holes coexist, but nevertheless thermal AdS dominates; the dot-dashed magenta line indicates a Hawking-Page transition; and in the green region large BCS black holes dominate the ensemble. The blue disks are the onset of scalar condensation; hairy black holes exist below this onset line and above the line $\mu=1$ but they are always sub-dominant. Finally, the dashed black line shows the location of the 0-th order phase transition briefly discussed in the main text.}
\end{figure}

Large charged AdS black holes can undergo Hawking-Page transitions \cite{Hawking:1982dh} so it should not come  as a surprise that the same is true for the large BCS black holes. To confirm this, first note that thermal AdS, by definition, has zero Gibbs free energy. Thus, all we need to do is to inspect the sign of $G$ for large AdS black holes (small BCS black holes are always sub-dominant). We summarise our findings in Fig.~\ref{fig:2D_Gibbs_BCS}. The light brown region indicates regions where BCS black holes do not exist, and thus the only available phase is thermal AdS. The light red region represents a region of the $(T,\mu)$ phase where thermal AdS and large BCS black holes coexist, but nevertheless thermal AdS dominates. The dot-dashed magenta line (the right boundary of the light red region) indicates a Hawking-Page transition similar to the one reported in \cite{Chamblin:1999tk} for standard five-dimensional Reissner-Nordstr\"om black holes with AdS asymptotics. In the green region, large BCS black holes dominate the grand-canonical ensemble. Note that the Hawking-Page phase transition from the red region to the green region (along the dot-dashed magenta line) is first order, but the transition from brown to green, along the dashed black line, is zeroth order. However, zeroth order phase transitions cannot occur in thermodynamically stable systems and are often an artefact of the thermodynamic approximation. Indeed, recall that in any thermodynamically stable phase in the grand-canonical ensemble, $-G$ must be a convex function of $T$ and $\mu$ \cite{LANDAU198079}, and this is not possible for zeroth order phase transition away from the strict thermodynamic limit. We interpret the presence of this ``forbidden'' zeroth order transition as indicating that we are missing a novel phase altogether for the single charge case. We will leave the construction of this new phase for future investigations.

To complete our discussion of the grand-canonical phase diagram, one might wonder where the hairy solutions fit in Fig.~\ref{fig:2D_Gibbs_BCS}. The blue solid disks in Fig.~\ref{fig:2D_Gibbs_BCS} represent the onset of the condensation instability of BCS black holes and we find that hairy solutions extend  from this curve down to the $\mu=1$ line. Computing the Gibbs free energy of the hairy solutions, we find it is always larger than that of large BCS black holes, making the new hairy solutions subdominant in the grand-canonical ensemble.

\subsection{Perturbative construction of hairy black holes}\label{sec:PerturbativeSingleQ}

In this section, we describe the basic strategy used to construct the single charged hairy black hole solutions (of Sections \ref{sec:AnsatzSingleQ}, \ref{sec:ThermoSingleQ}, \ref{sec:ResultsSingleQmicro} and \ref{sec:ResultsSingleQgrand}) in perturbation theory. It turns out that the equations of motion \eqref{SingleQ:EoM} in the gauge \eqref{SingleQ:ansatz} are difficult to solve analytically. It is more convenient to work in a slightly different gauge defined by the ansatz
\begin{equation}
\begin{split}\label{static-sol-ansatz-singlecharge}
\dt s^2 &= h^{\frac{1}{3}} \left( - \frac{f}{h} \dt t^2 +  \frac{\dt r^2}{f} + r^2 \dt \O_3^2 \right) , \\
\varphi_1 &= \sqrt{6} \varphi , \qquad \varphi_2 = 0 , \\
A^1 &= 0 , \qquad A^2 = 0 , \qquad A^3 = A_t dt , \\
\Phi_1 &= 0 , \qquad \Phi_2 = 0 , \qquad \Phi_3 = \Phi_3^\dagger = \Phi ,
\end{split}
\end{equation}
where all the quantities above are functions of $r$ only. In \eqref{SingleQ:ansatz}, $g_{rr}$ is an independent function and $\varphi$ is fixed in terms of $h$ whereas in this gauge, $g_{rr}$ is fixed in terms of $f$ and $\varphi$ is an independent function. The ansatz \eqref{static-sol-ansatz-singlecharge} has a leftover coordinate freedom $r \to \sqrt{r^2 + a}$ and we use this to fix\footnote{For completeness we note that the radial coordinate used here is related to the radial coordinate in \eqref{SingleQ:ansatz} by
$$r_{here}^2 =  \int \dt (r_{there}^2) \sqrt{\frac{f_{there}(r_{there})}{g_{there}(r_{there})}} + a.
$$
The integration constant $a$ is fixed by \eqref{phiconstraint}. The metric functions are related by
\begin{equation}
\begin{split}
\nonumber
r_{here}^6 h_{here} &= r_{there}^6 h_{there} , \quad r_{here}^4 f_{here} = r_{there}^4 f_{there} ,  \\
(A_t)_{here} &= (A_t)_{there} , \qquad \Phi_{here} = \Phi_{there} , \qquad \varphi_{here} = \frac{1}{\sqrt{6} } \varphi_{there} .
\end{split}
\end{equation}
}
\begin{equation}
\begin{split}\label{phiconstraint}
\varphi + \frac{1}{3} \ln h = \CO(r^{-4}) \quad \text{at large $r$.}
\end{split}
\end{equation}
Plugging the ansatz \eqref{static-sol-ansatz-singlecharge} into equations \eqref{GravEOM}--\eqref{ScalarsGaugeEOM}, we find the equations
\begin{align}
\label{singleQ:h2eq} 0 &= -\frac{3 A^2 h \Phi ^2}{f^2}+\frac{2 h'^2}{h^2}-\frac{3 \left(r h''+3 h'\right)}{r h}-\frac{3 \Phi '^2}{\Phi ^2+4}-18 \varphi '^2 , \\
0 &= \frac{3 r^2 \left(h^{5/3} e^{-4 \varphi } A'^2+f' h'+h^{4/3} \left(\Phi ^2 e^{4 \varphi }-8 \sqrt{\Phi ^2+4} e^{\varphi }-8 e^{-2 \varphi }\right)\right)}{h}+18 r f'  \nonumber \\
\label{singleQ:f1eq} &\qquad -\frac{3 r^2 A^2 h \Phi ^2}{f}+f \left(r^2 \left(-\frac{h'^2}{h^2}-\frac{3 \Phi '^2}{\Phi ^2+4}-18 \varphi '^2\right)+36\right)-36 , \\
\label{singleQ:A2eq} 0 &= \frac{f e^{-4 \varphi } \left(3 r h A''+A' \left(2 r h'+h \left(9-12 r \varphi '\right)\right)\right)}{3 r h^{4/3}}-A \Phi ^2 , \\
\label{singleQ:phi2eq}  0 &= -h^{\frac{1}{3}} e^{-5 \varphi } A'^2+\frac{3 e^{-\varphi } \left(\left(r f'+3 f\right) \varphi '+r f \varphi ''\right)}{r h^{\frac{1}{3}}}+2 \sqrt{\Phi ^2+4} - (\Phi^2+4)e^{3\varphi} ,  \\
0 &= \frac{4 A^2 h^{\frac{2}{3}} \Phi }{f}-\frac{\Phi  \left(\Phi ' \left(r \Phi  f'+f \left(r \Phi '+3 \Phi \right)\right)+r f \Phi  \Phi ''\right)}{r h^{\frac{1}{3}}\left(\Phi ^2+4\right)}  \nonumber \\
\label{singleQ:Phi2eq}  &\qquad +\frac{\left(r f'+3 f\right) \Phi '+r f \Phi ''}{r h^{\frac{1}{3}}}+\frac{f \Phi ^3 \Phi '^2}{h^{\frac{1}{3}} \left(\Phi ^2+4\right)^2}-4 \Phi \left(e^{4 \varphi }-\frac{4 e^{\varphi }}{\sqrt{\Phi ^2+4}}\right) . 
\end{align}
These are 5 coupled differential equations for 5 functions. The differential equations for $h$, $A$, $\varphi$ and $\Phi$ are second order whereas the one for $f$ is first order. The equations can then be solved up to 9 integration constants. Four of these are fixed by the AdS boundary conditions at $r=\infty$ and five are fixed by imposing regularity in the interior (either on the horizon $r=r_+$ or the origin $r=0$).

The asymptotic expansion of the fields in this gauge has the form
\begin{equation}
\begin{split}\label{singleQ:adsbdycondpert}
f(r) &= \frac{r^2}{L^2} + 1 + c_h + c_f \frac{L^2}{r^2} + \CO(L^4 r^{-4}) ,\\
h(r) &= 1 + c_h \frac{L^2}{r^2}  + \CO(L^4 r^{-4}) , \\
A_t(r) &= \mu + c_A \frac{L^2}{r^2} + \CO(L^4 r^{-4}) , \\
\varphi(r) &= - c_h \frac{L^2}{3r^2} + \CO(L^4  r^{-4}) , \\
\Phi(r) &= \e \frac{L^2}{r^2} + \CO(L^4r^{-4}) . \\
\end{split}
\end{equation}
Using holographic renormalization as described in section \ref{sec:ThermoSingleQ}, we find that the energy and electric charge are given by
\begin{equation}
\begin{split}\label{singleQ:EQlarger}
E = \frac{N^2}{L} \left( \frac{1}{2} c_h - \frac{3}{4} c_f  - \frac{3}{32} \e^2 \right)  , \qquad Q = \frac{N^2}{L } \left( - \frac{1}{2} c_A \right) . 
\end{split}
\end{equation}
Black hole solutions have a Killing horizon at $r=r_+$ where $f(r_+) = A_t(r_+) = 0$ ($r_+$ is the largest root of $f$). We will construct solutions which are regular on the horizon. The temperature, entropy and chemical potential of a regular black hole solution is given by
\begin{equation}
\begin{split}\label{singleQ:bhthermo}
T = \frac{f'(r_+)}{4\pi L \sqrt{h(r_+)}} , \qquad S = N^2 \pi r_+^3 \sqrt{h(r_+)}, \qquad \mu = \lim_{r\to\infty} A_t(r) . 
\end{split}
\end{equation}
These thermodynamic quantities must satisfy the first law of thermodynamics $\dt E = T \dt S + \mu \dt Q$.

In the rest of this section, we will set the AdS radius to unity, $L=1$.

\subsubsection{Hairy supersymmetric soliton}
\label{singleQ:hairysusysoliton-pertthy}

In this subsection, we describe the perturbative construction of the hairy supersymmetric soliton. This solution is of course known exactly and is described in Section \ref{sec:ansatzSingleQ-Solitons}. The purpose of this section then is to describe the qualitative features of the general perturbative construction in a simple setting where we can compare the perturbative approximation with the exact analitical result. The techniques introduced here can and will be generalized to the more complicated construction of hairy black hole solution in Section \ref{sec:singleQHBHpertthy}.

To initiate the perturbative construction, we expand the solitonic fields as
\begin{equation}
\begin{split}\label{soliton-singlecharge-exp}
f(r,\e) &= \sum_{n=0}^\infty \e^{2n} f_{(2n)}(r)  , \\
h(r,\e) &= \sum_{n=0}^\infty \e^{2n} h_{(2n)}(r)  , \\
A_t(r,\e) &= \sum_{n=0}^\infty \e^{2n} A_{(2n)}(r) , \\
\varphi(r,\e) &= \sum_{n=0}^\infty \e^{2n} \varphi_{(2n)}(r) , \\
\Phi(r,\e) &= \sum_{n=0}^\infty \e^{2n+1} \Phi_{(2n+1)}(r) ,
\end{split}
\end{equation}
where the leading order solution is vacuum $\ads_5$ given by
\begin{equation}
\begin{split}
f_\0(r) = 1 + r^2 , \qquad h_\0(r) = A_\0(r) = 1 , \qquad \varphi_\0(r)= 0 .
\end{split}
\end{equation}
The expansion parameter of the perturbation theory is $\e = \frac{\pi^2}{N^2} \avg{ \CO_\Phi }$ which is the expectation value of the operator dual to the scalar field $\Phi$. In terms of the bulk geometry, it is fixed as the leading coefficient of $r^{-2}$ in the near boundary expansion of $\Phi$ as
\begin{equation}
\begin{split}\label{singleQ:Phibc}
\Phi(r,\e) = \frac{\e}{r^2} + \CO(r^{-4}) . 
\end{split}
\end{equation}
We plug \eqref{soliton-singlecharge-exp} into the equations \eqref{singleQ:h2eq}--\eqref{singleQ:Phi2eq} and expand them in a power series in $\e$. At each order in $\e$, we have a set of linear differential equations, which we solve subject to AdS boundary condition \eqref{singleQ:Phibc} at $r=\infty$ and regularity at the origin $r=0$.\footnote{For the construction of the soliton solution, it is convenient to use the coordinate freedom $r \to \sqrt{r^2+a}$ to set the origin of the spacetime at $r=0$ instead of imposing \eqref{phiconstraint}.}

At $\CO(\e^{2n+1})$, only equation \eqref{singleQ:Phi2eq} is non-trivial and we obtain a single differential equation for $\Phi_{(2n+1)}(r)$ which has the form
\begin{equation}
\begin{split}
\td{}{r}  \left[ \frac{r^3}{1+r^2} \td{}{r} \left[  (1+r^2) \Phi_{(2n+1)}(r)  \right] \right] = \mfs^\Phi_{(2n+1)}(r) . 
\end{split}
\end{equation}
where the source $\mfs^\Phi_{(2n+1)}(r)$ is completely fixed by lower orders in perturbation theory and it should be thought of as a known function in terms of which we wish to determine $\Phi_{(2n+1)}(r)$. This equation is easily integrated and a solution for $\Phi_{(2n+1)}(r)$ can be obtained up to two integration constants which denote the source for the dual scalar operator and its response (expectation value). We are interested in solutions without any sources for the scalar field and the expectation value is defined by the boundary condition \eqref{singleQ:Phibc}. These two conditions fix both the integration constants.

At $\CO(\e^{2n})$, \eqref{singleQ:Phi2eq} is trivial but the remaining equations \eqref{singleQ:h2eq}--\eqref{singleQ:phi2eq} take the form
\begin{equation}
\begin{split}
\td{}{r} \left[ \frac{1+r^2}{r}  \td{}{r} \left[ r^2 \varphi_{(2n)}(r) \right] \right]  &= \mfs_{(2n)}^\varphi(r) , \\
\td{}{r} \left[ r^3 \td{}{r} A_{(2n)}(r) \right] &= \mfs_{(2n)}^A(r)   , \\
\td{}{r} \left[ r^3 \td{}{r}  h_{(2n)}(r) \right] &= \mfs_{(2n)}^h(r) , \\
\td{}{r} \left[ r^2 f_{(2n)}(r) \right]  &= \mfs_{(2n)}^f(r) - \frac{1}{3} r^8 \td{}{r} [ r^{-4} h_{(2n)}(r) ] . 
\end{split}
\end{equation}
As before, the source terms are all determined by lower orders in perturbation theory and known functions when we arrive at  $\CO(\e^{2n})$. These equations are easily integrated and the solutions are determined up to 7 integration constants. One of the integration constants in the first equation is fixed by regularity of $\varphi$ at $r=0$ and the other is fixed by AdS boundary conditions (namely, requiring that the source for dual operator $\CO_\varphi$ is zero.). One of the integration constants in the second equation is fixed by regularity of $A$ at $r=0$. The other is fixed at the next order in perturbation theory $\CO(\e^{2n+1})$ by requiring regularity of $\Phi$ at $r=0$. One of the integration constants in the third equation is fixed by AdS boundary conditions and the other is fixed by regularity at $r=0$. Finally, the integration constant in the last equation is fixed by regularity at $r=0$.

Explicit construction of the solution to $\CO(\e^3)$ is described in detail in Appendix \ref{App:PerturbativeSingleQSoliton} and the full solution to $\CO(\e^{15})$ is presented in the accompanying {\tt Mathematica} file. It is easily verified that the solution so constructed is supersymmetric and it satisfies equations \eqref{SingleQ:SUSYfunctions} and \eqref{SingleQ:solitonODE}. The energy (after removing the Casimir contribution) and charge of the soliton is given by 
\begin{equation}
\begin{split}
&E = Q = N^2 \left[ \frac{\e^2}{16}-\frac{\e ^4}{256}+\frac{\e^6}{2048}-\frac{5 \e^8}{65536}+\frac{7 \e^{10}}{524288} -\frac{21 \e^{12}}{8388608} + \frac{33 \e ^{14}}{67108864} + \CO (\e ^{16}) \right] . 
\end{split}
\end{equation}
It is also easy to verify the regular solution constructed here is precisely the perturbative expansion of the excat analytic soliton solution described in \eqref{SingleQ:RegularSoliton} once we identify $\e = 4 \sqrt{q(q+1)}$. 

\subsubsection{Hairy black hole}
\label{sec:singleQHBHpertthy}

The perturbative procedure described in the previous subsection can be generalized to construct hairy solutions with horizons as well, although we must introduce a second expansion parameter (the horizon radius in AdS radius units) and resort to a matched asymptotic expansion. So, in this section, we shall construct the single charge hairy black hole (BH) solution of Sections \ref{sec:AnsatzSingleQ}, \ref{sec:ThermoSingleQ}, \ref{sec:ResultsSingleQmicro} and \ref{sec:ResultsSingleQgrand} in a double perturbative expansion about the base BCS black hole \eqref{BCS:SingleQ}.
But before doing so, we start by deriving heuristically the leading order thermodynamic properties of such hairy solutions using a simple {\it non}-interacting thermodynamic model that does \emph{not} make use of the equations of motion.

\subsubsection*{Hairy BH as a noninteracting mix of BCS BH and supersymmetric soliton}
\label{singleQ:nonintmodel}

Before discussing the details of the perturbative construction, it is instructive to consider a toy model in which the hairy BH is treated as a non-interacting mix of the BCS black hole and the hairy supersymmetric soliton in thermodynamic equilibrium. That is to say, as a first approximation, the hairy BH can be found by placing a small bald BH (here, the BCS BH) on top of the soliton of the theory.
Although \`a priori crude, this model already proved to capture the correct leading order thermodynamic of many charged and/or rotating hairy black systems  
\cite{Basu:2010uz,Dias:2011at,Dias:2011tj,Cardoso:2013pza,Dias:2015rxy,Dias:2016pma,Bhattacharyya:2010yg,Markeviciute:2016ivy}, and this will also be the case in the present system.

The model assumes that, at leading order (and certainly only at this order), the energy (charge) of the non-interacting mix is given simply as a sum of the bald black hole energy (charge)  and the soliton energy (charge). Using the energy and charge \eqref{BCSsingleQ:EQ} of the bald BCS BH we can write in these conditions:
\begin{equation}
\begin{split}\label{singleQ:noninteractingmodel}
E &= \frac{N^2}{4} [3r_+^2(1+r_+^2)+ 2 \tilde{q}  (2+3r_+^2 )  ] + E_{sol} \,, \\
Q &= \frac{N^2}{2} \sqrt{ 2 {\tilde q} ( 1 + r_+^2 ) (  2 {\tilde q} + r_+^2 ) }  + E_{sol}\,,
\end{split}
\end{equation}
where we have also used the fact that the soliton is supersymmetric so $Q_{sol} = E_{sol}$. 

The soliton carries no entropy, $S_{sol}=0$, so the entropy $S$ of the hairy BH is simply the the BCS BH entropy~\eqref{BCSsingleQ:TS}:
\begin{align} \label{entropy}
S=S_{BCS}(E_{BCS},Q_{BCS})+S_{sol}(E_{sol},Q_{sol})=S_{BCS}(E-E_{sol},Q-Q_{sol}).
\end{align}

The hairy BH can partition its mass $E$ and charge $Q$ between the BCS BH and soliton components of the mixture. On physical grounds one expects this distribution to be such that, for fixed mass $E$ and charge $Q$, the entropy $S$ is maximised, $dS=dS_{BCS}=0$, while respecting the first law of thermodynamics $\mathrm d E=T \mathrm d S+\mu \mathrm d Q$. Not surprisingly, the maximisation of the entropy turns out to imply
\cite{Dias:2016pma} that the two mixed constituents of the system (and thus the  hairy BH) must be in thermodynamic (i.e. in chemical and thermal) equilibrium:
\begin{equation}
\begin{split}\label{singleQ:BHthermo}
\mu & \equiv \mu_{BCS}=\mu_{sol}=1  \quad \implies \quad  \sqrt{ \frac{ 2{\tilde q} ( 1 + r_+^2 ) }{2{\tilde q} + r_+^2 } }=1 \quad \implies \quad {\tilde q} = \frac{1}{2}\,, \\
T & \equiv T_{BCS}  \quad \implies \quad T = \frac{1+2{\tilde q}+2r_+^2}{2\pi \sqrt{2{\tilde q}+r_+^2} } = \frac{1}{\pi} \sqrt{ 1 + r_+^2 }\,,
\end{split}
\end{equation}
where we used~\eqref{BCSsingleQ:TS} for the chemical potential and temperature of the BCS BH and the fact that the supersymmetric soliton has $\mu_{sol}=1$.

We can now plug \eqref{singleQ:BHthermo} into \eqref{singleQ:noninteractingmodel} and solve for $r_+$ and $E_{sol}$ in terms of $E$ and $Q$, 
\begin{equation}
\begin{split}\label{singleQ:rpEsolnonintmix}
r_+^2 &= \frac{2}{3} \left( \left[ 1 + 3 \left( \frac{E-Q}{N^2}  \right) \right]^{\frac{1}{2}}- 1 \right)   , \\
E_{sol} &= Q  - \frac{N^2}{6} \left( 2  \left[ 1 + 3  \left( \frac{E-Q}{N^2}  \right)  \right]^{\frac{1}{2}} + 1 \right)  . 
\end{split}
\end{equation}
Since we must have $r_+^2 \geq 0$ and $E_{sol} \geq 0$, we find bounds on the total mass and charge of the mix,
\begin{equation}\label{singleQ:mixEbounds}
\begin{split}
\frac{N^2}{2} \leq Q \leq E \leq \frac{3Q^2}{N^2} - \frac{N^2}{4} .
\end{split}
\end{equation}
It follows that the hairy black hole solution (to the extent that it can be modeled as a non-interacting mix) can exist only in the parameter regime described  above. This domain of existence of the hairy black hole can be understood as follows. In one extremum, the BCS BH constituent is absent in the mixture and the supersymmetric soliton component with $E=Q\geq \frac{1}{2}$ carries all the mass and charge of the solution (this is the black dashed line above the red triangle in Fig.~\ref{fig:2D_single}). On the opposite extremum configuration, the soliton constituent is absent and all the mass and charge of the hairy BH is carried by the BCS component. This describes the hairy BH that merges with the BCS BH at the onset of the linear instability of the latter (this is the blue solid line in Fig.~\ref{fig:2D_single}). At leading order in $\{E,Q\}$ (i.e. in $r_+^2$) this is given by the upper bound in \eqref{singleQ:mixEbounds}.  
So this is indeed a very good approximation to the exact phase diagram shown in Fig.~\ref{fig:2D_single}.

\subsubsection*{Basic Setup for Perturbation Theory}

In this subsection, we construct the hairy black hole of the single charge truncation using a matched asymptotic expansion procedure that is a double expansion perturbation theory in the charged scalar condensate amplitude $\epsilon$ and on the adimensional horizon radius $r_+/L$. 

As with the perturbative construction of the soliton, we start by expanding all the fields of the theory in a power series in the charged scalar condensate $\e$,
\begin{equation}
\begin{split}\label{hbh-singlecharge-exp}
f(r,\e,r_+) &= \sum_{n=0}^\infty \e^{2n} f_{(2n)}(r,r_+)  , \\
h(r,\e,r_+) &= \sum_{n=0}^\infty \e^{2n} h_{(2n)}(r,r_+)  , \\
A_t(r,\e,r_+) &= \sum_{n=0}^\infty \e^{2n} A_{(2n)}(r,r_+) , \\
\varphi(r,\e,r_+) &= \sum_{n=0}^\infty \e^{2n} \varphi_{(2n)}(r,r_+) , \\
\Phi(r,\e,r_+) &= \sum_{n=0}^\infty \e^{2n+1} \Phi_{(2n+1)}(r,r_+) ,
\end{split}
\end{equation}
where the leading order solution is now the single charge BCS black hole of Section~\ref{sec:SingleQ-BCS},
\begin{equation}
\begin{split}\label{SingleQ:bhsol}
f_\0(r,r_+) &= \left( 1 - \frac{r_+^2}{r^2} \right) [ r^2 + 1 + 2{\tilde q}(\e,r_+) +  r_+^2 ] , \\
h_\0(r,r_+) &= \exp [ - 3 \varphi_\0(r,r_+) ] = 1 + \frac{2 {\tilde q}(\e,r_+) }{ r^2 } , \\
A_\0(r,r_+) &= \sqrt{ \frac{ 2 {\tilde q}(\e,r_+) ( 1 + r_+^2 ) }{ 2 {\tilde q} (\e,r_+)+ r_+^2 } } \left( 1 - \frac{2 {\tilde q}(\e,r_+) + r_+^2}{ 2 {\tilde q} (\e,r_+) + r^2} \right) . 
\end{split}
\end{equation}
The parameter ${\tilde q}$, defined in \eqref{BCS:singleQ-def-q}, is essentially related to the energy, charge and chemical potential of the solution; see \eqref{BCSsingleQ:EQ}. Thus, we expect that it also receives corrections as we climb the perturbation ladder. Therefore, we should also expand it in powers of $\e$,
\begin{equation}\label{hbh-singlecharge-exp2}
\begin{split}
{\tilde q}(\e,r_+) = \sum_{n=0}^\infty \e^{2n} {\tilde q}_{(2n)}(r_+) \,. 
\end{split}
\end{equation}

We must define precisely the expansion parameter $\e$. As for the soliton, we take it to be defined by the boundary condition
\begin{equation}\label{singleQ-pertBH:AdS-BC}
\begin{split}
\Phi(r,\e,r_+) = \frac{\e}{r^2} + \CO(r^{-4}) \,. 
\end{split}
\end{equation}
We substitute \eqref{hbh-singlecharge-exp} into the equations \eqref{singleQ:h2eq}--\eqref{singleQ:Phi2eq} and expand in a power series in $\e$. At each order in $\e$, we obtain linear differential equations for each of the coefficient functions in \eqref{hbh-singlecharge-exp}. Given the complicated structure of the base solution \eqref{SingleQ:bhsol}, these differential equations cannot be solved exactly. One might hope, however, that, at each order in $\epsilon$, they can be solved if we further do a power series expansion in $r_+$ (as done in \cite{Basu:2010uz}).

\begin{figure}[th]
\centering \includegraphics[width=0.6\textwidth]{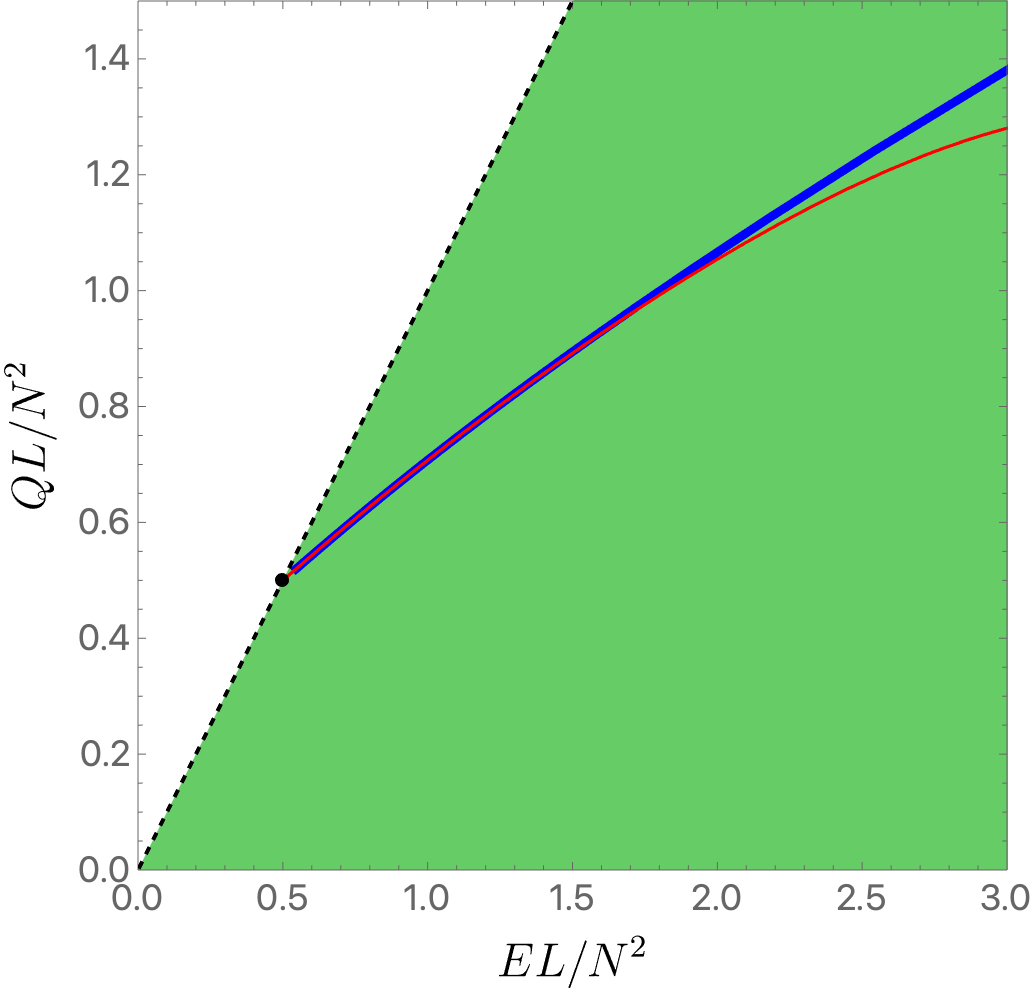}
\caption{\label{fig:comp_single} Comparing the analytic approximation given by~\eqref{eq:ana_sing} (represented as a solid red line) with the numerical output of Section \ref{sec:OnsetSingleQ}, given by the blue disks.}
\end{figure}

There is, however, an issue that arises immediately. The non-interacting model discussed earlier in this section clearly suggests that at leading order in $\e$ and $r_+$ (i.e. as $\e,r_+ \to 0$), we expect that ${\tilde q} \to \frac{1}{2}$; see \eqref{singleQ:BHthermo}. In this limit, the background BCS black hole solution \eqref{SingleQ:bhsol} reduces to the singular soliton solution described in \eqref{SingleQ:SingularSoliton} with $C_1 = 4 {\tilde q}({\tilde q}+1) \to 3$ and $C_2 = 2 ( 2 {\tilde q} + 1 ) \to 4$. It follows that the proposed double perturbative expansion in $\epsilon$ and $r_+$ is actually an expansion around a {\it singular} soliton solution. Perturbative expansions around singular backgrounds are typically ill-defined and this is also the case here. Indeed, the first signal of this problem arises at $\CO(\e^2r_+^4)$. The precise nature of the problem is discussed in Appendix \ref{singleQ:HBHsecondorderperthy}. Given these intricacies in the perturbative construction, we will limit ourselves here to perturbation theory at $\CO(\e)$. This will give us the linearized correction (in $\e$) to the thermodynamics and we find a very good fit with the numerical results displayed in Figs.~\ref{fig:2D_single}--\ref{fig:comp_sing}; see, in particular, the later Fig.~\ref{fig:comp_single}. We leave a complete analysis of the solution for future work. It is perhaps worth noting that the complications described here arise strictly for static singly charged solutions. Rotating single-charge supersymmetric solitons are perfectly regular at $r=0$ and the corresponding hairy black holes can be constructed in the usual way. The issues are also non-existent in the two-charge case discussed in section \ref{sec:PerturbativeTwoQ} where the hairy BH can be constructed to all orders in perturbation theory without issue.

At $\CO(\e)$, the only non-trivial equation is \eqref{singleQ:Phi2eq} (i.e. the backreaction of the charged scalar field on the other fields only kicks in at higher order in $\epsilon$) and this implies a second order differential equation for $\Phi_\1(r)$,
\begin{equation}
\begin{split}\label{SingleQ:scalareq1}
0&= (2 \tilde{q}_0(r_+)+r^2 ) r (r^2-r^2_+)  (2 \tilde{q}_0(r_+)+r_+^2+r^2+1 ) \Phi _\1''(r,r_+) \\
&~ + (2 \tilde{q}_0(r_+)+r^2 ) [ r^2  (6 \tilde{q}_0(r_+)+3 )-r_+^2  (2 \tilde{q}_0(r_+)+r_+^2+1)+5 r^4] \Phi_\1'(r,r_+) \\
&~ + \left[ 4  r^3  \left(4 \tilde{q}_0(r_+)+r^2\right) + \frac{8 (r^2-r^2_+) (r_+^2+1 ) r^3 \tilde{q}_0(r_+)  }{ (2 \tilde{q}_0(r_+)+r_+^2 ) (2\tilde{q}_0(r_+)+r_+^2+r^2+1 )} \right] \Phi_\1(r,r_+)\,,
\end{split}
\end{equation}
where $'$ denotes derivative w.r.t. $r$. As mentioned previously, this equation cannot be solved exactly. Solutions might however be constructed by expanding the scalar field further in $r^2_+$ as\footnote{The equation \eqref{SingleQ:scalareq1} is an analytic function of $r_+^2$ so it is clear that the perturbative expansion is one in $r_+^2$ and not $r_+$.}

\begin{equation}\label{hbh-singlecharge-exp3}
\begin{split}
\Phi_\1(r,r_+) = \sum_{n=0}^\infty r_+^{2n} \Phi_{(1,2n)}(r) , \qquad \tilde{q}_0(r_+) = \sum_{n=0}^\infty r_+^{2n} \tilde{q}_{(0,2n)}.
\end{split}
\end{equation}
Plugging this back into \eqref{SingleQ:scalareq1}, we find ODEs order-by-order in $r_+$ that we must solve for $\Phi_{(1,2n)}(r)$ and $\tilde{q}_{(0,2n)}$. However, typically we cannot solve these ODEs analytically unless we resort to a \emph{matched asymptotic expansion}, whereby we divide the outer domain of communications of our black hole into two regions. Restoring factors of $L$ for a moment, this is a \emph{near-field region} where $r_+\leq r\ll L$ (where we impose the horizon boundary condition), and a \emph{far-field region} where $r\gg r_+$ (and we impose the asymptotic boundary condition). Restricting the analysis to small black holes that have $r_+/L \ll 1$ (which is certainly the case since this quantity is one of our expansion parameters), the two regions then have an \emph{overlapping zone}, $r_+ \ll r\ll L$. In this overlapping region, we can match/relate the set of independent parameters that are generated by solving the perturbative ODEs in each of the two regions and that were not yet fixed by the two boundary conditions. (Onwards we set again $L\equiv 1$).

In this matched asymptotic expansion context, consider first the \emph{far-field region}, $r\gg r_+$. At $\CO(r_+^{2n})$, the equation for $\Phi_{(1,2n)}(r)$ and $\tilde{q}_{(0,2n)}$ takes the form\footnote{\label{foot:SingleQpert}The initial equation is written in terms of ${\tilde q}_{(0,0)}$ in \eqref{app:singleQ-ODE-0} or \eqref{app:singleQ-ODE-2} but after the matching asymptotic analysis at $\CO(r_+^2)$ we find that ${\tilde q}_{(0,0)} = \frac{1}{2}$ $-$ see \eqref{app:singleQ-ODE-q00} $-$ thus yielding the ODE \eqref{scalar-gen-eq-far} at any order in $r_+$.}
\begin{equation}
\begin{split}\label{scalar-gen-eq-far}
\td{}{r}  \left[ \frac{r^3}{2+r^2} \td{}{r} [ ( 2 + r^2 ) \Phi_{(1,2n)}(r) ] \right] = \mfs_{(1,2n)}^{\Phi}(r)\,,
\end{split}
\end{equation}
where, as stated before, the source $\mfs_{(1,2n)}^{\Phi}(r)$ is a known function of the solutions at lower orders $\CO(r_+^k)$, with $k\leq 2n-2$. This ODE is easily integrated at any order and the solution for $\Phi_{(1,2n)}(r)$ can be obtained up to two integration constants; typically, one which is fixed by AdS boundary condition \eqref{singleQ-pertBH:AdS-BC}, alike in the soliton construction, and the other by the matching procedure (if the boundary condition does not fix it also). For the latter, one needs to analyse the small $r$ behaviour of the far-field solution. It turns out that for small $r$, the scalar field $ \Phi_{(1,2n)}(r)$ diverges as a power of $\frac{r_+}{r}$. This indicates that the far-field analysis breaks down at $r\sim r_+$, which justifies why it is valid only for $r\gg r_+$.  It also follows from this observation that in the far-field region we can safely do a Taylor expansion in the expansion parameters $r_+ \ll 1$ and $\epsilon\ll 1$ since the large hierarchy of scales between the solution parameters $\{r_+,\epsilon\}$ and the distance $r$ guarantees that they do not compete.

Let us now move down to the near-field region, $r_+\leq r\ll L\equiv 1$.
This time, we should proceed with some caution when doing the Taylor expansion in $r_+ \ll 1$ and $\epsilon \ll 1$ since these small expansion parameters can now be of similar order as the radius $r$. This is closely connected with the fact that the far-field solution breaks down when  $r/r_+\sim \mathcal{O}(1)$. This suggests that, to proceed with the near-field analysis, we should define a new radial coordinate as
\begin{equation}\label{SingleQ:pertCoord-z}
z=\frac{r}{r_+}\,.
\end{equation}
The near-field region now corresponds to $1 \leq z \ll \frac{1}{r_+}$. If we further require that $r_+\ll 1$ (as we are doing in our double expansion) one sees that the near-field region corresponds to $z\geq 1 \gg r_+$ (and $z\gg \epsilon$). In particular, we can now safely do Taylor expansions in $r_+ \ll 1$ and $\varepsilon \ll 1$ since the radial coordinate $z$ and the black hole parameters $\{r_+,\epsilon\}$ have a large hierarchy of scales. At the heart of the matched asymptotic expansion procedure, note that a factor of $r_+$ (one of the expansion parameters) is absorbed in the new coordinate $z$!

To proceed with the near-field analysis further redefine the wavefunction as (onwards, we use the superscript $^\near$ to represent a near-field quantity) 
\begin{equation}
\begin{split}\label{SingleQ:scalarnear1}
\Phi_\1^\near ( z , r_+ ) \equiv \Phi_\1 ( z r_+ , r_+ ) . 
\end{split}
\end{equation}
The near-field expansion is performed by expanding $\Phi_\1^\near$ in a power series in $r_+^2$,
\begin{equation}
\begin{split}\label{SingleQ:scalarnear1exp}
\Phi_\1^\near ( z , r_+ ) = \sum_{n=0}^\infty r_+^{2n} \Phi_{(1,2n)}^\near ( z )  . 
\end{split}
\end{equation}
We now plug in \eqref{SingleQ:scalarnear1} and \eqref{SingleQ:scalarnear1exp} into \eqref{SingleQ:scalareq1} and then extract the equations order-by-order in $r_+^2$. At $\CO(r_+^{2n})$, the equation takes the form
\begin{equation}
\begin{split}\label{scalar-gen-eq-near}
\td{}{z} \left[ z ( z^2 - 1 ) \td{}{z} \Phi_{(1,2n)}^\near(z) \right] = \mfs_{(1,2n)}^{\Phi,\near}(z) .
\end{split}
\end{equation}
This ODE can easily be integrated and the solution for $\Phi_{(1,2n)}^\near(z)$ is fixed up to two integration constants. One of the integration constants is fixed by requiring regularity at the horizon $z=1$, and the other is fixed by matching the near-field solution to the far-field one as follows. At large $z$, the scalar field $\Phi_{(1,2n)}^\near(z)$ blows up as $\CO\left(z^{2n}\right)$. Consequently, the near-field expansion breaks down when $\CO(z)\sim \CO(r_+^{-1})$ and thus it is valid only when $1 \leq z \ll \frac{1}{r_+}$ or equivalently $r_+ \leq r \ll 1$, as we have claimed at the begin of our matched asymptotic analysis.

Since our expansion parameter satisfies $r_+ \ll L\equiv 1$,  there is a overlapping region $r_+ \ll r \ll L\equiv 1$ where both the near-field and far-field solutions are equally valid. Any near-field and far-field integration constants that were not yet determined by the horizon and the asymptotic boundary conditions are now fixed by the matching procedure of the near and far wavefunctions in this overlapping region. Typically, this matching procedure in the overlapping region also fixes ${\tilde q}_{(0,2n-2)}$.

The details of the perturbative construction including the matching process is described in Appendix \ref{app:singleQ:hbhpertthy} up to $\CO(r_+^2)$. Moreover, explicit results to $\CO(r_+^{10})$ are presented in the accompanying {\tt Mathematica} file. In the end of the day, we find that
\begin{equation}
\begin{split}
{\tilde q} &= \frac{1}{2} + \frac{r_+^2}{4} + r_+^4 \left( - \frac{3}{16} - \frac{1}{4} \ln \frac{r_+^2}{2}  \right) + r_+^6 \left(  \frac{7}{16} + \frac{5}{16} \ln \frac{r_+^2}{2} + \frac{1}{8} \ln^2  \frac{r_+^2}{2} \right)  \\
&\qquad + r_+^8 \left( - \frac{\zeta(3)}{8} - \frac{203}{256} - \frac{29}{32} \ln \frac{r_+^2}{2} - \frac{5}{16} \ln^2 \frac{r_+^2}{2} - \frac{1}{24} \ln^3 \frac{r_+^2}{2} \right) + \CO(r_+^{10}). 
\end{split}
\end{equation}
The thermodynamics of the hairy black hole at leading order in the charged scalar condensate is obtained by substituting the above expansion for ${\tilde q}$ into the thermodynamic quantities of the BCS black hole \eqref{BCSsingleQ:EQ} and \eqref{BCSsingleQ:TS},
\begin{equation}
\begin{split}
\frac{EL}{N^2}  &= \frac{1}{2} + \frac{7 r_+^2}{4L^2} + \frac{r_+^4}{16L^4} \left( -4 \ln \frac{r_+^2}{2L^2}+15 \right)+\frac{r_+^6}{32L^6} \left( 4 \ln ^2\frac{r_+^2}{2L^2}-2 \ln \frac{r_+^2}{2L^2} + 5 \right) \\
&\qquad +\frac{r_+^8 }{768L^8}\left(-32 \ln^3\frac{r_+^2}{2L^2} -96 \ln ^2\frac{r_+^2}{2L^2} -336 \ln \frac{r_+^2}{2L^2} -96 \zeta(3)-105 \right) \\
&\qquad + \CO\left( \frac{r_+^{10}}{L^{10}} \right) , \\
\frac{QL}{N^2} &= \frac{1}{2}+\frac{3 r_+^2}{4L^2}+\frac{r_+^4}{16L^4} \left( -4 \ln \frac{r_+^2}{2L^2}-1\right)+\frac{r_+^6}{32L^6}  \left(4 \ln ^2\frac{r_+^2}{2L^2}+6 \ln \frac{r_+^2}{2L^2}+11\right) \\
&\qquad +\frac{r_+^8}{768L^8}  \left(-32 \ln ^3\frac{r_+^2}{2L^2}-192 \ln ^2\frac{r_+^2}{2L^2}-576 \ln \frac{r_+^2}{2L^2}-96 \zeta(3)-453\right) \\
&\qquad + \CO\left( \frac{r_+^{10}}{L^{10}} \right) , \\
\mu &= 1+\frac{r_+^4}{4L^4}+\frac{r_+^6}{16L^6} \left( -4 \ln \frac{r_+^2}{2L^2}-9\right)\\
&\qquad +\frac{r_+^8}{32L^8} \left(4 \ln ^2\frac{r_+^2}{2L^2}+26 \ln \frac{r_+^2}{2L^2}+43\right)+ \CO\left( \frac{r_+^{10}}{L^{10}} \right) , \\
\frac{S}{N^2} &= \frac{\pi r_+^2}{L^2} + \frac{3 \pi  r_+^4}{4L^4} -\frac{ \pi r_+^6 }{32L^6} \left(8 \ln \frac{r_+^2}{2L^2}+15 \right) \\
&\qquad +\frac{\pi  r_+^8}{128L^8}  \left(16 \ln ^2\frac{r_+^2}{2L^2}+64 \ln \frac{r_+^2}{2L^2}+101\right)+ \CO\left( \frac{r_+^{10}}{L^{10}} \right) , \\
TL &= \frac{1}{\pi} +\frac{r_+^2}{2 \pi  L^2}-\frac{3 r_+^4}{32 \pi  L^4}-\frac{r_+^6}{64 \pi  L^6}  \left(4  \ln \frac{r_+^2}{2 L^2}+3\right) \\
&\qquad +\frac{r_+^8}{2048 \pi  L^8} \left( 128 \ln^2 \frac{r_+^2}{2 L^2}  + 544 \ln \frac{r_+^2}{2 L^2} + 611\right)+ \CO\left( \frac{r_+^{10}}{L^{10}} \right),  
\end{split}
\label{eq:ana_sing}
\end{equation}
where we have reinstated the AdS$_5$ radius $L$ in this final result. It is easy to verify that these quantities satisfy the first law of thermodynamics, $\mathrm{d}E=T \mathrm{d}S+\mu \mathrm{d}Q$, for the single charge system. 

Recall that the thermodynamics \eqref{eq:ana_sing} captures only the $\CO(\epsilon)$ contributions. Therefore, it should be a good approximation when the scalar condensate is small, i.e. in the region where the hairy black hole merges with the single charge BCS black hole. This occurs at the onset of the scalar condensation instability of the BCS black hole.
In Fig.~\ref{fig:comp_single}, we confirm that \eqref{eq:ana_sing} is indeed a good approximation. In this plot, we compare the the analytic approximation \eqref{eq:ana_sing}  for the onset (depicted as a red solid line) with the output of numerical procedure outlined in Section~\ref{sec:OnsetSingleQ} (blue disks). The match for small $r_+/L$ is reassuring and is a consistency test for both the numerical and matched asymptotic expansion analyses.

In addition to matching the numerical results, the expressions \eqref{eq:ana_sing} are also consistent with the non-interacting model of \ref{singleQ:nonintmodel} at leading order in $r_+$. Using \eqref{eq:ana_sing}, the RHS of \eqref{singleQ:rpEsolnonintmix} evaluates to
\begin{equation}
\begin{split}
\frac{2L^2}{3} \left( \left[ 1 + 3 \left( \frac{EL-QL}{N^2}  \right) \right]^{\frac{1}{2}}- 1 \right) = r_+^2 + \frac{r_+^4}{4L^2} - \frac{r_+^6}{16L^4} \left(9 + 4 \ln \frac{r_+^2}{2L^2} \right) + \CO \left( \frac{r_+^8}{L^6} \right)
\end{split}
\end{equation}
which precisely matches the non-interacting model \eqref{singleQ:rpEsolnonintmix} to leading order in $r_+$. Using \eqref{singleQ:rpEsolnonintmix}, we can also determine the mass of the BCS black hole in the non-interacting mix as
\begin{equation}
    \begin{split}
        E_{BCS} L &= E L - E_{sol} L  \\
        &= EL - QL + \frac{N^2}{6L} \left( 2  \left[ 1 + 3  \left( \frac{EL-QL}{N^2}  \right)  \right]^{\frac{1}{2}} + 1 \right) \\
        &= \frac{1}{2} + \frac{3r_+^2}{2L^2} + \frac{9r_+^4}{8L^4} - \frac{3r_+^6}{32L^6} \left( 5 + \ln \frac{r_+^2}{2L^2} \right) + \CO \left( \frac{r_+^8}{L^8} \right) .
    \end{split}
\end{equation}
This also matches -- to leading order in $r_+$ -- the energy of the ``bald'' BCS black hole \eqref{BCSsingleQ:EQ} at ${\tilde q}=\frac{1}{2}$.

\section{Consistent truncation with $ A^1=A^2\equiv A, A^3\equiv 0$}\label{sec:TwoQ}

\subsection{Setup the problem: Ansatz\"e and boundary conditions}\label{sec:AnsatzTwoQ}

We will denote this theory with  $A^1=A^2\equiv A, A^3\equiv 0$ and  $\Phi_1=\Phi_2\equiv \Phi, \Phi_3\equiv 0$ as the truncation with two equal charges.  Again motivated by the ansatz  \eqref{BCS:ansatz} we used for the BCS black hole, to find the static and spherically symmetric hairy solutions of this sector, we find convenient to use this time the ansatz: 
\begin{equation} \label{TwoQ:ansatz}
\begin{split}
\dt s^2&=h^{2/3}\left(-\frac{f}{h^2}\dt t^2+  \frac{\dt r^2}{g}+ r^2 \dt \Omega_3^2\right);\\
\varphi_1&=\sqrt{\frac{2}{3}}\ln h\,,\qquad \varphi_2=0;\\
A^1&=A_t \dt t\ \,,\qquad A^2= A_t \dt t\ \,,\qquad  A^3=0;\\
\Phi_1&=\Phi_1^\dagger= \Phi\,,\qquad \Phi_2= \Phi_2^\dagger= \Phi,\qquad \Phi_3=0 \,;\\
\end{split} 
\end{equation}
where $\dt \Omega_3^2$ is again the line element of a unit radius $S^3$ and we have selected the gauge where $h^{2/3} r^2$ measures the radius of the $S^3$. Moreover, we have fixed the $U(1)$ gauge freedom by taking $\Phi_1=\Phi_2= \Phi $ to be real, which implies that $A_t=0$ at the horizon location, $r=r_+$. Inserting this ansatz into the field equations \eqref{GravEOM}--\eqref{ScalarsGaugeEOM} we find that the system closes if the following five equations for $\{ h(r),f(r),g(r),A_t(r),\Phi(r)\}$ are satisfied:
\begin{equation} \label{TwoQ:EoM}
\begin{split}
0=& L^2 r g h \left(\Phi ^2+4\right) \left(r h'+3 h\right)f' 
+(\Phi^2+4)r^2  h^4 \left(L^2 g \left(A_t'\right)^2-A_t^2 \Phi^2\right)
\\
&-f  \left(\Phi ^2+4\right) \left[L^2 r^2  g\left(h'\right)^2+h^2 \left(6 L^2+4 r^2\right)\right]  \\
& +f h^2\left[ L^2  g\left(24-r^2 \left(\Phi '\right)^2+6 \Phi ^2\right)-4 h r^2 \left(\Phi ^2+4\right)^{3/2}\right],\\
0=&
L^2 r f h \left(\Phi ^2+4\right) \left(r h'+3 h\right)g' 
 +r^2  h^4 \left(\Phi ^2+4\right) \left[A_t^2 \Phi ^2-L^2 g \left(A_t'\right)^2\right]\\
&+ f \left(\Phi ^2+4\right) \left[ L^2 r^2 g \left(h'\right)^2+4 L^2 r (g-1) h h'+h^2 \left(4 r^2-6 L^2-4 r^3 \sqrt{\Phi ^2+4} \,h'\right)\right]\\
& +f h^2 \left[L^2 g\left(r^2 \left(\Phi '\right)^2+6 \Phi ^2+24\right)-8  r^2  h \left(\Phi ^2+4\right)^{3/2}\right],
\\
0=&h''-\frac{\left(h'\right)^2}{h}+\frac{h'}{L^2 r g} \left(L^2(g+2) +2 h r^2 \sqrt{\Phi ^2+4}\right)
+ \frac{2 h}{L^2 g}\left(h \sqrt{\Phi ^2+4}-2\right) +\frac{h^3 \left(A_t'\right)^2}{f} \,,\\
0=& L^2 r f g h \left(\Phi ^2+4\right) \left(r h'+3 h\right) A_t'' +L^2 r^2 g h^4 \left(\Phi ^2+4\right) \left(A_t' \right)^3-r A_t f h \Phi ^2 \left(\Phi ^2+4\right) \left(r h'+3 h\right)\\
& +A_t'\bigg\{ 
r \left(\Phi ^2+4\right)
 \bigg[f \left(L^2 r g \left(h'\right)^2+L^2 (7 g+2) h h'+2 h^2 r \left(r \sqrt{\Phi ^2+4}\, h'-2\right)\right)-r A_t^2 h^4 \Phi ^2 \bigg]\\
& + f h^2 \bigg[ L^2 g \left(36+9 \Phi ^2-r^2 \left(\Phi '\right)^2\right)+2 h r^2 \left(\Phi ^2+4\right)^{3/2} \bigg]
\bigg\}, \\
0= & L^2 g f \Phi'' -\frac{ L^2  g f \Phi  \left(\Phi '\right)^2}{\Phi ^2+4}  
+\left( \frac{L^2}{r}\,f(g+2)+2 r f h \sqrt{\Phi ^2+4} \right)\Phi' \\
  & +h  \left(A_t^2 h \left(\Phi ^2+4\right)+2 f \sqrt{\Phi ^2+4}\right)\Phi. 
\end{split} 
\end{equation}
This is a system of two first order ODEs for $\{f',g'\}$ plus three second order ODEs for $\{h'',A_t'',\Phi''\}$ very similar to the one discussed in the case \eqref{SingleQ:EoM} of the previous section.
As before we want to impose boundary conditions such that the solutions are asymptotically AdS$_5$ 
with normalizable fields. In particular, this requires that we set the sources of $\varphi_1$ and $\Phi$  to zero. After imposing these UV boundary conditions, a Frobenius analysis still yields the asymptotic expansion displayed in \eqref{SingleQ:expansionUV}, with the same 5 free UV parameters $\{h_2,f_2,\mu,\rho,\varepsilon\}$ not fixed by boundary conditions neither by the EoM.
For the same reasons as in Section \ref{sec:AnsatzSingleQ}, we require that at the horizon $r=r_+$, the  functions  $f$, $g$ and $A_t$ vanish linearly.
As for the consistent truncation sector of section \ref{sec:AnsatzSingleQ}, we will solve \eqref{TwoQ:EoM} with the above boundary conditions numerically or within perturbation theory. When solving the ODE system numerically, the above boundary conditions can be imposed efficiently if we introduce the same field redefinitions displayed in \eqref{SingleQ:FieldRedef} and look for solutions $q_j$, ($j=1,2,\cdots 5$) that are everywhere smooth. To find the numerical solutions it is again useful
to introduce the compact coordinate $y$ and dimensionless horizon radius $y_+$ defined in \eqref{SingleQ:compactRadius}. 

The auxiliary fields $q_j$ must satisfy boundary conditions that follow straightforwardly from the ones for the original fields $\{g,f,h,A_t,\Phi\}$ and from the field redefinitions  \eqref{SingleQ:FieldRedef}. 
The boundary conditions at the asymptotic boundary $y=1$ ($r\to \infty$) are the same as those already presented in \eqref{SingleQ:BCq}, where  $Q$ (this time $Q \equiv Q_1=Q_2$) is again the conserved electric charge of the solution (but this time associated to the gauge fields $A^1_\2$ and $A^2_\2$; see \eqref{TwoQ:EQ} later).
As before, the associated boundary condition effectively introduces $Q$ as an input parameter in our numerical code (which will allow us to run lines of constant $Q$).  On the other hand, at the horizon ($y=0$) the derived boundary conditions from the EoM are that 
$q_1$ must obey the Dirichlet $q_1\big|_{y=0}=\frac{1}{y_+^2}+2q_4\left(1+q_5^2\right)\big|_{y=0}$ and $q_{2,3,4,5}$ must obey mixed boundary conditions which are not enlightening to display.

\subsection{Thermodynamic quantities using holographic renormalization}\label{sec:ThermoTwoQ}

To find the thermodynamics of our solutions we implement {\it mutatis mutandis} the holographic renormalization procedure of \cite{Bianchi:2001de,Bianchi:2001kw}, as applied to our theory in Appendix \ref{app:holo-renorm}\footnote{See, in particular, expectations values, their conservation laws and associated anomalies in \eqref{app:VEVcurrent}--\eqref{app:VEVscalars} and \eqref{app:HoloStressTensor}--\eqref{app:TraceHoloStressTensor}.} and already discussed in section~\ref{sec:ThermoSingleQ}. 
The relation between the compact radial coordinate $y$ and the Fefferman-Graham radial coordinate $z$ is this time
 \begin{align}  \label{FGcoord}
y=& 1-y_+^2\frac{z^2}{L^2} - \frac{y_+^2}{6}\Big( 3-4y_+^2 q_4'(1)\Big)\frac{z^4}{L^4}  
- \frac{y_+^2}{144} \bigg[  27-12 y_+^2 \Big(3+2 q_4'(1) \Big) \nonumber \\
&+ 4y_+^4\left(  
\frac{9}{2}\,q_1''(1)+5 q_4'(1)^2+18 q_4'(1)+6 q_5(1)^2-9
\right) \bigg]\frac{z^6}{L^6} 
+\CO(L^{-8}z^8).
\end{align} 

Expanding the $U(1)$ gauge fields in FG coordinates off the boundary $z=0$, we find that the chemical potential $\mu$ and  holographic current $\langle \CJ_{a} \rangle=(\rho,0,\cdots,0)$ (where $\rho$ is the charge density) are given by
\begin{equation}\label{TwoQ:holoCurrent}
\mu= q_3(1)\,,\qquad \rho=\frac{N^2}{4 \pi ^2 L^4} y_+^2 \Big(q_3'(1)+q_3(1)\Big),
\end{equation}
where recall that this is the common source and common VEV of the dual operators of the fields $A^1=A^2\equiv A_t \dt t$.

Similarly, a FG expansion of the scalar fields $\varphi_1$ and $\Phi_1=\Phi_2 \equiv \Phi$ away from the boundary yields the expectation values for the operators dual to these fields,
\begin{eqnarray}\label{TwoQ:FGexpansion2} 
\langle {\cal O}_{\varphi} \rangle &= -\sqrt{\frac{2}{3}} \,\frac{y_+^2}{L^2}\,q_4'(1)\frac{N^2}{\pi^2} \,, \\\label{TwoQ:FGexpansion2b} 
 \langle {\cal O}_{\Phi} \rangle & = 2 \sqrt{2}\,\frac{y_+^2}{L^2}\,q_5(1)\frac{N^2}{\pi^2} \,.\label{TwoQ:FGexpansion2c} 
\end{eqnarray}
Recall that these  scalar fields have mass $m^2 L^2=-4$, \emph{i.e.} they saturate the BF bound in AdS$_5$, and we have set their sources to zero.
   
Finally, the relevant expansion of the gravitational field about the conformal boundary is
\begin{eqnarray}\label{TwoQ:FGexpansion1}
&& \dt s^2 = \frac{L^2}{z^2}\left[ \dt z^2 + \dt s^2_{\partial}+z^2 \,\dt s^2_\2+z^4\, \dt s^2_\4+\CO(z^6)\right]  \\[2mm]
&& \hbox{with}\nonumber\\[2mm]
&& \dt s^2_{\partial}=g_{ab}^\0\dt x^a \dt x^b=-\dt t^2+L^2 \dt \Omega_3^2,  \nonumber\\[2mm]
&& ds^2_\2=g_{ab}^\2\dt x^a \dt x^b=-\frac{1}{2 L^2}\,\left( \dt t^2+L^2 \dt \Omega_3^2
\right)\,,
\nonumber\\[2mm]
&&  ds^2_\4=g_{ab}^\4\dt x^a \dt x^b  \nonumber\\
&& \hspace{0.2cm} = -\frac{1}{144 L^4}
\bigg[ 9-36 y_+^2 \Big(3-4 q_4'(1)\Big)
+4 y_+^4 \left(\frac{27}{2} \left[q_1''(1)-2\right]-29q_4'(1)^2+54q_4'(1)+42q_5(1)^2\right)\bigg]\dt t^2\nonumber\\
&& \hspace{0.2cm}+ 
\frac{1}{144 L^2}
\bigg[ 9+12 y_+^2 \Big(3-4 q_4'(1) \Big)
-4 y_+^4 \left(\frac{9}{2}\,q_1''(1)-7q_4'(1)^2+18q_4'(1)+30q_5(1)^2-9 \right)
\bigg]
\dt \Omega_3^2 \nonumber
\,;
\end{eqnarray}
from which we can compute the expectation value $ \langle \CT_{ab} \rangle$ of the holographic stress tensor using \eqref{SingleQ:holoT}, this time  with
 $\langle {\cal O}_{\Phi_1} \rangle=\langle {\cal O}_{\Phi_2} \rangle\equiv \langle {\cal O}_{\Phi} \rangle$ and $\langle {\cal O}_{\Phi_3} \rangle=0$. We confirm that it is conserved, $\nabla^a\langle \CT_{ab} \rangle=0$, and its 
 trace $\langle \CT_{a}^{\:a} \rangle$ yields the expected Ward identity \eqref{SingleQ:traceAnomaly} associated to the gravitational conformal anomaly.

From \eqref{TwoQ:FGexpansion1} and \eqref{TwoQ:holoCurrent} we can compute the energy $E$ and electric charges $Q_1=Q_2\equiv Q$  of our solutions: 
\begin{eqnarray}\label{TwoQ:EQ}
E&=&\frac{N^2}{L}\,\frac{y_+^2 }{4}\left[ 
 3-4q_4'(1)  +3 y_+^2 \left( 1-\frac{1}{2}\,q_1''(1)-2q_4'(1)+q_4'(1)^2-2q_5(1)^2 \right) \right],
\nonumber\\
Q&=&  \frac{N^2}{L}\,\frac{y_+^2}{2} \Big(q_3'(1)+q_3(1)\Big),
\end{eqnarray}
where in $E$ we have already subtracted the Casimir energy $E_{\hbox{\tiny AdS}_5}= \frac{N^2}{L} \frac{3}{16}$.

The temperature $T$ and the entropy $S$ of the hairy black holes with two equal $U(1)$ charges can be read simply from the surface gravity and the horizon area of the solutions  \eqref{TwoQ:ansatz}, respectively:
\begin{eqnarray}\label{TwoQ:TS}
&& T=\frac{1}{L} \frac{y_+}{2\pi} \frac{q_1(0) q_2(0)} {q_4(0)}\,, \nonumber\\
&& S= N^2 \pi y_+^3 q_4(0)\,.
\end{eqnarray}

These thermodynamic quantities must obey the first law of thermodynamics \eqref{BCS:FirstLaw}
where for the theory of this subsection one has $Q_1=Q_2\equiv Q$, $\mu_1=\mu_2\equiv \mu$ and $Q_3=0$, $\mu_3=0$ and thus it reads $$\dt E= T \dt S +2\mu \dt Q.$$

From \eqref{TwoQ:EQ}--\eqref{TwoQ:TS} and \eqref{BCS:HelmoltzGibbs} we can also straightforwardly compute the Gibbs free energy $G=E - T S -2\mu Q$ which is useful to study the grand-canonical ensemble.

\subsection{Hairy supersymmetric solitons}\label{sec:ansatzTwoQ-Solitons}

In this subsection we describe the supersymmetric solitons of the consistent truncation of \eqref{OurCTaction} with two equal charges.
As for the single charge case of Section \ref{sec:ansatzSingleQ-Solitons}, the simplest way to find these solitons is to solve the first order Killing spinor equations. But these solutions are still described by the ansatz\"e \eqref{TwoQ:ansatz} and obey the associated  equations of motion \eqref{TwoQ:EoM}. 

The fields $\{f,g,A_t, \varphi_1,\Phi\}$ of supersymmetric solitons are given by
\begin{equation}\label{TwoQ:SUSYfunctions}
g=f=1+\frac{r^2}{L^2}\,h^2\,, \qquad A_t=\frac{1}{h}\,, \qquad  \varphi_1=\sqrt{\frac{2}{3}}\ln h\,,\qquad \Phi=2\sqrt{\left(h+\frac{1}{2}\,rh'\right)^2-1}\,, 
\end{equation}
\emph{i.e.} they are all a function of $h(r)$ which must solve the ODE
 \begin{equation}\label{TwoQ:solitonODE}
L^2 r\left(1+ \frac{r^2}{L^2}\,h^2\right)h''  +r^3 h \left(h'\right)^2
+\left(7 h r^2+3 L^2\right)h'  -4 r h \left(1-h^2\right)
=0.
\end{equation}
A Frobenius analysis of this ODE at the asymptotic boundary requires that  $h|_{r\rightarrow\infty}=1$ which also ensures that the fields $f$ and $g$ are asymptotically AdS$_5$ and $\varphi_1$ and $\Phi$ are normalizable (\emph{i.e.}, the scalar field sources are zero).  
On the other hand,  assuming that at the origin $h$ behaves as
\begin{equation}\label{TwoQ:solitonOrigin}
h\big|_{r\to 0} = \frac{h_\a}{r^\a}\,, 
\end{equation}
a Frobenius analysis of \eqref{TwoQ:solitonODE} yields two distinct  solutions: $\a=0$ and $\a=2$. The family with $\a=0$ is a regular supersymmetric soliton while the family with $\a=2$ is clearly an irregular supersymmetric soliton. 

Unfortunately, unlike in the single charged case of section \eqref{sec:ansatzSingleQ-Solitons}, it does not seem possible to solve \eqref{TwoQ:solitonODE} analytically. Therefore we resort to a full nonlinear numerical analysis to find the regular soliton with  $\a=0$. The soliton is also constructed perturbatively in section \ref{twoQ:solitonpert}.

We start by analysing the behaviour of $h$ at the origin and asymptotic boundary.
A series expansion of \eqref{TwoQ:solitonODE} about the origin yields 
 \begin{equation}\label{TwoQ:solitonIR}
 h\big|_{r= 0} \simeq \frac{h_2}{r^2}+h_0+\CO\left(r^2\right),
 \end{equation}
where $h_2$ and $h_0$ are two arbitrary constants  and all other coefficients of this expansion are fixed as a function of these two by the EoM. We want the regular soliton so we impose $h_2=0$ as a Dirichlet boundary condition. We are left with a single  IR free parameter $h_0$.
A similar series expansion of \eqref{TwoQ:solitonODE} but this time about the asymptotic boundary yields
 \begin{equation}\label{TwoQ:solitonUV}
 h\big|_{r\to \infty} \simeq 1+ \frac{c_2}{r^2}+ \frac{c_4}{r^4}+\CO\left(r^{-6}\right),
 \end{equation}
where $c_2$ and $c_4$ are the two arbitrary parameters, with all other coefficients of the expansion fixed as a function of these two by the EoM. There is no physical motivation to fix any of these two  parameters with a boundary condition since the solution is asymptoticaly AdS$_5$ and normalizable no matter their value. On way to conclude this is to note that, as it stands, we have 2 free UV parameters $\{c_2, c_4\}$ and 1 free IR parameter $h_0$ (after imposing regularity at origin). This is what we need to have for a 1-parameter soliton family since the difference between the number of UV  and IR free parameters is 1. 
Before further discussions, we find convenient to introduce the following compact coordinate and field redefinition:
\begin{align}\label{TwoQ:RegularSoliton}
\begin{split}
&y=\left(1+\frac{r^2}{L^2}\right)^{-1},\\
&h= 1+\left(1+\frac{r^2}{L^2}\right)^{-1}H = 1+y H\,. \qquad 
\end{split}
\end{align}
Now, the asymptotic AdS$_5$ boundary is at $y=0$ and the origin $r=0$ is at $y=1$. 
To justify these choices first note the the behaviour \eqref{TwoQ:solitonIR} of $h$ at the origin translates to $H\big|_{y= 1} \simeq \frac{h_2}{L^2(1-y)}+h_0-1+\cdots$. The numerical code can only capture smooth functions $H$ so the boundary condition $h_2=0$ that kills the divergence is automatically implemented. Next, note that the factor of 1 in the redefinition of $h$ absorbs the leading behaviour in \eqref{TwoQ:solitonUV}, and the UV expansion \eqref{TwoQ:solitonUV} for $h(r)$ translates into the following UV expansion for $H(y)$:
 \begin{equation}\label{TwoQ:solitonUVy}
 H\big|_{y\to 0} \simeq \eta_0 +\eta_1 y +\CO\left(y^{2}\right)
 \end{equation}
with the information of the free UV parameters $\{c_2,c_4\}$  of \eqref{TwoQ:solitonUV} now effectively transferred into $\{\eta_0,\eta_1\}$. The latter have the advantage that they can be read straightforwardly by evaluating $H(y)$ and $H'(y)$ at $y=0$, respectively.

The strategy to find the 1-parameter soliton is now much clear. We  have a well-posed elliptic problem if at the origin $y=1$ we impose a derived (because it follows directly from the EoM) mixed boundary condition. On the other hand, at $y=0$ we can impose a inhomogeneous Dirichlet boundary condition where we give the value $\eta_0$ of the function $H$.  
Concretely, the boundary conditions for the boundary-value problem are:
\begin{align}\label{TwoQ:RegularSolitonBCs}
\begin{split}
& H(0)=\eta_0\,,\\
& H'(1)=\frac{1}{2}H(1)^2 \Big(3+H(1)\Big)\,.
\end{split}
\end{align}
$\eta_0$ is an input parameter in our numerical code that fixes $H$ at $y=0$, and we let the EoM evolve subject to the IR condition to find the function $H$. We can then read the second UV parameter $\eta_1 =H'(0)$.
Then, we repeat the process, \emph{i.e.} we run the numerical code for several values of $\eta_0$ since this is the quantity that parametrizes the soliton. 

Finally, we can reconstruct the other functions from \eqref{TwoQ:SUSYfunctions} and compute the relevant thermodynamic quantities using holographic renormalization \cite{Bianchi:2001de,Bianchi:2001kw}. 
In the end of the day we  find that the energy $E$ (after removing the Casimir energy), electric charges $Q_K$, the chemical potential $\mu$, charge density $\rho$, and expectation values of the operators dual to $\varphi_1$ and $\Phi$ are, respectively, given by
\begin{align}\label{TwoQ:RegularSolitonHoloQuantities}
\begin{split}
& E=\frac{N^2}{L} \eta_0\,,\qquad  Q_1=Q_2=\frac{1}{2}\,\frac{N^2}{L} \eta_0\, \qquad Q_3=0\,;\\
& \mu_1=\mu_2=1, \quad \mu_3=0\,;\qquad \rho_1= \rho_2=\frac{N^2}{4\pi^2L^4}\,\eta_0\,,\quad  \rho_3=0;\qquad \\
& \langle \CO_{\varphi_1} \rangle=\frac{1}{L^2}\sqrt{\frac{2}{3}}\eta_0\frac{N^2}{\pi^2}\,;\qquad  
\langle \CO_{\varphi_2} \rangle=0\,;\qquad  \\
& \langle \CO_{\Phi_1} \rangle=\langle \CO_{\Phi_2} \rangle= \frac{2\sqrt{2}}{L^2}\,\sqrt{\eta_0-H'(0)}\frac{N^2}{\pi^2}\,, \qquad  \langle \CO_{\Phi_3} \rangle=0 \,;\\
\end{split}
\end{align}
where recall that we have set to zero the sources of the operators dual to $\varphi_1$ and $\Phi_K$.
Note that, as expected for our supersymmetric solution,  $E=Q_1+Q_2$, $\mu_1=\mu_2=1$ and the soliton satisfies the first law $\dt E=\sum_{K=1}^{3}\mu_K \dt Q_K=2\dt Q$.

The singular 2-parameter soliton (with $\a=2$) plays no role on the discussion of the hairy black hole solutions of the theory. Therefore we do not attempt to find it.

\subsection{Behrndt-Cveti\v c-Sabra black holes with  $A^1=A^2\equiv A, A^3\equiv 0$}\label{sec:TwoQ-BCS}

 The most general static Behrndt-Cveti\v c-Sabra black hole  \eqref{BCS:ansatz}  with three different charges $Q_K$ $(K=1,2,3)$ was presented in Section \ref{sec:BCS-BHs}. Here, we consider its special two charged case with $Q_1=Q_2\equiv Q$ and $Q_3=0$, which is a solution of the consistent truncation \eqref{OurCTaction} with $A^1=A^2\equiv A, A^3\equiv 0$ and no charged condensate, $\Phi_K=0$ $(K=1,2,3)$. 
The reason why we revisit this solution is because for our physical discussions of the hairy black holes with $\Phi_1=\Phi_2\equiv \Phi, \Phi_3=0$ it will be useful to display the two charged BCS black hole using the ansatz \eqref{SingleQ:ansatz} with the field redefinitions \eqref{SingleQ:compactRadius} and compact radial coordinate \eqref{SingleQ:FieldRedef}, \emph{i.e.} to present the auxiliary fields $q_j$ $(j=1,\cdots,5)$ for BCS. It will be useful because the hairy black hole family ultimately bifurcates from the BCS family at the onset of the scalar condensation instability. 

To study the properties of BCS black holes of the truncation with two charges set 
\begin{equation}\label{BCS:TwoQ-def-q}
\sinh \delta_1=\sinh \delta_2\equiv \frac{2 q}{r_0}\,, \qquad
\delta_3=0
\end{equation}
 in  \eqref{BCS:ansatz}, where $r_0$ is a function of the horizon radius $r_+$ and the charge parameter $q$ that follows from \eqref{BCS:def-r0}: $r_0=\frac{1}{L}\sqrt{(r_+^2+2q)^2+L^2 r_+^2}$. Further introduce $y_+=r_+/L$ and $\tilde{q}=q /L^2$ and choose a $U(1)$ gauge such that $A^1=A^2\equiv A$ vanish at the horizon $\mathcal{H}$. In these conditions the BCS black hole with  two equal charges sourced by $A^1=A^2\equiv A$   is described $-$ through \eqref{SingleQ:ansatz}, \eqref{SingleQ:compactRadius}  and \eqref{SingleQ:FieldRedef} $-$  by the functions 
\begin{align}\label{BCS:TwoQ}
\begin{split}
& q_1= 1+(1-y)\frac{y_+^2+4 \tilde{q}+1}{y_+^2}\,, \qquad 
q_2 = 1\,, \qquad 
q_3= \frac{\sqrt{2} \sqrt{\tilde{q}} \, y_+^2 \sqrt{y_+^2+2 \tilde{q}+1}}
{\left[y_+^2+2 \tilde{q} (1-y)\right]\sqrt{y_+^2+2 \tilde{q}} }\,, 
\\  
 & q_4 =1+ (1-y)\frac{2 \tilde{q}}{y_+^2}\,, \qquad q_5=0\,.
\end{split}
\end{align}
Using \eqref{BCS:TwoQ}, the thermodynamic quantities for the BCS black hole with two equal charges can now be read straightforwardly from \eqref{TwoQ:FGexpansion2b}--\eqref{TwoQ:holoCurrent} and \eqref{TwoQ:EQ} determined in the previous subsection. We conclude that the energy (after subtracting the Casimir energy), electric charges, chemical potentials and expectation values of the scalar fields are given by:
\begin{eqnarray}\label{BCStwoQ:EQ}
E&=&\frac{N^2}{L}\,\frac{1}{4}\Big[3y_+^2(1+y_+^2)+ 4 \tilde{q} \left(2+3y_+^2\right)+12 \tilde{q}^2 \Big],
\nonumber\\
Q_1&=&Q_2=\frac{N^2}{L}\,\frac{\sqrt{\tilde{q}}}{\sqrt{2}}\, \sqrt{y_+^2+2 \tilde{q}+1} \,\sqrt{y_+^2+2 \tilde{q}}\,,\qquad Q_3= 0\,;\\
 \mu_1&=&\mu_2=\frac{\sqrt{2} \sqrt{ \tilde{q} } \sqrt{y_+^2+2 \tilde{q}+1}}{\sqrt{y_+^2+2  \tilde{q}}}\,, \qquad \mu_3=0\,;\\
\langle \CO_{\varphi_1} \rangle &=& \frac{2}{L^2}\sqrt{\frac{2}{3}} \frac{N^2}{\pi^2}\, \tilde{q}\,;\qquad  \langle \CO_{\varphi_2} \rangle =0\,; \qquad \langle \CO_{\Phi_K} \rangle =0\:\:\:(K=1,2,3)\,;
\end{eqnarray}
and the temperature $T$ and the entropy $S$ of the single charged BCS black hole are:
\begin{eqnarray}\label{BCStwoQ:TS}
&& T=\frac{1}{L} \,\frac{2 y_+^2+4 \tilde{q}+1}{2 \pi  \left( y_+^2+2 \tilde{q}\right)}\,, \nonumber\\
&& S= N^2\,\pi  y_+^2   \left( y_+^2+2 \tilde{q}\right).
\end{eqnarray}
These quantities, of course, agree with \eqref{BCS:EQ}--\eqref{BCS:TS} in the appropriate limit.
It follows from \eqref{BCStwoQ:TS}  that there is no extremal configuration (\emph{i.e.} with $T\to 0$) in the $Q_1=Q_2$, $Q_3=0$ BCS family.

\subsection{The scalar condensation instability of Behrndt-Cveti\v c-Sabra black holes}\label{sec:GrowthTwoQ}

Just as we did for the single charge BCS solution, we first investigate the linear stability of the two charge BCS black hole with respect to the condensation of the charged scalar $\Phi$. The equation to solve turns out to be given by
\begin{equation}
D^a D_a \Phi+\frac{4}{L^2}\,e^{\frac{\varphi_1}{\sqrt{6}}}\Phi=0\,.
\label{eq:linear_two}
\end{equation}

Since the BCS backgrounds are static, we expand $\Phi$ in Fourier modes. Additionally, we will take the s-wave channel. All in all, we take $\Phi$ to be given by
\begin{equation}
\Phi(t,r)=e^{-i\omega t}\Phi_{\omega}(r)\,,
\end{equation}
where $\omega$ is the frequency of the perturbation mode.

At the horizon, we demand regularity in ingoing Eddinghton-Finkelstein coordinates $(v,r)$,
\begin{equation}
\dt v=\dt t+\frac{\dt r}{f(r)}\,,
\end{equation}
which in turn imposes
\begin{equation}
\widehat{\Phi}_{\omega}(r)\approx \left(1-\frac{r_+}{r}\right)^{-\i \frac{h(r_+)^{1/2}}{f^\prime(r_+)}\omega}\left[C^+_0+C^+_1\left(1-\frac{r_+}{r}\right)+\ldots\right]\,,
\end{equation}
near $r=r_+$ (where $C^+_0$ and $C^+_1$ are constants). At the conformal boundary, we choose standard quantisation for the scalar field $\Phi$, which in turn fixes the asymptotic behaviour of $\widehat{\Phi}_{\omega}$ to be
\begin{equation}
\widehat{\Phi}_\omega = \left(\frac{r_+}{r}\right)^2\left[C_0^-+C_1^-\left(\frac{r_+}{r}\right)^2+\ldots \right]
\end{equation}
as $r\to+\infty$ (where $C^-_0$ and $C^-_1$ are constants).

\begin{figure}[th]
\centering \includegraphics[width=\textwidth]{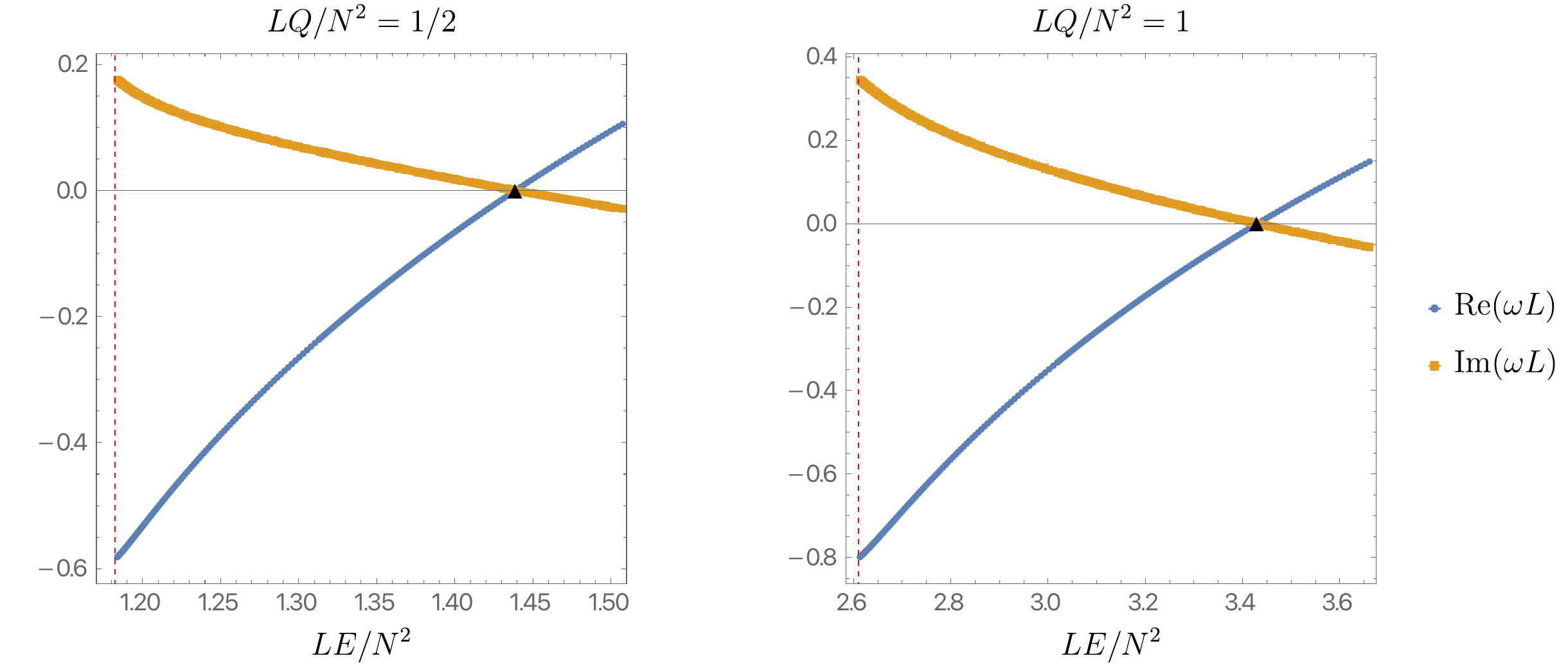}
\caption{\label{fig:growth_rates_double} Scalar condensation instability growth rate for the two charged BCS. The real (blue disks) and imaginary (orange squares) parts  of the frequency $\omega$ as a function of the energy $L E/N^2$ for $L Q/N^2=0.5$ (left panel) and $L Q/N^2=1$ (right panel). The dashed red lines show the singular extremal limit of the two charged BCS black hole for a given value of $L Q/N^2$ and the black triangle describes the onset of the instability with $\omega=0$.}
\end{figure}

To solve for $\widehat{\Phi}_\omega$ we change to a variable $q_{\omega}$ which is regular at $r=r_+$ and near the conformal boundary,
\begin{equation}
\widehat{\Phi}_\omega(r)=\left(1-\frac{r_+}{r}\right)^{-\i \frac{h(r_+)}{f^\prime(r_+)}\omega}\left(\frac{r_+}{r}\right)^2q_{\omega}(r)\,,
\label{eq:change_double}
\end{equation}
and introduce a compact coordinate just as in~ \eqref{eq:compact_linear}. We now solve for the eigenpair $\{\omega,q_{\omega}\}$ using familiar methods \cite{Dias:2015nua}. Our results are presented in Fig.~{\ref{fig:growth_rates_double}} where we plot the real (blue disks) and imaginary (orange squares) parts of the frequency $\omega$ as a function of the energy $E$ for two fixed values of $Q$. On the left panel we have $Q L/N^2=0.5$ whereas on the right panel we have $Q L/N^2=1$. Unlike the single charged BCS case of section~\ref{sec:GrowthSingleQ}, for the double charged case we find that $\mathrm{Im}(\omega L)$ and $\mathrm{Re}(\omega L)$ are non-zero at extremality. However, just like for the single charged case we also find that, for fixed charged, the instability ($\mathrm{Im}(\omega L)>0$) exists for $E\in[E_{\rm sing}(Q),E_{\rm onset}(Q)]$. We shall see this is the range where hairy black holes coexist with two charged BCS black holes and thus the latter provide a candidate for the instability endpoint (because for a given $\{E,Q\}$ they have higher entropy than the BCS) we just uncovered.

\subsection{Onset of scalar condensation instability}\label{sec:OnsetTwoQ}
Just as we did for the single charge case in section~\ref{sec:OnsetSingleQ}, we now proceed to determine the onset of the scalar condensation instability of the two charged BCS black hole directly, instead of computing $\omega$ and determine the onset \emph{a posteriori} as the configuration where $\omega=0$.

We take $\omega=0$ and rewrite~\eqref{eq:linear_two} just as in~\eqref{eq:quadratic_single}, but this time with
\begin{subequations}
\begin{equation}
L_2(y;\tilde{\l})=\left(1+2 \tilde{Q}\right) (1-y) y \left[1+2 \left(1+2 \tilde{Q}\right) \tilde{\l }-y \left(1+\tilde{\l }+4 \tilde{Q} \tilde{\l }\right)\right]^2\,,
\end{equation}
\begin{multline}
L_1(y;\tilde{\l})=\left(1+2 \tilde{Q}\right) \left[1+2 \left(1+2 \tilde{Q}\right) \tilde{\l }-y \left(1+\tilde{\l }+4 \tilde{Q} \tilde{\l}\right)\right] \bigg\{1+2\left(1+2 \tilde{Q}\right) \tilde{\l}
\\
+3 y^2\left(1+\tilde{\l }+4 \tilde{Q} \tilde{\l }\right)-2 y \left[2+\left(3+8 \tilde{Q}\right) \tilde{\l }\right]\bigg\}\,,
\end{multline}
and
\begin{multline}
L_0(y;\tilde{\l})=-16 \tilde{Q}^3 (1-y) (1-2 y) \tilde{\l }^2-(1-y) (1+\tilde{\l }) \left[1-y+(2-y) \tilde{\l }\right]
\\
-4 \tilde{Q}^2 \tilde{\l } \left[3+6 \tilde{\l }-y \left(7-4 y+8 (2-y)
   \tilde{\l }\right)\right]-2 \tilde{Q} \bigg\{[1+6 \tilde{\l } (1+\tilde{\l })
   \\
   +y^2 (1+\tilde{\l }) (1+5 \tilde{\l })-y \left[2+13 \tilde{\l }(1+\tilde{\l })\right]\bigg\}\,,
\end{multline}
\end{subequations}%
where we again defined $\tilde{\l}=y_+^2$ and $\tilde{Q}=\tilde{q}/y_+^2$. The above provides a quadratic eigenvalue problem in $\tilde{\l}$ for a given value of $\tilde{Q}$. Again, boundary conditions can be found at $y=0$ and $y=1$ by demanding that $q_0(y)$ admits a regular Taylor expansion at such regular singular points. The onset curve presented as a solid blue line in the figures of the subsections that follow is computed following the approach outlined in this section.

\subsection{Phase diagram in the microcanonical ensemble}\label{sec:ResultsMicroTwoQ}
We now discuss the phase diagram of the two charged system in the microcanonical ensemble. 

\begin{figure}[th]
\centering \includegraphics[width=0.6\textwidth]{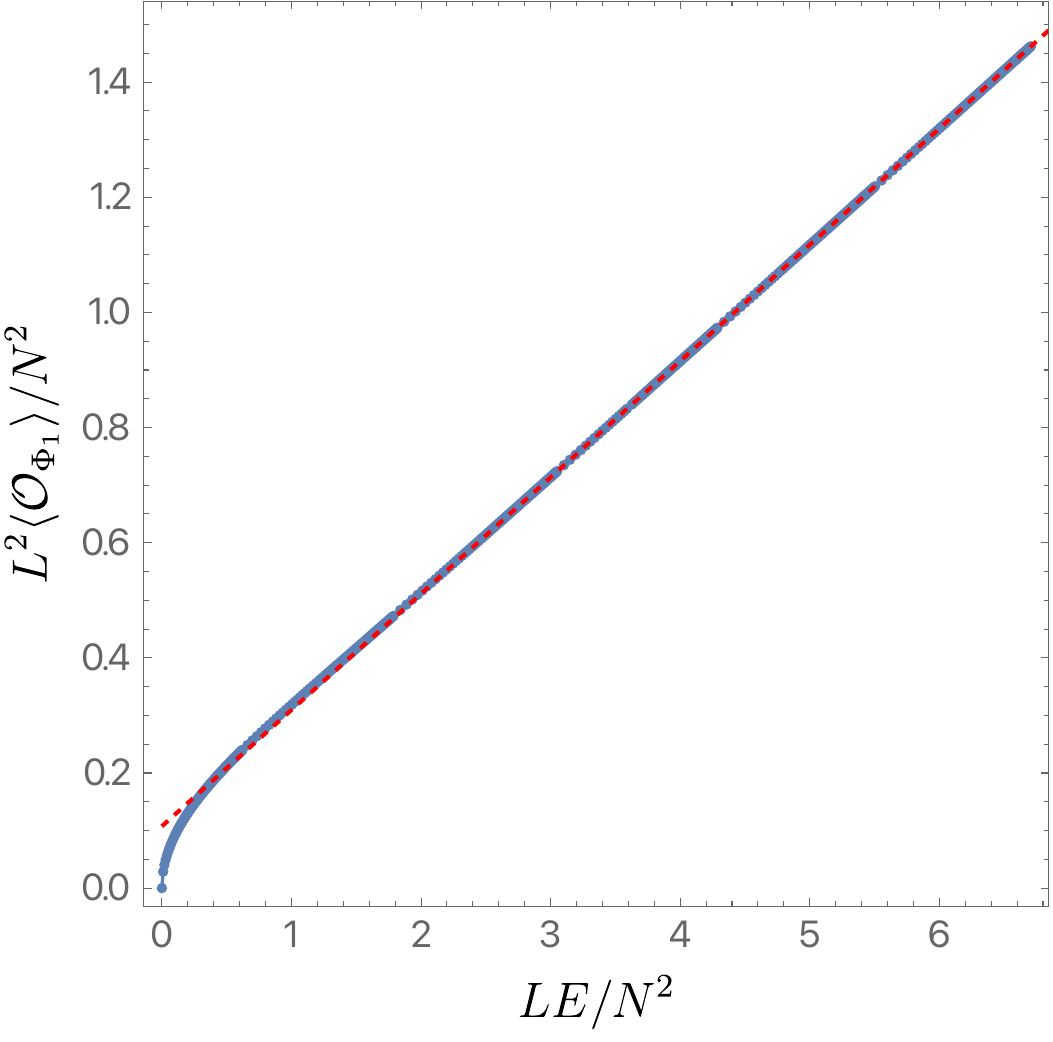}
\caption{\label{fig:soliton} Expectation value  $\langle \CO_{\Phi_1}\rangle=\langle \CO_{\Phi_2}\rangle$ as a function of $L E/N^2$ for the supersymmetric solitonic solution of the two charged system, with the numerical data being given by the blue disks and the dashed red line showing a linear best fit \eqref{TwoQ:bestfit} detailed in the main text.}
\end{figure}

We find convenient to start by discussing the supersymmetric solution of section \ref{sec:ansatzTwoQ-Solitons}. Unlike the single charge case, we were not able to find a closed form solution for the supersymmetric soliton, so we resort to numerical work.
It turns out that solving the EoM~\eqref{TwoQ:solitonODE} for the soliton numerically is harder than it might first appear. Naturally, the supersymmetric soliton will satisfy $M=2Q$. For this reason we plot instead $\langle \CO_{\Phi_1}\rangle$ as a function of $L E/N^2$ in Fig.~\ref{fig:soliton}, where the exact numerical data is represented by the blue dots and the dashed red line shows a best fit to a linear function
\begin{equation}\label{TwoQ:bestfit}
\frac{L^2}{N^2}\langle \CO_{\Phi_1}\rangle =a_0+\frac{b_0}{\pi^2} \frac{L E}{N^2}
\end{equation}
in the range $L E/N^2>3$ and we find $a_0=0.10746\pm0.00005$ and $b_0=1.9935\pm 0.0001$ from the numerical fit, where the error is estimated using a standard $\chi^2$ procedure. In particular, we see that there is no upper bound on the energy, that is to say, no Chandrasekhar limit, unlike in the three equal charge system of \cite{Bhattacharyya:2010yg}. Fig.~\ref{fig:soliton} will be important later on when we discuss the microcanonical phase space of solutions. In particular, it will provide an excellent \emph{norm} to understand whether our finite temperature hairy black hole solutions approach the soliton in the limit of vanishing temperature and area.

\begin{figure}[th]
\centering \includegraphics[width=0.6\textwidth]{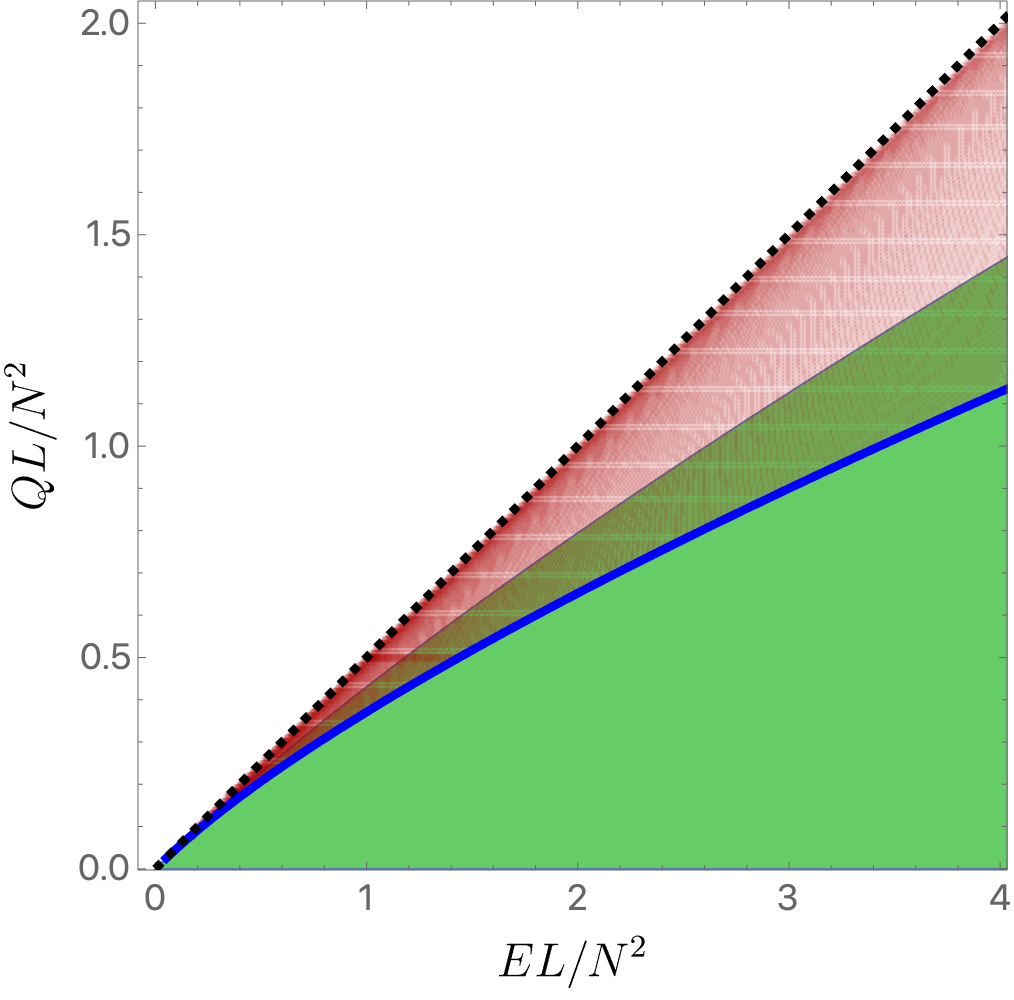}
\caption{\label{fig:microtwo} Phase diagram in the microcanonical ensemble for the two equal charge system. The green region indicates where BCS black holes with two equal charges exist, with the limiting upper boundary (the black solid thin curve) being the extremal line (where BCS black holes become singular); the black dashed line shows the supersymmetric bound, along which the supersymmetric solitons exist with no apparent upper bound on $E$; the blue disks describe the onset of the hairy black holes and the red disks (region) that extends from this onset all the way up to the BPS dashed black line  represent the hairy black holes of the system.}
\end{figure}

We now turn our attention to the full phase diagram of static solutions in the two charge system. We first note that, unlike the single charge case, in the two charge system the extremal limit of the BCS black holes does not saturate the supersymmetric bound $E=2Q$. Instead, we find that two equal charge extremal BCS black holes obey to\footnote{It might first appear that for $L Q/N^2\geq 1/(3\sqrt{3})$ the energy of extremal BCS black holes as given by \eqref{eq:trig} becomes complex. However, note that $X$ becomes purely imaginary in that limit, which then changes the $\cos$ in~\eqref{eq:trig} to $\cosh$, so that~\eqref{eq:trig} remains real for all $Q\geq0$.}
\begin{subequations}
\begin{equation}
\frac{L E}{N^2}=\frac{1}{12} \left\{2 \left[2 \cos \left(\frac{X(Q)}{3}\right)+\cos \left(\frac{2 X(Q)}{3}\right)\right]-1\right\}>2 \frac{L\,Q}{N^2}
\label{eq:trig}
\end{equation}
with
\begin{equation}
X(Q)=\arccos\left(54\frac{L Q}{N^2}-1\right)\,.
\end{equation}
\end{subequations}
This parametrizes the black solid thin line in Fig.~\ref{fig:microtwo}.

In Fig.~\ref{fig:microtwo} we plot the microcanonical phase diagram for the two equal charge system. In the microcanonical ensemble, the entropy is the relevant thermodynamic potential and the state variables are the charge $Q$ and energy $E$. Dominant solutions will maximise the energy $S$ at fixed $Q$ and $E$. The green region in Fig.~\ref{fig:microtwo} shows the region in moduli space where  two charged BCS black holes exist. As anticipated this region ends well before the supersymmetric bound $E=2Q$ is saturated. The extremal curve (black solid thin line) where this occurs has zero temperature, zero entropy and is singular. The black dashed curve with $Q=E/2$ represents the supersymmetric bound and the suspersymetric solitons exits along this curve for any value of $E=2Q$. Finally, the red disks are the hairy black holes we have determined numerically in this work. They exist in a region that extends from the blue curve (onset curve determined using the numerical method briefly outlined in section \ref{sec:OnsetTwoQ}) all the way up to the supersymmetric dashed black bound, possibly for arbitrarily large $E$ (and $Q\leq E/2$).

\begin{figure}[th]
\centering \includegraphics[width=0.6\textwidth]{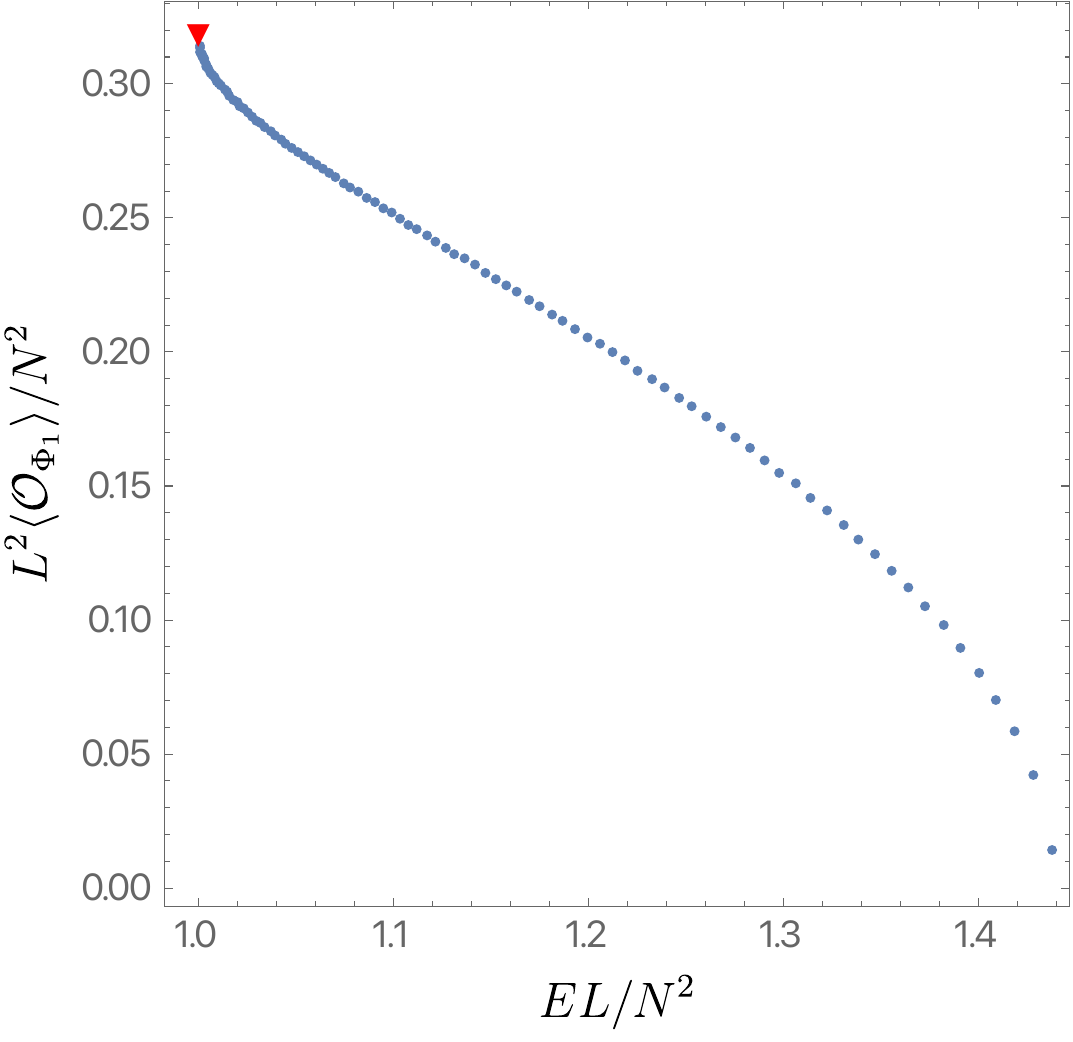}
\caption{\label{fig:comp_sol} Expectation value  $\langle \CO_{\Phi_1}\rangle=\langle \CO_{\Phi_2}\rangle$ as a function of $E L/N^2$ computed for a hairy black hole with fixed $Q L/N^2=0.5$. The solitonic red inverted triangle was retrieved from Fig.~\ref{fig:soliton} and the blue disks were computed directly using our novel hairy black holes. Clearly, the hairy black holes approach the supersymmetric soliton when the BPS limit is approached.}
\end{figure}

As the hairy black holes approach the supersymmetric bound, the geometry approaches that of the soliton. Perhaps the best way to see this is to plot the condensate $\langle \CO_{\Phi_1}\rangle=\langle \CO_{\Phi_2}\rangle$ as a function of $E L/N^2$ for a fixed value of $Q L/N^2$. We do this in Fig.~\ref{fig:comp_sol}, where the inverted red triangle was obtained from Fig.~\ref{fig:soliton} and the blue disks were computed using our hairy black holes. The agreement between the two methods as we approach the supersymmetric point backs up our claim.

\begin{figure}[th]
\centering \includegraphics[width=0.6\textwidth]{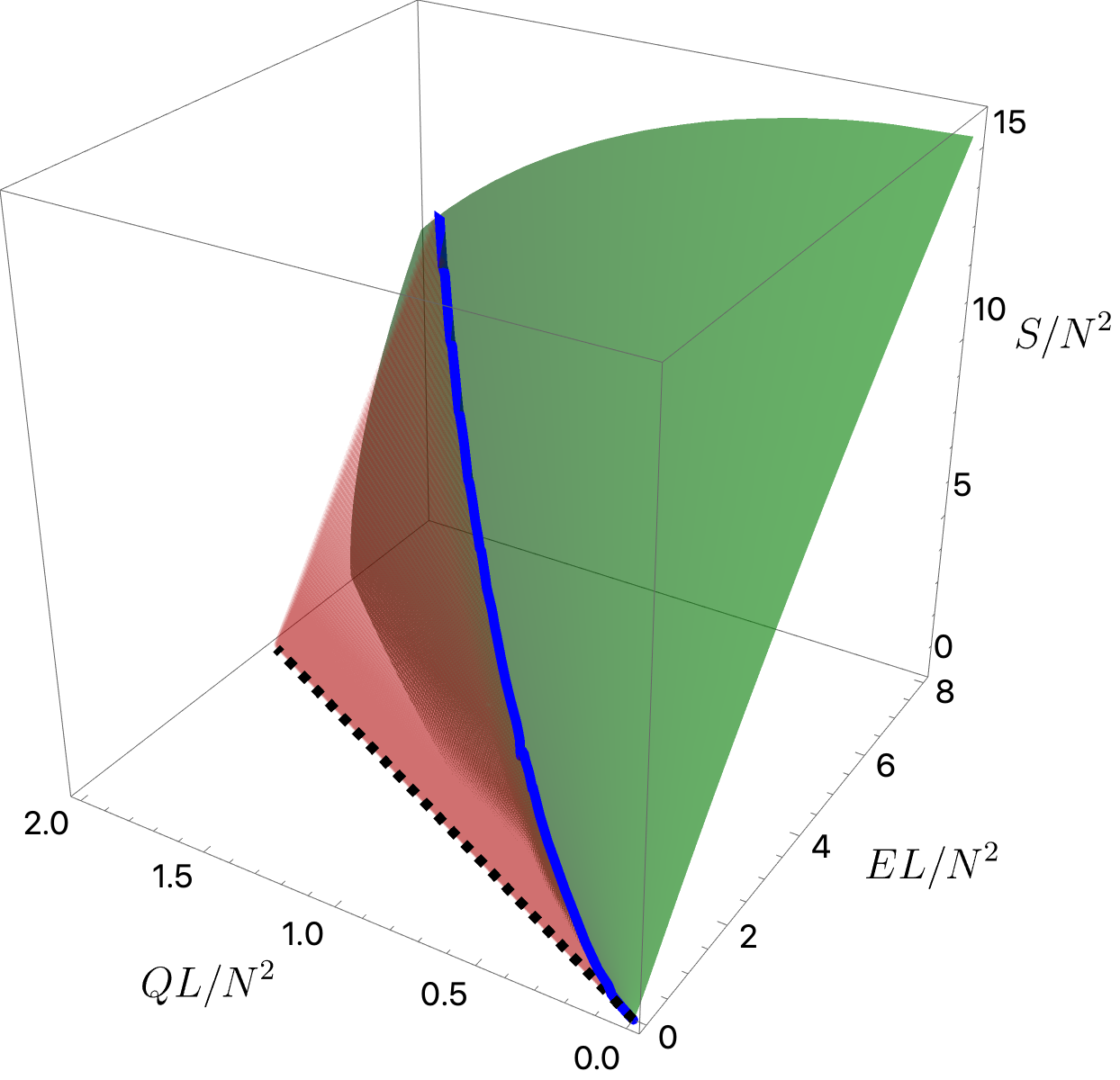}
\caption{\label{fig:3D_entropy_two} Microcanonical phase diagram for the two charged system: the entropy $S/N^2$ as a function of $E L/N^2$ and $Q L/N^2$. The colour coding used in this figure is the same as the one used in Fig.~\ref{fig:microtwo}. The red surface is always above the green surface in the $E-Q$ region where they coexist except at the solid blue line (onset curve) where the two families merge in a second order phase transition.}
\end{figure}

Next, we plot the entropy of the new hairy black hole solutions as a function of $E$ and $Q$ in Fig.~\ref{fig:3D_entropy_two} where we use the same colour coding as in Fig.~\ref{fig:microtwo}. In particular, this plot shows that the entropy of the hairy solutions is always larger than that of a BCS black hole with the same $E$ and $Q$, where the two families of black holes coexist. This fact, together with the results for the linear stability in section \ref{sec:GrowthTwoQ} provide very strong evidence that the hairy solutions we just found are should be the endpoint of the scalar condensation instability of the two charge BCS black hole uncovered in section \ref{sec:GrowthTwoQ}.

\begin{figure}[th]
\centering \includegraphics[width=0.6\textwidth]{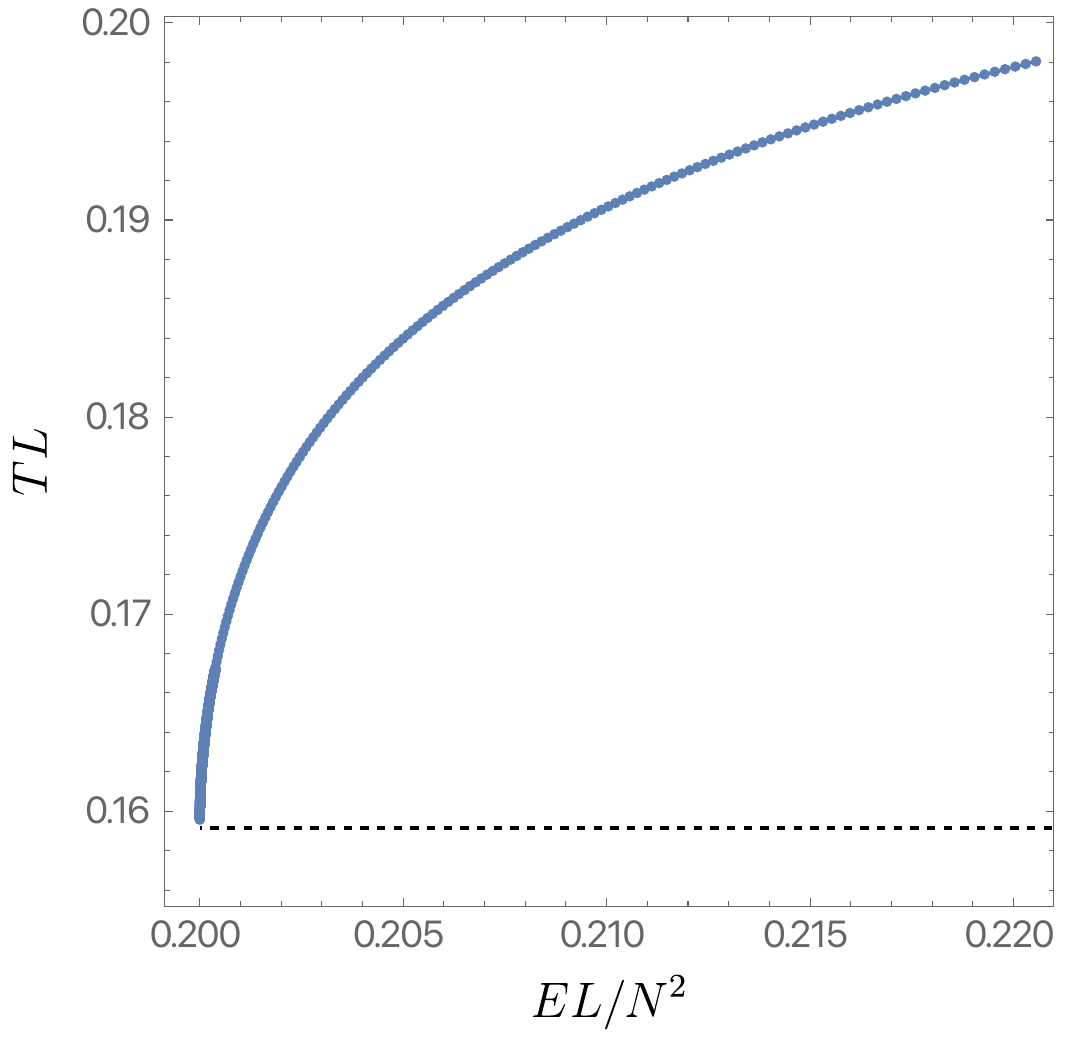}
\caption{\label{fig:two_temp} The temperature $T L$ as a function of $E L/N^2$ for a two charged hairy black hole family with fixed $Q L/N^2=0.1$. For reference, the horizontal dashed line marks the temperature $T L =1/(2\pi)$ and the hairy black hole temperature approaches this value in the BPS limit $E=2Q$.}
\end{figure}

Finally, we discuss the fate of the temperature of the hairy solutions as we approach the supersymmetric bound. One might think that the temperature will drop to zero as we approach this boundary. However, this turns out not to be the case. Indeed, we observe that, irrespectively of $E=2Q$, the temperature approaches the value $L T=1/(2 \pi)$ when we approach the supersymmetric bound $E=2Q$. At the moment, we have no understanding why this is the case.\footnote{Static supersymmetric solutions in \emph{four} dimensions exhibit a similar property wherein by taking two supercharges to zero simultaneously, it is possible to obtain a supersymmetric solution with finite temperature and zero entropy \cite{Kallosh:1993fj}. Our result seems to be the five dimensional analogue of this phenomenon. We hope to explore this further in future work.} To back up our claim, we plot in Fig.~\ref{fig:two_temp} the temperature of the hairy solutions as a function of their energy, for a particular value of $L Q/N^2=1/10$. We can clearly see the temperature reaching $T L=1/(2\pi)$ (represented as the dashed black line) as we approach the supersymmetric bound. We should also note that the perturbative scheme detailed in Section \ref{sec:PerturbativeTwoQ} also predicts such a limiting value for the temperature: see \eqref{twoQ:qtsol} and \eqref{eq:a_bit_crazy}.

\subsection{Phase diagram in the grand-canonical ensemble}\label{sec:ResultsCAnoTwoQ}
Having discussed the microcanonical ensemble, we now turn out attention to the grand-canonical ensemble. The relevant thermodynamic potential is now the Gibbs free energy $G=E-TS-2\mu Q$, and the state variables are $\mu$ and $T$. Dominant phases will minimise the Gibbs free energy at constant $\mu$ and $T$.

Just like for the single charge case (discussed in section~\ref{sec:ResultsSingleQgrand}), we find that for the two charged system the hairy solutions are also never dominant in the grand-canonical ensemble. We are thus left with discussing the two equal charge BCS black holes, a discussion that is missing in the literature. We will shortly argue that the two equal charge BCS black holes exhibit a critical point, similar to that of water!

\begin{figure}[th]
\centering \includegraphics[width=0.9\textwidth]{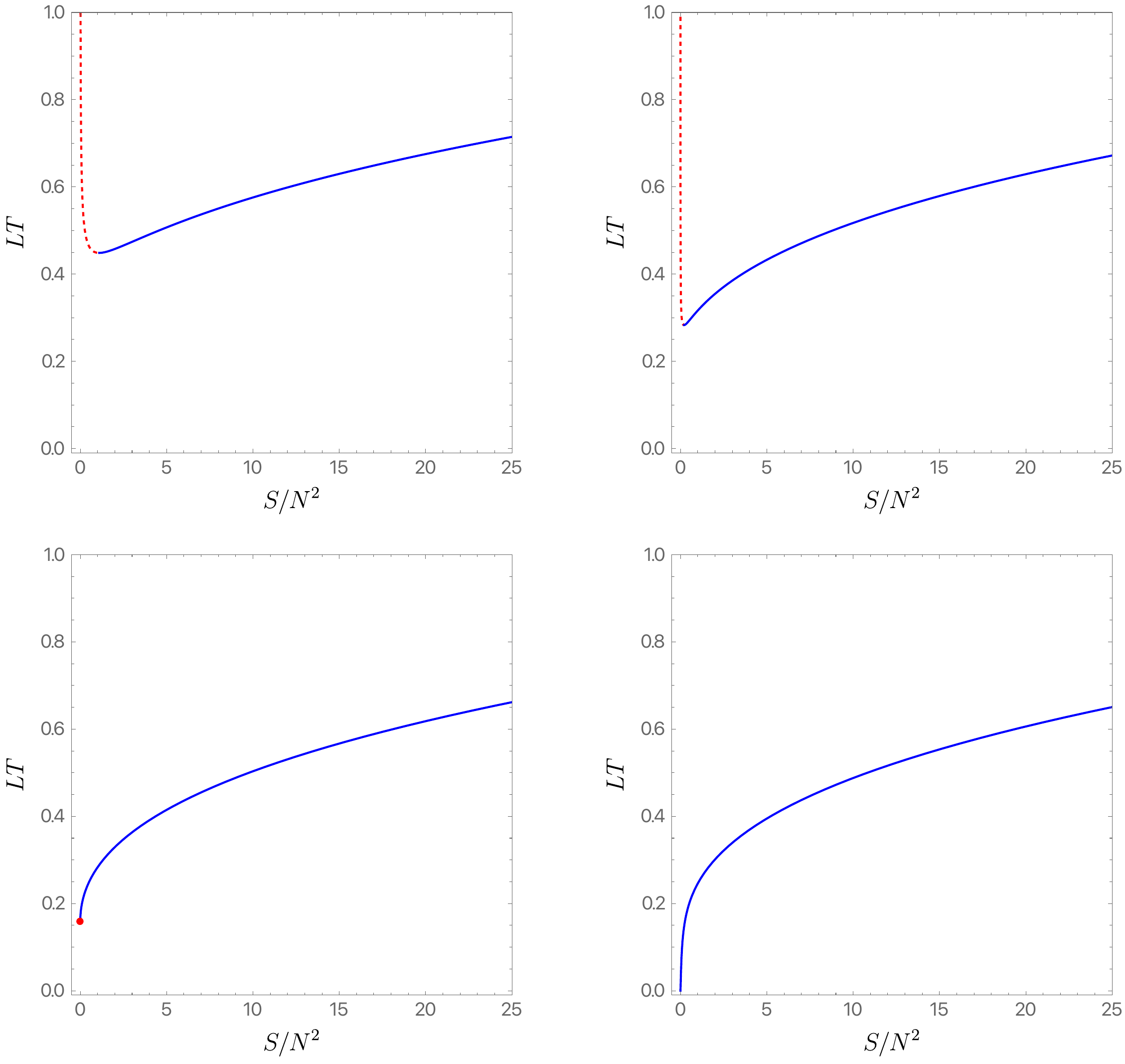}
\caption{\label{fig:temp_vs_entropy} The temperature $T L$ as a function of entropy $S/N^2$ for two equal charged BCS black holes with several fixed values of $\mu$. Using lexicographic ordering, we have $\mu=0.1,\,0.9,\,1, \,1.1$. The solid blue (dashed red) line describes large (small) BCS black holes.  In the bottom-left plot, the red dot on the left has $S=0$ and $L T=1/(2\pi)$.}
\end{figure}

For each value of $\mu^2\leq1$, there are two BCS black hole solutions at fixed temperature $T$. This is illustrated in Fig.~\ref{fig:temp_vs_entropy} where we plot the temperature of the two equal charge BCS black holes as a function of their entropy for several fixed values of $\mu$. Using lexicographic ordering, we have $\mu=0.1,\,0.9,\,1,\,1.1$. It is clear that for each value of $\mu^2<1$ and $T>T_{\min}(\mu)$ with
\begin{subequations}
\begin{equation}
T_{\min}(\mu)=\frac{1-\mu ^2+3 y_+^s(\mu )^2+\sqrt{\mu ^4+2 \mu ^2 [y_+^s(\mu )^2-1]+[1+y_+^s(\mu)^2]^2}}{4 \pi  y_+^s(\mu )}
\end{equation}
and
\begin{equation}
y_+^s(\mu)=\sqrt{\sqrt{1+8 \mu^2} \cos \left\{\frac{1}{3} \arccos \left[\frac{1-20 \mu ^2-8 \mu ^4}{\left(1+8 \mu ^2\right)^{3/2}}\right]\right\}-\frac{1}{2}-\mu^2}\,.
\end{equation}
\end{subequations}
there are two BCS black hole solutions. The one with larger entropy we coin as ``\emph{large}'' BCS black holes (solid blue curves), whereas the one with smaller entropy we label as the ``\emph{small}'' BCS black holes (dashed red curves). The entropy of the large black hole scales as $S/N^2\simeq \pi^4 (L T)^3$, whereas for the small black hole branch we have $S/N^2\simeq (1-\mu ^2)^2/[8 \pi ^2 (L T)^3]$ (at large $T$).

Note that for $\mu^2>1$ there is only \emph{one} black hole solution at fixed $T$ and $\mu$, with $T_{\min}=0$ and $S/N^2=0$ (bottom-right panel of Fig.~\ref{fig:temp_vs_entropy}). For $\mu^2=1$, we find that $L T_{\min}=1/(2\pi)$ (bottom-left plot in Fig.~\ref{fig:temp_vs_entropy}). Note that the behaviour when $\mu^2>1$ is markedly different from the single charged BCS black hole. Indeed, in the single charge case (recall Fig.~\ref{fig:entropy_single_BCS}), two solutions existed for fixed $T$ and $\mu>1$, whereas in the two charge case a single black hole solution exists for fixed $\mu>1$ and $T>0$.

The small black hole branch turns out to be locally thermodynamically unstable, whereas the large black hole branch is locally thermodynamically stable. To see this we could again study the Hessian of the Gibbs free energy $G$ as a function of $T$ and $\mu$, just like we did in section \ref{sec:ResultsSingleQgrand}. Here, instead, we will follow \cite{Monteiro:2008wr} where the local thermodynamic stability of charged black holes in the grand-canonical ensemble was shown to be equivalent to the positivity of the specific heat $C_Q$ at constant charge $Q$, defined as
\begin{equation}
C_Q = T\,\left(\frac{\partial S}{\partial T}\right)_Q
\end{equation}
and positivity of the isothermal permittivity (or capacitance) $\epsilon_T$ defined as
\begin{equation}
\epsilon_T = \left(\frac{\partial Q}{\partial \mu}\right)_T\,.
\end{equation}
These quantities can be readily computed in terms of $\tilde{q}$ and $y_+$ defined in~\eqref{BCS:TwoQ} and turn out to be given by
\begin{subequations}
\begin{equation}
C_Q/N^2=\frac{\pi  \left(2 \tilde{q}+y_+^2\right) \left(y_++4 \tilde{q} y_++2 y_+^3\right) \left[12 \tilde{q}^2+4 \tilde{q} \left(1+3 y_+^2\right)+3 y_+^2\left(1+y_+^2\right)\right]}{48 \tilde{q}^3+28 \tilde{q}^2 \left(1+2
   y_+^2\right)+y_+^2 \left(2 y_+^4+y_+^2-1\right)+4 \tilde{q} \left(1+4 y_+^2+5 y_+^4\right)}
\end{equation}
and
\begin{equation}
\epsilon_T/N^2=\frac{\left(2 \tilde{q}+y_+^2\right) \left[48 \tilde{q}^3+28 \tilde{q}^2 \left(1+2 y_+^2\right)+y_+^2 \left(2 y_+^4+y_+^2-1\right)+4 \tilde{q} \left(1+4 y_+^2+5 y_+^4\right)\right]}{2 L \left[16 \tilde{q}^3+4
   \tilde{q}^2 \left(1+6 y_+^2\right)+y_+^2 \left(2 y_+^4+y_+^2-1\right)+4 \tilde{q} y_+^2\left(1+3 y_+^2\right)\right]}\,.
\end{equation}
\end{subequations}%
It is a simple exercise to show that the product $C_Q \epsilon_T$ is negative on the small BCS black hole branch, thus indicating a local thermodynamic unstable phase. On the other hand both $C_Q$ and $\epsilon_T$ are positive on the large BCS black hole branch, thus indicating a local thermodynamically stable phase. For $\mu^2\geq1$, where only one BCS black hole phase exists at fixed $T$ and $\mu$, both $C_Q$ and $\epsilon_T$ are positive, thus indicating a local thermodynamically stable phase.

\begin{figure}[th]
\centering \includegraphics[width=0.6\textwidth]{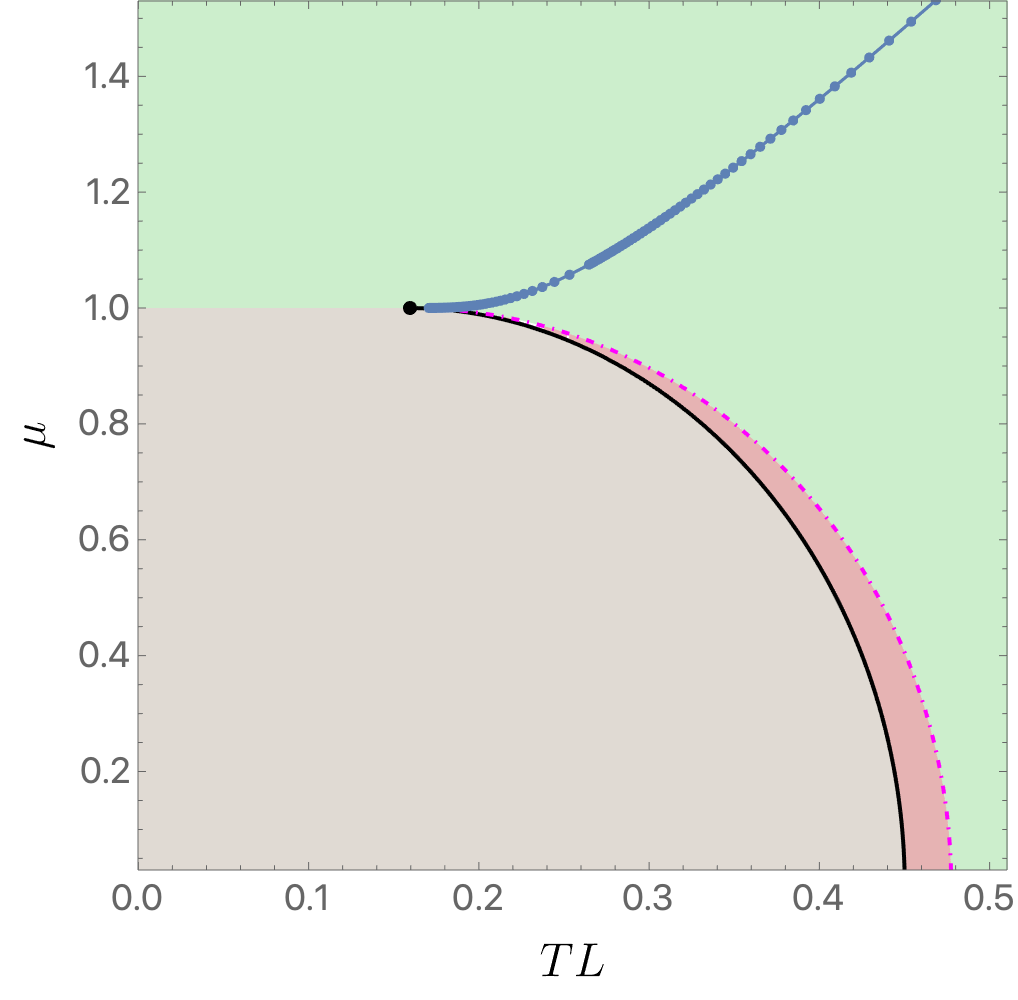}
\caption{\label{fig:two_grand} Phase space of the two equal charge BCS black holes in the grand-canonical ensemble. The light brown region indicates regions where only thermal AdS exists; the light red region is a region where thermal AdS and large BCS black holes coexist, but nevertheless thermal AdS dominates; the dot-dashed magenta line indicates a Hawking-Page transition; and in the green region large BCS black holes dominate the ensemble. The blue disks are the onset of scalar condensation. The black disk at $(L T,\mu)=(1/(2\pi),1)$ marks the location of a critical point, similar to the critical point of water (ending the phase transition between vapor and liquid water). Hairy black holes exist below the blue disk onset line and above $\mu=1$ but they never dominate the garnd-canonical ensemble.}
\end{figure}

We now turn to the issue of global thermodynamic stability in the grand-canonical ensemble, \emph{i.e.} which phase has lower Gibbs free energy at fixed $T$ and $\mu$. The associated phase diagram appropriated for this discussion is plotted in Fig.~\ref{fig:two_grand} where we investigate the phase space of solutions in the grand-canonical ensemble. The light brown region indicates regions where two charge BCS black holes do not exist neither hairy black holes (thermal AdS dominates here); the light red region is a region where thermal AdS and large BCS black holes coexist, but nevertheless thermal AdS dominates; the dot-dashed magenta line (on the right boundary of the red region) indicates a Hawking-Page transition between thermal AdS and large BCS; and in the green region large BCS black holes dominate the ensemble. The blue disks are the onset of scalar condensation of the BCS black hole. The black disk, where the solid black line and the blue disk onset line merge at $TL=1/(2\pi)$ and $\mu=1$,  marks the location of a critical point, similar to the critical point of water (ending the phase transition between vapor and liquid water). We reiterate that the hairy black holes numerically constructed in this manuscript never dominate the grand-canonical ensemble. They exist below the blue disk onset line and $\mu=1$. 

The existence of a critical point at $(L T,\mu)=(1/(2\pi),1)$ $-$ the black point in Fig.~\ref{fig:two_grand} $-$ is intriguing. One might ask what happens to the capacitance $\epsilon_T$ and specific heat $C_Q$ as one approaches that point along the dot-dashed magenta Hawking-Page transition line. This is similar to finding the critical exponents for water around the critical point ending the phase transition between vapor and liquid water. For our system we find
\begin{subequations}
\begin{equation}
C_Q\simeq \pi ^3 \left(L T-\frac{1}{2 \pi }\right)^2+\CO\left[\left(L T-\frac{1}{2 \pi }\right)^3\right]
\end{equation}
and
\begin{equation}
\epsilon_T\simeq \frac{1}{\displaystyle 2 \pi  L \left(L T-\frac{1}{2 \pi }\right)}+\frac{1}{L}+\CO\left[\left(L T-\frac{1}{2 \pi }\right)^3\right]\,.
\end{equation}
\end{subequations}

\subsection{Perturbative construction of hairy solitons and black holes}\label{sec:PerturbativeTwoQ}

In this section, we describe the basic strategy used to construct the hairy black hole solutions in perturbation theory with two expansion parameters and using matched asymptotic expansion. This is largely identical to the one described in section \ref{sec:PerturbativeSingleQ} for the one-charge hairy black hole so we shall be brief and only highlight the differences. The equations of motion \eqref{TwoQ:EoM} in the gauge \eqref{TwoQ:ansatz} are difficult to solve analytically. It is more convenient to work in a different gauge defined by
\begin{equation}
\begin{split}\label{twoQ:ansatz}
\dt s^2 &= h^{\frac{2}{3}} \left[ - \frac{f}{h^2} \dt t^2 +  \frac{\dt r^2}{f}  + r^2 \dt \O_3^2 \right]  , \\
\varphi_1 &= \sqrt{6} \varphi , \qquad \varphi_2 = 0  , \\
A^1 &= A_t \dt t , \qquad A^2 = A_t \dt t , \qquad A^3 = 0 , \\
\Phi_1 &= \Phi_1^\dagger = \Phi , \qquad \Phi_2  = \Phi_2^\dagger = \Phi , \qquad \Phi_3 = 0 . 
\end{split}
\end{equation}
where all the quantities above are functions of $r$. In \eqref{TwoQ:ansatz}, $g_{rr}$ and $\varphi_1$ is fixed in terms of $h$ whereas in this gauge, $g_{rr}$ is fixed in terms of $f$ and $\varphi_1$ is an independent field. We can consistently set $A^3=0$ only in the static case considered here. More generally, $A^3$ is sourced by the angular momentum of the black holes and is generically non-vanishing for rotating black holes. The ansatz \eqref{twoQ:ansatz} has a residual coordinate freedom $r \to \sqrt{r^2 +a }$ and we use this to fix\footnote{The radial coordinate used here is related to the one in \eqref{TwoQ:ansatz} by
$$
r_{here}^2 = \int \dt ( r_{there}^2 ) \sqrt{ \frac{ f_{there} ( r_{there} )} { g_{there} ( r_{there} ) } } + a  . 
$$
The integration constant $a$ is fixed by \eqref{twoQ:residualcoordinatefix}. The metric functions are related by
\begin{equation}
\begin{split}
\nonumber
r_{here}^3 h_{here} &= r_{there}^3 h_{there} , \quad r_{here}^4 f_{here} = r_{there}^4 f_{there} ,  \\
(A_t)_{here} &= (A_t)_{there} , \qquad \Phi_{here} = \Phi_{there} , \qquad \varphi_{here} = \frac{1}{\sqrt{6} } \varphi_{there} .
\end{split}
\end{equation}
}
\begin{equation}
\begin{split}\label{twoQ:residualcoordinatefix}
\varphi - \frac{1}{3} \ln h = \CO(L^4 r^{-4}) \quad \text{at large $r$.} 
\end{split}
\end{equation}
Plugging the ansatz \eqref{twoQ:ansatz} into the equations \eqref{GravEOM}--\eqref{ScalarsGaugeEOM}, we find the equations
\begin{align}
\label{twoQ:E11} 0 &=  h'^2-\frac{3 h \left(r h''+3 h'\right)}{r}+h^2 \left(-\frac{3 \Phi '^2}{\Phi ^2+4}-9 \varphi '^2 \right) - \frac{3 A^2 h^4 \Phi^2}{f^2}  , \\
\label{twoQ:E22}  0 &=r^2 h^{\frac{4}{3} } e^{2 \varphi } A'^2+\frac{r^2 f' h'}{h}+3 r f'-4 r^2 h^{\frac{2}{3}} e^{-2 \varphi } \left( e^{3 \varphi } \sqrt{\Phi^2+4} +1\right)-6  \nonumber \\
&\qquad \qquad \qquad \qquad - \frac{r^2 A^2 h^2 \Phi^2}{f} +f \left(-\frac{2r^2 h'^2}{3 h^2}-\frac{r^2 \Phi '^2}{\Phi ^2+4}-3 r^2 \varphi '^2+6\right) , \\
\label{twoQ:M1}  0 &= f h^{-\frac{2}{3}} e^{2 \varphi } \left(3 r h A''+A' \left(4 r h'+h \left(6 r \varphi '+9\right)\right)\right)  - 3 r A h \Phi ^2 , \\
\label{twoQ:p1} 0 &= h^{\frac{4}{3}} e^{2 \varphi } A'^2+3 f' \varphi '+3 f \left(\varphi'' +\frac{3 \varphi '}{r}\right)+h^{\frac{2}{3}} \left(2 e^{\varphi } \sqrt{\Phi ^2+4} -4 e^{-2 \varphi }\right) , \\
\label{twoQ:P1} 0 &= \frac{A^2 h^2 \Phi }{f}r  (\Phi ^2+4 )^2 + \Phi '  ( (\Phi ^2+4 )  (r f'+3 f )-r f \Phi  \Phi ' ) \nonumber  \\
&\qquad \qquad \qquad \qquad \qquad  + r f  (\Phi ^2+4 ) \Phi ''+2 r h^{\frac{2}{3}} \Phi (\Phi^2+4)^{\frac{3}{2}} e^{\varphi} . 
\end{align}
These are 5 coupled differential equations for 5 functions. The equations for $h$, $A$, $\varphi$ and $\Phi$ are second order whereas the one for $f$ is first order. The equations can then be solved up to 9 integration constants. Four of these are fixed by the AdS boundary conditions at $r=\infty$ and five are fixed by imposing regularity in the interior (either on the horizon $r=r_+$ or the origin $r=0$).

The asymptotic expansion of the fields in this gauge has the form
\begin{equation}
\begin{split}\label{AdSbdycond}
f(r) &= \frac{r^2}{L^2} + 1 + 2 c_h + c_f \frac{L^2}{r^2} + \CO(L^4 r^{-4}) , \\
h(r) &= 1 + c_h \frac{L^2}{r^2} + \CO(L^4 r^{-4}) , \\
A_t(r) &= \mu + c_A \frac{L^2}{r^2} + \CO(L^4 r^{-4}) , \\
\varphi(r) &= c_h \frac{L^2}{3r^2} + \CO(L^4 r^{-4}) , \\
\Phi(r) &= \e \frac{L^2}{r^2} + \CO(L^4 r^{-4}) . 
\end{split}
\end{equation}
Using the holographic renormalization procedure described in section \ref{sec:ThermoTwoQ}, we find that the mass and charge is then given by 
\begin{equation}
\begin{split}\label{mass-charge-calc}
E = \frac{N^2}{L} \left( c_h - \frac{3}{4} c_f  + \frac{3}{4} c_h^2  - \frac{3}{16} \e^2 \right)  , \qquad Q &= \frac{N^2}{L} \left( - \frac{1}{2} c_A \right) . 
\end{split}
\end{equation}
Black hole solutions have a Killing horizon at $r=r_+$ where $f(r_+)=A_t(r_+)=0$. The temperature, entropy and chemical potential of the black hole is
\begin{equation}
\begin{split}
T = \frac{f'(r_+)}{4\pi L h(r_+)} , \qquad S = N^2 \pi r_+^3 h(r_+) , \qquad \mu =  \lim_{r\to\infty} A_t(r)  .  
\end{split}
\end{equation}
The thermodynamic quantities must satisfy the first law of thermodynamics, $\dt E = T \dt S + 2 \mu \dt Q$.

In the rest of this section, we will set the AdS radius to unity, $L=1$.

\subsubsection{Hairy supersymmetric soliton}
\label{twoQ:solitonpert}

Unlike in the single charge case, the two equal charge hairy supersymmetric soliton is not known exactly and we must construct it in perturbation theory. The process is identical to the one described in section \ref{singleQ:hairysusysoliton-pertthy}. We expand the metric functions as 
\begin{equation}
\begin{split}
\label{twoQ:soliton-exp}
f(r,\e) &= \sum_{n=0}^\infty \e^{2n} f_{(2n)}(r)  , \\
h(r,\e) &= \sum_{n=0}^\infty \e^{2n} h_{(2n)}(r)  , \\
A_t(r,\e) &= \sum_{n=0}^\infty \e^{2n} A_{(2n)}(r) , \\
\varphi(r,\e) &= \sum_{n=0}^\infty \e^{2n} \varphi_{(2n)}(r) , \\
\Phi(r,\e) &= \sum_{n=0}^\infty \e^{2n+1} \Phi_{(2n+1)}(r) ,
\end{split}
\end{equation}
The base solution is taken to be empty AdS,
\begin{equation}
\begin{split}
f_\0(r) = r^2 + 1  , \qquad h_\0(r) = A_\0(r) =  1 , \qquad \varphi_\0(r) = 0 .
\end{split}
\end{equation}
The perturbative parameter is the charged scalar condensate VEV $\e = \frac{\pi^2}{N^2} \avg{ \CO_\Phi }$. We plug \eqref{twoQ:soliton-exp} into equations \eqref{twoQ:E11}--\eqref{twoQ:P1} and solve the equations so obtained order-by-order in $\e$ with AdS boundary conditions at $r=\infty$ and regularity at $r=0$.

At $\CO(\e^{2n+1})$, only \eqref{twoQ:P1} is non-trivial and the equation takes the form
\begin{equation}
\begin{split}\label{twoQ:solitoneq1}
\td{}{r} \left[ \frac{r^3}{1+r^2} \td{}{r} [ ( 1 + r^2 ) \Phi_{(2n+1)}(r) ] \right] &= \mfs^\Phi_{(2n+1)}(r) . 
\end{split}
\end{equation}
where $\mfs^\Phi$ denotes a source term that is fixed by lower orders in perturbation theory (we will continue to use this notation for the rest of this section). The equation is easily integrated and the solution for $\Phi_{(2n+1)}(r)$ up to two integration constants which are fixed using AdS boundary conditions. 

At $\CO(\e^{(2n)})$, the differential equations for $f_{(2n)}$, $h_{(2n)}$, $\varphi_{(2n)}$ and $A_{(2n)}$ take the form
\begin{equation}
\begin{split}\label{twoQ:solitoneq2}
[ r^3 h_{(2n)}'(r) ]' &= \mfs^h_{(2n)}(r) , \\
[ r^3 [ r^2 f_{(2n)}(r) ]' ]' &= \mfs^f_{(2n)}(r)  + 16 r^5 h_{(2n)}(r)  , \\
[ r^3 A_{(2n)}'(r) ]' &= \mfs^A_{(2n)}(r) , \\
\left( \frac{1+r^2}{r} [ r^2 \varphi_{(2n)}(r)]' \right)' &= \mfs^\varphi_{(2n)}(r) . \\
\end{split}
\end{equation}
These equations are easily fixed up to 7 integration constants. Two of these are fixed by the AdS boundary conditions and five are fixed by regularity at $r=0$ (one of these is fixed at $\CO(\e^{2n+1})$).

The explicit construction of the solution to $\CO(\e^3)$ is described in Appendix \ref{app:twoQsolitonpert} and the explicit solution to $\CO(\e^{13})$ can be found in the accompanying {\tt Mathematica} file. It is easily verified that the solution is supersymmetric and satisfies \eqref{TwoQ:SUSYfunctions} and \eqref{TwoQ:solitonODE}. The mass and charge of the soliton is
\begin{equation}\label{TwoQ:pert:solitonThermo}
\begin{split}
E = 2Q = N^2 \left( \frac{\e^2}{8}-\frac{\e^4}{384}+\frac{11 \e^8}{4423680}-\frac{47 \e^{12}}{8918138880}+\CO  (\e^{13} ) \right) . 
\end{split}
\end{equation}

\subsubsection{Hairy black hole}
\label{twoQ:HBHpert}

In this section, we construct analytically the hairy black hole (BH) solution of Sections \ref{sec:AnsatzTwoQ}, \ref{sec:ThermoTwoQ}, \ref{sec:ResultsMicroTwoQ} and \ref{sec:ResultsCAnoTwoQ} in a double perturbative expansion (in the scalar condensate $\epsilon$ and in the adimensional horizon radius $r_+/L$) about the base BCS two charge black hole \eqref{BCS:TwoQ}. This requires that we also resort to a matched asymptotic expansion analysis with three zones.

But before doing so, we first find heuristically the leading order thermodynamic properties of such hairy solutions using a simple {\it non}-interacting thermodynamic model that does \emph{not} make use of the equations of motion.

\subsubsection*{Hairy BH as a noninteracting mix of BCS BH and supersymmetric soliton}
\label{TwoQ:non-int-model}

Alike for the single charge case of Section~\ref{singleQ:nonintmodel}, the leading order thermodynamics of the two charge hairy BH can be obtained if we take the latter to be a non-interacting mixture (in thermodynamic equilibrium) of the two charge BCS black hole and the hairy supersymmetric soliton. 

The non-interacting model assumes that the total energy (charge) of the mixture is simply a linear sum of the bald black hole mass (charge) and soliton mass (charge). Using the energy and charge \eqref{BCStwoQ:EQ} of the bald BCS BH, we can thus write the energy and charge of the hairy BH as 
\begin{equation}
\begin{split}\label{twoQ:EQnon-int}
E &=  \frac{N^2}{4} [ {\tilde q}  + 3 ( {\tilde q} + r_+^2 ) ( 1 + {\tilde q} + r_+^2 ) ]   + E_{sol} \, , \\
Q &= \frac{N^2}{2} \sqrt{{\tilde q}({\tilde q}+r_+^2)(1+{\tilde q}+r_+^2)}  + \frac{1}{2} E_{sol}  \,,
\end{split}
\end{equation}
where we have also used the fact that the soliton is supersymmetric so $Q_{sol} = \frac{1}{2} E_{sol}$. Naturally, the model further assumes that the BCS BH and soliton are in chemical and thermal equilibrium and thus the chemical potential and temperature of the hairy BH and of its BCS and soliton constituents must be the same
\begin{equation}
\begin{split}\label{twoQ:qtsol}
\mu & \equiv \mu_{BCS}=\mu_{sol}=1 \quad \implies \quad {\tilde q} = \frac{r_+}{2} \left( \sqrt{4 + r_+^2 } - r_+ \right) \,, \\
T & \equiv T_{BCS}  \quad \implies \quad T = \frac{r_+}{2\pi} \left( 2 + \frac{1}{{\tilde q}+r_+^2} \right) = \frac{ 1 + r_+ \sqrt{4 + r_+^2 } + r_+^2  }{ \pi \left( \sqrt{4 + r_+^2 } + r_+  \right) } \, , 
\end{split}
\end{equation}
where we used~\eqref{BCStwoQ:TS} for the chemical potential and temperature of the BCS BH and the fact that the supersymmetric soliton has $\mu_{sol}=1$.

Actually, these conditions also follow from requiring that the system distributes the energy and charge among the two components in such a way that the entropy of the system is maximized while respecting the first law of thermodynamics $\mathrm d E=T \mathrm d S+2\mu \mathrm d Q$. The entropy of the hairy BH is simply the entropy~\eqref{BCStwoQ:TS} of the BCS BH since the soliton is horizonless.

Substituting~\eqref{twoQ:qtsol} into \eqref{twoQ:EQnon-int}, we find
\begin{equation}
\begin{split}
E &= N^2 \left( r_+^2 + \frac{3r_+^4}{8} + \frac{r_+}{8} ( 3 r_+^2 + 4 ) \sqrt{4+r_+^2} \right) + E_{sol}  , \\
Q &= \frac{N^2 r_+}{4} \left( r_+ + \sqrt{ 4 + r_+^2 } \right)  + \frac{1}{2} E_{sol} . 
\end{split}
\end{equation}
Next, we wish to solve for $E_{sol}$ and $r_+$ in terms of $E$ and $Q$. This is difficult to do exactly, but we can find a solution in a perturbative expansion in $x = \frac{1}{N} \sqrt{2(E - 2 Q)}$,
\begin{equation}
\begin{split}\label{eq:rEsolexp}
r_+ &= x-\frac{3 x^2}{4}+\frac{33 x^3}{32}-\frac{57 x^4}{32}+\frac{7119 x^5}{2048}-\frac{3741 x^6}{512}+\CO(x^7) , \\
E_{sol} &= \frac{2Q}{N^2} -x+\frac{x^2}{4}-\frac{13 x^3}{32}+\frac{3 x^4}{4}-\frac{3095 x^5}{2048}+\frac{207 x^6}{64} + \CO(x^7). 
\end{split}
\end{equation}
Since we must have $r_+ \geq 0$ and $E_{sol} \geq 0$, we find bounds on $x$
\begin{equation}
\begin{split}\label{eq:xboundnonint}
0 \leq x \leq \frac{2Q}{N^2} + \frac{Q^2}{N^4} -\frac{9 Q^3}{4N^6}+\frac{41 Q^4}{8N^8}-\frac{757 Q^5}{64N^{10} }+\frac{3543 Q^6}{128N^{12} } + \CO((Q/N^2)^8) . 
\end{split}
\end{equation}
We can recast this into a bound for $E$,
\begin{equation}\label{twoQ:mixEbounds}
\begin{split}
2 Q  \leq E \leq 2 Q \left( 1 + \frac{Q}{N^2}  + \frac{Q^2}{N^2( N^2 + 2 Q )}  \right) .
\end{split}
\end{equation}
The lower bound describes the energy $E=2Q$ of the two charge hairy soliton, i.e. the energy and charge partition of the mixture in which they are all stored in the soliton component. On the other extremum, the energy and charge partition is such they are stored in the BCS BH constituent of the mixture. This happens when the hairy BH merges with the BCS BH at the onset of the instability of the latter. To leading order in a small perturbation in $E$ and $Q$ this merger/onset is described by the upper bound of~\eqref{twoQ:mixEbounds}. That is to say, the lower bound of \eqref{twoQ:mixEbounds} describes the dashed black line (the supersymmetric soliton) in Fig.~\ref{fig:microtwo}, while the upper bound of ~\eqref{twoQ:mixEbounds} describes, within a good approximation valid for small $E$ and $Q$, the solid blue curve of Fig.~\ref{fig:microtwo}.

\subsubsection*{Basic Setup for Perturbation Theory}

We are ready to construct analytically the hairy black hole of the two charge truncation using a double expansion perturbation theory (in the charged scalar condensate amplitude $\epsilon$ and on the adimensional horizon radius $r_+/L$), supplemented with a matched asymptotic expansion procedure with three zones (not with the two zones used in the single charge case). 

Given the intricacy of the discussion and analysis that will follow it is perhaps better to first give a panoramic overview of the matched asymptotic expansion procedure we will employ.
To be able to solve analytically the inhomogeneous ODEs of the problem at each order in perturbation theory, we need to employ a matched asymptotic expansion (at each order) whereby we divide the outer domain of communications of our hairy black hole into three regions, namely the \emph{near-field}, \emph{intermediate} and \emph{far-field} regions (we leave the reasons that justify them for later and next we simply introduce/define them). 
The \emph{far-field} region is the zone $r\gg \sqrt{r_+}$ where the fields are expanded around global AdS$_5$ and we impose the asymptotic boundary conditions.
The \emph{intermediate-region} spans the range $r_+ \ll r\ll L\equiv 1$ where the fields are essentially expanded around a neutral BH \eqref{TwoQ:pert:BCSintermediate} (with perturbatively weak $U(1)$ gauge fields) written in isotropic coordinates  $\{\tau=t/\sqrt{r_+},y=r/\sqrt{r_+}\}$.
Finally, the \emph{near-field} region  covers the domain $r_+\leq r\ll \sqrt{r_+}$ where the fields are expanded around the AdS$_2\times S^3$ near-horizon geometry \eqref{TwoQ:pert:BCSnear} of the two charge BCS BH (moduli a conformal factor), we use the radial coordinate $z=r/r_+$ and we impose the horizon boundary conditions. Restricting the analysis to small black holes that have $r_+/L \ll 1$ (which is necessarily the case by construction since this is one of our expansion parameters), we see that the intermediate and far-field regions overlap in the zone $\sqrt{r_+}\ll r \ll L\equiv 1$, while the near-field and intermediate regions have an overlapping zone in $r_+\ll r \ll \sqrt{r_+}$. In each of the 3 zones we have a ODE system of order 8 which means that, at each order $\CO(\epsilon^{2n}r^k)$ we have 8 integration constants in each of the three zones, i.e. a total of 24 integration constants plus the the parameter $\eta_{(2n,2k)}$ to determine. These are all fixed by the horizon and asymptotic boundary conditions and by the matching conditions in the near-intermediate and intermediate-far overlapping zones. It is worth emphasizing two important aspects of this perturbation method. Firstly, note that when moving along the far$\to$intermediate$\to$near regions we work successively with the radial coordinates $\{r\}\to \{y=\frac{r}{\sqrt{r_+}}\}\to \{z=\frac{y}{\sqrt{r_+}}=\frac{r}{r_+}\}$, i.e. at each step we absorb a factor of $r_+^{-1/2}$ in the new radial coordinate. This is keystone of the matched asymptotic expansion. This choice of radial coordinate redefinitions is not arbitrary; instead it is selected unequivocally by the system as we shall see later. Secondly, this is also a systematic unambiguous matched asymptotic expansion in the sense that the only quantities that are taken to be small are the expansion parameters $\epsilon$ and $r_+/L$ of the double perturbation theory and thus, at each order in the expansion, we keep \emph{all} contributions in the perturbative equations of motion (with no single exception). Nowhere at any stage of the analysis do we make any further assumption neither do we neglect some contribution in the perturbative equations of motion.\footnote{Note that when studying linear perturbations about black holes there are many studies in the literature, starting in the 70's, that use a less systematic matched asymptotic expansion where, besides taking the expansion parameter to be small, it is also assumed that some other quantity (usually the frequency of the perturbation mode in horizon radius units) is small and it is argued that some contributions in the near-field and far-field regions can then be neglected when compared with other terms. In this sense, such an analysis has approximations that are not implied by the smalleness of the expansion parameter and as such it less systematic, robust and accurate. Typically, any such problems (although time dependent) can be solved using the systematic exact approach we employ here after identifying the natural novel radial coordinates of the system (which absorb powers of $r_+$) in each matched asymptotic region.}   
 
After these introductory remarks, we are ready to initiate the perturbative construction of the hairy BH solution, we first expand all the fields in the charged scalar condensate amplitude $\epsilon$ as
\begin{equation}
\begin{split}\label{twoQ:hbh-exp}
f(r,\e,r_+) &= f_\0(r,\e,r_+) + \sum_{n=1}^\infty \e^{2n} f_{(2n)}(r,r_+)  , \\
h(r,\e,r_+) &= h_\0(r,\e,r_+) +  \sum_{n=1}^\infty \e^{2n} h_{(2n)}(r,r_+)  , \\
A_t(r,\e,r_+) &= A_\0(r,\e,r_+) +  \sum_{n=1}^\infty \e^{2n} A_{(2n)}(r,r_+) , \\
\varphi(r,\e,r_+) &= \varphi_\0(r,\e,r_+) + \sum_{n=1}^\infty \e^{2n} \varphi_{(2n)}(r,r_+) , \\
\Phi(r,\e,r_+) &= \sum_{n=0}^\infty \e^{2n+1} \Phi_{(2n+1)}(r,r_+) ,
\end{split}
\end{equation}
where the leading order solution is the two charge BCS black hole of Section~\ref{sec:TwoQ-BCS},
\begin{equation}
\begin{split}\label{twoQ:basebh}
f_\0(r,\e,r_+) &= \left( 1 - \frac{r_+^2}{r^2} \right) ( r^2 + 1 + 2 {\tilde q}(\e,r_+) + r_+^2 )  , \\
h_\0(r,\e,r_+) &= \exp [ 3 \varphi_\0(r,\e,r_+) ] = 1 + \frac{{\tilde q}(\e,r_+)}{r^2}  , \\
A_\0(r,\e,r_+) &= \sqrt{\frac{{\tilde q}(\e,r_+)(1+{\tilde q}(\e,r_+)+r_+^2)}{{\tilde q}(\e,r_+)+r_+^2}} \left( 1 - \frac{{\tilde q}(\e,r_+)+r_+^2}{{\tilde q}(\e,r_+)+r^2} \right) .
\end{split}
\end{equation}
Based on our preliminary analysis in the previous section, we see that in the hairy BH solution ${\tilde q} = \CO(r_+)$, where $q$ is originally defined in \eqref{BCS:TwoQ-def-q} and recall that ${\tilde q} =q/L^2\equiv q$ (since $L=1$ in this section). Consequently, we are invited to write

\begin{equation}\label{twoQ:hbh-exp2}
\begin{split}
{\tilde q}(\e,r_+) = r_+ \eta (\e,r_+)\,, 
\end{split}
\end{equation}
where $\eta(\e,r_+)$ is also expected to get corrections at each order in perturbation theory (since it determines the chemical potential, energy and charge of the solution). Thus, we also expand $\eta(\e,r_+)$ in powers of $\e$,
\begin{equation}\label{twoQ:hbh-exp2b}
\begin{split}
\eta (\e,r_+)  = \sum_{n=0}^\infty \e^{2n} \eta_{(2n)}(r_+).
\end{split}
\end{equation}
The perturbative parameter $\e$ is unambiguously defined to be the expectation value of the dual scalar operator $\e = \avg{ \CO_\Phi }$.

We substitute \eqref{twoQ:hbh-exp} into the equations \eqref{twoQ:E11}--\eqref{twoQ:P1} and obtain differential equations at each order in $\e$. These are too complicated to solve exactly. We proceed by introducing a second expansion parameter (the adimensional horizon radius). That is we further expand the functions in powers of $r_+$,
\begin{equation}
\begin{split}\label{twoQ:farfieldexp}
f_{(2n)}(r) &= \sum_{k=0}^\infty r_+^k f_{{(2n,k)}}(r) , \\
h_{(2n)}(r) &= \sum_{k=0}^\infty r_+^k h_{{(2n,k)}}(r) , \\
A_{(2n)}(r) &= \sum_{k=0}^\infty r_+^k A_{{(2n,k)}}(r) , \\
\varphi_{(2n)}(r) &= \sum_{k=0}^\infty r_+^k \varphi_{{(2n,k)}}(r)  , \\
\Phi_{(2n+1)}(r) &= \sum_{k=0}^\infty r_+^k \Phi_{{(2n+1,k)}}(r),
\end{split}
\end{equation}
and
\begin{equation}
\begin{split}
\eta_{(2n)}(r_+) = \sum_{k=0}^\infty r_+^k \eta_{(2n,2k)} . 
\end{split}
\end{equation}
Note that in the limit $r_+ \to 0$, the background two charge BCS black hole \eqref{twoQ:basebh} (with ${\tilde q} = \eta r_+$) reduces to empty global AdS$_5$. Consequently, the perturbative expansion constructed here is actually an expansion around vacuum AdS$_5$. This is in stark contrast with the one charge case where the perturbative expansion was around a singular soliton solution (with associated issue).

At each order $\CO(\e^m r_+^n)$, we obtain differential equations for the component functions which are exactly off the form \eqref{twoQ:solitoneq1} and \eqref{twoQ:solitoneq2} (the soliton solution is also constructed in a perturbative expansion around global AdS$_5$ so we get the same set of equations). Four of the nine integration constants are fixed by AdS boundary conditions at $r=\infty$. The leftover constants are to be fixed by matching conditions with the interior region. To define the latter, start by noting that at small $r$, all the fields at $\CO(r_+^k)$ blow up as $\frac{1}{r^{2k}}$ which implies that the expansion \eqref{twoQ:farfieldexp} is really an expansion in $r_+/r^2$ and the far-field analysis breaks down at $\CO(\sqrt{r_+}/r)\sim 1$. Consequently, the expansion is valid only for $r \gg \sqrt{r_+}$. We refer
to this region as the \emph{far-field region}. 

This discussion also indicates unequivocally that, to move further down towards the horizon, we should introduce an \emph{intermediate-field region} governed by the new radial coordinate (and time coordinate rescaled to have the same dimensions as $y$)
\begin{equation}\label{TwoQ:pert:defy}
y=\frac{r}{\sqrt{r_+}}\,,\qquad \tau=\frac{t}{\sqrt{r_+}}.    
\end{equation}
The \emph{intermediate-field region} is the region spanned by $\sqrt{r_+} \ll y \ll \frac{L}{\sqrt{r_+}}\equiv \frac{1}{\sqrt{r_+}}$ or equivalently by $r_+ \ll r \ll L \equiv 1$, for reasons that will be understood soon.
To construct the solution in the \emph{intermediate-field region} one introduces the intermediate-fields (denoted by the superscript $^{\rm int}$) that are now a function of $y$:
\begin{equation}
\begin{split}
f_{(2n)}^\intt(y,r_+) &\equiv f_{(2n)}(y \sqrt{r_+}  ,r_+) , \\
h_{(2n)}^\intt(y,r_+) &\equiv h_{(2n)}(y \sqrt{r_+}  ,r_+) , \\
A_{(2n)}^\intt(y,r_+) &\equiv A_{(2n)}(y \sqrt{r_+}  ,r_+) , \\
\varphi_{(2n)}^\intt(y,r_+) &\equiv \varphi_{(2n)}(y \sqrt{r_+}  ,r_+) , \\
\Phi_{(2n+1)}^\intt(y,r_+) &\equiv \Phi_{(2n+1)}(y \sqrt{r_+}  ,r_+) , 
\end{split}
\end{equation}
and expand each of these fields in powers of $r_+$ keeping $y$ fixed,
\begin{equation}
\begin{split}\label{twoQ:intfieldexp}
f_{(2n)}^\intt(y,r_+) &= \sum_{k=0}^\infty r_+^k f^\intt_{(2n,k)}(y) , \\
h_{(2n)}^\intt(y,r_+) &=  \sum_{k=0}^\infty r_+^k h^\intt_{(2n,k)}(y ) , \\
A_{(2n)}^\intt(y,r_+) &=  \sum_{k=0}^\infty r_+^k A^\intt_{(2n,k)}(y) , \\
\varphi_{(2n)}^\intt(y,r_+) &=  \sum_{k=0}^\infty  r_+^k  \varphi^\intt_{(2n,k)}(y) , \\
\Phi_{(2n+1)}^\intt(y,r_+) &=  \sum_{k=0}^\infty  r_+^k  \Phi^\intt_{(2n+1,k)}(y) . 
\end{split}
\end{equation}
In this intermediate region, we are zooming closer to the horizon of the black hole (but not too close!) and the perturbative expansion is no longer around global AdS$_5$. Indeed, introducing the intermediate coordinates \eqref{TwoQ:pert:defy} into the two charge BCS background \eqref{twoQ:basebh} and taking $r_+ \to 0$, we find that the metric and $U(1)$ gauge fields of the base solution in this intermediate region are
\begin{equation}\label{TwoQ:pert:BCSintermediate}
\begin{split}
r_+^{-1} \dt s_\0^2 &= V_0^{-\frac{4}{3}}  [ - \dt \tau^2 + V_0^2  (  \dt y^2 + y^2 \dt \O_3^2  )  ]  , \qquad V_0(y) = 1 + \frac{1}{y^2} \,, \\
A^1=A^2 &= \sqrt{r_+} A_{(0)}\mathrm{d}\tau\,,
\end{split}
\end{equation}
where we have used the fact that $\eta \to 1$ as $r_+ \to 0$. 
So, in the \emph{intermediate-field region} the $U(1)$ gauge fields $A^1=A^2$ have now an explicit factor of $\sqrt{r_+} \ll 1$ (also relative to the gravitational fields), and thus they are weak (i.e. negligible when compared with the mass scale set by the horizon radius and thus with the gravitational fields) and are to be seen as a small perturbation around a \emph{neutral} bald BH whose geometry is described in isotropic coordinates by \eqref{TwoQ:pert:BCSintermediate}. 

We substitute \eqref{twoQ:intfieldexp} into the equations of motion to obtain differential equations at each $\CO(\e^m r_+^k )$. It turns out that the solutions at $\CO(\e^m r_+^k)$ behave as $y^{2k}$ at large $y$ and as $y^{-2k}$ at small $y$. Consequently, at small $y$ the expansion \eqref{twoQ:intfieldexp} is really an expansion in $r_+/y^2$ (and thus breaks dow at $\CO(y)\sim \CO(\sqrt{r_+})$), and at large $y$ it is really an expansion in $y^2 r_+$ (and thus breaks down at $\CO(y)\sim \CO\big(\frac{1}{\sqrt{r_+}}\big)$). This justifies why the \emph{intermediate-field } perturbative expansion is valid only for $\sqrt{r_+} \ll y \ll \frac{1}{\sqrt{r_+}}$ or equivalently when $r_+ \ll r \ll 1$, as we have been claiming. The intermediate-field region overlaps with the far-field region when $\sqrt{r_+} \ll r \ll 1$ and the intermediate and far solutions may be matched here which fixes the far-field region integration constants that were not fixed by the asymptotic boundary conditions. 

As stated above, the intermedate-field expansion breaks down at $\CO(y)\sim \CO(\sqrt{r_+})$. This unequivocally suggests that, to move further towards the horizon, we should introduce an \emph{near-field region} spanning the zone $\sqrt{r_+}\leq y \ll 1$ (i.e. $r_+\leq r\ll \sqrt{r_+}$, for reasons that will be understood soon) governed by the new radial and time coordinates 
\begin{equation}\label{TwoQ:pert:defz}
z=\frac{y}{\sqrt{r_+}}=\frac{r}{r_+}\,,\qquad \tilde{\tau} = \sqrt{r_+} \tau = t .    
\end{equation}
That is to say, the near-field region is described by $1\leq z \ll \frac{1}{\sqrt{r_+}}$.

To construct the solution in the \emph{near-field region}, one introduces the near-fields (denoted by the superscript $^{\rm near}$) that are a function of $z$:
\begin{equation}
\begin{split}
f_{(2n)}^\near(z,r_+) &\equiv f_{(2n)}(zr_+  ,r_+) , \\
h_{(2n)}^\near(z,r_+) &\equiv h_{(2n)}(zr_+  ,r_+) , \\
A_{(2n)}^\near(z,r_+) &\equiv A_{(2n)}(zr_+  ,r_+) , \\
\varphi_{(2n)}^\near(z,r_+) &\equiv \varphi_{(2n)}(zr_+  ,r_+) , \\
\Phi_{(2n+1)}^\near(z,r_+) &\equiv \Phi_{(2n+1)}(zr_+  ,r_+) , 
\end{split}
\end{equation}
and expand each of these fields in powers of $r_+$. Looking at the base solution \eqref{twoQ:basebh} in the limit $r_+ \to 0$ (keeping $z=r/r_+$ fixed), we find that $f_\0,\varphi_\0 = \CO(1)$, $h_\0 = \CO(\frac{1}{r_+})$ and $A_\0 = \CO(r_+)$. We must therefore expand the near-field functions in powers of $r_+$ as
\begin{equation}
\begin{split}\label{twoQ:nearfieldexp}
f_{(2n)}^\near(z,r_+) &= \sum_{k=0}^\infty r_+^k f^\near_{(2n,k)}(z) , \\
h_{(2n)}^\near(z,r_+) &=  \frac{1}{r_+} \sum_{k=0}^\infty r_+^k h^\near_{(2n,k)}(z ) , \\
A_{(2n)}^\near(z,r_+) &=  r_+ \sum_{k=0}^\infty r_+^k A^\near_{(2n,k)}(z) , \\
\varphi_{(2n)}^\near(z,r_+) &=  \sum_{k=0}^\infty  r_+^k  \varphi^\near_{(2n,k)}(z) , \\
\Phi_{(2n+1)}^\near(z,r_+) &=  \sum_{k=0}^\infty  r_+^k  \Phi^\near_{(2n+1,k)}(z) . 
\end{split}
\end{equation}
In this near-region, we are zooming extremely close to the horizon of the black hole. The background geometry for the perturbative expansion in this region is given by
\begin{equation}\label{TwoQ:pert:BCSnear}
\begin{split}
\dt s^2 &= r_+^{\frac{4}{3}} z^{\frac{2}{3}} \left[ - (z^2-1) \dt \tilde{\tau}^2 +  \frac{ \dt z^2}{z^2 -1}  + \dt \O_3^2 \right]  , \\
\end{split}
\end{equation}
We see that up to the conformal factor this is precisely the $\ads_2 \times S^3$ geometry which describes the near-horizon geometry of the two charge BCS black hole.

We substitute \eqref{twoQ:nearfieldexp} into the equations of motion and obtain ODEs at each order in $\CO(\e^m r_+^k)$. We can solve them analytically, and some of its integration constants are fixed by regularity on the horizon $z=1$. The leftovers are fixed by the matching conditions with the intermediate-field region. For that note that at large $z$, the solutions at $\CO(\e^m r_+^k)$ diverge as $z^{2k}$ and thus perturbation theory in this near-region breaks down at $\CO(z)\sim\CO\big( \frac{1}{\sqrt{r_+}} \big)$ which implies that the near-field perturbative expansion is valid only for $1 \leq z \ll \frac{1}{\sqrt{r_+}}$ or equivalently for $r_+ \leq r \ll \sqrt{r_+}$, as claimed before \eqref{TwoQ:pert:defz}. Since our expansion parameter is $r_+\ell 1$, this near-field region overlaps with the intermediate-field region when $r_+ \ll r \ll \sqrt{r_+}$ and the matching of the near-field and intermediate-field solutions provide conditions to fix the near-field integration constants that were not fixed by the horizon boundary conditions.

The detailed forms of all the differential equations, their solutions and the matching process is described in Appendix \ref{app:twoQ:perthy-hbh}. Explicit results are shown in the Appendix to leading order in $r_+$. The full explicit solution can be found in the accompanying \emph{\tt Mathematica} file. We can extract all the thermodynamic quantities from the full solution and they are given by
\begin{subequations}
\begin{equation}
\begin{split}
\frac{EL}{N^2} &= \bigg[ \frac{r_+}{L} +\frac{r_+^2}{L^2} +\frac{11 r_+^3}{8L^3}  +\frac{r_+^4}{8 L^4}  \left[ -16 \ln \frac{r_+}{L}-3\right] \\
&\qquad \qquad  + \frac{r_+^5}{128 L^5}  \left[ 512 \ln ^2\frac{r_+}{L}+768 \ln \frac{r_+}{L}+483\right] + \CO \left( \frac{r_+^6}{L^6} \right) \bigg] \\
&  +\e^2 \left[ \frac{1}{8} +\frac{r_+^2}{48 L^2}  \left(15-\pi ^2\right)  +\frac{r_+^3}{32 L^3}   \left(-8 \ln \frac{r_+}{L}-12 \zeta (3)+3\pi ^2-26\right) \right. \\
&\left.  + \frac{r_+^4}{8640 L^4} \left(4320 \ln^2\frac{r_+}{L}+16560 \ln \frac{r_+}{L}+18780 \zeta (3)  -87 \pi ^4-3405 \pi ^2\right. \right. \\
&\left. \left.  +32980\right) + \CO \left( \frac{r_+^5}{L^5} \right) \right]  +\e^4 \left[ -\frac{1}{384} -\frac{r_+}{288 L} + \frac{r_+^2}{69120 L^2}  (-6965+300 \pi ^2\right. \\
&\left.   +42 \pi^4 ) + \CO \left( \frac{r_+^3}{L^3} \right) \right] + \CO(\e^6) , \\
\frac{QL}{N^2} &= \bigg[ \frac{r_+}{2 L}  + \frac{r_+^2}{4 L^2} +\frac{5 r_+^3}{16L^3} +\frac{r_+^4 }{2 L^4} \left(-2 \ln \frac{r_+}{L}-1\right) \\
&\qquad \qquad + \frac{r_+^5}{256 L^5}  \left[ 512 \ln ^2\frac{r_+}{L}+896 \ln \frac{r_+}{L}+519\right] + \CO \left( \frac{r_+^6}{L^6} \right)  \bigg] \\
& +\e^2 \left[ \frac{1}{16} +\frac{r_+^2}{96 L^2} \left(15-\pi ^2\right) +\frac{r_+^3 }{192 L^3} \left(-24 \ln \frac{r_+}{L}-36 \zeta(3)+11 \pi ^2-111\right) \right. \\
&\left.  + \frac{r_+^4}{17280 L^4}  \left(4320 \ln^2\frac{r_+}{L}+20880 \ln \frac{r_+}{L}+22020 \zeta (3)  -87 \pi ^4-4170 \pi ^2\right. \right. \\
&\left. \left.  +39550\right) + \CO \left( \frac{r_+^5}{L^5} \right) \right] +\e^4 \left[ -\frac{1}{768} -\frac{r_+}{576 L} + \frac{r_+^2}{138240 L^2}  (-6965+300 \pi^2 \right. \\
&\left.+42 \pi ^4 ) + \CO \left( \frac{r_+^3}{L^3} \right)  \right] + \CO(\e^6) , \\
\mu &= \bigg[ 1+\frac{r_+^3}{2L^3}  +\frac{r_+^4}{4 L^4}  \left(-8 \ln \frac{r_+}{L}-7\right)  + \frac{r_+^5}{16 L^5}  \left(64 \ln ^2\frac{r_+}{L}+160 \ln \frac{r_+}{L}+101\right) \\
&+ \CO \left( \frac{r_+^6}{L^6} \right) \bigg] + \e^2 \left[\frac{r_+^3}{12 L^3} + \frac{r_+^4 }{288 L^4} \left(-168 \ln \frac{r_+}{L}-9 \pi ^2-103\right) + \CO \left( \frac{r_+^5}{L^5} \right) \right] \\
& + \e^4 \left[ \CO \left( \frac{r_+^3}{L^3} \right)   \right]  + \CO(\e^6) , \\
\end{split}
\end{equation}

\begin{equation}
\begin{split}
TL &= \bigg[ \frac{1}{2 \pi } +\frac{3 r_+}{4 \pi  L} -\frac{3r_+^2}{16 \pi  L^2} +\frac{r_+^3}{\pi  L^3}  \left(\ln \frac{r_+}{L}+1\right) \\
&\qquad \qquad + \frac{r_+^4}{256 \pi  L^4}  \left(-512 \ln ^2\frac{r_+}{L}-1408 \ln \frac{r_+}{L}-905\right)  + \CO \left( \frac{r_+^5}{L^5} \right)  \bigg] \\
& +\e^2 \left[ \frac{3 r_+}{32 \pi  L}  +\frac{r_+^2 }{192 \pi  L^2} \left(-96 \ln \frac{r_+}{L}-3 \pi ^2-89\right) + \frac{r_+^3 }{576 \pi  L^3} \left(720 \ln ^2\frac{r_+}{L} \right. \right. \\
&\left. \left. +2364 \ln \frac{r_+}{L}-162 \zeta (3)+63 \pi ^2+1366\right) + \CO \left( \frac{r_+^4}{L^4} \right)  \right] \\
&+ \e^4 \left[ \frac{3 r_+}{512 \pi  L} + \CO \left( \frac{r_+^2}{L^2} \right) \right]  + \CO(\e^6) , \\
\frac{S}{N^2} &= \bigg[ \frac{\pi  r_+^2}{L^2} +\frac{\pi  r_+^3}{2 L^3}  +\frac{5\pi  r_+^4}{8 L^4}  -\frac{\pi  r_+^5}{2 L^5}  \left(4 \ln \frac{r_+}{L}+3\right) \\
&\qquad \qquad  + \frac{\pi  r_+^6}{128 L^6} \left(512 \ln ^2\frac{r_+}{L}+1152 \ln \frac{r_+}{L}+711\right)   + \CO \left( \frac{r_+^7}{L^7} \right)  \bigg]  \\
&+ \e^2 \left[ -\frac{\pi  r_+^3}{48 L^3}  \left(2 \pi ^2-27\right) +\frac{\pi  r_+^4}{96 L^4}  \left(-48 \ln \frac{r_+}{L}-72 \zeta(3)+23 \pi ^2-185\right) \right. \\
&\left.  - \frac{\pi  r_+^5}{4320 L^5}  \left(-6480 \ln^2\frac{r_+}{L}-13140 \ln \frac{r_+}{L}-22830 \zeta (3)+87 \pi ^4+4710 \pi ^2 \right. \right. \\
&\left. \left. -34840\right)  + \CO \left( \frac{r_+^6}{L^6} \right)  \right] + \e^4 \left[ \frac{\pi r_+^3}{5760 L^3}   \left(-1070+50 \pi ^2+7 \pi ^4\right)  +  + \CO \left( \frac{r_+^4}{L^4} \right)  \right] \\
&+ \CO(\e^6) . 
\end{split}
\end{equation}
\label{eq:a_bit_crazy}
\end{subequations}
Here, we have reinstated the AdS radius $L$. It can be verified that these quantities satisfies the first law of thermodynamics, $\mathrm{d}E=T \mathrm{d}S+2\mu \mathrm{d}Q$, for the two charge system. As it could not be otherwise, when we set $r_+=0$ in these expressions we recover the thermodynamic expansion \eqref{TwoQ:pert:solitonThermo} for the two charge supersymmetric soliton (i.e. the dashed black curve of Figs.~\ref{fig:microtwo} and \ref{fig:3D_entropy_two}) . On the other hand, when we set $\epsilon=0$, \eqref{eq:a_bit_crazy} yields the scalar onset curve where the hairy black hole family merges with the two charge BCS black hole family (i.e. the perturbative description of the solid blue curve of Figs.~\ref{fig:microtwo} and \ref{fig:3D_entropy_two}). For finite $\epsilon$ and $r_+/L$, \eqref{eq:a_bit_crazy} gives the perturbative description of the hairy black holes (red surface of Figs.~\ref{fig:microtwo} and \ref{fig:3D_entropy_two} for small $E$ and $Q$).    
  
\begin{figure}[th]
\centering \includegraphics[width=\textwidth]{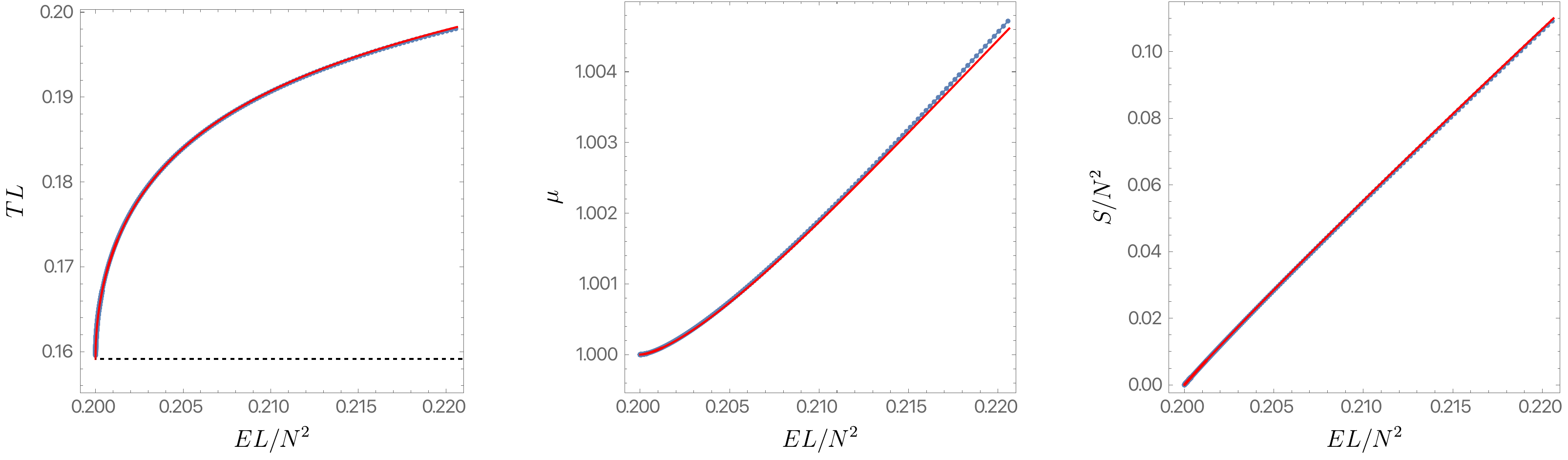}
\caption{\label{fig:comp_tute} Comparison between the analytic expressions given by~\eqref{eq:a_bit_crazy} (depicted as the solid red lines) and the exact numerical results (represented by the blue disks). All data in these three figures was collected at fixed $L Q/N^2 =0.1$. The horizontal dashed line on the left most plot represents $L T= 1/(2\pi)$.}
\end{figure}

In Fig.~\ref{fig:comp_tute}, we compare the perturbative expressions \eqref{eq:a_bit_crazy} $-$ represented by the solid red curves $-$ for the adimensional temperature, chemical potential and entropy  with our exact numerical calculations (described by the blue disks). For small energy and charge, the agreement is striking and reassures that both our analytic expansions and numerical methods are working as expected. These plots where generated at fixed $L Q/N^2=0.1$.

As in the single charge case, it can be verified that the exact expressions \eqref{eq:a_bit_crazy} are consistent with the non-interacting model of \eqref{TwoQ:non-int-model} at leading order in the perturbative expansion. To see this, we first note that the perturbative expansion of $x = \frac{1}{N} \sqrt{2(EL-2QL)}$ in terms of $r_+$ and $\e$ is
\begin{equation}
\begin{split}
x &= \left[ \frac{r_+}{L}  + \frac{3 r_+^2}{4L^2} + \frac{11 r_+^3}{32L^3 } + \CO \left( \frac{r_+^4}{L^4} \right) \right]  + \e^2 \left[ \frac{r_+^2}{L^2} \frac{33-2 \pi^2}{96}  \right. \\
&\qquad \qquad \left. +\frac{r_+^3}{384L^3} \left(-192 \ln \frac{r_+}{L} -144 \zeta (3)+40 \pi^2-391\right) + \CO\left( \frac{r_+^4}{L^4} \right)\right] \\
&\qquad \qquad \qquad + \e^4 \left[ -\frac{ (33-2 \pi^2)^2 r_+^3}{18432L^3} + \CO\left( \frac{r_+^4}{L^4} \right) \right] + \CO ( \e^6) . 
\end{split}
\end{equation}
Comparing the expansions, we find that $x$ and $\frac{QL}{N^2}$ are perturbatively of the same order. Using this, we can invert the expansions and determine $r_+$ and $\e$ in terms of $Q$ and $x$,
\begin{equation}
\begin{split}
\frac{r_+}{L} &= x - \frac{3x^2}{4} + \CO \left( x^3 , x^2 \frac{QL}{N^2} , x \bigg( \frac{QL}{N^2} \bigg)^2 , \bigg( \frac{QL}{N^2} \bigg)^3 \right)  , \\
\e^2 &= 8 \left( \frac{2QL}{N^2} - x \right) + \frac{2}{3} \left[ 8 \left( \frac{QL}{N^2} \right)^2 - 8 x \frac{QL}{N^2} + 5 x^2 \right] \\
&\qquad \qquad \qquad \qquad \qquad \qquad + \CO \left( x^3 , x^2 \frac{QL}{N^2} , x \bigg( \frac{QL}{N^2} \bigg)^2 , \bigg( \frac{QL}{N^2} \bigg)^3 \right)  . 
\end{split}
\end{equation}
These results are consistent with those of the non-interacting model at leading order in $x$ and $Q$. The match with the expression for $r_+$ in \eqref{eq:rEsolexp} is clear. To match to $E_{sol}$, we use \eqref{TwoQ:pert:solitonThermo} which to leading order in $x$ implies
\begin{equation}
\begin{split}
\frac{E_{sol} L}{N^2} = \frac{2 QL}{N^2} - x + \frac{x^2}{4} + \CO \left( x^3 , x^2 \frac{QL}{N^2} , x \bigg( \frac{QL}{N^2} \bigg)^2 , \bigg( \frac{QL}{N^2} \bigg)^3 \right)  . 
\end{split}
\end{equation}
which matches the expression for $E_{sol}$ in \eqref{eq:rEsolexp}. Using the fact that $r_+ \geq 0$ and $\e^2 \geq 0$, we obtain a bound for $x$
\begin{equation}
\begin{split}
0 \leq x \leq \frac{2QL}{N^2} + \left( \frac{QL}{N^2} \right)^2 - \frac{34}{8} \left( \frac{QL}{N^2} \right)^3 + \CO \left( \left( \frac{QL}{N^2} \right)^4 \right) ,
\end{split}
\end{equation}
which matches \eqref{eq:xboundnonint} to leading order.

\section{Summary and outlook\label{sec:disc}}

In this manuscript, we have studied $U(1)^3$ gauged supergravity in five dimensions, i.e. for the truncation of  $SO(6)$ gauged supergravity that retains a $U(1)^3 \cong SO(6) / \mzz_2^3$ gauge symmetry (associated to the Cartan subgroup of $SO(6)$) and associated chemical potentials. In the absence of sources for the scalar and gauge fields, solutions are described by 6 conserved asymptotic charges -- energy $E$, two angular momenta $J_1$ and $J_2$ and three $U(1)$ charges $Q_1$, $Q_2$ and $Q_3$. We constructed new static ($J_1=J_2=0$) hairy solutions in two different consistent truncations of this theory, adding to the solutions already known for a third truncation \cite{Bhattacharyya:2010yg,Markeviciute:2016ivy,Markeviciute:2018yal}.

The truncation with a single non-vanishing $U(1)$ charge has $Q_1=Q_2=0$ and $Q_3\equiv Q$. In the microcanonical ensemble (Figs.~\ref{fig:2D_single} and \ref{fig:3D_single}), solutions are parameterized by the energy $E$ and charge $Q$ and satisfy the BPS bound $E \geq Q$. Previously known solutions in this truncation were the supersymmetric soliton ($E=Q$) and the static single charge BCS black hole (without charged scalar fields) which exists for all energies $E \geq Q \geq 0$. We have shown that these black holes are unstable to (charged) scalar condensation when $\frac{N^2}{2L} < Q < E \leq E_{\text{onset}}(Q)$. The endpoint of this instability are the  hairy black holes which have been constructed in this manuscript both numerically and to first order in perturbation theory. The single charge hairy solutions exist precisely in the region where the single charge BCS black holes are unstable and are the dominant phase in this region (i.e. have higher entropy for a given $E$ and $Q$). In contrast, the hairy black holes are never the dominant phase in the grand canonical ensemble, where the ensemble is dominated either by the ``large'' BCS black holes or by the thermal AdS$_5$ bath of gravitons.

The truncation with two equal non-vanishing $U(1)$ charges has $Q_1=Q_2\equiv Q$ and $Q_3=0$. In the microcanonical ensemble (see Figs.~\ref{fig:microtwo} and \ref{fig:3D_entropy_two}), solutions are parameterized by energy $E$ and charge $Q$ and satisfy the BPS bound $E \geq 2 Q$. Previously known solution in this truncation is the (bald) static two-charge BCS black hole family which exists when $E \geq E_{\text{min}}(Q) > 2 Q$. Note that in contrast to the single charge case, the regime of existence of two charge BCS black holes in an $E-Q$ phase diagram does extend up to the BPS line (see Fig.~\ref{fig:microtwo}).
This truncation also admits a supersymmetric solution ($E=2Q$) which we constructed numerically and in perturbation theory. We have shown that the two charge BCS black holes are unstable to (charged) scalar condensation when $ E_{\text{min}}(Q) \leq E \leq  E_{\text{onset}}(Q)$ and as in the single charge case, the hairy black holes constructed in this manuscript should be the endpoint of the instability. The hairy solutions exist in the region $2Q \leq E \leq  E_{\text{onset}}(Q)$. Note that this region overlaps with the instability region of the two charge BCS black holes but (unlike BCS) extends all the way up to the BPS bound (see Fig.~\ref{fig:microtwo}). In the region where they coexist, the hairy black holes are always the dominant phase over the two BCS black holes (i.e. they have higher entropy for given $E$ and $Q$). In the grand canonical ensemble, the hairy black holes are never dominant.  the grand-canonical ensemble is dominated either by the ``large'' BCS black holes or by thermal AdS$_5$. 

In addition to the phase diagrams of the two $SO(6)$ truncations studied in this manuscript, we also know the phase diagram of the  truncation with three equal non-vanishing $U(1)$ charges with has $Q_1=Q_2=Q_3\equiv Q$ \cite{Bhattacharyya:2010yg,Markeviciute:2016ivy,Markeviciute:2018yal}. The (static) microcanonical phase diagram of the equal three charge truncation is qualitatively very similar to the phase diagram of the two equal charge case of Figs.~\ref{fig:microtwo} and \ref{fig:3D_entropy_two} (so we crudely borrow these figures to complement the summary of the results of \cite{Bhattacharyya:2010yg,Markeviciute:2016ivy} that we give next). Namely, static three charge BCS solutions (which are literally described by the Reissner-Nordstr\"om$-$AdS$_5$ black hole since in this case the neutral scalars $\varphi_{1,2}$ vanish) exist for $E,Q\geq 0$ in the $E-Q$ phase diagram all the way up to the extremal regular configuration where their temperature vanishes (the equivalent of the black solid thin line in Fig.~\ref{fig:microtwo}). Three charge BCS black holes are however unstable to charged scalar condensation of $\Phi_1=\Phi_2=\Phi_3$ when they are in between this extremal configuration and the instability onset curve (the equivalent of the solid blue line in Fig.~\ref{fig:microtwo}). Three charge hairy black holes merge with the BCS black hole along this onset curve and exist all the way up to the supersymmetric line $E=3Q$ (the equivalent of the black dashed curve in Fig.~\ref{fig:microtwo}). This BPS line $E=3Q$, in its totality, also describes collectively the four hairy supersymmetric solitons of the theory (here, the three equal charge case differs significantly from the two equal charge truncation since the latter has a single regular soliton). In short, for the three charge truncation, the system has a regular supersymmetric soliton family that has a Chandrasekhar limit in the sense that it exists from $E=Q=0$ all the way up to a critical $E=Q=Q_c\neq 0$ (where the central scalar density blows up). Then there is a singular supersymmetric soliton family that departs from $E=Q=Q_c$ and extends all the way to $E=Q\to \infty$ (for a detailed account of the solitons of this truncation, including the other two, see discussion below \eqref{SingleQ:SingularSolitonHoloQuantities})\footnote{Although we have not attempted to prove this, we believe that whenever the three charges $Q_1, Q_2,Q_3$ are non-zero, the solitonic spectra of the system should be similar to the three charge case reviewed here \eqref{SingleQ:SingularSolitonHoloQuantities}. However, when at least one of the $U(1)$ charges vanishes, the regular solitons of the system have no Chandrasekhar limit, i.e. they extend to arbitrarly large energy and charges obeying the BPS relation.}. 
When the hairy black holes terminate on the BPS line below the critical point $Q_c$ they do so smoothly and at $T\to 0$, while when they terminate at the BPS line above $Q_c$ they do approach a singular configuration with $T\to\infty$.
Altogether, the three charge hairy black holes are described by surfaces similar to those displayed in the red region/surface of Figs.~\ref{fig:microtwo} and \ref{fig:3D_entropy_two}. Three charge hairy black holes dominate the microcanonical ensemble; in particular for values of $E,Q$ where they coexist with the BCS black hole, hairy black holes always have higher entropy than BCS (very much like for the single charge and two equal charge truncations studied in this manuscript). Finally, for the three charge truncation the hairy black holes also never dominate the grand-canonical ensemble (alike for the two truncations studied here). The phase diagram in this ensemble is qualitatively similar to the one for the two charge case but with a fundamental difference: the analogue of the  point of Fig.~\ref{fig:two_grand} where the three curves meet is at $T=0$ (and $\mu=1$). 

Altogether, the supergravity solutions that we found and those of  \cite{Bhattacharyya:2010yg,Markeviciute:2016ivy,Markeviciute:2018yal},
permit a good understanding of the full phase space of hairy solutions with three arbitrary $U(1)$ charges $Q_{1,2,3}$. Via the original AdS$_5$/CFT$_4$ correspondence \cite{Maldacena:1997re,Gubser:1998bc,Witten:1998qj,Aharony:1999ti}, this phase diagram of hairy black holes is dual to the phase space o thermal states at finite chemical potential of $\mathcal{N}=4$. 
In future work \cite{DMS2021}, we plan to extend the present study of the supergravity system to include rotation $J_1,J_2$ along the two independent rotation planes of AdS$_5$ with $SO(4)$ symmetry. In the dual CFT, these will describe thermal states at finite chemical potential with weights $J_L \equiv J_1 + J_2$ and $J_R \equiv J_1 - J_2$ of  $SO(4) \sim SU(2)_L \times SU(2)_R$ \cite{Kunduri:2006ek}.

Our discovery of new solutions to gauged supergravity with $\CO(N^2)$ entropy and which are dominant in the microcanonical ensemble is particularly fascinating in light of recent developments in the microstate counting of the entropy of supersymmetric black holes via a twisted index computation on the CFT side (see \cite{Zaffaroni:2019dhb} and references therein). These calculations have reproduced exactly the entropy of the general supersymmetric Kunduri-Lucietti-Reall black holes \cite{Kunduri:2006ek}. The hairy black holes constructed in this manuscript do not have a (smooth) BPS limit with $\CO(N^2)$ entropy. However, we expect such a limit to exist when rotations are turned on $J_1,J_2 \neq 0$ and give rise to new supersymmetric hairy black holes. Further, we also expect such black holes to dominate (in the microcanonical ensemble) over their bald counterparts in the region of phase space where they overlap. Existence of such dominant hairy black hole solutions has already been established in the equal charge and equal angular momentum case $Q_1=Q_2=Q_3$, $J_1=J_2$ \cite{Markeviciute:2018yal} and it is important to explore the phase space of such solutions in the single and two-charge case as well \cite{DMS2021}. 

Additionally, given the new machinery that has been established in the microstate counting of supersymmetric black holes, it is important to understand the interpretation of such new hairy solutions on the CFT side. Indeed, a key puzzle here is that since the hairy solutions are often the dominant phase in the regions of phase space where they exist, the index should be computing the entropy of \emph{these} black holes and \emph{not} the bald ones. A second related puzzle is regarding the existence of supersymmetric $1/8$-BPS black holes. It has long been established that in the absence of charged scalar hair ($\Phi_K=0$), all supersymmetric black holes are $1/16$-BPS. However, recent computations of a twisted SYM index in the so-called Macdonald limit ($Q_3+J_2=0$) suggest an $\CO(N^2)$ entropy for $1/8$-BPS states \cite{Choi:2018hmj}. It is unclear what solutions this index corresponds to in gauged supergravity. We leave such explorations for future work.

\smallskip

\subsection*{Acknowledgments}

We would like to thank Daniel Kapec, R. Loganayagam, Abhishek Pathak and Andrew Strominger for many useful conversations. We would like to especially thank Shiraz Minwalla for extensive discussions in the early stages of this work and Nicholas Warner for useful comments on a draft of our manuscript. The authors acknowledge the use of the IRIDIS High Performance Computing Facility, and associated support services at the University of Southampton, in the completion of this work.
 O.~C.~D. acknowledges financial support from the STFC ``Particle Physics Grants Panel (PPGP) 2018'' Grant No.~ST/T000775/1.  P.~M. and J.~E.~S. work have been partially supported by STFC consolidated grant ST/T000694/1. P.~M. acknowledges support from Birla Institute of Technology and Science and Tata Institute of Fundamental Research in the early stages of this research.

\appendix


\section{Holographic renormalization of $U(1)^3$ gauged supergravity}
\label{app:holo-renorm}

In this appendix, we present all the details of holographic renormalization (following \cite{Bianchi:2001de,Bianchi:2001kw} for a similar theory; see convention discussion of footnote~\ref{footHoloRen}) for $U(1)^3$ gauged supergravity, i.e. for the truncation of  $SO(6)$ gauged supergravity that retains a $U(1)^3 \cong SO(6) / \mzz_2^3$ gauge symmetry (associated to the Cartan subgroup of $SO(6)$) with associated gauge fields $\{A_\1^K\}$. We applied the results of this appendix in Sections \ref{sec:ThermoSingleQ} and \ref{sec:ThermoTwoQ} (among others). $U(1)^3$ gauged supergravity is described by the bulk action \eqref{OurCTaction}, which we can rewrite in differential form as
\begin{equation}
\begin{split}\label{app:action-trunc}
S &= \frac{1}{16\pi G_5} \int_\CM \bigg[  ( R - V ) \star 1 - \frac{1}{2} \sum_{K=1}^3 \frac{1}{X^2_K}  F^K  \wedge  \star F^K  -  \frac{1}{2} \sum_{r=1}^2 \dt \varphi_r \wedge  \star \dt \varphi_r \\
&\qquad \qquad \qquad \qquad - \frac{1}{8} \sum_{K=1}^3 \left( D \Phi_K \wedge   \star D \Phi_K^\dagger - \frac{ \dt \l_K \wedge  \star \dt \l_K}{4(4+\l_K)}  \right) \\
&\qquad \qquad \qquad \qquad  - \frac{1}{6} \sum_{I,J,K=1}^3 C_{IJK} F^I  \wedge F^J \wedge  A^K  \bigg] + S_{\text{bdy}} , \\
\end{split}
\end{equation}
where $\{I,J,K\}=1,2,3$; $\{r,s\}=1,2$, $\star$ is the Hodge dual and $\dt $ is the exterior derivative (we use the differential form conventions listed in appendix of \cite{Dias:2019wof}), $S_{\text{bdy}}$ is a boundary term that we add to \eqref{OurCTaction}, and
\begin{equation}
\begin{split}
F^K = \dt A^K , \quad D \Phi_K = \dt \Phi_K - i \frac{2}{L} A^K \Phi_K , \quad \l_K = | \Phi_K |^2 , \quad C_{IJK} = | \ve_{IJK} | . 
\end{split}
\end{equation}
The scalar potential $V$ is
\begin{equation}
\begin{split}
V &= \frac{1}{2L^2}  \sum_I \bigg( X_I^2 \l_I - \sum_{JK} C_{IJK} X_J X_K \sqrt{ 4 + \l_J } \sqrt{ 4 + \l_K } \bigg).
\end{split}
\end{equation}
where
\begin{equation}
\begin{split}
X_1 &= \exp \left( - \frac{\varphi_1}{\sqrt{6}} - \frac{ \varphi_2 }{ \sqrt{2} } \right) , \\
X_2 &= \exp \left( - \frac{\varphi_1}{\sqrt{6}} + \frac{ \varphi_2 }{ \sqrt{2} } \right) , \\
X_3 &= \exp \left( \sqrt{ \frac{2}{3}} \varphi_1 \right). 
\end{split}
\end{equation}
Again, note that the action \eqref{app:action-trunc} is completely equivalent to \eqref{OurCTaction} up to boundary terms. The boundary action $S_{bdy}$ will be determined via holographic renormalization in this appendix.

Varying the action \eqref{app:action-trunc} and keeping careful track of the boundary terms, we find
\begin{equation}
\begin{split}\label{app:action-var}
\delta S &= \frac{1}{16\pi G_5} \int_\CM (\text{EoM}) \delta (\text{fields}) +  \frac{1}{16\pi G_5} \int_{\p\CM} [ {\bs \Theta}_{GR} (\delta) + {\bs \Theta}_{M} (\delta) ] + \delta S_{\text{bdy}}   . \\
\end{split}
\end{equation}
where the equations of motion (EoM) are given explicitly in \eqref{GravEOM}--\eqref{ScalarsGaugeEOM}. The boundary terms in $\delta S_{bulk}$ are the integral ($\int_{\p\CM}$) over the symplectic potential current density ${\bs \Theta}$ which has the form
\begin{equation}
\begin{split}\label{app:spcd-sugra}
[ \star {\bs \Theta}_{GR} (\delta) ]^a &= \n_b \delta g^{ab} - g_{bc} \n^a \delta g^{bc} , \\
\star {\bs \Theta}_{M} (\delta) &=  \sum_{K=1}^3  \star \bigg[ \frac{1}{X^2_K} \star F^K \wedge  \delta A^K  + \frac{1}{3} \sum_{I,J=1}^3 C_{IJK}  F^I  \wedge A^J \wedge \delta A^K  \bigg]  \\
& +    \sum_{r=1}^2 \dt \varphi_r \delta \varphi_r  + \frac{1}{8} \sum_{K=1}^3   \bigg[ \delta \Phi_K^\dagger   D \delta \Phi_K +   \delta \Phi_K  D \Phi_K^\dagger   - \frac{  \delta \l_K  }{2(4+\l_K)} \dt \l_K  \bigg]   ,
\end{split}
\end{equation}
In this manuscript, we impose Dirichlet boundary conditions on all the fields. This implies that in the variational principle \eqref{app:action-var}, we require
\begin{equation}
\begin{split}
\delta \Phi_K |_{\p\CM} = \delta \varphi_r |_{\p\CM}  = \delta A^K  |_{\p\CM} = \delta h  |_{\p\CM} = 0\,,
\end{split}
\end{equation}
where $h$ is the induced metric on $\p\CM$. It is clear that under these conditions ${\bs \Theta}_M$ vanishes identically. However, ${\bs \Theta}_{GR}$ does not vanish since it depends both on $\delta h$ (which vanishes under Dirichlet boundary conditions) and its normal derivative $\cl_n \delta h$. This can be resolved by adding to this action the famous Gibbons-Hawking-York boundary term namely
\begin{equation}
\begin{split}\label{app:GHbdyterm}
S_{bdy} = \frac{1}{16\pi G_5} \int_{\p \CM} d^4 y \sqrt{-h} ( 2 \CK ) + S_{\text{ct}} , 
\end{split}
\end{equation}
where $\CK$ is the extrinsic scalar curvature on $\p\CM$ and $y^\mu$ are intrinsic coordinates on $\p\CM$. The proof that this boundary term precisely cancels the contribution of ${\bs \Theta}_{GR}$ is standard and will not be repeated here.

Of particular interest to us is the counterterm action $S_{\text{ct}}$. This term is needed to make the on-shell action finite. However, it must not affect the cancellation of the boundary terms in the variational principle so it must satisfy
\begin{equation}
\begin{split}
\delta S_{\text{ct}} |_{\text{Dirichlet b.c.}} = 0 
\end{split}
\end{equation}
It follows that $S_{\text{ct}}$ can depend only on quantities which are intrinsically defined on $\p\CM$. The precise counterterm action needed can be determined via the method of holographic renormalization \cite{Bianchi:2001kw}.

\subsection{Asymptotic expansions}

To renormalize the on-shell action, we need to determine the asymptotic structure of solutions in the theory. To do this, it is convenient to work in Fefferman-Graham (FG) gauge where the metric and gauge fields take the form
\begin{equation}\label{app:FGmetric}
\begin{split}
\dt s^2 = \frac{L^2}{z^2} \dt z^2 + g_{\mu\nu}(z,y) \dt y^\mu \dt y^\nu , \qquad A^K = A^K_\mu(z,y) \dt y^\mu . 
\end{split}
\end{equation}
To regulate the divergences which arise due to the infinite volume of AdS, we set the boundary at $z=\e$. Once all the divergences are cancelled, we will take $\e \to 0$. The outward-pointing unit normal vector on $\p\CM$ is
\begin{equation}
\begin{split}
n = - \frac{1}{L} z \p_z . 
\end{split}
\end{equation}
In these coordinates, the induced metric (first fundamental form) on $\p\CM$ is $h_{\mu\nu}(y) = g_{\mu\nu}(\e,y)$. The extrinsic curvature (second fundamental form) on $\p\CM$ is given by
\begin{equation}
\begin{split}
\CK_{\mu\nu} = \n_\mu n_\nu |_{\p\CM} = - \frac{1}{2L} \e \p_\e g_{\mu\nu} . 
\end{split}
\end{equation}

In $U(1)^3$ gauged supergravity, all five scalar fields have $m^2 L^2 = - 4$. It follows that the asymptotic behavior (near $z=0$) of the scalars is\footnote{In $\ads_{d+1}$, a scalar field with mass $m^2$ has the behaviour $\phi = z^{d-\D} (\cdots ) + z^\D(\cdots)$ near $z=0$ where $\Delta = \frac{d}{2} + \sqrt{ \frac{d^2}{4} + m^2 L^2 }$ is the dimension of the dual operator. There are additional $\log$ terms in the expansion if $\Delta \in \frac{d}{2} + \mzz$ which is the case here.}
\begin{equation}
\begin{split}\label{app:phiexp}
\Phi_K (z,y) &= \frac{z^2}{L^2} \left( {\wt \Phi}^\0_K (y) + \Phi_K^\0 (y) \ln \frac{z}{L} \right) + \CO \left( \frac{z^4}{L^4} \right)  , \\
\varphi_r (z,y) &= \frac{z^2}{L^2} \left( {\wt \varphi}^\0_r (y) + \varphi_r^\0 (y) \ln \frac{z}{L} \right) + \CO \left( \frac{z^4}{L^4} \right)  .
\end{split}
\end{equation}
$\Phi_K^\0$ and $\varphi_r^\0$ are the sources (free data) for the scalar fields. As we will see shortly, under the AdS/CFT dictionary the sources couple to the dual operators in the CFT. In the main text of this manuscript, we have set these sources to zero but in this section, we consider the most general case.

The gauge field is massless  and admits the following asymptotic expansion\footnote{In $\ads_{d+1}$, a vector field with mass $m^2$ has the behaviour $A_\mu = z^{d-1-\D}(\cdots) + z^{\D-1}(\cdots)$ near $z=0$ where $\Delta = \frac{d}{2}  + \sqrt{  ( \frac{d-2}{2}  )^2 + m^2 L^2}$ is the dimension of the dual operator. There are additional log terms when $\Delta = \frac{d}{2} + \mzz$. In the case of a massless gauge field, $m^2 = 0 \implies \Delta=d-1$ and the dual operator is a conserved current.}
\begin{equation}
\begin{split}\label{app:Aexp}
A_\mu^K (z,x)&= A_\mu^{\0K}(y) + \frac{z^2}{L^2} \left[ A_\mu^{\2K}(y) + {\wt A}_\mu^{\2K}(y)\ln \frac{z}{L} \right] + \CO \left( \frac{z^4}{L^4} \right)  , \\
\end{split}
\end{equation}
The scalar and gauge fields source the metric through Einstein's equations and the induced metric $h_{\mu\nu}$ has the following asymptotic expansion\footnote{Note that the definitions of FG coordinate $z$ in the main text and in this Appendix differ slightly by a power of $L$: compare \eqref{SingleQ:FGexpansion1} or \eqref{TwoQ:FGexpansion1} with \eqref{app:FGmetric} and \eqref{app:gexp}. The two agree when we set $L\equiv 1$.}
\begin{equation}
\begin{split}\label{app:gexp}
g_{\mu\nu} ( z , x ) &= \frac{L^2}{z^2} g^\0_{\mu\nu}(x) + g^\2_{\mu\nu} (x) + \frac{z^2}{L^2} \left[ g^\4_{\mu\nu}(x) + \wt{g}^\4_{\mu\nu}(x) \ln\frac{z}{L}  + \wt{\wt{g}}^\4_{\mu\nu}(x) \ln^2 \frac{z}{L} \right] \\
&\qquad \qquad + \CO \left( \frac{z^4}{L^4} \right)  , \\
\end{split}
\end{equation}

We now substitute these expansions \eqref{app:phiexp}--\eqref{app:gexp} into the equations of motion \eqref{GravEOM}--\eqref{ScalarsGaugeEOM} and solve order-by-order in small $z/L$. This is a tedious exercise and we only reproduce the final results here. At each order in $z$, we obtain algebraic equations which fixes most of the coefficients of the fields in the small $z$ expansion in terms of the sources $\Phi^\0_K$, $\varphi^\0_r$, $A^{\0K}_\mu$ and $g^\0_{\mu\nu}$. 

More precisely, ${\wt \Phi}^\0_K$ and ${\wt \varphi}^\0_r$ are \emph{not} fixed by all subleading terms in the expansion are. Similarly, $A_\mu^{\2K}$ is not fixed but ${\wt A}_\mu^{\2K}$ and all further subleading terms in the expansion are,
\begin{equation}
\begin{split}\label{app:Csol}
{\wt A}_\mu^{\2K} &= - \frac{L^2}{2} (\n^\0)^\nu F_{\nu\mu}^{\0K} , \qquad F^{\0K} = \dt A^{\0K} . 
\end{split}
\end{equation}
Here, $\n^\0$ is the covariant derivative w.r.t. $g^\0_{\mu\nu}$. The equations of motion also fix the divergence of $A_\mu^{\2K}$, 
\begin{equation}
\begin{split}\label{app:Bdiv}
( \n^\0 )^\mu A^{\2K}_\mu = - \frac{L}{16} (\ve^\0)^{\mu\nu\rho\s} \sum_{IJ} C_{IJK} F^{\0I}_{\mu\nu} F^{\0J}_{\rho\s} - \frac{1}{4L} \text{Im}  \left[ \Phi_K^{\0\dagger} {\wt \Phi}^\0_K \right] . \\ 
\end{split}
\end{equation}
Here and in the rest of this appendix, Greek indices are raised and lowered w.r.t. the metric $g^\0_{\mu\nu}$. 

Finally all coefficients in the metric expansion apart from $g^\4_{\mu\nu}$ are fixed by the equations of motion as
\begin{equation}
\begin{split}\label{app:metriccoeff}
g^\2_{\mu\nu} &= - \frac{L^2}{2} \left( R^\0_{\mu\nu}  - \frac{1}{6} g^\0_{\mu\nu} R^\0 \right) , \\
\wt{g}^\4_{\mu\nu} &= \frac{1}{4} ( T^{\0\text{grav}}_{\mu\nu} + T^{\0\text{gauge}}_{\mu\nu} )  - \frac{1}{24} g^\0_{\mu\nu}   \bigg[ \sum_{K=1}^3 \text{Re} \left[ (\Phi_K^\0)^\dagger {\wt \Phi}^\0_K \right]  + 4 \sum_{r=1}^2 \varphi_r^\0 \wt{\varphi}^\0_r \bigg] , \\
\wt{\wt{g}}^\4_{\mu\nu} &= - \frac{1}{48} g^\0_{\mu\nu} \bigg[ \sum_{K=1}^3 | \Phi^\0_K |^2  + 4 \sum_{r=1}^2 ( \varphi^\0_r )^2 \bigg] .
\end{split}
\end{equation}
Here, $T^{\0\text{gauge}}_{\mu\nu}$ and $T^{\0\text{grav}}_{\mu\nu}$ are the gauge and gravity anomaly stress tensors respectively,
\begin{equation}
\begin{split}\label{app:anomalyst}
T^{\0\text{grav}}_{\mu\nu} &= L^4 \bigg[ R^\0_{\mu\rho\nu\s} (R^\0)^{\rho\s} - \frac{1}{4} g^\0_{\mu\nu}   ( R^\0_{\rho\s} )^2   - \frac{1}{3} R^\0 ( R^\0)^\tf_{\mu\nu} \\
&\qquad \qquad \qquad \qquad  - \frac{1}{6} ( \n^\0_\mu \n^\0_\nu R^\0 )^\tf + \frac{1}{2} (\n^\0)^2  (R^\0)^\tf_{\mu\nu} \bigg] , \\
T^{\0\text{gauge}}_{\mu\nu} &= - L^2 \sum_{K=1}^3 \bigg[ F^{\0K}_{\mu \rho} (F^{\0K})_\nu{}^\rho - \frac{1}{4} g^\0_{\mu\nu} ( F^{\0K}_{\rho\s}  )^2 \bigg] .
\end{split}
\end{equation}
The superscript TF denotes the trace-free part of the tensor. Both of these stress tensors can be obtained by varying an action
\begin{equation}
\begin{split}\label{app:anomalyst-actvar}
T^{\0\text{grav}}_{\mu\nu} + T^{\0\text{gauge}}_{\mu\nu} &= \frac{2}{ \sqrt{ - g^\0 } } \frac{\delta}{\delta (g^\0)^{\mu\nu} } \int_{\p\CM} \dt^4 y \sqrt{ - g^\0 } \\
&\qquad\times \bigg[ \frac{L^4}{4} \bigg( ( R^\0_{\mu\nu} )^2  - \frac{1}{3}  (R^\0)^2  \bigg) - \frac{L^2}{4} \sum_{K=1}^3 ( F^{\0K}_{\mu\nu} )^2 \bigg] . 
\end{split}
\end{equation}
This action is Weyl invariant so both the stress tensors are traceless (this can be checked explicitly as well). This also implies that $T^{\0\text{grav}}$ is conserved whereas $T^{\0\text{gauge}}$ satisfies
\begin{equation}
\begin{split}
(\n^\0)^\nu T^{\0\text{gauge}}_{\mu\nu} = - L^2 \sum_{K=1}^3 F^{\0K}_{\mu\rho} \n^\0_\nu ( F^{\0K} )^{\nu\rho} .
\end{split}
\end{equation}
In addition to \eqref{app:metriccoeff}, the equations of motion also fix the trace and divergence of $g^\4$. 
\begin{equation}
\begin{split}\label{app:qprop}
\text{tr} [ g^\4 ]  &= \frac{1}{4} ( g^\2_{\mu\nu})^2 + \frac{L^2}{48} \sum_{K=1}^3 ( F^{\0K}_{\mu\nu}  )^2  \\
& - \frac{1}{96}  \bigg[ \sum_{K=1}^3 \left( | \Phi^\0_K |^2 + 8  | \wt{\Phi}^\0_K |^2 \right)   + 4 \sum_{r=1}^2 \left( (\varphi^\0_r)^2 + 8 (\wt{\varphi}_r^\0)^2 \right)  \bigg]  , \\
(\n^\0)^\nu (g^\4)^\tf_{\mu\nu}  &= ( \n^\0 )^\nu \bigg[  \frac{1}{2} [ (g^\0)^2 ]^\tf_{\mu\nu} - \frac{1}{4}  \text{tr} [ g^\2 ] (g^\2)_{\mu\nu}^\tf  \bigg] \\
&\qquad  - \frac{1}{16} \p_\mu \bigg[ \frac{L^4}{4} \bigg( R^\0_{\mu\nu})^2 - \frac{1}{3} ( R^\0 )^2 \bigg) - \frac{L^4}{4} \sum_{K=1}^3 ( F_{\rho\s}^{\0K} )^2  \bigg]  \\
&\qquad  - \frac{1}{128} \p_\mu  \bigg[ \sum_{K=1}^3 | \Phi^\0_K |^2  + 4 \sum_{r=1}^2  (\varphi^\0_r)^2 \bigg]  \\
&\qquad  + \frac{1}{8}  \sum_{K=1}^3 F^{\0K}_{\mu\nu} \left[ 4 (A^{\2K})^\nu  +  (\wt{A}^{\2K})^\nu \right] \\
&\qquad -  \frac{1}{32} \bigg[ \sum_{K=1}^3 \text{Re} \left[  ( \wt{\Phi}_K^\0 )^\dagger \overleftrightarrow{D^\0_\mu} \Phi^\0_K \right] + 4 \sum_{r=1}^2 \wt{\varphi}_r^\0 \overleftrightarrow{\p_\mu} \varphi^\0_r \bigg]  .
\end{split}
\end{equation}

\subsection{Counterterm action}

As mentioned previously, the counterterm action must cancel all the divergences in the on-shell action and at the same time not spoil the variational principle. This is done by taking the counterterm action to depend only on quantities which are intrinsically defined on $\p\CM$. In the present case, it turns out that the following counterterm action does the job:
\begin{equation}
\begin{split}\label{app:ctaction}
S_{\text{ct}} &= \frac{1}{16\pi G_5} \int_{\p\CM} \dt^4 y \sqrt{-h} \left[ - \frac{6}{L} - \frac{L}{2} \CR - \frac{1}{4L} \bigg( 1 + \frac{1}{2\log\frac{\e}{L}} \bigg) \bigg(   \sum_{K=1}^3 |\Phi_K|^2 + 4 \sum_{r=1}^2 \varphi_r^2 \bigg) \right. \\
&\left. \qquad \qquad \qquad\qquad + \frac{1}{L} \log \frac{\e}{L} \bigg[ \frac{L^4}{4} \bigg( ( \CR_{\mu\nu} )^2   - \frac{1}{3} \CR^2 \bigg) - \frac{L^2}{4} \sum_{K=1}^3 ( F_{\mu\nu}^K  )^2 \bigg] \right] ,
\end{split}
\end{equation}
where $\CR_{\mu\nu}$ is the Ricci tensor of $\p\CM$ and $F_{\mu\nu}$ is the induced field strength on $\p\CM$. Note that $S_{\text{ct}}$ explicitly depends on $\log \frac{\e}{L}$ (note that the boundary is located at $z=\e$) which is not covariant. The presence of such terms breaks conformal symmetry and are responsible for the conformal anomaly.

In this section, instead of showing that the full on-shell action is finite with this choice of counterterm action, we will show that the on-shell variation is finite. Importantly, in this calculation, we are no longer imposing Dirichlet boundary conditions on the variations and all divergences must cancel out for all on-shell variations.

Varying the action and imposing equations of motion (and \emph{not} imposing Dirichlet boundary conditions), we find
\begin{equation}
\begin{split}
\delta S_{\text{on-shell}} = \delta S_{\text{grav}} + \delta S_{\text{gauge}} + \delta S_{\text{scalar}} ,
\end{split}
\end{equation}
where
\begin{equation}
\begin{split}
\delta S_{\text{grav}} &= \frac{1}{16\pi G_5} \int_{\p\CM} \dt^4 y \sqrt{-h}  \delta h^{\mu\nu} \bigg( \CK_{\mu\nu} - h_{\mu\nu} \CK  +  \frac{3}{L} h_{\mu\nu} - \frac{L}{2} \CR_{\mu\nu} + \frac{L}{4} h_{\mu\nu} \CR \\
&  + \frac{1}{8L} h_{\mu\nu}  \bigg[ 1 + \frac{1}{2\log\frac{\e}{L}} \bigg] \bigg[ \sum_{K=1}^3 |\Phi_K|^2 + 4 \sum_{r=1}^2 \varphi_r^2 \bigg] + \frac{1}{2L} \log \frac{\e}{L}  \left[ T^{\text{grav}}_{\mu\nu} + T^{\text{gauge}}_{\mu\nu} \right]  \bigg) , \\
\delta S_{\text{gauge}} &= \frac{1}{16\pi G_5} \int_{\p\CM} d^4 y \sqrt{-h} \sum_{K=1}^3 \bigg[ \bigg(  \frac{1}{X^2_K} ( F^K )^{ab}  n_b + \frac{1}{6} \sum_{I,J=1}^3 C_{IJK} \ve^{abcde} F^I_{bc} A^J_d n_e  \bigg)   \delta A^K_a \\
&\qquad \qquad \qquad \qquad \qquad \qquad \qquad \qquad \qquad  +  L \log \frac{\e}{L}  \n_\mu (F^K)^{\mu\nu} \delta A^K_\nu   \bigg] ,   \\
\delta S_{\text{scalar}} &= \frac{1}{16\pi G_5} \int_{\p\CM} d^4 y \sqrt{-h}   \sum_{r=1}^2 \bigg[ - \bigg( n^a \p_a  +  \frac{2}{L} \bigg)  \varphi_r  - \frac{\varphi_r }{L\log\frac{\e}{L}}  \bigg]  \delta \varphi_r    \\
&\qquad +  \frac{1}{16\pi G_5} \int_{\p\CM} d^4 y \sqrt{-h} \sum_{K=1}^3   \bigg[ - \frac{1}{4} \text{Re} \bigg[ \delta \Phi_K^\dagger \left( n^a D_a + \frac{2}{L} \right) \Phi_K \bigg]  \\
&\qquad \qquad \qquad \qquad \qquad \qquad  + \bigg( \frac{ n^a \p_a \l_K  }{16(4+\l_K)}  - \frac{1}{8L\log\frac{\e}{L}} \bigg)   \delta \l_K   \bigg]  . 
\end{split}
\end{equation}
Plugging in the asymptotic expansions \eqref{app:phiexp}--\eqref{app:gexp} into the above, one can explicitly check that all the divergences cancel on-shell and each of these variations are independently finite in the $\e \to 0$ limit. Explicitly, these limits work out to be
\begin{equation}
\begin{split}\label{app:deltaScheck}
\delta S_{\text{scalar}} &= \frac{1}{16\pi G_5} \int_{\p\CM} d^4 y \sqrt{-g^\0} \bigg( - \frac{1}{4L} \bigg[ \sum_{K=1}^3 \text{Re} \left( {\wt \Phi}^{\0\dagger}_K \delta \Phi^\0_K \right) + 4  \sum_{r=1}^2 \wt{\varphi}_r^\0 \delta \varphi^\0_r \bigg]   \bigg) , \\
\delta S_{\text{gauge}} &= \frac{1}{16\pi G_5} \int_{\p\CM} d^4 y \sqrt{-g^\0} \sum_{K=1}^3 \bigg[ \frac{1}{L} \left[ 2 ( A^{\2K} )^\mu + ( {\wt A}^{\2K} )^\mu \right] \\
&\qquad \qquad \qquad \qquad \qquad -  \frac{1}{6} \sum_{I,J=1}^3 C_{IJK}  (\ve^\0)^{\mu\nu\rho\s} F^{\0I}_{\nu\rho} A^{\0J}_\s  \bigg]\delta A^{\0K}_\mu , \\
\delta S_{\text{grav}} &= \frac{1}{2} \int \dt^4 y \sqrt{-g^\0} \avg{ \CT^{\mu\nu} } \delta g^\0_{\mu\nu} \,,
\end{split}
\end{equation}
where
\begin{equation}
\begin{split}
\avg{ \CT_{\mu\nu} } &= - \lim_{z \to 0}\frac{1}{8\pi G_5}  \frac{L^2}{z^2} \bigg[ \CK_{\mu\nu} - h_{\mu\nu} \CK  +  \frac{3}{L} h_{\mu\nu} - \frac{L}{2} \CR_{\mu\nu} + \frac{L}{4} h_{\mu\nu} \CR \\
& + \frac{1}{8L} h_{\mu\nu}  \bigg[ 1 + \frac{1}{2\log\frac{\e}{L}} \bigg] \bigg[ \sum_{K=1}^3 |\Phi_K|^2 + 4 \sum_{r=1}^2 \varphi_r^2 \bigg] + \frac{1}{2L} \log \frac{\e}{L}  \left[ T^{\text{grav}}_{\mu\nu} + T^{\text{gauge}}_{\mu\nu} \right] \bigg] \\
&=  \frac{1}{4\pi G_5 L} \bigg[   (g^\4)^\tf_{\mu\nu} + \frac{3}{16} T^{\0\text{grav}}_{\mu\nu} + \frac{1}{16} T^{\0\text{gauge}}_{\mu\nu} - \frac{1}{2} ( ( g^\2 )^2 )_{\mu\nu}^\tf  \\
& + \frac{1}{4} \text{tr} [ g^\2 ]  ( g^\2 ) ^\tf_{\mu\nu}  - g^\0_{\mu\nu}  \bigg(  \frac{1}{16}  \text{tr} [ g^\2 ]^2 - \frac{1}{16}  \text{tr} [ ( g^\2 )^2 ]  + \frac{L^2}{64}  \sum_{K=1}^3 ( F^{\0K}_{\mu\nu}  )^2 \\
& - \frac{1}{128} \bigg[ \sum_{K=1}^3 | \Phi_K^\0  |^2  + 4 \sum_{r=1}^2 ( \varphi_r^\0)^2 \bigg]  + \frac{1}{32} \bigg[ \sum_{K=1}^3 \text{Re} \left( {\wt \Phi}_K^{\0\dagger}  \Phi_K^\0 \right)  + 4 \sum_{r=1}^2 \wt{\varphi}_r^\0 \varphi_r^\0 \bigg]  \bigg) \bigg]  .
\end{split}
\end{equation}

\subsection{VEV of dual CFT operators}

AdS/CFT duality conjectures that the bulk partition function with boundary conditions determined by the sources (classically, this is given by the on-shell action) is equal to the CFT partition function where sources couple to their dual operators,
\begin{equation}\label{dictionary}
\begin{split}
&e^{ i S_{\text{on-shell}} \left[ g^\0 , A^{\0K} , \Phi^\0_K , \varphi^\0_r \right] }  \\
&\qquad= \left\langle e^{ i \int \dt^4 y \sqrt{-g^\0} \left[ \sum\limits_{K=1}^3 A_\mu^{\0K} \CJ_K^\mu - \frac{1}{64L^2} \sum\limits_{K=1}^3 \left(  \Phi_K^{\0\dagger} \CO_{\Phi_K} +  \Phi_K^\0 \CO_{\Phi_K^\dagger}  \right)  - \frac{1}{8L^2} \sum\limits_{r=1}^2 \varphi_r^\0 \CO_{\varphi_r} \right]} \right\rangle_{\text{CFT}} . 
\end{split}
\end{equation}
The metric source $g^\0$ describes the background geometry to which the CFT couples. Note also the unusual normalization for the scalar operators.

Differentiating both sides of \eqref{dictionary} w.r.t. the sources, we can find the VEV of all the operators
\begin{equation}
\begin{split}
\left\langle   \CJ_K^\mu \right\rangle_{\text{CFT}}  &= \frac{1}{\sqrt{-g^\0}}\frac{\delta}{\delta A^{\0K}_\mu} S_{\text{on-shell}} \left[ g^\0 , A^{\0K} , \Phi^\0_K , \varphi^\0_r \right] , \\
 \left\langle \CO_{\varphi_r}   \right\rangle_{\text{CFT}}  &= - 8 L^2 \frac{1}{\sqrt{-g^\0}} \frac{\delta}{\delta \varphi_r^\0} S_{\text{on-shell}} \left[ g^\0 , A^{\0K} , \Phi^\0_K , \varphi^\0_r \right]  , \\
\left\langle   \CO_{\Phi_K}  \right\rangle_{\text{CFT}}  &=- 64L^2 \frac{1}{\sqrt{-g^\0} } \frac{\delta}{\delta \Phi_K^{\0\dagger}  } S_{\text{on-shell}} \left[ g^\0 , A^{\0K} , \Phi^\0_K , \varphi^\0_r \right]  , \\
\left\langle   \CO_{\Phi^\dagger_K}  \right\rangle_{\text{CFT}}  &= - 64L^2 \frac{1}{\sqrt{-g^\0} } \frac{\delta}{\delta \Phi_K^{\0}  } S_{\text{on-shell}} \left[ g^\0 , A^{\0K} , \Phi^\0_K , \varphi^\0_r \right] .
\end{split}
\end{equation}
Using \eqref{app:deltaScheck}, we can determine the VEV for the $U(1)$ currents as
\begin{equation}\label{app:VEVcurrent}
\begin{split}
\left\langle   \CJ_K^\mu \right\rangle_{\text{CFT}} &=  \frac{1}{16\pi G_5L} \bigg[ 2 ( A^{\2K} )^\mu + ( {\wt A}^{\2K} )^\mu -  \frac{L}{6} \sum_{I,J=1}^3 C_{IJK}  (\ve^\0)^{\mu\nu\rho\s} F^{\0I}_{\nu\rho} A^{\0J}_\s  \bigg]  .
\end{split}
\end{equation}
The equations of motion \eqref{app:Bdiv} imply a divergence constraint for the current
\begin{equation}
\begin{split}
\n^\0_\mu \left\langle  \CJ_K^\mu \right\rangle_{\text{CFT}} &= - \frac{1}{32\pi G_5L^2} \bigg[ \text{Im}  \left[ \Phi_K^{\0\dagger} {\wt \Phi}^\0_K \right] + \frac{5L^2}{12} (\ve^\0)^{\mu\nu\rho\s} \sum_{IJ} C_{IJK} F^{\0I}_{\mu\nu} F^{\0J}_{\rho\s}  \bigg]  .
\end{split}
\end{equation}
The VEVs of the scalar operators are
\begin{equation}\label{app:VEVscalars}
\begin{split}
\left\langle \CO_{\varphi_r}   \right\rangle_{\text{CFT}}  &=   \frac{L }{2\pi G_5} \wt{\varphi}_r^\0  , \\
\left\langle   \CO_{\Phi_K}  \right\rangle_{\text{CFT}} & =  \frac{L}{2\pi G_5}  {\wt \Phi}^{\0}_K   , \\
\left\langle   \CO_{\Phi^\dagger_K}  \right\rangle_{\text{CFT}} &= \frac{L}{2\pi G_5}  {\wt \Phi}^{\0\dagger}_K .  \\
\end{split}
\end{equation}

Finally, differentiating the partition function w.r.t. the background metric  $g^\0$ we obtain the VEV of the CFT stress tensor (a.k.a. the holographic stress tensor),
\begin{equation}\label{app:HoloStressTensor}
\begin{split}
\avg{ \CT_{\mu\nu} }_{\text{CFT} }  &=  \frac{1}{4\pi G_5 L} \bigg[   (g^\4)^\tf_{\mu\nu} + \frac{3}{16} T^{\0\text{grav}}_{\mu\nu} + \frac{1}{16} T^{\0\text{gauge}}_{\mu\nu} - \frac{1}{2} ( ( g^\2 )^2 )_{\mu\nu}^\tf  \\
&\qquad + \frac{1}{4} \text{tr} [ g^\2 ]  ( g^\2 ) ^\tf_{\mu\nu}  - g^\0_{\mu\nu}  \bigg(  \frac{1}{16}  \text{tr} [ g^\2 ]^2 - \frac{1}{16}  \text{tr} [ ( g_{\rho\s}^\2 )^2 ]   \\
&\qquad + \frac{L^2}{64}  \sum_{K=1}^3 ( F^{\0K}_{\rho\s}  )^2  - \frac{1}{128} \bigg[ \sum_{K=1}^3 | \Phi_K^\0  |^2  + 4 \sum_{r=1}^2 ( \varphi_r^\0)^2 \bigg] \\
&\qquad + \frac{1}{32} \bigg[ \sum_{K=1}^3 \text{Re} \left( {\wt \Phi}_K^{\0\dagger}  \Phi_K^\0 \right)  + 4 \sum_{r=1}^2 \wt{\varphi}_r^\0 \varphi_r^\0 \bigg]  \bigg) \bigg]  .
\end{split}
\end{equation}
Using \eqref{app:qprop}, we can fix the divergence and trace of the stress tensor
\begin{equation}\label{app:DivHoloStressTensor}
\begin{split}
(\n^\0)^\nu \avg{ \CT_{\mu\nu} }_{\text{CFT} }  &=  \frac{1}{16\pi G_5 L} \bigg[ \sum_{K=1}^3 F^{\0K}_{\mu\nu} \left[ 2 (A^{\2K})^\nu  +    (\wt{A}^{\2K})^\nu \right] \\
&\qquad \qquad - \frac{1}{4} \sum_{K=1}^3 \text{Re} \left[  ( \wt{\Phi}_K^\0 )^\dagger D^\0_\mu \Phi^\0_K \right] - \sum_{r=1}^2 \wt{\varphi}_r^\0 \p_\mu \varphi^\0_r \bigg]  ,
\end{split}
\end{equation}
and
\begin{equation}\label{app:TraceHoloStressTensor}
\begin{split}
\avg{ \CT^\mu{}_\mu }_{\text{CFT} }  &=    \frac{L^3}{64\pi G_5} \bigg[ ( R^\0_{\mu\nu} )^2 - \frac{1}{3} (R^\0)^2 - \frac{1}{L^2} \sum_{K=1}^3 ( F^{\0K}_{\mu\nu}  )^2  \bigg] \\
&\qquad \qquad \qquad   -  \frac{1}{32\pi G_5 L}  \bigg[ \frac{1}{4} \sum_{K=1}^3 | \Phi_K^\0  |^2  +  \sum_{r=1}^2 ( \varphi_r^\0)^2 \bigg]  \\
&\qquad \qquad \qquad + \frac{1}{32\pi G_5 L}  \bigg[ \sum_{K=1}^3 \text{Re} \left( {\wt \Phi}_K^{\0\dagger}  \Phi_K^\0 \right)  + 4 \sum_{r=1}^2 \wt{\varphi}_r^\0 \varphi_r^\0 \bigg] . 
\end{split}
\end{equation}
The last equation is precisely the conformal anomaly in a holographic CFT (with $a=c$). This receives contributions from three sources -- one from the background metric, one from the gauge field and one from the scalar fields. The latter is due to the fact that the dual operators have dimension $\D = 2$ so their squares have dimension $\D = 4 + \CO(1/N^2)$. In this manuscript, we have turned off all the sources and the background geometry is $\mrr \times S^3$ so all the anomaly contributions vanish and the stress tensor is exactly conserved. That is to say, with vanishing sources and the holographic dictionary entry \eqref{G5}, \eqref{app:VEVcurrent}--\eqref{app:VEVscalars} and \eqref{app:HoloStressTensor}--\eqref{app:TraceHoloStressTensor} reduce to 
\eqref{SingleQ:holoCurrent}, \eqref{SingleQ:vevs} and \eqref{SingleQ:holoT}--\eqref{SingleQ:traceAnomaly} in the single charge truncation of Section~\ref{sec:ThermoSingleQ}, and also give the VEVs of the two charge truncation of Section~\ref{sec:ThermoTwoQ}.

\section{Perturbative construction of hairy solitons and black holes  with $A^1=A^2\equiv 0, A^3\equiv A$} \label{App:PerturbativeSingleQ}

In this appendix, we present the explicit solutions for the supersymmetric soliton and hairy black holes obtained in perturbation theory for the single charge truncation. Additional details of the construction which are omitted in Section~\ref{sec:PerturbativeSingleQ} of the main text are also presented here. 

\subsection{Hairy supersymmetric soliton}
\label{App:PerturbativeSingleQSoliton}

Recall that we initiate the perturbative construction with the field expansion~\eqref{soliton-singlecharge-exp}.
The equation for the scalar field $\Phi_\1$ at $\CO(\e)$ is given by
\begin{equation}
\begin{split}
\td{}{r}  \left[ \frac{r^3}{1+r^2} \td{}{r} \left[  (1+r^2) \Phi_\1(r)  \right] \right] = 0 . 
\end{split}
\end{equation}
The general solution to this equation is
\begin{equation}
\begin{split}
\Phi_\1(r) &= c_1 \frac{1}{1+r^2}  + c_2 \frac{ 2r^2 \ln r - 1 }{r^2(1+r^2)}  . 
\end{split}
\end{equation}
$c_1$ ($c_2$) denotes the source (response) for the dual scalar operator $\CO_\Phi$. The AdS boundary condition \eqref{singleQ:Phibc} implies that $c_1=1$ and $c_2=0$.

At $\CO(\e^2)$, the differential equations for $\varphi_\2$, $A_\2$, $h_\2$ and $f_\2$ take the form
\begin{equation}
\begin{split}
\td{}{r} \left[ \frac{1+r^2}{r}  \td{}{r} \left[ r^2 \varphi_\2(r) \right] \right]  &=  \frac{r}{6(1+r^2)^2}   , \\
\td{}{r} \left[ r^3 \td{}{r} A_\2(r) \right] &=\frac{r^3}{(1+r^2)^3}   , \\
\td{}{r} \left[ r^3 \td{}{r}  h_\2(r) \right] &= - \frac{r^3}{(1+r^2)^3}  , \\
\td{}{r} \left[ r^2 f_\2(r) \right]  &=  \frac{r^3}{3(1+r^2)^2}  - \frac{1}{3} r^8 \td{}{r} [ r^{-4} h_\2(r) ] . 
\end{split}
\end{equation}
These equations are solved to
\begin{equation}
\begin{split}
\varphi_\2(r) &= \frac{c_3}{r^2} + c_4 \frac{\ln ( 1 + r^2 )}{r^2}  - \frac{1}{24(1+r^2)}  , \\
A_\2(r) &= \mu_\2 + \frac{c_5}{r^2}   - \frac{1}{8(1+r^2)}  , \\
h_\2(r) &= c_6 + \frac{c_7}{r^2} + \frac{1}{8(1+r^2)}  , \\
f_\2(r) &= \frac{c_8}{r^2} + \frac{c_6r^2}{3}  + c_7 + \frac{r^2}{8(1+r^2)}  .
\end{split}
\end{equation}
AdS boundary conditions \eqref{singleQ:adsbdycondpert} fixes $c_4 = c_6 = 0$ and regularity at $r=0$ fixes $c_3=c_5=c_7=c_8=0$. The integration constant $\mu_\2$ is unfixed at this order. It will be fixed by imposing regularity at $r=0$ at $\CO(\e^3)$.

At $\CO(\e^3)$, the equation for $\Phi_\3$ is
\begin{equation}
\begin{split}
\label{eqPhi}
\td{}{r}  \left[ \frac{r^3}{1+r^2} \td{}{r} \left[  (1+r^2) \Phi_\3(r)  \right] \right] =  - \frac{r^3( 2 - 5 r^2 + r^4 )}{2(1+r^2)^5}  - \frac{8 \mu_\2 r^3}{(1+r^2)^3} \, .
\end{split}
\end{equation}
Its solution is
\begin{equation}
\begin{split}
\Phi_\3(r) = c_9 \frac{1}{1+r^2}  + c_{10} \frac{ 2r^2 \ln r - 1 }{r^2(1+r^2)}  - \frac{r^2}{8(1+r^2)^3} - \mu_\2 \frac{ 1+ r^2\ln ( 1 + \frac{1}{r^2} ) }{ r^2 ( 1 + r^2 ) }\, .
\end{split}
\end{equation}
The AdS boundary condition \eqref{singleQ:Phibc} fixes $c_9=c_{10}=0$ and regularity at $r=0$ fixes $\mu_\2=0$.

Proceeding in this fashion, we can construct the solution to all orders in $\e$. The explicit solution to $\CO(\e^{15})$ can be found in the accompanying {\tt Mathematica} file. 

\subsection{Hairy black hole at $\CO(\e)$}
\label{app:singleQ:hbhpertthy}

Recall that we initiate the black hole double expansion perturbative construction with the field expansion~\eqref{hbh-singlecharge-exp}, \eqref{hbh-singlecharge-exp2} and \eqref{hbh-singlecharge-exp3}.
At $\CO(r_+^0)$, the differential equation for $\Phi_{(1,0)}(r)$ is given by
\begin{equation}\label{app:singleQ-ODE-0}
\begin{split}
\td{}{r} \left[ \frac{r^3}{1+2{\tilde q}_{(0,0)}+r^2}\td{}{r}   [ ( 1+2{\tilde q}_{(0,0)}+r^2 ) \Phi_{(1,0)}(r) ] \right] = 0 . 
\end{split}
\end{equation}
The solution satisfying the AdS boundary condition \eqref{singleQ-pertBH:AdS-BC} is given by
\begin{equation}
\begin{split}\label{singleQ:scalarfield1sol}
\Phi_{(1,0)} = \frac{1}{1+2{\tilde q}_{(0,0)} + r^2} . 
\end{split}
\end{equation}
Using this and moving to $\CO(r_+^2)$, we find the differential equation
\begin{equation}\label{app:singleQ-ODE-2}
\begin{split}
&\td{}{r} \left[ \frac{r^3}{1+2{\tilde q}_{(0,0)}+r^2} \td{}{r}  [ ( 1+2{\tilde q}_{(0,0)}+r^2 ) \Phi_{(1,2)}(r) ] \right] \\
&= \frac{r}{{\tilde q}_{(0,0)} (2 {\tilde q}_{(0,0)} + 1+r^2 )^4}  \left[ 2 r^4  ({\tilde q}_{(0,0)}  (4 {\tilde q}_{(0,2)}+2 )-1 ) \right. \\
&\left. \qquad -2 r^2  (2 {\tilde q}_{(0,0)}+1 )  ({\tilde q}_{(0,0)}  (8 {\tilde q}_{(0,2)}+2 )+1 )+4 {\tilde q}_{(0,0)} (2 {\tilde q}_{(0,0)}+1)^2 \right] .
\end{split}
\end{equation}
The solution satisfying the AdS boundary condition \eqref{singleQ-pertBH:AdS-BC} is given by
\begin{equation}
\begin{split}
\Phi_{(1,2)}(r) &= \frac{1}{4 {\tilde q}_{(0,0)} r^2  (2 {\tilde q}_{(0,0)} +1  + r^2 )^2} \left[ -2 {\tilde q}_{(0,0)}  (2 {\tilde q}_{(0,0)}+(4 {\tilde q}_{(0,2)}+3) r^2 ) \right. \\
&\left. \qquad \qquad \qquad + r^2  (2 {\tilde q}_{(0,0)}+ 1 + r^2 ) \ln  \left( 1+\frac{2 {\tilde q}_{(0,0)}+1}{r^2} \right)+r^2+1 \right] . 
\end{split}
\end{equation}
This describes the solution in the \emph{far-field} region to $\CO(r_+^2)$. 

Recalling the near-region radial coordinate \eqref{SingleQ:pertCoord-z}, $z=r/r_+$, the leading order \emph{near-field} equation is
\begin{equation}
\begin{split}
\td{}{z} \left[ z ( z^2 - 1 ) \td{}{z} \Phi_{(1,0)}^\near(z) \right]  = 0. 
\end{split}
\end{equation}
The general solution to this equation is given by
\begin{equation}
\begin{split}
\Phi_{(1,0)}^\near(z) = c_1 + c_2 \ln \frac{z^2}{z^2-1} . 
\end{split}
\end{equation}
Regularity at the horizon $z=1$ requires that $c_2=0$. Using this and moving to $\CO(r_+^2)$, we find the equation
\begin{equation}
\begin{split}
\td{}{z} \left[ z ( z^2 - 1 ) \td{}{z} \Phi_{(1,2)}^\near(z) \right]  = - \frac{8 c_1 z^3 } {1+2{\tilde q}_{(0,0)} } . 
\end{split}
\end{equation}
The general solution is
\begin{equation}
\begin{split}
\Phi_{(1,2)}^\near(z) = c_3 +  c_4 \ln \frac{z^2}{z^2-1} - \frac{c_1}{1+2{\tilde q}_{(0,0)} } \left( z^2  + \ln \frac{r_+^2z^2}{1+2{\tilde q}_{(0,0)}} \right) . 
\end{split}
\end{equation}
Regularity at $z=1$ sets $c_3=0$.

The integration constants $c_1$ and $c_3$ are fixed by matching the near- and far-field solutions. Let us now describe this procedure. We start by setting $r = z r_+$ in the far-field solution and then expand the solution at small $z$ and $r_+$. We find
\begin{equation}
\begin{split}\label{singleQ:far-smallr}
\Phi_\1(z r_+,r_+) &= \Phi_{(1,0)}(zr_+) + r_+^2 \Phi_{(1,2)} ( z r_+ ) + \CO(r_+^4)  \\
&= \left[\frac{1-2 {\tilde q}_{(0,0)}}{4 {\tilde q}_{(0,0)} (2 {\tilde q}_{(0,0)}+1 ) z^2}+\frac{1}{2 {\tilde q}_{(0,0)}+1}+\CO(z^4) \right]  \\
&\qquad  + r_+^2 \left[\frac{-8 {\tilde q}_{(0,2)} {\tilde q}_{(0,0)} -2 {\tilde q}_{(0,0)} -1 - ( 2 {\tilde q}_{(0,0)}  + 1 ) \ln \frac{ r^2_+ z^2 }{2 {\tilde q}_{(0,0)}+1}  }{4 {\tilde q}_{(0,0)} (2 {\tilde q}_{(0,0)}+1)^2} \right. \\
&\left. \qquad \qquad \qquad \qquad -\frac{z^2}{(2 {\tilde q}_{(0,0)}+1)^2}+\CO(z^4) \right]+\CO(r_+^4). 
\end{split}
\end{equation}
We next expand the near-field solution at large $z$,
\begin{equation}
\begin{split}\label{singleQ:near-larger}
\Phi^\near_\1(z,r_+) &= \Phi^\near_{(1,0)}(z) + r_+^2 \Phi^\near_{(1,2)} (z) + \CO(r_+^4) \\
&= c_1+r_+^2 \left[- \frac{c_1 z^2}{2 {\tilde q}_{(0,0)}+1}+c_3 - \frac{c_1 \ln \frac{r_+^2 z^2}{2 {\tilde q}_{(0,0)}+1}}{2
   {\tilde q}_{(0,0)}+1}+ \CO(z^{-4})  \right] + \CO(r_+^4). 
\end{split}
\end{equation}
The expansion \eqref{singleQ:far-smallr} is valid when $r \ll 1$ or $z \ll \frac{1}{r_+}$ whereas the expansion \eqref{singleQ:near-larger} is valid when $z \gg 1$. When $r_+\ll1$, the two expansions have an overlapping region of validity and we can match the expansions exactly in this region. It is clear that the matching requires us to set
\begin{equation}\label{app:singleQ-ODE-q00}
\begin{split}
{\tilde q}_{(0,0)} = \frac{1}{2} , \qquad c_1 = \frac{1}{2} , \qquad c_3 = - \frac{1}{4} ( 1 + 2 {\tilde q}_{(0,2)}  ) . 
\end{split}
\end{equation}
Note that after this matching we can replace ${\tilde q}_{(0,0)} = \frac{1}{2}$ back in \eqref{app:singleQ-ODE-0} or \eqref{app:singleQ-ODE-2} to get the ODE \eqref{scalar-gen-eq-far} that we present in the main text (see also the associated footnote~\ref{foot:SingleQpert}).

Proceeding in this fashion, we can construct the solution to all orders in $r_+^2$. The solution to $\CO(r_+^8)$ is given in the {\tt Mathematica} file.

\subsection{Hairy black hole at $\CO(\e^2)$}
\label{singleQ:HBHsecondorderperthy}

As mentioned in section \ref{sec:singleQHBHpertthy}, the perturbative construction of the single charge hairy black hole solution is intricate due to the fact that the solution is constructed as a perturbation around a singular solution, namely the singular supersymmetric soliton \eqref{SingleQ:SingularSoliton}. The first indication of these intricacies show up at $\CO(\e^2)$ in perturbation theory and we describe these in this section. We will find that this complicates the perturbative construction significantly. We leave a resolution of the issues discussed here for future work.

The issue arises since we are effectively perturbing around the singular soliton. To understand this, we strip off all the complications of the BCS black hole solution and consider a simpler perturbative expansion around the singular soliton. In the gauge \eqref{static-sol-ansatz-singlecharge}, the general singular soliton \eqref{SingleQ:SingularSoliton} is given by
\begin{equation}
\begin{split}
f = 1 + r^2 h , \qquad A_t = \frac{1}{h} , \qquad  \Phi = \frac{\sqrt{ C_2^2 - 4 ( 1 + C_1 ) }}{ 1 + r^2 h } , \qquad \varphi = - \frac{1}{3} \ln h , 
\end{split}
\end{equation}
where
\begin{equation}
\begin{split}
h = \sqrt{1 + \frac{C_2}{r^2} + \frac{1 + C_1}{r^4} } - \frac{1}{r^2}\,. 
\end{split}
\end{equation}
The regular solution occurs for $C_1=0$. The $r_+ \to 0$ limit of the BCS black hole is the singular soliton satisfying $C_2 = 2 \sqrt{1+C_1} = 2 + 4 {\tilde q} \equiv 2 \eta $. The hairy BH solution is a perturbation around this singular solution with ${\tilde q}=\frac{1}{2} \Rightarrow \eta=2$ whereas the regular supersymmetric soliton solution is a perturbation around vacuum AdS with ${\tilde q}=0\Rightarrow \eta=1$. We consider the perturbative expansion for arbitrary $\eta$ to illustrate the key differences between the two cases.

To initiate the perturbative expansion, we set
\begin{equation}
\begin{split}
f(r,\e) &= \sum_{n=0}^\infty \e^{2n} f_{(2n)}(r)  , \\
h(r,\e) &= \sum_{n=0}^\infty \e^{2n} h_{(2n)}(r)  , \\
A_t(r,\e) &= \sum_{n=0}^\infty \e^{2n} A_{(2n)}(r) , \\
\varphi(r,\e) &= \sum_{n=0}^\infty \e^{2n} \varphi_{(2n)}(r) , \\
\Phi(r,\e) &= \sum_{n=0}^\infty \e^{2n+1} \Phi_{(2n+1)}(r) ,
\end{split}
\end{equation}
where
\begin{equation}
\begin{split}
f_\0 = 1 + r^2 h_\0 , \qquad A_\0 = \frac{1}{h_\0} , \qquad \varphi_\0 = - \frac{1}{3} \ln h_\0 , \qquad h_\0 = 1 + \frac{\eta-1}{r^2} .  
\end{split}
\end{equation}
We plug this into the equations \eqref{singleQ:h2eq}--\eqref{singleQ:Phi2eq} and solve them order-by-order in $\e$. At $\CO(\e)$, only the scalar equation is non-trivial and the solution satisfying the AdS boundary condition is given by
\begin{equation}
\begin{split}
\Phi_\1(r) = \frac{1}{\eta + r^2 } . 
\end{split}
\end{equation}
This precisely matches the solution \eqref{singleQ:scalarfield1sol} obtained at $\CO(\e r_+^0)$. We use this solution and move on to $\CO(\e^2)$. Here, we find it convenient to define new functions
\begin{equation}
\begin{split}\label{singleQ:secondorderfunctions}
\varphi_\2(r) &\equiv X_\2(r)  + \frac{r^2 ( \eta - 1 + 3 r^2 ) h_\2(r)}{ 6 ( \eta - 1 ) ( \eta - 1 + r^2 ) }  - \frac{[ r^{-1} [ r^4 f_\2(r) ]' ]'  }{16(\eta-1)r} , \\
h_\2(r) &\equiv Y_\2(r) - \frac{[ r^4 f_\2(r) ]'}{2r^5}  , \\
A_\2(r) &\equiv Z_\2 (r)+ \frac{r^6  (\eta+1+r^2)Y_\2(r) }{2 (\eta -1)  (\eta -1+r^2 )^2} -\frac{r^4  (\eta +r^2 )}{8 (\eta -1)  (\eta-1+r^2)} f_\2''(r)  \\
&\qquad -\frac{r^3[ 3 \eta ^2+\eta +r^4+(4 \eta -1) r^2-4 ]  }{8 (\eta -1)  (\eta-1+r^2)^2} f_\2'(r) \\
&\qquad + \frac{r^2 \left(-2 \eta +r^4+(\eta +1) r^2+2\right)}{2 (\eta -1)  (\eta-1+r^2)^2} f_\2(r) .
\end{split}
\end{equation}
Note that these definitions are valid only if $\eta\neq1$. The $\eta=1$ background is vacuum $\ads_5$ and the corresponding perturbative construction reproduces the regular hairy supersymmetric soliton as described in \eqref{singleQ:hairysusysoliton-pertthy}. In any case, we are interested in the case $\eta=2$. The differential equations for $X_\2$, $Y_\2$ and $Z_\2$ take a particularly simple form,
\begin{equation}
\begin{split}\label{singleQ:XYZeq}
X_\2(r) &= \frac{r^2(r^2+2\eta)}{16(\eta-1)(r^2+\eta)^3} , \\
\left[ \frac{1}{r} [ r^4 Y_\2(r) ]' \right]' &= - \frac{r^3}{(\eta+r^2)^3}  - 8 ( \eta-1)X_\2'(r)  , \\
Z_\2'(r) &= \frac{4r(\eta-1)X_\2(r)}{(\eta-1+r^2)}   - \frac{2r^2(\eta+r^2)X_\2'(r)}{\eta-1+r^2}  - \frac{r^5}{4(\eta-1)(\eta+r^2)^3} . 
\end{split}
\end{equation}
$X_\2$ satisfies an algebraic equation which has already been solved above. The last two differential equations are easily integrated up to 3 integration constants.
\begin{equation}
\begin{split}
Y_\2(r) &= \frac{c_1}{r^2} + \frac{c_2}{r^4} + \frac{\eta}{8r^2(\eta + r^2)^2} , \\
Z_\2(r) &=  c_3 + \frac{(\eta+1)r^4}{ 8 ( \eta - 1 ) ( \eta + r^2 )^2 ( \eta - 1 + r^2 ) } .
\end{split}
\end{equation}
Using these solutions, we then find a 4th order differential equation for $f_\2$. It is convenient to write
\begin{equation}
\begin{split}
f_\2(r) = - c_1 - \frac{1}{8(\eta+r^2)} + {\tilde f}_\2(r) . 
\end{split}
\end{equation}
${\tilde f}_\2(r)$ satisfies a homogeneous differential equation
\begin{equation}
\begin{split}\label{singleQ:f2deq}
0&= r^3 ( \eta + r^2 ) {\tilde f}_\2''''(r) + 2 r^2 ( 6 r^2 + 5 \eta ) {\tilde f}_\2'''(r) + r ( 29 r^2 + 23 \eta-8) {\tilde f}_\2''(r) \\
&\qquad \qquad + (-13r^2+9\eta-24) {\tilde f}_\2'(r) - 32 r {\tilde f}_\2(r) .
\end{split}
\end{equation}
The general solution to this equation is
\begin{equation}
\begin{split}\label{singleQ:ft2homogeneoussol}
{\tilde f}_\2(r) &= \frac{c_4}{r^2} + c_5 ( 2 \eta - 2 + r^2 )  + c_6 \g_+(r) + c_7 \g_-(r) ,
\end{split}
\end{equation}
where
\begin{equation}
\begin{split}\label{singleQ:gammapm}
\g_\pm(r) \equiv r^{-1\pm\a} \, _2F_1\left( \frac{1\pm\a}{2} ,  \frac{-3\pm\a}{2}  ; 1 \pm \a ; -\frac{r^2}{\eta}\right) , \qquad \a \equiv \sqrt{1+8/\eta} . 
\end{split}
\end{equation}
We immediately notice a qualitative different structure for the solution depending on whether $\a \in \mnn$ or not. When $\a \in \mnn$, the hypergeometric reduces to a simple rational function of $r^2$. Consequently, in this case the solution for all $\varphi_\2$, $h_\2$, $A_\2$ are all rational functions of $r^2$ and asymptotically AdS boundary conditions are easy to impose. Further, we recall that the solutions constructed here are valid in the \emph{far-field} region. To match the solutions to the near-field expansion at small $r$, we set $r=r_+z$ and expand in small $z$. Since, the functions are rational functions of $r^2$, only terms of the form $\CO(r^{2n})$ appear in its small $r$ expansion. Such terms are matched to terms of the form $\CO(r_+^{2n} z^{2n})$ in the large $z$ expansion of the near-field solution. This process is identical to the one described for the scalar field $\Phi_\1$ at $\CO(\e)$ in section \ref{app:singleQ:hbhpertthy}. Note that this is exactly what happens for $\eta=1\Rightarrow \a=3\in\mnn$.

On the other hand, the hairy BH is a perturbation around the $\eta=2\Rightarrow \a=\sqrt{5} \not\in \mnn$ singular soliton. In this case, $\g_\pm(r)$ have problematic terms in their expansion near $r=\infty$ and $r=0$. It's expansion near $r=\infty$ is
\begin{equation}
\begin{split}
\g_\pm(r) ~~\propto ~~  \frac{\a^2-1}{\a^2-9} r^2 - 2 + \frac{1}{ r^2}  \left[ 3-4H_{\frac{1}{2}(\pm\a-1)} + 2 \ln \frac{r^2}{\eta}   \right]  + \CO(r^{-4}) . 
\end{split}
\end{equation}
The $\log$ term in this expansion violates the AdS boundary conditions \eqref{singleQ:adsbdycondpert}. The resolution for this is simple - we simply choose the integration constants $c_6$ and $c_7$ so that the $\log$ terms cancel out from $f_\2(r)$.

Near $r=0$, the hypergeometric function admits an expansion of the form
\begin{equation}
\begin{split}
\g_\pm(r) = r^{-1\pm\a} [ 1 + \CO(r^2) ] . 
\end{split}
\end{equation}
Such terms are matched to terms of the form $\CO(r_+^{-1\pm\a} z^{-1\pm\a})$ in the large $z$ expansion of the near-field solution. However, this immediately implies that the near-field solution \emph{does not} admit an analytic expansion in $r_+^2$, e.g. the scalar field at $\CO(\e)$ in \eqref{SingleQ:scalarnear1exp}. A possible resolution to this is to set both $c_6=c_7=0$. The problem is that choice may clash with the imposition of AdS boundary conditions and this is indeed what happens at higher orders in perturbation theory (in the construction of the hairy BH, this first shows up at $\CO(\e^2r_+^4)$). At this order in perturbation theory and beyond, AdS boundary conditions will force at least one of $c_6$ or $c_7$ (or both) to be non-vanishing. This will then force us to introduce to non-analytic terms in the near-field expansion of $f_\2$. A general small $r_+$ expansion for $f_\2$ then takes the form
\begin{equation}
\begin{split}
f_\2^\near(z,r_+) = \sum_{m,n,p=0}^\infty r_+^{2p+m(-1+\a)+n(-1-\a)} f_{(2,2p,2m,2n)}^\near(z).
\end{split}
\end{equation}
Similar expansions will also exist for $h_\2$, $\varphi_\2$ and $A_\2$. However, now, the presence of such terms in the near-field expansion will backreact and introduce similar non-analytic terms in the far-field expansion as well so we will have to modify that expansion as well. Of course, it will further backreact on to the scalar field at $\CO(\e^3)$.

Another complication arises due to the fact that $\a+1=\sqrt{5}+1>2$ which implies that the far-field expansion actually breaks down when $r \sim r_+^{\frac{1}{2} ( \sqrt{5}-1)}$ which for small $r_+$ is parametrically larger than $r_+$. Consequently, we would need to introduce a new intermediate-field expansion to construct the solution in this region.

Based on the discussion in this section, it is clear that the construction of the hairy BH solution is significantly more intricate than previously presumed. We hope to resolve these issues in future work.

\subsubsection*{Explicit Results for the hairy BH solution at $\CO(\e^2)$}

Having the described the issue qualitatively in the previous section, we present explicit results up to $\CO(\e^2 r_+^4)$ at which point the perturbative construction breaks down.

We start by plugging in the expansion \eqref{hbh-singlecharge-exp} into equations \eqref{singleQ:h2eq}--\eqref{singleQ:phi2eq} and extracting the equations at $\CO(\e^2)$. To solve these equations, we further expand in $r_+$
\begin{equation}
\begin{split}
f_\2(r,r_+) &= \sum_{n=0}^\infty r_+^{2n} f_{(2,2n)}(r) , \\
h_\2(r,r_+) &= \sum_{n=0}^\infty r_+^{2n} h_{(2,2n)}(r)  , \\
A_\2(r,r_+) &= \sum_{n=0}^\infty r_+^{2n} A_{(2,2n)}(r)  , \\
\varphi_\2(r,r_+) &= \sum_{n=0}^\infty r_+^{2n} \varphi_{(2,2n)}(r) . 
\end{split}
\end{equation}
The differential equations at each order in $r_+^2$ take a simpler form if we work instead with the functions $X_{(2,2n)}$, $Y_{(2,2n)}$ and $Z_{(2,2n)}$ which are defined as in \eqref{singleQ:secondorderfunctions} with $\eta=2$. The differential equations these functions and $f_{(2,2n)}$ have exactly the same form as \eqref{singleQ:XYZeq} and \eqref{singleQ:f2deq}, now with $\eta=2$ and additional source terms. At $\CO(r_+^0)$, the solutions are
\begin{equation}
\begin{split}
X_{(2,0)}(r) &= \frac{r^2(4+r^2)}{16(2+r^2)^3} , \\
Y_{(2,0)}(r) &= \frac{c_1}{r^2} + \frac{c_2}{r^4} - \frac{4+r^2}{16(2+r^2)^2} , \\
Z_{(2,0)}(r) &= c_3 + \frac{1+r^2}{2(2+r^2)^2} , \\
f_{(2,0)}(r) &= c_4 ( 2 + r^2 ) + \frac{c_5}{r^2} - c_1 + \frac{r^2}{16(2+r^2)} . 
\end{split}
\end{equation}
Here we have set $c_6=c_7=0$ as this is consistent with AdS boundary conditions. Using this, we find
\begin{equation}
\begin{split}\label{singleQ:far-field-sol-2}
\varphi_{(2,0)}(r) &= \frac{c_2 + c_5}{6r^2} + \frac{\frac{1}{16} + c_1 + c_2  - c_4 + c_5}{ 3(1+r^2)} - \frac{1}{24(2+r^2)} , \\
h_{(2,0)}(r) &= 3 c_4 - \frac{c_1 - 4 c_4}{r^2} + \frac{c_2+c_5}{r^4} + \frac{1}{16(2+r^2)} , \\
A_{(2,0)}(r) &=c_3 + \frac{2 r^6+7 r^4+10 r^2+4}{16  (r^2+1 )^2  (r^2+2 )} \\
&\qquad \qquad +  \frac{   (2c_1+2c_5+ 3 c_2-4c_4)  r^2+(c_2 - 2 c_4 ) r^4  }{2(1+r^2)^2} , \\
f_{(2,0)}(r) &=  c_4 ( 2 + r^2 ) + \frac{c_5}{r^2} - c_1 + \frac{r^2}{16(2+r^2)} . 
\end{split}
\end{equation}
AdS boundary conditions imply $c_4=0$. All other constants are fixed by matching with the near-field solution.

To construct the solution in the near-field region, we define
\begin{equation}
\begin{split}
f_\2^\near(z,r_+) &= f_\2(zr_+,r_+) ,\\
h_\2^\near(z,r_+) &= h_\2(zr_+,r_+) ,\\
A_\2^\near(z,r_+) &= A_\2(zr_+,r_+), \\
\varphi_\2^\near(z,r_+) &= \varphi_\2(zr_+,r_+) . \\
\end{split}
\end{equation}
Note that as $r_+ \to 0$, the base solution has the behaviour
\begin{equation}
\begin{split}
f_\0^\near(z,r_+) &= \CO(r_+^0) , \\
h_\0^\near(z,r_+) &= \CO(r_+^{-2}) , \\
A_\0^\near(z,r_+) &= \CO(r_+^2) , \\
\varphi_\0^\near ( z , r_+) &= \CO(r_+^0) . 
\end{split}
\end{equation}
Consequently, the small $r_+$ expansion of the fields are
\begin{equation}
\begin{split}
f^\near_\2(z,r_+) &= \sum_{n=0}^\infty r_+^{2n} f^\near_{(2,2n)}(z) , \\
h^\near_\2(z,r_+) &= \frac{1}{r_+^2} \sum_{n=0}^\infty r_+^{2n} h^\near_{(2,2n)}(z)  , \\
A^\near_\2(z,r_+) &= r_+^2 \sum_{n=0}^\infty r_+^{2n} A^\near_{(2,2n)}(z)  , \\
\varphi^\near_\2(z,r_+) &= \sum_{n=0}^\infty r_+^{2n} \varphi^\near_{(2,2n)}(z) . 
\end{split}
\end{equation}
The differential equations at each order in $r_+^2$ takes a simpler form if we define
\begin{equation}
\begin{split}
X^\near_{(2,2n)}(z) &\equiv \varphi^\near_{(2,2n)}(z) +  \frac{1}{8}z^{1/3} [ z^{-7/3} [  z^{8/3} h^\near_{(2,2n)}(z)  ] ]'   , \\
Y^\near_{(2,2n)}(z)  &\equiv  h^\near_{(2,2n)}(z)  -  \frac{1}{2z^5} [ z^4 f_{(2,2n)}^\near(z)]'   , \\
Z^\near_{(2,2n)}(z) &\equiv  A^\near_{(2,2n)}(z) + \frac{1}{2z} [ z^4 f_{(2,2n)}^\near(z)]'    . 
\end{split}
\end{equation}
The differential equations then take the form
\begin{equation}
\begin{split}\label{near-2-eq}
\mfs_{(2,2n)}^{X,\near}(z) &= [ X_{(2,2n)}^\near(z)]', \\
\mfs_{(2,2n)}^{Y,\near}(z) &= [ z^4 Y_{(2,2n)}^\near(z)]' - 8y X_{(2,2n)}^\near(z) , \\
\mfs_{(2,2n)}^{Z,\near}(z) &= [ z^{-1} [Z_{(2,2n)}^\near(z)]']' ,\\
\mfs^{f,\near}_{(2,2n)}(z)  &= z^2(z^2-1) [f_{(2,2n)}^\near(z)]''' + 7 y ( z^2 - 1 ) [f_{(2,2n)}^\near(z)]''   \\
&\qquad + 5 ( z^2 - 1 ) [f_{(2,2n)}^\near(z)]'  - 8 z f_{(2,2n)}^\near(z) + 4 [ Z_{(2,2n)}^\near(z)]' \\
&\qquad   +  2  [ z (z^2-1)^{\frac{1}{2}}  [ z^2  (z^2-1)^\frac{1}{2} Y_{(2,2n)}^\near(z) ]'  ]'  \\
&\qquad  -16 (z^2-1)^{\frac{1}{2}}   [  (z^2-1)^{\frac{1}{2}}  X_{(2,2n)}^\near(z)  ]' . 
\end{split}
\end{equation}
As always, the sources are fixed by lower orders in perturbation theory. It is easy to integrate these differential equations and obtain the solution up to 7 integration constants. The general solution has the form
\begin{equation}
\begin{split}
\varphi^\near_{(2,2n)}(z) &= \frac{4b_1  - b_4}{6} + \frac{ b_2 + b_5 }{6z^2}  + \text{source}, \\
h^\near_{(2,2n)}(z) &= \frac{b_2+b_5}{z^4} + \frac{2(2 b_1+b_4)}{z^2}  + \text{source}  , \\
A^\near_{(2,2n)}(z) &=  b_3  - b_5 - b_4 z^2 + \text{source} , \\
f_{(2,2n)}^\near(z) &= \frac{b_5}{z^2} - \frac{b_2 + b_5}{z^4}  + b_4 + b_6 \rho_+(z) + b_7 \rho_-(z) + \text{source} ,
\end{split}
\end{equation}
where the ``source'' terms are obtained by integration the sources $\mfs$ in \eqref{near-2-eq} and 
\begin{equation}
\begin{split}
\rho_\pm(z) =  (z^2-1)^{-\frac{1}{2} ( \pm \sqrt{5}+1)} \,_2 F_1 \left( \frac{1\pm\sqrt{5}}{2} , \frac{5\pm\sqrt{5}}{2}  ; 1 \pm \sqrt{5} ; \frac{1}{1-z^2} \right) . 
\end{split}
\end{equation}
The hypergeometric functions $\rho_\pm$ are the near-field analogue of the far-field hypergeometrics \eqref{singleQ:gammapm}. Their presence in the near-field solution would imply non-analytic terms in the small $r_+$ expansion of the far-field solution and this is precisely what we expect will happen at a sufficiently high order.

At $\CO(r_+^0)$, all the near-field sources are identically zero and we find the solution
\begin{equation}
\begin{split}
\varphi_{(2,0)}^\near(z) &= \frac{4 b_1 - b_4}{6} + \frac{b_2 + b_5}{6z^2} , \\
h_{(2,0)}^\near(z) &= \frac{2(2b_1+b_4)}{z^2} + \frac{b_2+b_5}{z^4} , \\
A_{(2,0)}^\near(z) &=  b_3 - b_5 - b_4 z^2 , \\
f_{(2,0)}^\near(z) &=  - \frac{b_2}{z^4}  + b_4+ b_5 \frac{z^2-1}{z^4} .
\end{split}
\end{equation}
Here, we have consistently set $b_6=b_7=0$. $f$ and $A$ must vanish on the horizon $z=1$. This implies $b_4=b_2$ and $b_5=b_3-b_2$. The remaining constants are fixed by matching to the far-field solution \eqref{singleQ:far-field-sol-2} in the usual way. As a result of the matching, we have $c_2 = c_5 = b_1 = b_2 = 0$ and $c_3 = - \frac{1}{8}$. The remaining unfixed integration constants are $c_1$ and $b_3$. $b_3=x_{(2,0)}$ will will be fixed at $\CO(r_+^2)$ in perturbation theory. On the other hand, $c_1$ is a redundant integration constant. In the full solution this constant appears alongside ${\tilde q}_{(2,0)}$ in the combination $c_1 - 2 {\tilde q}_{(2,0)}$. Since only this combination appears in the final solution, we can set $c_1 = 0$ without loss of generality.

The far- and near-field solution at $\CO(r_+^2)$ is given by
\begin{equation}
\begin{split}
\varphi_{(2,2)}(r) &= -\frac{r^2 \log \left(\frac{2}{r^2}+1\right)}{48  (r^2+1 ) (r^2+2 )}+\frac{-r^6+5 r^4+8 r^2}{192  (r^4+3 r^2+2 )^2} \\
h_{(2,2)}(r) &=  \frac{r^2-4}{64  (r^2+2 )^2}+\frac{\log \left(\frac{2}{r^2}+1\right)}{16  (r^2+2 )} , \\
A_{(2,2)}(r) &=  -\frac{r^4 \log \left(\frac{2}{r^2}+1\right)}{16  (r^2+1 )^2  (r^2+2 )}+\frac{-r^8+11r^6+24 r^4+8 r^2}{64  (r^2+1 )^3  (r^2+2 )^2} , \\
f_{(2,2)}(r) &= \frac{ (r^2-4 ) r^2}{64  (r^2+2 )^2}+\frac{r^2 \log \left(\frac{2}{r^2}+1\right)}{16  (r^2+2 )} , \\
\end{split}
\end{equation}
\begin{equation}
\begin{split}
\varphi_{(2,2)}^\near(z) &= \frac{x_{(2,2)}}{6 z^2}-\frac{z^2}{96} , \\
h_{(2,2)}^\near(z) &= \frac{x_{(2,2)}}{z^4}+\frac{1}{32} , \\
A_{(2,2)}^\near(z) &=  -\frac{1}{32} z^2 \left(z^2-1\right) , \\
f_{(2,2)}^\near(z) &=  \frac{\left(z^2-1\right) \left(32 x_{(2,2)}+z^4+z^2\right)}{32 z^4}.
\end{split}
\end{equation}
The matching of solutions also sets $x_{(2,0)} = 0$. 

Finally, we turn to the solution at $\CO(r_+^4)$. The far-field solution has the form
\begin{equation}
\begin{split}
X_{(2,4)}(r) &= \frac{r^2}{384 \left(r^2+1\right)^3 \left(r^2+2\right)^5} \left[55 r^{10}+607 r^8+2779 r^6+5207 r^4+4214 r^2 \right.\\
&\qquad \qquad \left. +12  (r^2+1 )^3  (r^2+2 )^2  (r^2+4 ) \ln ^2\left(\frac{2}{r^2}+1\right) \right.\\
&\qquad \qquad \left.  +4 \left(r^2+1\right) \left(r^2+2\right) \left(2 r^6+24r^4+39 r^2+16\right) \ln \frac{r_+^2}{2} \right.\\
&\qquad \qquad \left.  -2  (r^2+1 )  (r^2+2 )  (15 r^8+148 r^6+531 r^4 \right.\\
&\qquad \qquad \left. +672 r^2+272 ) \ln \left(\frac{2}{r^2}+1\right) +1232 \right] , \\
Y_{(2,4)}(r) &= \frac{c_2}{r^2}+c_1+\frac{1}{64 r^2  (r^2+2 )^4} \left[ r^2 \left( (r^2+2 ) \ln \left(\frac{2}{r^2}+1\right) \left(5 r^6+38 r^4  \right. \right. \right. \\
&\left. \left. \left. \qquad \qquad  +84 r^2-2 \left(r^4+6 r^2+8\right) r^2 \ln \left(\frac{2}{r^2}+1\right)+8\right) \right. \right. \\
&\left. \left. \qquad \qquad - 4  (r^4-4 ) \ln \frac{r_+^2}{2}  \right)  -r^2  (10 r^6+77 r^4+192 r^2+56 ) -16 \right]  , \\
Z_{(2,4)}(r) &= c_3 - \frac{1}{192  (r^4+3 r^2+2 )^4} \left[ 46 r^{14}+394 r^{12}+1027 r^{10}+1088 r^8 \right. \\
&\left. \qquad+296 r^6 - 370 r^4-360 r^2+12  r^4 (r^2+2 )^2  (r^2+1)^4 \ln ^2\left(\frac{2}{r^2}+1\right) \right. \\
&\left. \qquad +4  (5 r^{10}+30 r^8+73 r^6+94 r^4+66 r^2+20 ) r^4 \ln \frac{r_+^2}{2} \right. \\
&\left. \qquad -2  (15 r^{14}+154 r^{12}+600 r^{10}+1190 r^8+1337 r^6+876 r^4\right. \\
&\left. \qquad +316 r^2 +48 ) r^2 \ln \left(\frac{2}{r^2}+1\right)-96 \right]. 
\end{split}
\end{equation}
Finally, the solution for $f_{(2,4)}$ is given by
\begin{equation}
\begin{split}
f_{(2,4)}(r) &= \frac{c_4}{r^2}+c_5 (r^2+2 ) + c_6 \g_+(r) + c_7 \g_-(r) -c_1-\frac{\ln\frac{r_+^2}{2}}{16  (r^2+2 )^2} \\
&\qquad +\frac{r^2 \ln ^2\left(\frac{2}{r^2}+1\right)}{32(r^2+2 )}  -\frac{ (r^6+16 r^4+48 r^2+16 ) \ln \left(\frac{2}{r^2}+1\right)}{128  (r^2+2 )^2} \\
&\qquad \frac{5 r^8+38 r^6+124 r^4+194 r^2+80}{128  (r^2+2 )^3} \\
&\qquad - \frac{r^4}{320(2+r^2)} \, _3F_2\left(1,1,3;\frac{7}{2}-\frac{\sqrt{5}}{2},\frac{7}{2}+\frac{\sqrt{5}}{2};\frac{r^2}{r^2+2}\right) . 
\end{split}
\end{equation}
The hypergeometric $\,_3F_2$ has a logarithm in its large $r$ expansion which must be cancelled in order to impose asymptotically AdS boundary conditions. In particular, this implies that we cannot consistently set $c_6=c_7=0$ anymore which in turn implies non-analytic terms in the near-field expansion as explained in the previous section.

\section{Perturbative construction of hairy solitons and black holes with $ A^1=A^2\equiv A, A^3\equiv 0$} \label{App:PerturbativeTwoQ}

In this appendix, we present the details of the perturbative construction of the supersymmetric soliton and hairy black holes that were omitted in Section~\ref{sec:PerturbativeTwoQ} of the main text and the explicit solutions to the order that we have derived them.

\subsection{Hairy supersymmetric soliton}
\label{app:twoQsolitonpert}

 Recall that we initiate the two charge soliton perturbative construction with the field expansion~\eqref{twoQ:soliton-exp}.
The equation for $\Phi_\1$ at $\CO(\e)$ is
\begin{equation}
\begin{split}
\td{}{r} \left[ \frac{r^3}{1+r^2} \td{}{r} [ ( 1 + r^2 ) \Phi_\1(r) ] \right] &= 0. 
\end{split}
\end{equation}
The general solution this is
\begin{equation}
\begin{split}
\Phi_\1(r) = c_1 \frac{1}{1+r^2} + c_2 \frac{2r^2 \ln r - 1 }{ r^2 ( 1 + r^2 ) } 
\end{split}
\end{equation}
AdS boundary conditions \eqref{AdSbdycond} imply that $c_1=1$ and $c_2=0$. 

At $\CO(\e^2)$, the differential equations are
\begin{equation}
\begin{split}\label{app:twoQ:solitoneq2}
[ r^3 h_\2'(r) ]' &= - \frac{r^3}{(1+r^2)^2}  , \\
[ r^3 [ r^2 f_\2(r) ]' ]' &=  \frac{2r^5 ( 2 + r^2 )}{  ( 1 + r^2 )^3 } + 16 r^5 h_\2(r)  , \\
[ r^3 A_\2'(r) ]' &= \frac{r^3}{(1+r^2)^3}  , \\
\left( \frac{1+r^2}{r} [ r^2 \varphi_\2(r)]' \right)' &= - \frac{r}{6(1+r^2)^2} . \\
\end{split}
\end{equation}
The general solution is
\begin{equation}
\begin{split}
h_\2(r) &= c_1 + \frac{c_2}{r^2} + \frac{1}{8(1+r^2)}  , \\
f_\2(r) &= \frac{c_3}{r^2} + \frac{c_4}{r^4} + \frac{2c_1r^2 }{3} + 2 c_2 + \frac{r^2}{4(r^2+1)} , \\
A_\2(r) &= \mu_\2 + \frac{c_5}{r^2} - \frac{1}{8(r^2+1)} , \\
\varphi_\2(r) &= \frac{c_6}{r^2} + c_7 \frac{\ln(1+r^2)}{2r^2} + \frac{1}{24(1+r^2)} . 
\end{split}
\end{equation}
AdS boundary conditions implies $c_1=c_7=0$ and regularity at $r=0$ fixes $c_2=c_3=c_4=c_5=c_6=0$. $\mu_\2$ is fixed at $\CO(\e^3)$.

At $\CO(\e^3)$, the differential equation for $\Phi_\3$ is
\begin{equation}
\begin{split}
\td{}{r} \left[ \frac{r^3}{1+r^2} \td{}{r} [ ( 1 + r^2 ) \Phi_\3(r) ] \right] &=  \frac{r^3(4-7r^2+r^4)}{2(1+r^2)^5} - \frac{8\mu_\2 r^3}{(1+r^2)^3}  . 
\end{split}
\end{equation}
The general solution to this is
\begin{equation}
\begin{split}
\Phi_\1(r) = c_1 \frac{1}{1+r^2} + c_2 \frac{2r^2 \ln r - 1 }{ r^2 ( 1 + r^2 ) } + \frac{1-2r^2}{16(1+r^2)^3} - \mu_\2 \frac{1+r^2 \ln ( 1 + \frac{1}{r^2} ) }{ r^2 ( 1 + r^2 ) } .
\end{split}
\end{equation}
AdS boundary conditions sets $c_1=c_2=0$ and regularity at the origin sets $\mu_\2 = 0$.

Proceeding in this fashion, we can construct the solution to all orders in $\e$. The solution to $\CO(\e^{13})$ can be found in the {\tt Mathematica} file.

\subsection{Hairy black hole}
\label{app:twoQ:perthy-hbh}

Recall that we initiate the black hole double expansion perturbative construction with the field expansion~\eqref{twoQ:hbh-exp}, \eqref{twoQ:hbh-exp2}, \eqref{twoQ:hbh-exp2b} and \eqref{twoQ:farfieldexp}.

\subsubsection*{General structure of the differential equations at $\CO(\e^{2n+1}r^k)$}

At this order, the only non-trivial equation is for the scalar field. In the far-, intermediate- and near-field regions, the equations take the form
\begin{equation}
\begin{split}\label{app:twoQ:solitoneq1}
\left[ \frac{r^3}{1+r^2} [ ( 1 + r^2 ) \Phi_{(2n+1)}(r) ]' \right]' &= \mfs^\Phi_{(2n+1,k)}(r) , \\
[ y^3 [ \Phi^\intt_{(2n+1,k)}(y)]' ]' &= \mfs^{\Phi,\intt}_{(2n+1,k)}(y) , \\
[ z(z^2-1) [ \Phi^\near_{(2n+1,k)}(z)]' ]' &= \mfs^{\Phi,\near}_{(2n+1,k)}(z) . \\
\end{split}
\end{equation}
Each of these differential equations are easily solved up to two integration constants each which are fixed by the AdS boundary condition, regularity on the horizon and matching.

The general solutions for the scalar field at $\CO(\e r_+^0)$ and $\CO(\e r_+)$ is
\begin{equation}
\begin{split}
\Phi_{\ttbr{1,0}}(r) &= \frac{1}{1+r^2} , \\
\Phi_{\ttbr{1,1}}(r) &=  \frac{1+r^2- (5 r^2+1 ) \eta _{\ttbr{0,0}}^2+ r^2( r^2 + 1 )  (\eta _{\ttbr{0,0}}^2+1 ) \ln\left(\frac{1}{r^2}+1\right)}{2\eta_{\ttbr{0,0}}r^2 (r^2+1)^2}  , \\
\Phi^\intt_{\ttbr{1,0}}(y)  &= a_1 + \frac{a_2}{y^2} , \\
\Phi^\intt_{\ttbr{1,1}}(y)  &= - a_1 y^2 + \frac{a_2}{2y^4} + a_3 + \frac{a_4}{y^2}  - \left( a_2 + a_1 \eta_{\ttbr{0,0}} - \frac{a_2 \eta_{\ttbr{0,0}}}{y^2} \right) \ln (r_+ y^2)   , \\
\Phi^\near_{\ttbr{1,0}}(z)  &= b_1  , \\
\Phi^\near_{\ttbr{1,1}}(z)  &=  b_2 - 2 b_1 \eta_{\ttbr{0,0}} \ln  (r_+z) .
\end{split}
\end{equation}
The integration constants in the far-field solution have been fixed by asymptotically AdS boundary conditions and one of the integration constants in the near-field solution is fixed by regularity on the horizon.

Consider now the matching of the far- and intermediate-field solutions. Setting $r=\sqrt{r_+}y$ into the far-field solution and expanding in small $y$, we find
\begin{equation}
\begin{split}
\Phi_\1(\sqrt{r_+}y,r_+) &= \bigg[ \frac{1-\eta_{\ttbr{0,0}}^2}{2\eta_{\ttbr{0,0}} y^2} + 1 + \CO(y^2) \bigg]  \\
&\qquad + \bigg[ - \frac{1+3\eta_{\ttbr{0,0}}^2}{2\eta_{\ttbr{0,0}}} - \frac{1+\eta_{\ttbr{0,0}}^2}{2\eta_{\ttbr{0,0}}} \ln ( r_+ y^2 ) - y^2 + \CO(y^4) \bigg] r_+  \\
&\qquad + \CO(r_+^2)
\end{split}
\end{equation}
Expanding the intermediate-field solution at large $y$, we find
\begin{equation}
\begin{split}
\Phi^\intt_\1(y,r_+) &= \bigg[ a_1 + \frac{a_2}{y^2} + \CO(y^{-4}) \bigg] \\
&\qquad + \bigg[ - a_1 y^2 + [ a_3 - 2 ( a_1 \eta_{\ttbr{0,0}} + a_2 ) \ln y ] + \CO(y^{-2}) \bigg] r_+ + \CO(r_+^2) . 
\end{split}
\end{equation}
Matching the expansions, we find
\begin{equation}
\begin{split}
a_1 = 1 , \qquad a_2 = \frac{1-\eta_{\ttbr{0,0}}^2}{2\eta_{\ttbr{0,0}}}   , \qquad a_3 = - \frac{1+3\eta_{\ttbr{0,0}}^2}{2\eta_{\ttbr{0,0}}} . 
\end{split}
\end{equation}
Next, we turn to the matching of the intermediate-field solution to the near-field solution. To do this, we substitute $y=\sqrt{r_+}z$ into the intermediate field solution and expand in small $z$,
\begin{equation}
\begin{split}
\Phi^\intt_\1(\sqrt{r_+}z,r_+) &= \frac{1}{r_+} \bigg[ \frac{1-\eta_{\ttbr{0,0}}^2}{4\eta_{\ttbr{0,0}}z^4}  +  \frac{1-\eta_{\ttbr{0,0}}^2}{2\eta_{\ttbr{0,0}}z^2} + \CO(z^4)  \bigg] \\
&\qquad + \bigg[  \frac{a_4 + ( 1 - \eta_{\ttbr{0,0}}^2 ) \ln (r_+ z ) }{z^2} + 1 + \CO(z^2) \bigg] \\
&\qquad + \bigg[ - \frac{1+3\eta_{\ttbr{0,0}}^2 + 2 ( 1 + \eta_{\ttbr{0,0}}^2 ) \ln (r_+ z ) }{ 2 \eta_{\ttbr{0,0}} } + \CO(z^2) \bigg] r_+ \\
&\qquad + \CO(r_+^2) . 
\end{split}
\end{equation}
Expanding the near-field solution at large $z$, we find
\begin{equation}
\begin{split}
\Phi^\near_\1(z,r_+) &= b_1 + \bigg[ b_2 - 2 b_1 \eta_{\ttbr{0,0}} \ln ( r_+ z ) + \CO(z^{-3}) \bigg] r_+ + \CO(r_+^2) . 
\end{split}
\end{equation}
Matching the expansions, we find
\begin{equation}
\begin{split}
\eta_{\ttbr{0,0}} = 1 , \qquad b_1 = 1 , \qquad b_2 = - 2 , \qquad a_4 = 0 . 
\end{split}
\end{equation}
The rest of the solutions to higher orders is constructed in exactly the same way. The explicit solutions are presented at the end of this section.

\subsubsection*{General structure of the differential equations at $\CO(\e^{2n}r^k)$}

At this order, we obtain non-trivial differential equations for $f_{(2n,k)}$, $h_{(2n,k)}$, $A_{(2n,k)}$ and $\varphi_{(2n,k)}$. The equations in the far-field region are
\begin{equation}
\begin{split}
[ r^3 h_{(2n,k)}'(r) ]' &= \mfs^h_{(2n,k)}(r) , \\
[ r^3 [ r^2 f_{(2n,k)}(r) ]' ]' &= \mfs^f_{(2n,k)}(r)  + 16 r^5 h_{(2n,k)}(r)  , \\
[ r^3 A_{(2n,k)}'(r) ]' &= \mfs^A_{(2n,k)}(r) , \\
\left( \frac{1+r^2}{r} [ r^2 \varphi_{(2n,k)}(r)]' \right)' &= \mfs^\varphi_{(2n,k)}(r) . \\
\end{split}
\end{equation}
The differential equations in the intermediate-field region take a simpler form if we set
\begin{equation}
\begin{split}
\varphi^\intt_{(2n,k)}(y) &= X_{(2n,k)}^\intt(y) + \frac{y^3}{4} V_0(y)^{2/3} \td{}{y} \bigg( \frac{h_{(2n,k)}^\intt(y)}{V_0(y)^{-2/3}} \bigg) , \\
A^\intt_{(2n,k)}(y) &= Y_{(2n,k)}^\intt(y) - \frac{X_{(2n,k)}^\intt(y)}{V_0(y)} + \frac{y^5}{8} V_0(y)^2 \bigg( \frac{f_{(2n,k)}^\intt(y)}{y^2 V_0(y)^2} \bigg)'  \\
&\qquad \qquad \qquad - \frac{y^3}{4} V_0(y) \bigg( \frac{h_{(2n,k)}^\intt(y)}{V_0(y)^2} \bigg)'
\end{split}
\end{equation}
where $V_0(y) = 1 + y^{-2}$. The equations are then
\begin{equation}
\begin{split}
[ X_{(2n,2k)}^\intt(y)]' &= \mfs_{(2n,2k)}^{X,\intt}(y) , \\
[ Y_{(2n,2k)}^\intt(y)]' &= \mfs_{(2n,2k)}^{Y,\intt}(y) , \\
[ y^3 [ y^2 f_{(2n,2k)}^\intt(y)]' ]' &= \mfs_{(2n,2k)}^{f,\intt}(y) , \\
[ y^3 [ y^3 [h_{(2n,k)}^\intt(y)]' ]' ]' &= \mfs^{h,\intt}_{(2n,k)}(y)  + \frac{1}{2} \bigg( y^7 V_0(y)^2  \bigg[ \frac{ [ y^2 f_{(2n,k)}^\intt(y) ]' }{ y^3 V_0(y)  } \bigg]' \bigg)'  \\
&\qquad \qquad - \frac{4}{V_0(y)^2}   \left( y^3 V_0(y)^2 [ X_{(2n,k)}^\intt(y) ]' \right)' \\
&\qquad \qquad + \frac{4}{V_0(y)} \left( y^3 V_0(y)^2  [ Y^\intt_{(2n,k)}(y)]' \right)' . 
\end{split}
\end{equation}
Finally, the differential equations in the near-field region take on a simple form if we write
\begin{equation}
\begin{split}
\varphi^\near_\ttbr{2n,k}(y) &= X_\ttbr{2n,k}^\near(y) + \frac{z^2}{3} h_\ttbr{2n,k}^\near(y)  + \frac{z^3}{4} [ h_\ttbr{2n,k}^\near(y) ]' , \\
h_\ttbr{2n,k}^\near(y) &= Y_\ttbr{2n,k}^\near(y) - \frac{1}{z^4} A_\ttbr{2n,k}^\near(y) . 
\end{split}
\end{equation}
The equations are then
\begin{equation}
\begin{split}
[ X_\ttbr{2n,k}^\near(z)]' &= \mfs_\ttbr{2n,k}^{X,\near}(z) , \\
[ z^3 [ z^2 Y_\ttbr{2n,k}^\near(z) ]' ]' &= \mfs_\ttbr{2n,k}^{Y,\near}(z) , \\
[ z^3 [ z^2 f_\ttbr{2n,k}^\near(z) ]' ]'  &= \mfs_\ttbr{2n,k}^{f,\near}(z),
\end{split}
\end{equation}
and
\begin{equation}
\begin{split}
\mfs^{A,\near}_\ttbr{2n,k}(z)  &= \bigg( z(z^2-1)^2 \bigg[ \frac{A_\ttbr{2n,k}^\near(z)}{z^2(z^2-1)} \bigg]' - 4 (z^2 - 1 ) X_\ttbr{2n,k}^\near(z) \\
&\qquad \qquad \qquad \qquad \qquad  + \frac{z}{2} [ z^2 f_\ttbr{2n,k}^\near(z) - 2 z^2 ( z^2 - 1 ) Y_\ttbr{2n,k}^\near(z) ]' \bigg)'.
\end{split}
\end{equation}
All the differential equations above can be easily solved. Let us now describe the matching process.

At $\CO(\e^2r_+^0)$, the far-, intermediate- and near-field solutions are given by
\begin{equation}
\begin{split}
h^\far_\ttbr{2,0}(r) &= \frac{a_1}{r^2}  + \frac{1}{8(1+r^2)} , \\
f^\far_\ttbr{2,0}(r) &=  \frac{a_2}{r^2} + a_1 \left( 2 + \frac{4}{r^2} \right) + \frac{r^2}{4(1+r^2)} , \\
A^\far_\ttbr{2,0}(r) &= a_3 + \frac{a_4}{r^2} - \frac{1}{8(1+r^2)} , \\
\varphi^\far_\ttbr{2,0}(r) &= \frac{a_1}{3r^2} + \frac{1}{24(1+r^2)} , \\
X_\ttbr{2,0}^\intt(y) &= b_1 , \\
Y_\ttbr{2,0}^\intt(y) &= b_2 ,\\
f_\ttbr{2,0}^\intt(y) &= \frac{b_3}{y^2} + \frac{b_4}{y^4} , \\ 
h_\ttbr{2,0}^\intt(y) &= \frac{b_4}{6y^6} + \frac{b_5}{y^4} + \frac{b_6}{y^2} + b_7 , \\
X_\ttbr{2,0}^\near(z) &= c_1 , \\
Y_\ttbr{2,0}^\near(z) &= \frac{c_2}{z^2} + \frac{c_3}{z^4} , \\
f_\ttbr{2,0}^\near(z) &= - 2 ( 2 c_1 + c_2 - c_3 ) \frac{z^2-1}{z^4} , \\
A_\ttbr{2,0}^\near(z) &= - \frac{1}{2} ( z^2 - 1 ) ( 4 c_1 + 2 c_2 + c_4 z^2 ) . 
\end{split}
\end{equation}
Here, we have used AdS boundary conditions and regularity at the horizon in addition to the requirement $f^\near(1) = A^\near(1)=0$ to fix some of the integration constants. The remaining ones are fixed by matching.

Setting $r=\sqrt{r_+}y$ in the far-field solution and expanding at small $r_+$ and $y$, we find
\begin{equation}
\begin{split}
f_\2(\sqrt{r_+}y,r_+) &= \frac{1}{r_+} \left[ \frac{4a_1+a_2}{y^2}  + \CO(y^2) \right] + 2a_1 + \CO(r_+) , \\
h_\2(\sqrt{r_+}y,r_+) &= \frac{1}{r_+} \left[ \frac{a_1}{y^2}  + \CO(y^2) \right] +  \frac{1}{8}  + \CO(r_+) , \\
A_\2(\sqrt{r_+}y,r_+) &= \frac{1}{r_+} \left[ \frac{a_4}{y^2}  + \CO(y^2) \right] + a_3 - \frac{1}{8}  + \CO(r_+) , \\
\varphi_\2(\sqrt{r_+}y,r_+) &= \frac{1}{r_+} \left[ \frac{a_1}{3y^2}  + \CO(y^2) \right] + \frac{1}{24}  + \CO(r_+) . \\
\end{split}
\end{equation}
We next expand the intermediate field solution at small $r_+$ and large $y$,
\begin{equation}
\begin{split}
f^\intt_\2(y,r_+) &= \CO(y^{-2}) + \CO(r_+) , \\
h^\intt_\2(y,r_+) &= b_7 + \CO(r_+) , \\
A^\intt_\2(y,r_+) &= - b_1 + b_2 - \frac{b_3}{2} + \frac{b_6}{2} - b_7 + \CO(r_+) , \\
\varphi^\intt_\2(y,r_+) &= b_1 - \frac{b_6}{2} + \frac{b_7}{3}  + \CO(r_+) . 
\end{split}
\end{equation}
Matching the two expansions, we find
\begin{equation}
\begin{split}
a_1 = a_2 = a_4 = 0 , \qquad a_3 = b_2 - \frac{b_3}{2} , \qquad b_6 = 2 b_1 , \qquad b_7  = \frac{1}{8} . 
\end{split}
\end{equation}
To match the intermediate-field solution to the near-field one, we set $y=\sqrt{r_+}z$ in the intermediate-field solution and expand at small $r_+$ and $z$ and we find
\begin{equation}
\begin{split}
f_\2^\intt(\sqrt{r_+}z,r_+) &= \frac{1}{r_+^2} \left[ \frac{b_4}{z^4}  + \CO(z^2) \right] + \frac{1}{r_+} \left[ \frac{b_3}{z^2} + \CO(z^2) \right] + \CO(r_+) , \\
h_\2^\intt(\sqrt{r_+}z,r_+) &= \frac{1}{r_+^3} \left[ \frac{b_4}{6z^6} + \CO(z^2) \right] + \frac{1}{r_+^2} \left[ \frac{b_5}{z^4} + \CO(z^2) \right] \\
&\qquad + \frac{1}{r_+} \left[ \frac{2b_1}{z^2} + \CO(z^2) \right] + \CO(r_+^0) , \\
A_\2^\intt(\sqrt{r_+}z,r_+) &= \frac{1}{r_+} \left[ - \frac{b_4}{6z^2} + \CO(z^2) \right] + b_2 - \frac{5b_4}{12} \\
&\qquad  + \left[ \frac{1}{4} ( - 8 b_1 - 2 b_3 + b_4 + 4 b_5 ) z^2 + \CO(z^4) \right] r_+ + \CO(r_+^2) , \\
\varphi_\2^\intt(\sqrt{r_+}z,r_+) &= \frac{1}{r_+^2} \left[ - \frac{7b_4}{36z^4} + \CO(z^2) \right] + \frac{1}{r_+} \left[ \frac{1}{z^2} \left( - \frac{b_4}{18} - \frac{2b_5}{3} \right) + \CO(z^2) \right] \\
&\qquad + \frac{1}{18} ( 12 b_1 + b_4 - 6 b_5 ) + \CO(r_+) . \\
\end{split}
\end{equation}
Finally, we expand the near-field solution at small $r_+$ and large $z$,
\begin{equation}
\begin{split}
f_\2^\near(z,r_+) &= \left[ - \frac{2(2c_1+c_2-c_3)}{z^2} + \CO(z^{-4}) \right] + \CO(r_+) , \\
h_\2^\near(z,r_+) &= \frac{1}{r_+} \left[ \frac{c_4}{2} + \frac{4(c_1+c_2) - c_4 }{ 2 z^2 } + \CO(z^{-4}) \right]  + \CO(r_+^0), \\
A_\2^\near(z,r_+) &= \left[ - \frac{c_4 z^4}{2} + \frac{1}{2} ( - 4 c_1 - 2c_2 + c_4 ) z^2 + ( 2 c_1 + c_2 ) + \CO(z^{-2}) \right] r_+  \\
&\qquad + \CO(r_+^2) , \\
\varphi_\2^\near(z,r_+) &= \left[ \frac{c_4}{6} z^2 + \frac{1}{12} ( 8 c_1 - 4 c_2 + c_4 ) + \CO(z^{-2}) \right] + \CO(r_+) . \\
\end{split}
\end{equation}
The expansions are matched by setting
\begin{equation}
\begin{split}
b_1 = c_1 , \qquad b_2 = b_3 = b_4 = b_5 = c_2 = c_4 = 0 , \qquad c_3 = 2 c_1 + x_\ttbr{2,0} . 
\end{split}
\end{equation}
We find that the matching procedure fixes all constants except $x_\ttbr{2,0}$ and $c_1$. The former is fixed by the matching procedure at $\CO(r_+^2)$ which sets
\begin{equation}
\begin{split}
x_\ttbr{2,0} = 0. 
\end{split}
\end{equation}
On the other hand, $c_1$ is a redundant parameter as the full hairy BH solution only depends on the combination $\eta_\ttbr{2,0} + 2 c_1$. We can then set $c_1=0$ without loss of generality.

This procedure can be continued on to any order in perturbation theory. The explicit solution can be found in the accompanying {\tt Mathematica} file.

\bibliography{refsSO6sugra}{}
\bibliographystyle{JHEP}

\end{document}